\documentclass[a4paper,oneside]{article}

\usepackage[a4paper,hmargin={3cm,3cm}]{geometry}

\usepackage[utf8]{inputenc}
\usepackage[T1]{fontenc}

\usepackage{graphicx}
\usepackage{textcomp}
\usepackage{booktabs}
\usepackage{xcolor}
\usepackage[binary-units]{siunitx}
\usepackage{xspace}
\usepackage{url}
\usepackage{tabularx}
\usepackage[version=4]{mhchem}

\usepackage[algpseudocodecmds,natbibpackage]{pauli_new}
\usepackage{mathtools}
\usepackage{redescriptions}

\newcommand{\ownalgo}{\algname{MCMCTreePair}\xspace}
\newcommand{\sampling}{\algname{AltMCMC}\xspace}
\newcommand{\expmech}{\algname{AltExpM}\xspace}
\newcommand{\splittrees}{\algname{SplitT}\xspace}
\newcommand{\layeredtrees}{\algname{LayeredT}\xspace}

\newcommand{\mimic}{\datasetname{MIMIC-III}\xspace}
\newcommand{\mimicX}[1]{\datasetname{MIMIC-III-#1}\xspace}
\newcommand{\medicare}{\datasetname{CMS}\xspace}
\newcommand{\medicX}[2]{\datasetname{CMS-#1-#2}\xspace}
\newcommand{\nerdy}{\datasetname{NPAS}\xspace}
\newcommand{\mammals}{\datasetname{MAMMAL}\xspace}

\newcommand{\parameter}[1]{\mathit{#1}}
\newcommand{\MCIter}{\parameter{MCIter}}
\newcommand{\RMIter}{\parameter{RMIter}}
\newcommand{\InitTrials}{\parameter{InTr}}
\newcommand{\varArray}{S}
\newcommand{\runInd}{i}

\newcommand{\VarTh}{\sigma}

\newcommand{\classvar}{\mathbf{c}}

\newcommand{\function}[1]{\mathit{#1}}
\newcommand{\score}{\function{score}}

\newcommand{\attrs}{\text{\texttt{attrs}}}
\newcommand{\aview}{\text{\texttt{view}}}
\newcommand{\treeroot}{\mathit{root}}

\newcommand{\val}{\mathtt{val}}

\newcommand{\qual}{\score} %
\newcommand{\quality}{g} %
\newcommand{\pbdg}{\varepsilon}

\newcommand{\client}{\mathrm{[cl]}}
\newcommand{\server}{\mathrm{[ser]}}
\newcommand{\inv}{\mathrm{[inv]}}

\makeatletter
\newcommand\footnoteref[1]{\protected@xdef\@thefnmark{\ref{#1}}\@footnotemark}
\makeatother

\newcommand{\infourl}{\url{https://doi.org/10.5281/zenodo.7599160}}

\usepackage{hyperref}
\hypersetup{
  colorlinks=true,
  breaklinks=true,
  linkcolor=blue,
  citecolor=blue,
  urlcolor=blue,
  pdftitle={Differentially Private Tree-Based Redescription Mining},
  pdfauthor={Matej Mihelčić and Pauli Miettinen},
  pdfkeywords={redescription mining, differential privacy, health care informatics}
}

\begin{document}

\title{Differentially Private Tree-Based Redescription Mining}

\author{Matej Mihel\v{c}i\'{c}\\
   Department of Mathematics\\
  University of Zagreb  \\
  Zagreb, Croatia \\
\texttt{matmih@math.hr}
  \and
  Pauli Miettinen\\
   School of Computing \\
  University of Eastern Finland \\
  Kuopio, Finland \\
  \texttt{pauli.miettinen@uef.fi}
  }

  \date{}

\maketitle
  
\begin{abstract}
Differential privacy provides a strong form of privacy and allows preserving most of the original characteristics of the dataset. Utilizing these benefits  requires one to design specific differentially private data analysis algorithms. In this work, we present three tree-based algorithms for mining redescriptions while preserving differential privacy. Redescription mining is an exploratory data analysis method for finding connections between two views over the same entities, such as phenotypes and genotypes of medical patients, for example. It has applications in many fields, including some, like health care informatics, where privacy-preserving access to data is desired. Our algorithms are the first differentially private redescription mining algorithms, and we show via experiments that, despite the inherent noise in differential privacy, it can return trustworthy results even in smaller datasets where noise typically has a stronger effect.

\end{abstract}

\section{Introduction}
\label{sec:introduction}

Modern data analysis methods have had a huge impact on analysing health and bioinformatics data, with personalised medicine~\citep{frohlich18from}, computer-aided diagnoses~\citep{vanginneken01computer-aided}, and new drugs~\citep{napolitano13drug} being just some examples. Exploratory data analysis has still much more to contribute to these fields, but the access to the data is often restricted due to the sensitivity of health information. The standard procedure for requesting access to such data typically involves submitting a detailed research plan, which is often required to take the form of a confirmatory study, meaning that the researchers must present a hypothesis to be tested. This standard procedure is ill-suited for exploratory analysis, which relies on trying out a number of different approaches in order to see whether the data reveals some novel knowledge. As a consequence, requests for permission to access data for exploratory research often get rejected.

The particular method of exploratory data analysis that we focus on in this paper is \emph{redescription mining}. It aims at finding pairs of queries over the data attributes such that the queries describe approximately the same sets of entities. It can be used, among other things, to find connections between patient diagnoses and lab test results; for instance, finding that patients with following laboratory results
\begin{enumerate}
  \item normal total iron binding capacity, total \ce{CO2} and hemoglobin not tested; or%
  \item abnormal total \ce{CO2} and normal eosinophils levels with triglycerides not tested; or%
  \item normal \ce{CO2} level and blood potassium level, polys (neutrophils) levels not tested
\end{enumerate}
suffer either from chronic pulmonary heart diseases, hypovolemia (decreased blood volume) or hypertension (and vice versa). Information like this -- obtained from real-world data with a method proposed in this paper -- can be used to guide doctors to best practices (which lab tests should be ordered), to identify particularly strong or weak combinations of tests, or to improve diagnoses.

The problem with the above data analysis is that it is often impossible to do, as the data are confidential. And even if it would be done by a trusted party on the hospital data, sharing the results would still be a problem, as it could lead to leaking private patient information.

The solution to these problems lies in the well-known method of \emph{differential privacy} \citep{dwork06calibrating}. By giving researchers differentially private access to the dataset, the data owner can rest assured that they will not be able to infer information about any individual in the dataset, providing much stronger privacy guarantees than standard anonymisation methods. Consequently, the data owner can give researchers differentially private access to the data under a more lenient application process, thus allowing them to experiment with new methods more easily. Also, the results can be shared without fear of leaking sensitive information.

In addition to providing strong theoretical privacy guarantees, differential privacy has the advantage of allowing the use of unchanged, original data via different mechanisms that guarantee the privacy of the subjects in all information obtained and potentially publicly released using these mechanisms. A very important part of this information can be lost using other techniques for preserving privacy. For example, perturbation-based techniques affect the range and values of the attributes \citep{Agrawal01DdataPrivacy}, projection-based perturbation techniques in addition cause losing information about the meaning and number of the original attributes  \citep{kenthapadi2013privacy,liu06RPprivacy}, and using fuzzy logic-based perturbation techniques potentially obfuscate the numerical boundaries and ranges of the original attributes  \citep{jahan2014comparative}.

On the other hand, developing a differentially-private algorithm for a pattern-based data mining task, such as redescription mining, is perhaps even harder than many other applications of differential privacy, such as classification. A small perturbation can break a pattern, while it would not typically have any major impact on classification.

In this paper, we propose three tree-based differentially private redescription mining algorithms and study their properties. Two algorithms closely follow alternations of the well known CARTwheels algorithm \citep{ramakrishnan04turning} using differentially private decision trees: the \expmech algorithm is based on the exponential mechanism \citep{Friedman10} and the \sampling algorithm uses embedding of differential privacy into decision tree construction \citep{Xuanyu17}. The third algorithm, 
 denoted \ownalgo, is a differentially private redescription mining algorithm that uses a completely novel technique for embedding differential privacy into the construction of pairs of decision trees.

Concretely, we present the following contributions: 
\begin{itemize}
\item A novel methodology for the construction of differentially private redescriptions (\texttt{extractReds}) using a pair of differentially private decision trees. This technique allows using much lower noise levels for differential privacy, and consequently, produces more accurate results.
\item A novel algorithm (\texttt{sampleTreePair}) for the creation of a differentially private pair of decision trees using MCMC sampling. This extends the ideas of  \citet{Xuanyu17} to simultaneously create pairs of differentially private trees with matching leaves so that they are suitable for redescription construction. 
\item An adaptation of the alternating procedure for redescription construction \citep{ramakrishnan04turning} to the setting where differential privacy needs to be preserved (\texttt{createRedsAlt}). %
\item An adaptation of an existing differentially private decision tree algorithm based on MCMC sampling \citep{Xuanyu17} for creating differentially private redescriptions (\texttt{createDTMCMC}). Our approach utilizes a fixed tree depth and avoids unnecessary computations. %
\item A similar adaptation of an existing differentially private decision tree algorithm based on the exponential mechanism \citep{Friedman10} for creating differentially private redescriptions (\texttt{createDTExpMech}).
\end{itemize}

We performed a thorough experimental evaluation of the three developed approaches for differentially private redescription mining. The obtained results show that the \ownalgo algorithm outperforms the \sampling and \expmech algorithms on smaller datasets (although not always significantly), but \sampling and \expmech significantly outperform the \ownalgo algorithm on medium-sized datasets. As one allows more and more redescriptions to be generated during the experiments (which necessitates stricter privacy setting),  \ownalgo outperforms \sampling and \expmech on an increasing number of medium-sized datasets as well. Additionally, it is possible that \sampling and \expmech do not produce any redescriptions in some runs. This occurs with relatively small probability ($\leq 0.1$ on our test datasets), but it entirely depends on the quality of the input data, which can only be influenced by the data providers. For this reason, we propose a way to significantly reduce the probability of such failure (to $0$ in our experiments). This procedure reduces the overall performance of the \sampling and \expmech algorithms, reducing their advantage compared to the \ownalgo algorithm.

To conclude, all three approaches have their advantages and there are scenarios in which they excel. The \ownalgo algorithm excels on small, potentially unprepared data, or where a strict privacy setting is required, the robust versions of the \sampling and \expmech algorithms excel on medium-sized or large data, potentially not preprocessed, where a medium or small number of repeated differentially private experiments needs to be performed, and the non-robust \sampling and \expmech algorithms excel on well preprocessed medium-sized or large datasets where very few, or just one, differentially private experiments needs to be performed.

The rest of the paper is organized as follows. As the paper combines two topics, we start with brief primers and reviews on related work of either of them: Section~\ref{sec:prim-redescr-mining} is about redescription mining and Section~\ref{sec:prim-diff-priv} about differential privacy. The tree-based differentially private algorithms are presented in Section~\ref{sec:algorithms}, the experimental evaluation in Section~\ref{sec:exper-eval}, and conclusions in Section~\ref{sec:conclusions}.

\section{Primer on Redescription Mining}
\label{sec:prim-redescr-mining}

Redescription mining aims at finding two different ways to describe almost the same observations. In this paper, the dataset consists of a pair of tables, denoted $\Table_{\iLHS}$ and $\Table_{\iRHS}$, i.e.\ $\Table = (\Table_{\iLHS}, \Table_{\iRHS})$. Both tables are over the same entities (rows) but have disjoint collections of attributes (columns). The set of attributes of a table $\Table_X$ is denoted $\attrs(\Table_X)$ while $\aview(a)$ indicates which table a given attribute $a$ belongs to, so that for any $a \in \attrs(\Table_{X})$, $\aview(a) = X$.

The pair of data tables, together with user-defined constraints such as, in particular, minimum support and minimum accuracy, constitute the input of redescription mining.
In this context, a redescription consists of a pair of logical \emph{queries}, one over either data table, respectively denoted $\lquery$ and $\rquery$.
More specifically, a query consists of literals over Boolean, categorical or numerical attributes, combined with logical conjunction and disjunction operators, possibly involving negations.
For instance, the following query
\[
\query = (\neg\,\text{Hypertension} \land [\text{State} = \text{CA}]) \lor [\text{BirthY} \leq 1950 ]
\]
describes entities, in this case individuals, who either do not suffer from hypertension and reside in California, or are born before $1950$.
The collection of entities described by the queries is called the \emph{support} of the query, denoted $\supp(\query)$.
In other words, $\supp(\query)$ is the set of entities that satisfy the query $\query$.

The main quality of a redescription is that the queries have similar supports.
The similarity of supports is typically measured using the Jaccard similarity index.
Hence, the accuracy of a redescription $(\lquery, \rquery)$ is defined as
\[
\jacc(\lquery, \rquery) = \frac{\abs{\supp(\lquery) \cap \supp(\rquery)}}{\abs{\supp(\lquery) \cup \supp(\rquery)}}\;.
\]

In addition to having a high accuracy, to be of interest a redescription should cover neither too few nor too many entities. This is typically ensured by setting minimum and maximum support thresholds. %
Finally, the size of the intersection of the queries' supports should not be a direct consequence of their respective supports. For instance, if both queries cover almost the entire dataset, it is not at all surprising, and hence not interesting, that they have a large overlap. This is controlled by computing \pvalue{s} representing the probability that the intersection of the supports of two randomly chosen queries is as large as or larger than observed in the studied redescription. Queries are randomly chosen with a priori probabilities equal to the fraction of entities described by redescription's queries. The assumption is that all entities can be sampled with equal probability.  Redescriptions with \pvalue{s} over a chosen significance threshold (typically $0.01$) are then deemed uninteresting.
For details about the computation of \pvalue{s} and a more extensive account of redescription mining, we direct the reader to the work of \citet{galbrun18redescription} and references therein.

\paragraph{Related Work}
\label{sec:related-work-rm}

Since redescription mining was introduced by \citet{ramakrishnan04turning}, various algorithms have been proposed for the task.
Some algorithms are based, for instance, on iteratively growing the queries in a greedy fashion \citep{gallo08finding,galbrun12black}, or on mining and combining frequent itemsets \citep{gallo08finding}.

Several algorithms, including the pioneering one \citep{ramakrishnan04turning}, are based on decision trees, be it classification trees \citep{zinchenko15mining}, predictive clustering trees (PCT) \citep{mihelcic17framework} or random forests of PCTs \citep{mihelcic17redescription}.
These algorithms build the queries by inducing a decision tree over one side of the data, using the current query on the other side or an initialization vector as the target, and alternating between the sides. 
Different algorithms are obtained depending on precisely how the trees are induced, growing entire trees to full depth at once \citep{ramakrishnan04turning,mihelcic17framework}, progressively increasing the depth of the trees as in the \splittrees algorithm \citep{zinchenko15mining}, or growing subtrees to expand existing branches as in the \layeredtrees algorithm  \citep{zinchenko15mining}.
For reasons that will become clear, methods that rely on tree induction are the most relevant to us, and the approach we propose also belongs to this family.

Initially limited to Boolean data, the task was later extended to numerical attributes by \citet{galbrun12black} and \citet{zinchenko15mining} (following the greedy and tree-based approaches, respectively).

Redescription mining has been used in various application domains, including political analysis \citep{galbrun16analysing}, econometrics \citep{galbrun17computational}, bio-in\-for\-mat\-ics \citep{ramakrishnan09redescription} and medicine \citep{Mihelcic17using}.

\section{Primer on Differential Privacy}
\label{sec:prim-diff-priv}

Differential privacy ensures that it cannot be inferred, within a fixed probability, whether or not a particular individual is included in a dataset -- irrespective of what other information the adversary might have. This is done by withholding direct access to the data and only returning randomized answers.

Formally, if $\Table$ and $\Table'$ are two datasets that differ in exactly one row, $A$ is a randomized algorithm operating on these tables,  $S$ is any set of potential results $A$ might return, and $\pbdg>0$ is the privacy parameter (or budget), then $A$ is \emph{$\pbdg$-differentially private} (later also simply $\pbdg$-private) if
\begin{equation}
  \label{eq:dp:def}
  \Pr[A(\Table)\in S] \leq e^{\pbdg} \Pr[A(\Table')\in S]\;,
\end{equation}
where the probability is over the randomness of $A$ \citep[see][]{DworkFoundations}. This property ensures that a change in one row of a dataset has a limited impact on the output of $A$. Thus, very little information about the individual can be obtained if a sufficiently strict privacy level is enforced. 
Two important features of differential privacy are \emph{composability} and \emph{resistance to post-processing}. Composability means that if the user runs multiple algorithms on the data (or the same algorithm multiple times), then assuming the source of randomness is independent, the results are still differentially private: if the user runs $k$ algorithms that are $\pbdg$-private, the results are $k\pbdg$-private \citep{mcsherry09privacy} and conversely, if one wants to preserve $\pbdg$-privacy over $k$ runs, each run must be $\pbdg/k$-private.
Resistance to post-processing means that the results of an $\pbdg$-differentially private algorithm will stay (at least) $\pbdg$-differentially private after any post-processing that does not have access to the original data. 

Composability has been a topic of intense study in differential privacy. The aforementioned $k\pbdg$-privacy bound cannot be improved in ``pure'' differential privacy, but if one weakens the privacy definition, one can obtain different trade-off, using, e.g.\ the composition theorem of \citet{dwork10boosting} or concentrated differential privacy \citep{dwork16concentrated}. These methods are straight-forward to apply to $\pbdg$-differentially private mechanisms, but they do require knowledge of the acceptable privacy trade-off in the application domain. Therefore, further analysis using these weaker forms of privacy is out of scope for this manuscript.

When considering differentially private algorithms, an important concept is the sensitivity of the underlying functions. Assume that the differentially private algorithm $A$ computes some deterministic function $f$ of the data (e.g.\ counts a particular type of observations). The \emph{sensitivity} of $f$ is defined as $\Delta f = \max_{\Table, \Table'}\norm{f(\Table) - f(\Table')}_1$, where the maximum is taken over all pairs of datasets $\Table$ and $\Table'$ that differ in exactly one row.

When $A$ is an algorithm that returns a number (e.g.\ a counting query), a typical way to ensure differential privacy is to use the \emph{Laplace mechanism} \citep{dwork06calibrating}: noise drawn from the Laplace distribution $\texttt{Laplace}(0; \Delta f/\pbdg)$ is added to the correct answer. %

For more complex queries, the \emph{exponential mechanism} \citep{mcsherry07mechanism} can be used. Here, instead of adding noise to the answer, the algorithm samples the final answer from the set of all answers. The probability of sampling a particular answer $r$ is proportional to
\begin{equation}
  \label{eq:exp_mec}
  \exp(\pbdg \cdot \qual_\Table(r))\cdot \mu(r)\;,
\end{equation}
where $\qual_\Table(r)$ is the quality of answer $r$ on data $\Table$ and $\mu$ is a measure assigning a base probability to $r$ ($\mu$ is often uniform).
That is, the exponential mechanism requires a way to assign a quality to each answer. The level of privacy, on the other hand, depends on the sensitivity of the quality measure, that is, on how much the quality can change with a change to a single record in the dataset. 

\paragraph{Related Work}
\label{sec:related-work-dp}

After the initial formulation by \citet{dwork06calibrating}, building on their earlier work on the SuLQ framework \citep{blum05practical}, and the introduction of the exponential mechanism by \citet{mcsherry07mechanism}, differential privacy has seen significant interest in academia, and recent years have brought uptake by the industry as well \citep[e.g.][]{bolin17collecting}. McSherry's \citeyearpar{mcsherry09privacy} PINQ is a framework for implementing differentially private data analysis using smaller building blocks. For example, \citet{Friedman10} presented a differentially private ID3 decision tree algorithm using PINQ operations.

Since then, differential privacy has been integrated in many fields of data mining and machine learning. Many classification algorithms, such as decision trees \citep{Friedman10, Jagannathan12, Xuanyu17,Gong20}, random forests \citep{Rana15}, naive Bayes \citep{Vaidya13, Gong20}, SVM \citep{Rubinstein12, Li14, Gong20}, $k$ nearest neighbours \citep{Gursoy17}, artificial bee colonies, and differential evolution \citep{Zorarpaci13} have been adapted to preserve differential privacy. Differential privacy has also been used in a semi-supervised setting \citep{Jagannathan13semi} where private data is augmented by publicly available data to create classifiers of improved accuracy. It has been integrated into (distributed) online learning \citep{Gong20} and regression algorithms (linear and logistic regression) \citep{Gong20}.

Various types of neural and deep neural networks have been privatized using differential privacy (autoencoders, convolutional deep belief networks, long short-term networks, and generative neural networks) \citep{Fan20,Gong20}. Clustering \citep{Nissim07, Gong20} and dimensionality reduction \citep{blum05practical, Gong20} are the most prominent unsupervised techniques privatized using differential privacy. In addition to PCA, various matrix factorization approaches  \citep{Berlioz15, Balu16} and more complex tensor decompositions \citep{Wang16, Imtia18} have been modified to preserve privacy using methods and techniques developed in the field of differential privacy. The creation of differentially private matrix factorization approaches is largely motivated by their application to the construction of differentially private recommender systems \citep{Mcsherry09Rescs}, where it can be very important to preserve the full anonymity of users. Differentially private association rule mining \citep{Tsou20}, frequent itemset mining \citep{Zeng12}, and frequent subgraph mining \citep{Xu16} are examples of the integration of differential privacy with data mining techniques.

After the submission of this work, \citet{karjalainen23serenade} published a greedy differentially private redescription mining method. Their method follows the \algname{ReReMi} approach of \citet{galbrun12black} and hence, the queries in their redescriptions are not based on decision trees and have a very different structure to the ones presented here. Evaluating the relative benefits of these two approaches is left for future work. 

\section{Algorithms}
\label{sec:algorithms}

Fully differentially private redescription mining algorithms require parts of the computation to be performed on the secured server (where all data is stored). We use the notation $\mathit{object}^{\client}$ to denote an object that is stored in the memory of the client computer and $\mathit{object}^{\server}$ to denote an object that is stored in the memory of the secured server (thus not directly available to the client). Similarly, we use $\mathtt{function}^{\client}$ for functions that are completely computed on the client computer, $\mathtt{function}^{\server}$ for functions completely computed on the server, and $\mathtt{function}^{\inv}$ for functions that are started on the client but initiate (invoke) some computations on the server, potentially using data from both server and client (client data is sent to the server) and potentially obtaining some results that are differentially private and stored in the memory of the client computer.

\subsection{Motivation}
\label{sec:motivation}

From the various approaches used to build redescriptions, in this work we focus on decision-tree based methods. Decision trees are easy to interpret, but can express complex relations. Also, building differentially-private decision trees is a well-studied problem \citep{fletcher10decision}, which allowed us to build on existing approaches. Mining redescriptions (especially differentially private ones) imposes specific requirements on the decision tree induction algorithms, preventing any existing approach to be efficiently used directly.

Of the different existing approaches, we decided to base one approach on the work of \citet{Friedman10}, which presents the creation of differentially private decision trees using a standard exponential mechanism to select the appropriate split at each step of the decision tree induction, and to base the other approach on the work of \citet{Xuanyu17}, which implements the exponential mechanism on the level of decision trees of a predefined depth using MCMC sampling. Using MCMC sampling, we do not need to instantiate all trees in the sampling space, but can still benefit from the efficient use of the privacy budget provided by the exponential mechanism.

Next, we will explain how the alternating redescription mining approach of \citep{ramakrishnan04turning} must be adapted to work in differentially private manner. This approach can use arbitrary differentially private decision tree creation procedures in a direct way to compute differentially private redescriptions. We use modified versions of two differentially private decision tree algorithms \citep{Friedman10,Xuanyu17} as the building blocks of our algorithms called \expmech and \sampling, respectively. In Section~\ref{sec:outline}, we will present our third algorithm, \ownalgo, that extends the ideas of \citet{Xuanyu17} to allow sampling of tree-pairs instead of individual trees.
In Section~\ref{sec:gener-redescr}, we explain the process of creating redescriptions using pairs of trees and computing the statistics (support and \pvalue) for them in a differentially private manner. This part is shared by all tree-based redescription mining algorithms. Section \ref{sec:redprun} explains our redescription pruning strategy, an important step for obtaining accurate differentially private redescriptions in the presence of potentially high levels of noise.

The asymptotic time complexities of the algorithms are analyzed in Appendix~\ref{sec:time_complexity}.

\subsection{Main Building Blocks}
\label{sec:buildBlocks}
All redescription mining algorithms that will be presented in the following sections consist of three main parts: 
\begin{enumerate}
\item Creation of an initial target variable.
\item Creation of differentially private trees.
\item Creation of differentially private redescriptions. 
\end{enumerate}

Being supervised algorithms, decision trees need a target variable. As redescription mining is unsupervised, we need to create an initial target variable.
It is created from an attribute of one of the data tables.  For Boolean and categorical attributes, each distinct value forms a target class. Numerical attributes, on the other hand, are discretized using a standard equal-width binning approach, where the number of bins is derived using the Freedman--Diaconis rule \citep{Freedman1981OnTH}. Each bin forms one target class. 

\begin{figure}[tb]
  \begin{center}
  \includegraphics{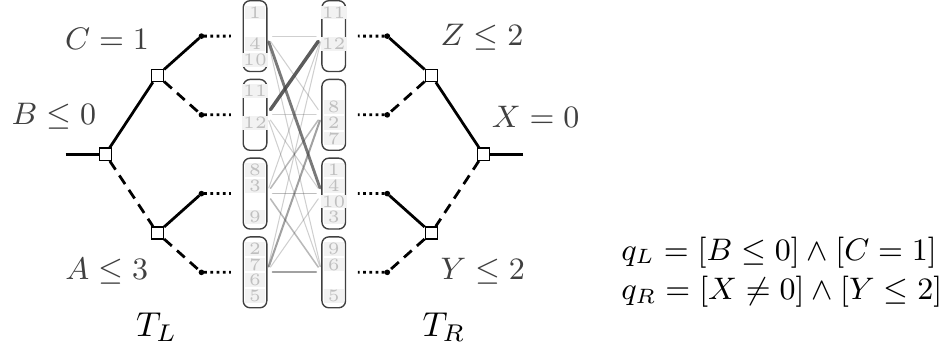}
  \end{center}
  \caption{(Left) A pair of trees forming a redescription. Solid and dashed lines respectively represent ``yes'' and ``no'' branches, while boxes depict the leaves. Numbers in the leaves indicate the entities. They are rendered for illustrative purposes only and very faint to emphasize the fact that they are \emph{not visible} to the algorithm. The size of the intersection of entities for every leaf pair is indicated by the thickness and shade of the line between the leaves. The algorithm obtains these values using the Laplace mechanism. Redescriptions are obtained from matching paths to pairs of leaves that have a high (noisy) Jaccard index. (Right) Example redescription
    $(q_L, q_R)$ with $\lvert\supp(q_L) \cap \supp(q_R)\rvert = 3$ and $\jacc(q_L, q_R) = 3/4$.
  }
  \label{fig:treepair}
\end{figure}

After the target variable is created, a process of differentially private tree construction begins. Different algorithms use different mechanisms for building trees, however the main goal of this step is to create a sequence of tree-pairs with matching leaves (see Figure~\ref{fig:treepair}).

Pairs of differentially private trees are utilized to obtain differentially private redescriptions, with a specifically developed procedure (see Section~\ref{sec:gener-redescr} and Figure~\ref{fig:treepair}) that preserves differential privacy. The branch leading to a leaf in a decision tree can straightforwardly be turned into a logical conjunctive query, by combining the tests encountered along the branch (see Figure~\ref{fig:treepair}).
A simple redescription is obtained by pairing the queries that correspond to a pair of leaves from opposite trees. Such simple redescriptions can be extended with logical conjunctive queries, obtained in either tree from the pair, using disjunctions and negations, to produce more complex, hopefully more accurate, redescriptions.

In the following sections we provide a detailed description of all three main parts of differentially private redescription creation. %

\subsection{Alternation-Based Algorithms}
\label{sec:alternationmethods}
In this section, we describe two differentially private redescription mining algorithms: the first, \expmech, uses individually created differentially private decision trees based on the exponential mechanism \citep{Friedman10} and the second, \sampling, embeds the sampling into the construction of the decision trees \citep{Xuanyu17}. We do not consider differentially private decision trees using noisy counts since prior work shows that these are outperformed by trees constructed using exponential mechanism \citep{Friedman10}.

The algorithms presented in this section follow a well-known alternation scheme used in the \algname{CARTwheels} algorithm \citep{ramakrishnan04turning}. In this scheme, the initial tree is created on one table (usually $\Table_{\iLHS}$) using a set of initial target labels (derived from data). Consecutive steps use information about entity membership in the leaves of the tree from the previous iteration to create target labels which allow the construction of a matching tree on the other table ($\Table_{\iRHS}$). Redescriptions are formed based on the paths from the root of one tree to the root of a second tree. Since decision tree construction cannot match created target labels exactly, each alternation creates slightly different trees, allowing the creation of new redescriptions.

We first present the general framework before explaining how individual trees are built.

\subsubsection{Alternation-Based Framework for Differentially Private Redescription Mining}
\label{sec:difPRIT}

Algorithm~\ref{algo:dprmSimple} builds differentially private redescriptions using differentially private decision trees provided by the \texttt{createDPTree} algorithm. When \texttt{createDPTree} uses Algorithm \ref{algo:dtem} to build the decision trees, the underlying redescription mining algorithm is called \expmech; when it uses Algorithm \ref{algo:dtMCMC} instead, it is called \sampling. The next two subsections describe these two different algorithms to build decision trees. All tree-specific parameters are explained in Appendix \ref{sec:param-used-exper}.

\begin{algorithm}[tb]
\caption{\texttt{createRedsAlt}: Create differentially private redescriptions using differentially private decision trees}\label{algo:dprmSimple}
\begin{algorithmic}[1]
  \small
\Require{Dataset with two views $(\Table_{L}, \Table_{R})^{\server}$, privacy budget $\pbdg^{\client}$, maximum number of initial trials $\InitTrials^{\client}$, maximum number of mining iterations $\RMIter^{\client}$, maximum tree depth $d^{\client}$, other tree-specific parameters $\mathit{params}^{\client}$, additional redescription constraints $\Gamma^{\client}$}
\Ensure{Set of redescriptions $P^{\client}$}
\State $P^{\client} \leftarrow \emptyset$
\State $\varepsilon'^{\client} = \pbdg^{\client}/\bigl(\InitTrials^{\client}(2\RMIter^{\client}+1)\bigr)$
\For{$\mathit{initTry}^{\client}$ \textbf{from} $1$ \textbf{to} $\InitTrials^{\client}$} \label{algo:dprmSimple:forStart}
\State $[\mathit{class}^{\server}, \mathit{side}^{\client}]\leftarrow \mathtt{createInitialClass}^{\inv}\bigl(\mathtt{randomAttribute}^{\server}\bigl($
\par \hfill$\mathtt{attributes}^{\server}(\Table_L^{\server}\cup \Table_R^{\server})\bigr)\bigr)$ \label{algo:dprmSimple:createInitialClass}
\If{$\mathit{side}^{\client} == 0$}
\State $T_L^{\client}\leftarrow\mathtt{createDPTree}^{\inv}(\Table_{L}^{\server}, d^{\client}, \varepsilon'^{\client}, \mathit{params}^{\client}, \mathit{class}^{\server})$ \label{algo:dprmSimple:createDPTree:L}
\Else{} 
\State $T_R^{\client}\leftarrow\mathtt{createDPTree}^{\inv}(\Table_{R}^{\server}, d^{\client}, \varepsilon'^{\client}, \mathit{params}^{\client}, \mathit{class}^{\server})$ \label{algo:dprmSimple:createDPTree:R}
\EndIf 
\For{$\runInd$ \textbf{from} $1$ \textbf{to} $\RMIter^{\client}$} \label{algo:dprmSimple:whl}
\If{$\mathit{side}^{\client} ==0$} \label{algo:dprmSimple:targetsStart}
\State $\mathit{class}^{\server}\leftarrow \mathtt{createClass}^{\inv}(\Table_{L}^{\server}, T_L^{\client})$
\State  $T_R^{\client}\leftarrow\mathtt{createDPTree}^{\inv}(\Table_{R}^{\server}, d^{\client}, \varepsilon'^{\client}, \mathit{params}^{\client}, \mathit{class}^{\server})$
\Else{}
\State $\mathit{class}^{\server}\leftarrow \mathtt{createClass}^{\inv}(\Table_{R}^{\server}, T_R^{\client})$
\State  $T_L^{\client}\leftarrow\mathtt{createDPTree}^{\inv}(\Table_{L}^{\server}, d^{\client}, \varepsilon'^{\client}, \mathit{params}^{\client}, \mathit{class}^{\server})$
\EndIf \label{algo:dprmSimple:targetsEnd}
\State $\mathit{side}^{\client}\leftarrow \mathtt{flip}^{\client}(\mathit{side}^{\client})$
\State $O^{\client} \leftarrow \mathtt{extractReds}^{\inv}(\Table_{L}^{\server},\Table_{R}^{\server} ,T_{L}^{\client}, T_{R}^{\client}, \varepsilon'^{\client}, \Gamma^{\client})$ \label{algo:dprmSimple:red}
\State $P^{\client} \leftarrow P^{\client} \cup O^{\client}$ \label{algo:dprmSimple:store}
\EndFor
\EndFor
\State \textbf{return} $P^{\client}$ \label{algo:dprmSimple:return}
\end{algorithmic}
\end{algorithm} 

Algorithm \ref{algo:dprmSimple} iterates for $\mathit{InTr}$ iterations (line~\ref{algo:dprmSimple:forStart}), chooses a random attribute at each iteration and creates an initial target variable using values of the selected attribute (line~\ref{algo:dprmSimple:createInitialClass}). 
The target variable creation step is entirely performed on the secured server (the one hosting the data) thus it does not require protection (the end-user does have access to this data). This operation is conceptually similar to the \texttt{Partition} operation, commonly used in differential privacy. This operation groups the data on the server based on some criterion (usually the value of a target label). Since this grouping is performed entirely on the secured server, it does not reveal any information to the end user. Then, the secured server returns the corresponding noisy counts of the number of entities contained in these groups to the end-user (thus privacy is preserved).  

Once an initial target variable is constructed, it is used either to construct a differentially private decision tree using the exponential mechanism (see Section~\ref{subalt:trees}) or to perform decision tree construction by sampling the space of decision trees (see Section~\ref{subalt:trees}). The resulting tree is differentially private and returned to the client. Which protection mechanism is chosen depends on the choice of a differentially private decision tree algorithm used during the process of differentially private redescription creation (lines \ref{algo:dprmSimple:createDPTree:L} and \ref{algo:dprmSimple:createDPTree:R}). 

At each consecutive step of the loop starting in line~\ref{algo:dprmSimple:whl}, the successive targets are created directly from the leaves of a tree constructed in the previous iteration (lines \ref{algo:dprmSimple:targetsStart} to \ref{algo:dprmSimple:targetsEnd}). Each leaf represents one class and entities are assigned a class label based on their membership in a leaf; the labels in turn are used to create a new tree using attributes from the other view. The main idea is to obtain pairs of trees having very similar or identical sets of entities in their leaves. 

Target creation is again performed entirely on the secure server and the corresponding data is accessed by the differentially private algorithms using the exponential mechanism or by sampling. The fact that MCMC sampling upon convergence produces trees as would be selected by applying the exponential mechanism to the space of trees of a predefined depth has been proven by \citet{Xuanyu17}. The obtained pairs of trees (without information about their node sizes) are used to create redescriptions in differentially private manner using noisy counts, obtained by making two passes over the dataset, to compute noisy redescription support, accuracy and significance (line~\ref{algo:dprmSimple:red}; see also Section~\ref{sec:gener-redescr}).

To summarize, after creating a target variable (class attribute, see Section~\ref{sec:buildBlocks}) in the first of $\mathit{RMIter}$ iterations, redescriptions are obtained by iteratively creating pairs of differentially private decision trees with matching leaves and then combining their leaves into redescriptions in a differentially private manner described in Section \ref{sec:gener-redescr}.

The total budget of Algorithm~\ref{algo:dprmSimple} is $\varepsilon$. In each of the $\mathit{initTry}$ iterations, Algorithm \ref{algo:dprmSimple} calls \texttt{createDPTree}, which is $\varepsilon/(\InitTrials(2\cdot \RMIter+1))$ differentially private (see Section \ref{subalt:trees}). Then, in each of the $\RMIter$ iterations, it calls \texttt{createDPTree} and \texttt{extractReds}, both of which are $\varepsilon/(\InitTrials(2\cdot \RMIter+1))$ differentially private (see Sections \ref{subalt:trees} and \ref{sec:gener-redescr}). %
Hence, the overall algorithm is $\InitTrials\bigl(\varepsilon/\bigl(\InitTrials(2\cdot \RMIter+1)\bigr) + \RMIter\cdot 2\cdot \varepsilon/\bigl(\InitTrials(2\cdot \RMIter+1)\bigr)\bigr) =
\varepsilon$ differentially private.

\subsubsection{Types of Differentially Private Decision Trees}
\label{subalt:trees}
We use two different types of differentially private decision trees:  differentially private decision trees using the exponential mechanism \citep{Friedman10}, presented in Algorithm~\ref{algo:dtem}, and differentially private decision trees obtained using MCMC sampling \citep{Xuanyu17}, presented in Algorithm~\ref{algo:dtMCMC}.

Algorithm~\ref{algo:dtem} first computes the available budget for every level of the tree (line~\ref{algo:dtem:budget}). It uses the exponential mechanism (line~\ref{algo:dtem:expmech}) to choose the split point out of all possible split points (attribute-value pairs) \citep[see][]{Friedman10}. Next, it partitions the dataset based on the values of the split point attribute using the \texttt{Partition} operation (line~\ref{algo:dtem:partition}) and then recursively calls itself to create the next level of the tree (line~\ref{algo:dtem:recursive}). Finally, it connects subtrees into the final tree and returns it.

\begin{algorithm}[tb]
\caption{\texttt{createDTExpMech}: Create differentially private decision trees using exponential mechanism}\label{algo:dtem}
\begin{algorithmic}[1]
  \small
\Require{Dataset $\Table^{\server}$, maximum tree depth $d^{\client}$, privacy budget $\pbdg^{\client}$, level $l^{\client}$}
\Ensure{Tree $T^{\client}$}
\If{$l^{\client}==0$}
\State $\varepsilon_{\mathit{level}}^{\client} = \pbdg^{\client}/d^{\client}$ \label{algo:dtem:budget}
\EndIf
\If{$d^{\client} == 0$} 
\State \textbf{return}
\EndIf
\State $\overline{A}^{\client} = \mathtt{ExpMech}^{\inv}(\attrs(D)^{\server}, \varepsilon_{\mathit{level}}^{\client} , q^{\client}$) \label{algo:dtem:expmech}
\State $\Table_i^{\server} \leftarrow \mathtt{Partition}^{\server}(\Table^{\server}, \val(\overline{A})^{\server})$ for all $i\in\val(\overline{A}^{\server})$ \label{algo:dtem:partition}
\State $\mathit{subtree}_i^{\client} \leftarrow \mathtt{createDTExpMech}^{\inv}(\Table_i^{\server}, (d-1)^{\client},  \varepsilon_{level}^{\client}, (l+1)^{\client})$ for all $i\in val(\overline{A})^{\client}$ \label{algo:dtem:recursive} 
\State $\mathit{root}(T)^{\client} = \overline{A}^{\client}; \mathit{subtrees}(T)^{\client} = \{\mathit{subtree}_i^{\client}: i\in \val(\overline{A})^{\client}\}$ \label{algo:dtem:treebuild}
\State \textbf{return} $T^{\client}$
\end{algorithmic}
\end{algorithm} 

At each level, calls to the exponential mechanism are $\varepsilon/d$ differentially private. Applying the exponential mechanism on siblings does not necessitate increasing the privacy level thanks to the property of parallel composition \citep{mcsherry09privacy}. The \texttt{Partition} function does not spend any budget, as no result is returned to the user.
The function calls itself recursively at most $d$ times, thus by the property of sequential composition \citep{mcsherry09privacy}, Algorithm~\ref{algo:dtem} is $d\cdot \varepsilon/d = \varepsilon$ differentially private.  The parameter $q$ in a call of the exponential mechanism in line~\ref{algo:dtem:expmech} represents the Gini index, which has sensitivity $2$ \citep{Friedman10}; the larger sensitivity is taken care of in the exponential mechanism calculations. The exponential mechanism is executed on a secured server by calculating all splitting choices and returning a random solution sampled from a distribution depending on the desired privacy level and the corresponding score function.

Algorithm \ref{algo:dtem} is based on the algorithm developed by \citet{Friedman10} with the following modifications: a) computations used to determine the noisy counts of entities at each node of a tree are removed and b) trees are built until a predefined depth is reached. The main reason for these modifications is to save budget, and they are justified by the fact that redescription mining algorithms usually use shallow trees (depth $\leq 8$). For various reasons (budget savings, reduction of false positive redescriptions), this depth is even smaller in differentially private redescription mining algorithms. This allows using $\varepsilon/d$ of the budget for the exponential mechanism. Noisy count computations are not required because  target variables in the alternation process are created on the secured server (see Algorithm \ref{algo:dprmSimple}) and pairs of differentially private trees without information on target value distribution in their nodes are sufficient to compute differentially private redescriptions (see Section \ref{sec:gener-redescr}).  

The second approach uses a deep embedding of differential privacy into the decision tree construction (Algorithm~\ref{algo:dtMCMC}). The embedding is achieved by sampling the tree space using Markov Chain Monte Carlo (MCMC) \citep{Hastings}. This process has been shown to generate, upon convergence, trees as would be obtained if selected by the exponential mechanism \citep{Xuanyu17}. The advantage of this approach compared to Algorithm~\ref{algo:dtem} is that it does not have to divide the available budget between the levels of a tree. Instead, it can use the whole budget for sampling. The construction of a differentially private object of this complexity cannot be achieved easily by applying the exponential mechanism once to the vast space of all trees of depth $d$. 

\begin{algorithm}[tb]
\caption{\texttt{createDTMCMC}: Create differentially private decision trees using decision tree sampling with MCMC}\label{algo:dtMCMC}
\begin{algorithmic}[1]
  \small
\Require{Dataset $\Table^{\server}$, maximum tree depth $d^{\client}$, privacy budget $\pbdg^{\client}$,  maximum number of MCMC iterations $\MCIter^{\client}$, MCMC iteration termination variance threshold $\sigma^{\client}$}
\Ensure{Tree $T^{\client}$}
\State $T^{\server}\leftarrow \mathtt{generateRandomTree}^{\server}(d^{\client})$ \label{algo:dtMCMC:randomTree}
\State $S^{\server}\leftarrow []$; $i^{\server}\leftarrow 0$
\While{$\runInd^{\server} < \MCIter^{\client}$ \textbf{and not} $\mathtt{stabilizedVar}^{\server}(\varArray^{\server}, \VarTh^{\client})$}
\State $\varArray^{\server} \leftarrow \varArray^{\server}\mathtt{.append}^{\server}( g(T)_1^{\server} )$
\State $n^{\server}\leftarrow \mathtt{randomNode}^{\server}(N_i(T)^{\server})$ \label{algo:dtMCMC:randomNode}
\State $n'^{\server}\leftarrow \mathtt{randomSplit}^{\server}(\mathtt{randomAttr}^{\server}(\Table^{\server}))$
\State $T'^{\server}\leftarrow \mathtt{replaceNode}^{\server}(T^{\server}, n^{\server}, n'^{\server})$ \label{algo:dtMCMC:replaceNode}
\State $\alpha^{\server}\leftarrow \exp^{\server}\bigl(\varepsilon^{\client}\cdot (g(T')_1/4)^{\server}\bigr)/\exp^{\server}\bigl(\varepsilon^{\client}\cdot (g(T)_1/4)^{\server}\bigr)$
\If{$\mathtt{random}^{\server}() < \alpha^{\server}$}
\State $T^{\server}\leftarrow T'^{\server}$ \label{algo:dtMCMC:replaceTree}
\EndIf
\State $i^{\server}\leftarrow (i+1)^{\server}$
\EndWhile
\State \textbf{return} $T^{\client}$
\end{algorithmic}
\end{algorithm}  

Algorithm \ref{algo:dtMCMC} is a version of the approach presented by \cite{Xuanyu17}, modified by removing the noisy count computation. This allows using the whole available budget $\varepsilon$ for tree sampling.
The reason why noisy counts are not required is the same as for the \texttt{createDTExpMech}: 
the initial targets are created on the secure server (see Algorithm~\ref{algo:dprmSimple}) and noisy counts are not required as input to the construction of differentially private redescriptions (see Section \ref{sec:gener-redescr}).

The work of \citet{Xuanyu17} demonstrates that trees sampled using MCMC outperform trees based on the selection of split points using the exponential mechanism, which strongly motivates the use of the former method in an algorithm for differentially private redescription creation.

The score function for the tree construction is as defined by \cite{Xuanyu17}:

\begin{equation}
  \label{eq:quality:1}
    \quality(n)_1 = 
     \begin{cases}
       \sum_{m \in C(n)} \quality(m)_1  & \text{if }n \in N_i(T)\;,\\
      -\tau_n\bigl(1-\sum_{c \in \classvar} (\frac{\tau_{n,c}}{\tau_n})^2\bigr) & \text{if }n\in N_l(T)\;,
     \end{cases}
   \end{equation}
where $C(n)$ denotes the set of children of the tree node $n$, while $\tau$, $\tau_n$, and $\tau_{n,c}$ denote respectively the total number of entities in the dataset, the number of entities in $n$ and the number of entities of class $c$ in $n$.  $N_l$ denotes a set of leaf nodes, whereas $N_i$ denotes a set of internal (non-leaf) nodes. The quality of a tree, $\quality(T)_1$, is computed as the quality of its root node.
 
Algorithm \ref{algo:dtMCMC} is completely performed on the secured server. It first generates a completely random tree of depth $d$ (line~\ref{algo:dtMCMC:randomTree}), then it iterates for $\MCIter$ iterations, selects a random node of a tree $T$ and computes the score of a newly constructed tree where all other nodes are equal to the nodes of $T$ but a selected node is replaced with a node corresponding to randomly chosen split point (lines~\ref{algo:dtMCMC:randomNode} to~\ref{algo:dtMCMC:replaceNode}). The quality of the new tree is compared to the old one in a similar fashion as in the exponential mechanism, and if the change is accepted, the old tree is replaced (line~\ref{algo:dtMCMC:replaceTree}). If changes in a tree score during the iteration drop below a predefined threshold (determined by the function \texttt{stabilizedVar}), iterations are terminated and the resulting tree is returned to the client.

\subsection{The \ownalgo Algorithm}
\label{sec:outline}

In this section, we present an algorithm whose main ingredient is MCMC sampling of tree-pairs instead of individual trees.

This algorithm uses the privacy budget more efficiently than the alternation-based algorithms. Assuming one has a budget $\varepsilon_t$ to construct one pair of trees in a differentially private manner, applying the algorithm described here (that samples pairs of trees) allows using this whole budget $\varepsilon_t$ to sample tree-pairs. Constructing a tree-pair by first creating single trees necessarily requires using $\varepsilon_t/2$ budget to create each tree (regardless of the technique used to obtain individual trees).  However, sampling pairs of trees is more complicated than sampling decision trees or split points, thus there is a trade-off between the budget spent and the complexity of the sampling.

The algorithm creates the initial target variable (line~\ref{algo:own:splA} in Algorithm \ref{algo:own}) as described in Sections \ref{sec:buildBlocks} and \ref{sec:difPRIT}. Next, a random pair of decision trees of a predefined depth is created (line~\ref{algo:own:pair} in Algorithm \ref{algo:own}, lines \ref{algo:tp:initL} and \ref{algo:tp:initR} in Algorithm \ref{algo:tp}). This pair is iteratively improved by selecting a random node in either tree and replacing it with a node created by making a split on a randomly selected split point (attribute--category or attribute--numerical value pair) on the data contained in this node (while-loop starting in line \ref{algo:tp:whl} in Algorithm \ref{algo:tp}). Potentially newly gained branches are completed to random subtrees until a predefined depth. %
As a result, we obtain a pair of decision trees, respectively over the variables from either data tables, which are matched at the leaves through the entities they contain (line \ref{algo:own:red} of Algorithm~\ref{algo:own}). %

\begin{algorithm}[tb]
\caption{\ownalgo: Differentially private redescription mining}\label{algo:own}
\begin{algorithmic}[1]
  \small
\Require{Dataset with two views $(\Table_{L}^{\server}, \Table_{R}^{\server})$, privacy budget $\pbdg^{\client}$, maximum number of initial trials (mining iterations) $\InitTrials^{\client}$, maximum number of MCMC iterations $\MCIter^{\client}$, MCMC iteration termination variance threshold $\VarTh^{\client}$, maximum tree depth $d^{\client}$, additional redescription constraints $\Gamma^{\client}$, budget re-distribution weight $\omega^{\client}\in [0,1]$}
\Ensure{Set of redescriptions $P^{\client}$}
\State $P^{\client} \leftarrow \emptyset$
\For{$\runInd^{\client}$ \textbf{from} $1$ \textbf{to} $\InitTrials^{\client}$} \label{algo:own:whl}
\State $a^{\server} \leftarrow \mathtt{randomAttribute}^{\inv}\bigl(\attrs^{\server}(\Table_{L}^{\server})\cup\attrs^{\server}(\Table_{R}^{\server})\bigr)$ \label{algo:own:splA}
\State $(T_L,\ T_R)^{\client} \leftarrow \mathtt{sampleTreePair}^{\inv}\bigl(\Table_{L}^{\server}, \Table_{R}^{\server}, a^{\server}, d^{\client},\omega^{\client}\cdot\pbdg^{\client}/\InitTrials^{\client}, \MCIter^{\client}, \VarTh^{\client}\bigr)$ \label{algo:own:pair}
\State $O^{\client} \leftarrow \mathtt{extractReds}^{\inv}\bigl(\Table_L^{\server}, \Table_R^{\server}, T_{L}^{\client}, T_{R}^{\client},(1-\omega^{\client})\cdot \pbdg^{\client}/\InitTrials^{\client}, \Gamma^{\client}\bigr)$ \label{algo:own:red}
\State $P^{\client} \leftarrow P^{\client} \cup O^{\client}$ \label{algo:own:store}
\EndFor
\State \textbf{return} $P^{\client}$ \label{algo:own:return}
\end{algorithmic}
\end{algorithm}

This process is repeated, starting with different attributes from either data table, to produce a collection of redescriptions (for-loop starting in line~\ref{algo:own:whl} of Algorithm \ref{algo:own}). These redescriptions -- the pairs of queries together with their corresponding support statistics (support sizes, Jaccard index, and $\pvalue{}$) -- constitute the output of the algorithm.

Thus, two important steps are: a) building pairs of matching trees (line \ref{algo:own:pair} of Algorithm \ref{algo:own}) and b) evaluating and combining queries corresponding to the branches of the trees to form redescriptions and compute their support statistics (line \ref{algo:own:red} of Algorithm \ref{algo:own}). The main challenge we need to tackle is to perform these steps in a differentially private manner while still being able to produce high-quality redescriptions.

Specifically, in the first step we use an MCMC method similar to the one used in \sampling. However, we have to extend the method to simultaneously sample pairs of matching trees. We show that this process, upon convergence, produces a tree-pair as would be produced if the exponential mechanism was applied to the very large space of all possible pairs of trees of a fixed depth $d$. Thus, upon convergence, we may consider the produced pairs of trees to be differentially private.
In the second step we use noisy counts, i.e.\ apply the Laplace mechanism, in a targeted manner to compute the support statistics of obtained redescriptions. This guarantees that the returned redescriptions are differentially private.
The methodology allows balancing budget expenditures between the two steps.

In each of the $\InitTrials$ iterations, procedure \texttt{sampleTreePair} is $\omega\varepsilon/\InitTrials$ differentially private (see Section~\ref{sec:sampling-pair-trees}) whereas procedure \texttt{extractReds} is  $(1-\omega)\varepsilon/\InitTrials$ differentially private (see Section~\ref{sec:gener-redescr}). Thus, by the sequential composition property of differential privacy \citep{mcsherry09privacy}, each iteration of \ownalgo is
$\varepsilon/\InitTrials$ differentially private. As a result,
the \ownalgo algorithm is $\InitTrials\cdot \varepsilon/\InitTrials = \varepsilon$ differentially private. 

\subsubsection{Sampling a Pair of Trees}
\label{sec:sampling-pair-trees}

The MCMC-based procedure for sampling a pair of trees is described in Algorithm~\ref{algo:tp}. It is an extension of the MCMC sampling procedure for trees developed by \cite{Xuanyu17} and used in Algorithm~\ref{algo:dtMCMC}. In addition to the extensions in MCMC sampling iterations, development of the tree-pair sampling procedure necessitated development of a novel tree-pair evaluation function. This necessitated creating a normalized tree evaluation score \eqref{eq:quality}, which is a modified version of the score from \eqref{eq:quality:1}, and incorporating this modified score into the novel tree-pair evaluation function \eqref{eq:score}.

For simplicity of exposition, we assume that the initialization attribute $a$ comes from table $\Table_{R}$. If attribute $a$ comes from $\Table_{L}$ instead, the roles of the two views are simply inverted.
We let $N_i(T)$ and $N_l(T)$ denote the sets of inner nodes and of leaf nodes of the tree $T$, respectively.
The procedure relies on a score function to evaluate the quality of pairs of trees, denoted $\score(T_L,T_R)$, to which we will come back later.

The procedure in Algorithm~\ref{algo:tp} starts by randomly generating trees $T_L$ and $T_R$ of depth $d$ (lines \ref{algo:tp:initL} and \ref{algo:tp:initR}). Function \texttt{makeTarget} is used to create a target variable from a given attribute or tree as explained above, and function \texttt{randomSplitTree} is used to build a tree of chosen depth by selecting splits at random. Then, in each iteration, a node $n$ is selected at random from the current tree-pair (line \ref{algo:tp:splN}). A new node $n'$ is created, by selecting at random an attribute from the same view as the attribute in $n$, and a random split value in its range (line \ref{algo:tp:split}).

\begin{algorithm}[tb]
\caption{\texttt{sampleTreePair}: MCMC sampling of a tree-pair}\label{algo:tp}
\begin{algorithmic}[1]
  \small
\Require{Dataset with two views $(\Table_{L}^{\server}, \Table_{R}^{\server})$, initialization attribute $a^{\server}$, maximum tree depth $d^{\client}$, privacy budget $\pbdg^{\client}$, maximum number of MCMC iterations $\MCIter^{\client}$, MCMC iteration termination variance threshold $\VarTh^{\client}$}
\Ensure{Tree-pair $(T_L,\ T_R)^{\client}$}
\State $T_L^{\server} \leftarrow \mathtt{randomSplitTree}^{\server}(\Table_{L}^{\server}, \mathtt{makeTarget}^{\server}(a^{\server}), d^{\client})$ \label{algo:tp:initL}
\State $T_R^{\server} \leftarrow \mathtt{randomSplitTree}^{\server}(\Table_{R}^{\server}, \mathtt{makeTarget}^{\server}(T_L^{\server}), d^{\client})$ \label{algo:tp:initR}
\State $\varArray^{\server} \leftarrow []\,;\quad \runInd^{\server} \leftarrow 0$ \label{algo:tp:initvrs}
\While{$\runInd^{\server} < \MCIter^{\client}$ \textbf{and not} $\mathtt{stabilizedVar}^{\server}(\varArray^{\server}, \VarTh^{\client})$} \label{algo:tp:whl}
\State $\varArray^{\server} \leftarrow \varArray^{\server}\mathtt{.append}^{\server}( \score(T_L,\ T_R)^{\server} )$ \label{algo:tp:appSc}
\State $n^{\server} \leftarrow \mathtt{randomNode}^{\server}(N_i(T_L)^{\server} \cup N_i(T_R)^{\server})$\label{algo:tp:splN} 
\State $n'^{\server} \leftarrow \mathtt{randomSplit}^{\server}( \mathtt{randomAttr}^{\server}(\Table_{\aview(n)}^{\server}))$ \label{algo:tp:split}
\State $(T'_L, T'_R)^{\server}  \leftarrow \mathtt{replaceNode}^{\server}((T_L, T_R)^{\server}, n^{\server}, n'^{\server})$ \label{algo:tp:subtree}
\State $\alpha^{\server} \leftarrow \frac{\exp^{\server}(\varepsilon^{\client}\cdot \score(T'_L,\ T'_R)^{\server} /2)}{ \exp^{\server}(\varepsilon^{\client}\cdot\score(T_L,\ T_R)^{\server} /2)}$ \label{algo:tp:scratio}
\If{$\mathtt{random}^{\server}() < \alpha^{\server}$} \label{algo:tp:finScA}
\State $(T_L, T_R)^{\server}  \leftarrow (T'_L, T'_R)^{\server}$ \label{algo:tp:replaceT}
\If{$\aview(n)^{\server} == L^{\server}$} \label{algo:tp:updateA} 
\State $T_{R}^{\server} \leftarrow \mathtt{updateStats}^{\server}(T_R^{\server}, \Table_{R}^{\server},  \mathtt{makeTarget}^{\server}(T_L^{\server}))$ \label{algo:tp:upstatsT}
\EndIf \label{algo:tp:updateZ}
\EndIf \label{algo:tp:finScZ}
\State $\runInd^{\server} \leftarrow (\runInd+1)^{\server}$ \label{algo:tp:incremIt}
\EndWhile
\State \textbf{return} $(T_L,\ T_R)^{\client}$ \label{algo:tp:return}
\end{algorithmic}
\end{algorithm}  

The score for the tree-pair obtained by replacing node $n$ with node $n'$ (line \ref{algo:tp:subtree}) is computed and compared to the score of the original tree-pair to decide whether to accept the replacement (line \ref{algo:tp:scratio}). Bad choices for replacement (e.g. internal nodes which affect several leaves negatively) will significantly reduce the score of a newly constructed tree, thus the change will be accepted with smaller probability.

If the replacement is accepted and the modified node is in $T_L$, the statistics of $T_R$ must be updated to account for the modified target (line \ref{algo:tp:upstatsT}).
Such node replacement attempts are performed for a predefined number of iterations ($\MCIter$) or until the variance of the scores in the $k$ previous iterations drops below a predefined threshold, as checked by function \texttt{stabilizedVar} (line \ref{algo:tp:whl}). The last obtained tree-pair is returned (line \ref{algo:tp:return}).

Clearly, the crux of this procedure is the score function to evaluate the quality of pairs of trees, $\score(T_L,T_R)$, and the replacement probability $\alpha$ that is derived from it.
The replacement probability should be such that the procedure is able to traverse the whole space of tree-pairs and, upon convergence, samples a tree-pair $(T_L,\ T_R)$ with the probability assigned to it by the exponential mechanism
\begin{equation}
  \label{eq:sampling_probability}
  \frac{\exp(\pbdg\cdot \score(T_L,\ T_R)/(2\cdot \Delta \score))}{\sum_{(T'_L, T'_R)\in \mathit{AllTreePairs}_d} \exp(\pbdg \cdot \score(T'_L, T'_R)/(2\cdot \Delta \score))}\; ,
\end{equation}
where $\mathit{AllTreePairs}_d$ is the set of all pairs of trees of depth at most $d$ and $\Delta \score$ is the sensitivity of the score function.
Assuming the tree-pairs have uniform base probability, the above probability is clearly in the form of \eqref{eq:exp_mec}. We show that the sampling follows the above probability in Appendix \ref{sect:appendix:correctness}.

Given a decision tree $T$ and target variable $\classvar$, we measure the quality of the subtree rooted in node $n$ as the quality of the associated split with respect to the target, using the recursive formula
\begin{equation}
  \label{eq:quality}
    \quality(n) = 
     \begin{cases}
       \sum_{m \in C(n)} \quality(m)  & \text{if }n \in N_i(T)\;,\\
       \frac{\tau_n}{\tau} \sum_{c \in \classvar} \bigl(\frac{\tau_{n,c}}{\tau_n}\bigr)^2 & \text{if }n\in N_l(T)\;,
     \end{cases}
   \end{equation}
   where, as in \eqref{eq:quality:1}, $C(n)$ denotes the set of children of $n$, while $\tau$, $\tau_n$, and $\tau_{n,c}$ denote respectively the total number of entities in the dataset, the number of entities in $n$, and the number of entities of class $c$ in $n$. 
Unlike \citet{Xuanyu17} and in \eqref{eq:quality:1}, here we use a normalized quality measure that does not depend on the data size. A value of $1$ indicates a pure split, i.e.\ a perfect match between the (sub)tree and the target.
Computing $\quality(\treeroot(T))$ gives the quality of tree $T$.

The score of a pair of trees is measured by combining their respective scores
\begin{equation}
  \label{eq:score}
   \score(T_L,T_R) = \quality(\treeroot(T_L))\cdot \big(1+ \quality(\treeroot(T_R))\big)/2\;.
\end{equation}
The quality score of tree $T_R$ is weighted by the quality score of tree $T_L$ since a fairly accurate tree $T_L$ must be built first, to ensure that the resulting tree-pair adequately models the initial target. Then, as tree $T_L$ becomes more accurate, the overall score depends increasingly on the accuracy of tree $T_R$, forcing both trees to match and to model the initial target.

\begin{lemma}
  \label{lemma:score_sensitivity}
  The function $\score(T_L, T_R)$ from \eqref{eq:score} has sensitivity $1$.
\end{lemma}
The proof of the above lemma is presented in Appendix~\ref{sect:appendix:sensitivity}. Note that the score used by \citet{Xuanyu17} has sensitivity $2$ instead. Thus, our score reduces the level of noise needed in the differentially-private mechanism.

To conclude, Algorithm~\ref{algo:tp} uses the whole initial budget $\varepsilon$ for tree-pair sampling. The produced tree-pair, upon convergence of the MCMC sampling procedure, is as if produced by the exponential mechanism applied to the space of all tree-pairs of depth $d$ and can thus be considered $\varepsilon$-differentially private. Sampling tree-pairs is much more complex than sampling individual trees. For this procedure to work, the underlying quality score must be normalized to the $[0,1]$ interval. Although this allows reducing sensitivity, it also changes the acceptance probabilities of the exponential mechanism (random choices can be accepted with substantially larger probability compared to the non-normalized version).

\subsection{Generating Redescriptions}
\label{sec:gener-redescr}

Having obtained a pair of matching trees, the next step consists in generating redescriptions from it by evaluating and combining queries corresponding to the branches of the trees.
In particular, doing this in a differentially private way requires obtaining noisy counts of the entities shared by the leaves from either tree, from which we can then compute all necessary support statistics. A na\"ive way to compute the support cardinalities would be to use the Laplace mechanism to obtain the size of the support $\abs{\supp(q)}$ for both conjunctive queries of every simple redescription, and then use the mechanism again for every attempted combination of these queries via disjunctions or negations. This, however, would use too much of the privacy budget, and we show how we can compute the required counts at a significantly lower privacy budget. 

Recall that the leaves of a given tree contain disjoint sets of entities. All necessary information to evaluate redescriptions generated from the pair of trees $T_L$ and $T_R$ with sets of leaf nodes $N_l(T_L)$ and $N_l(T_R)$, respectively, can be computed using two passes over the data:
first computing $\abs*{\supp(n_L) \cap \supp(n_R)}$ for all $(n_L, n_R) \in N_l(T_L) \times N_l(T_R)$ in the first pass over the dataset (line~\ref{algo:extractReds:firstPass} in Algorithm \ref{algo:extractReds}),
and then computing $\abs*{\supp(n_L)}$ for all $n_L \in N_l(T_L)$ in the second pass (line~\ref{algo:extractReds:secondPass}). As the entities in $\supp(n_L) \cap \supp(n_R)$ are disjoint over the leaves, learning the (noisy) cardinality of one of these reveals nothing about the rest of the data, and hence we can use the same $\varepsilon$ for computing each of these cardinalities. The same holds true for computing $\abs*{\supp(n_L)}$.

\begin{algorithm}[tb]
\caption{\texttt{extractReds}: Extract redescriptions from decision trees}\label{algo:extractReds}
\begin{algorithmic}[1]
  \small
\Require{Dataset with two views $(\Table_{L}^{\server}, \Table_{R}^{\server})$, a pair of differentially private trees  $T_{L}^{\client}, T_{R}^{\client}$, privacy budget $\pbdg^{\client}$, additional redescription constraints $\Gamma^{\client}$}
\Ensure{Set of redescriptions $P^{\client}$}
\State $P^{\client} \leftarrow \emptyset$
\State $\mathit{CntInt}^{\client}\leftarrow \mathtt{computeNLeafInterCards}^{\inv}(\Table_{L}^{\server}, \Table_{R}^{\server}, T_{L}^{\client}, T_{R}^{\client}, \pbdg^{\client}/2)$ \label{algo:extractReds:firstPass}
\State $\mathit{CntL}^{\client}\leftarrow \mathtt{computeNLLeafCards}^{\inv}(\Table_{L}^{\server}, T_{L}^{\client}, \pbdg^{\client}/2)$ \label{algo:extractReds:secondPass}
\State $\mathit{CntR}^{\client}\leftarrow  \mathtt{computeNRLeafCards}^{\client}(\mathit{CntInt}^{\client})$ \label{algo:extractReds:CntR}
\State $\mathit{dSize}\leftarrow \mathtt{computeDSize}(\mathit{CntL}^{\client})$ \label{algo:extractReds:dSize}
\State $[\mathit{SQ}_L, \mathit{SQ}_R]^{\client}\leftarrow \mathtt{createSimpleQueries}^{\client}(T_{L}^{\client}, T_{R}^{\client}, \mathit{CntL}^{\client}, \mathit{CntR}^{\client})$ \label{algo:extractReds:SQs}
\State $\mathit{SReds}^{\client}\leftarrow \mathtt{createSimpleReds}^{\client}(\mathit{SQ}_L^{\client}, \mathit{SQ}_R^{\client}, \Gamma^{\client}, \mathit{CntInt}^{\client}, \mathit{dSize})$
\For{$R_s\in \mathit{SReds}$}
\For{$i$ \textbf{from} $1$ \textbf{to} $\mathit{maxClauses}$}\label{algo:extractReds:forMaxClauses}
\State $\mathit{bestExt}_L^{\client}\leftarrow \argmax_{p\in \mathit{SQ}_L} \mathtt{evaluate}\bigl((q_{\mathit{sL}} \vee p, q_{\mathit{sR}}), \mathit{dSize}\bigr)$
\If{$\jacc(q_{\mathit{sL}} \vee p, q_{\mathit{sR}})>\jacc(R_s)$ and $\mathtt{sat}\bigl((q_{\mathit{sL}} \vee p, q_{\mathit{sR}}) , \mathit{dSize}, \Gamma^{\client}\bigr)$}
\State $R_s \leftarrow (q_{\mathit{sL}} \vee p, q_{sR})$
\EndIf
\State $\mathit{bestExt}_R^{\client}\leftarrow \argmax_{p\in \mathit{SQ}_R} \mathtt{evaluate}\bigl((q_{\mathit{sL}}, q_{\mathit{sR}} \vee p), \mathit{dSize}\bigr)$
\If{$\jacc(q_{\mathit{sL}}, q_{\mathit{sR}} \vee p) > \jacc(R_s)$ and $\mathtt{sat}\bigl((q_{\mathit{sL}}, q_{\mathit{sR}} \vee p), \mathit{dSize}, \Gamma^{\client}\bigr)$}
\State $R_s \leftarrow (q_{\mathit{sL}}, q_{\mathit{sR}} \vee p)$
\EndIf
\EndFor
\If{$\mathtt{sat}(R_s,  \mathit{dSize}, \Gamma^{\client})$}
\State $P^{\client} \leftarrow P^{\client} \cup \{R_s\}$\label{algo:extractReds:growP}
\EndIf
\EndFor
\State \textbf{return} $P^{\client}$ 
\end{algorithmic}
\end{algorithm} 

Using the computed noisy cardinalities, we can also compute
\begin{align*}
  \abs{\supp(n_R)} &= \sum_{n_L \in N_l(T_L)} \abs{\supp(n_R)\cap \supp(n_L)}\quad\text{ for $n_R \in N_l(T_R)$ and}\\
  \abs{\Table} &= \sum_{n_L \in N_l(T_L)} \abs{\supp(n_L)} \; .
\end{align*}
The procedures described above are used in lines~\ref{algo:extractReds:CntR} and \ref{algo:extractReds:dSize} of Algorithm \ref{algo:extractReds}.
For $I_L \subseteq N_l(T_L)$ and $I_R \subseteq N_l(T_R)$, we have that 
\begin{align*}
  \abs*[bigg]{\supp\Bigl(\bigvee_{I_L} \query_L\Bigr)\cap \supp\Bigl(\bigvee_{I_R} \query_R\Bigr)}
  &= \sum_{n_L \in I_L}\sum_{n_R \in I_R}\!\abs{\supp(n_L) \cap \supp(n_R)}\\
  \shortintertext{and}
    \abs*[bigg]{\supp(\neg\query_m)\cap \supp\Bigl(\bigvee_{I_R} \query_R\Bigr)} &= 
  \sum_{n_L \in I_L\setminus \{n_m\}}\sum_{n_R \in I_R} \abs{\supp(n_L) \cap \supp(n_R)}\; .
\end{align*}
The formulae to compute noisy counts described above are used inside the inner for-loop of Algorithm~\ref{algo:extractReds}.

Based on these differentially private values, we can compute the support sizes of the simple queries corresponding to single branches from the trees, as well as of the extended queries, that combine different branches, or their negations, from either tree. We can also compute the accuracy and the \pvalue of every query pair, as this only requires knowledge of the support sizes of the queries, their intersection, and the size of the data.

Simple redescriptions, generated by function \texttt{createSimpleReds}, consist of query-pairs $(\bar{q}_L, \bar{q}_R)$ obtained from $(n_L, n_R)\in N_l(T_L) \times N_l(T_R)$ (line~\ref{algo:extractReds:SQs} in Algorithm~\ref{algo:extractReds}), where $\bar{q}_X$ denotes the simple query associated to $n_X$ or its negation. These simple redescriptions are iteratively extended by at most three other simple queries (or their negations) obtained from each view, using the disjunction operator (the use of disjunctions and their properties are described by \cite{mihelcic17redescription}). An extension is accepted if it increases the Jaccard index compared to the simple redescription (checked by function \texttt{evaluate}) and if it satisfies conditions from $\Gamma^{\client}$ (checked by function \texttt{sat}). The resulting redescriptions after this step can be, for example, $(n_{L_1}\lor n_{L_2}, n_{R_1})$, $(n_{L_1} \lor n_{L_2} \lor n_{L_3}, n_{R_1})$, $(n_{L_1} \lor n_{L_2}, n_{R_1} \lor \neg  n_{R_2})$, $(n_{L_1} \lor n_{L_2} \lor n_{L_3}, n_{R_1} \lor n_{R_2} \lor n_{R_3})$. Extended redescriptions are often more accurate than simple redescriptions. Such redescriptions also have larger support set size compared to corresponding simple redescriptions. All redescriptions satisfying the constraints from $\Gamma^{\client}$ are added to the redescription set (line~\ref{algo:extractReds:growP}) that is returned to the user.

When the data contains missing values, we have 
    \[
      \abs{\supp(\neg\,\query_m)} \geq \sum_{l \in I_L\setminus \{m\}}\abs{\supp(n_l)}\; .
    \]
This is because entities for which the value of the split attribute(s) is missing cannot be sorted into the leaves. In such cases, we propose to estimate the support of negated queries with a heuristic count $\abs{\supp(\neg\,\query_m)} \approx \sum_{l \in I_L\setminus \{m\}}\abs{\supp(n_l)}$. Detailed justification for this choice is provided in in Appendix \ref{sect:appendix:missing}.

Procedures \texttt{computeNLeafInterCards} and \texttt{computeNLLeafCards} from Algorithm \ref{algo:extractReds} are $\varepsilon/2$ differentially private, since both procedures apply  the Laplace mechanism with budget $\varepsilon/2$, thus Algorithm \ref{algo:extractReds} is $\varepsilon/2 + \varepsilon/2 = \varepsilon$ differentially private.

\subsection{Redescription Pruning Strategy}
\label{sec:redprun}

The redescription mining process on data where differential privacy must be preserved is significantly harder than the regular task. Redescription mining algorithms have traditionally had a number of parameters that require tuning, usually through experimentation. However, in a setting where differential privacy needs to be preserved, such experimentation is not an option. Thus, constraints applied to filter redescriptions during the mining process need to be loose so that regardless of data characteristics some usable redescriptions can be obtained. This necessitates using relatively low noisy Jaccard threshold, as setting this parameter to very high values might remove many useful redescriptions and does not guarantee removing inaccurate artefact redescriptions.

The Support of redescriptions is an important quality measure. In the differentially private setting, we do not know the exact size of the data, and hence cannot set the minimum support threshold exactly. In our experiments, we set a coarse minimum support threshold (see Appendix~\ref{sec:param-used-exper}). Noisy support calculation has the biggest relative effect on redescriptions with small support, having also a noticeable impact on the accuracy of such redescriptions. As the real support of redescriptions increases, the noisy support size and accuracy of redescriptions are relatively closer to the real values. 

To keep the mining process general but also allow obtaining usable sets of redescriptions, we apply a redescription pruning strategy after the redescription mining step. The aim of this post-processing step is to use the information about noisy support and noisy accuracy to determine a noisy support threshold that will be used to filter out redescriptions that are probably artefacts from the introduced noise. The easiest way to determine such threshold is to study a noisy support vs. noisy redescription accuracy scatter plot. In a differentially private setting, there will always be a very dense block in the upper left corner of such a scatter plot (low noisy support, very high accuracy). Redescriptions making up such blocks should be pruned as most of them are inaccurate.

Notice that this step can be applied only once the redescription mining step is complete. Although this step drastically reduces the number of noise-induced redescriptions, it is still possible to obtain redescriptions with strongly overestimated accuracy or insignificant redescriptions. However, as we demonstrate in Section~\ref{sec:exper-eval}, the redescription sets obtained after pruning are usable in practice.

\section{Experimental Evaluation}
\label{sec:exper-eval}

\subsection{The Setup}
\label{sec:setup}

All presented algorithms are implemented in Java.\!\footnote{The source code is available at \infourl .} The experiments were conducted on a server with two 64-core AMD EPYC 2 processors at \SI{2}{\giga\hertz} and with \SI{1}{\tera\byte} of main memory. All experiments used less than \SI{25}{\giga\byte} of main memory.

\paragraph{Datasets}
We conducted experiments on seven real-world datasets based on four different types of data. The datasets are selected to cover a variety of data sizes and types, with an emphasis on the kind of data where confidential access would be important. \medicare is a family of datasets based on synthetic health insurance claims;  \nerdy is based on online psychological surveys;  \mammals contains mammal occurrences and climate information; and \mimic is based on hospital records. \medicare, \nerdy, and \mimic are examples of sensitive data with different properties, while  \mammals is a standard dataset in the redescription mining literature.  All datasets are publicly available. Basic properties of the datasets are presented in Table~\ref{tab:datasets}; further information is available in Appendix \ref{sec:datas-prop-pre}.

\begin{table}
  \centering
  \caption{Properties of the datasets}
  \label{tab:datasets}
  \begin{tabular}{@{}
    l
    S[table-format=6.0,group-minimum-digits=4,table-comparator]
    @{}L
    r
    S[table-format=4.0,group-minimum-digits=4]
    r
    S[table-format=4.0,group-minimum-digits=4]
    @{}}
    \toprule
    Name & {Rows} && \multicolumn{2}{c}{Left side} & \multicolumn{2}{c}{Right side} \\
    \cmidrule(r){4-5}\cmidrule(l){6-7}
         & & & type & {cols} & type & {cols} \\
    \midrule
    \medicX{2}{k}& \approx 7200 & \cdot k&  Boolean & 452 & numeric & 152 \\
    \nerdy & 1221 && mixed & 17 & numeric & 47 \\
    \mammals & 2576 && Boolean & 194 & numeric & 48\\
     \mimic & 46065 && Boolean & 107  & $\{-1,0,1\}$ & 77\\
    \bottomrule
  \end{tabular}
\end{table}

\paragraph{Baseline Methods}
We compared all three differentially private algorithms to two existing general (not privacy-preserving) tree-based redescription mining algorithms, \splittrees and \layeredtrees~\citep{zinchenko15mining,galbrun18mining} using the implementations from the \texttt{python-clired} package.\!\footnote{\url{https://pypi.org/project/python-clired/} v6.0.4, accessed 18 September 2020} We used only tree-based methods for the comparison to be fair; the purpose of these experiments is to show the effects of differential privacy, not to compare different types of redescriptions.

\paragraph{Parameters}
Setting the parameters for differentially private algorithms requires some care. Unlike in a normal setting, with differential privacy the user cannot tune the parameters. We provide the parameters used in the experiments, as well as some general guidelines on how to set the parameters when the data is not accessible, in Appendix~\ref{sec:param-used-exper}.

\paragraph{Evaluation Setup}

We conducted five types of experiments. First, we studied the convergence properties of the MCMC simulations used to create tree-pairs by the \ownalgo algorithm (see Appendix~\ref{sec:add-exp-res}). This is studied because sampling tree-pairs in this manner has not been previously reported in the literature. Second, we studied the differences between the quality of the redescriptions as reported by the differentially private algorithms (i.e.\ the noisy Jaccard and \pvalue) vs.\ the true values. Here, we used $\InitTrials = 1$ to have the largest budget for tree creation in \expmech and \sampling. Noise is required to preserve privacy, but it obviously also means that the results the user will obtain will not be entirely accurate. The aim of this experiment was to assess the error introduced by adding noise. 

In the next experiment, we compared the redescriptions found with differentially private algorithms to redescriptions obtained by standard (non-differentially private) tree-based algorithms \splittrees and \layeredtrees. This experiment measured the overall quality of differentially private redescriptions against standard redescriptions. Then, we compared the performance of differentially private algorithms among $10$ runs with budget $1.0$ each (we perform repeated runs to measure variation between executions), on $10$ differentially private runs with budget $0.1$ each, and on $100$ differentially private runs with budget $0.01$ each. We compared differentially private algorithms based on whether any redescriptions were obtained in a run, the percentage of significant redescriptions discovered, the distribution of true Jaccard index values among the obtained redescriptions, and the distribution of the absolute difference between the true and reported, noisy Jaccard index values of the redescriptions. Finally, we tested the scalability of the proposed differentially private algorithms.

\subsection{Effects of Differential Privacy on Redescription Quality}
\begin{figure*}[tbp]
  \centering
  \begin{tabularx}{\textwidth}{@{}X@{}X@{}X@{}}
    \multicolumn{1}{c}{\nerdy}  & \multicolumn{1}{c}{\mammals} & \multicolumn{1}{c}{\mimicX{Ter}}\\
    \includegraphics[width=\linewidth]{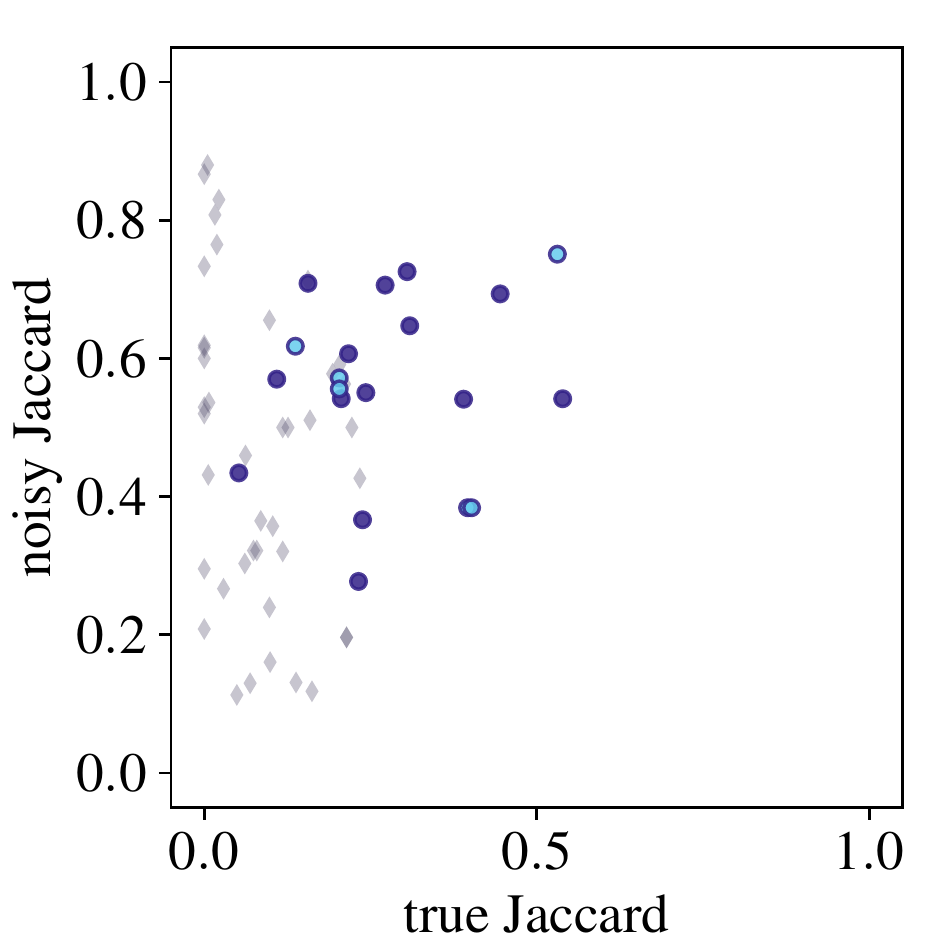} &
    \includegraphics[width=\linewidth]{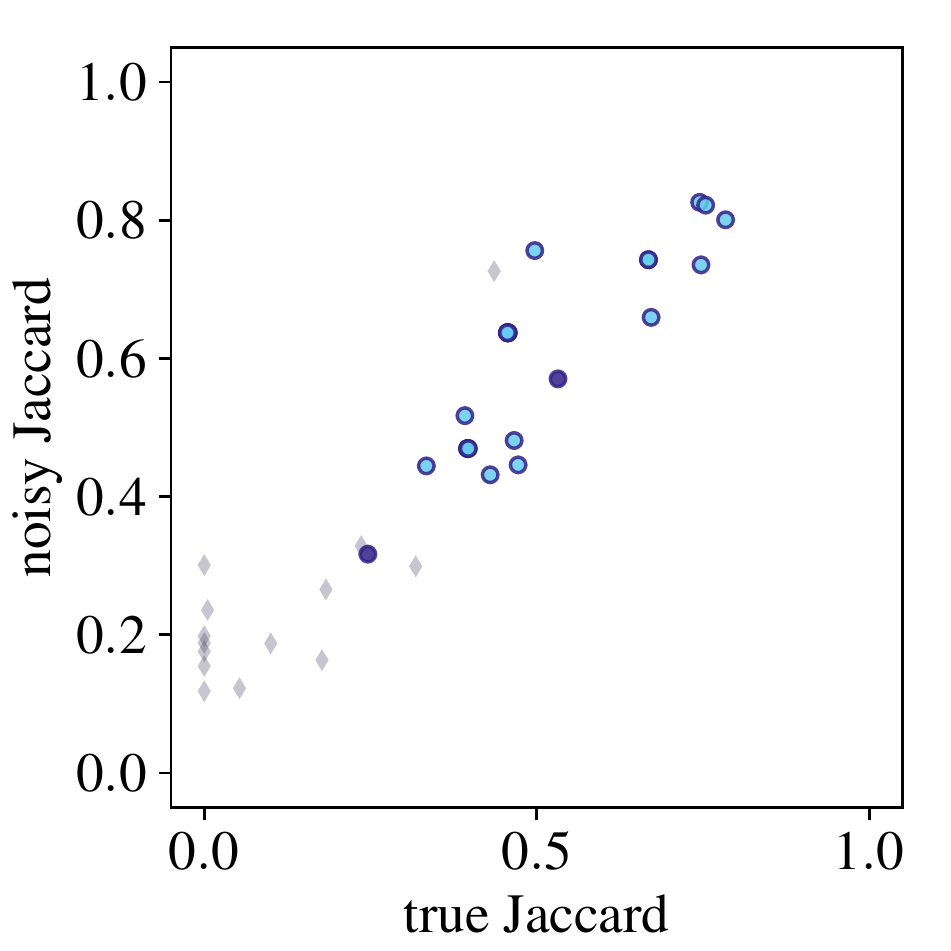} &
                                                                \includegraphics[width=\linewidth]{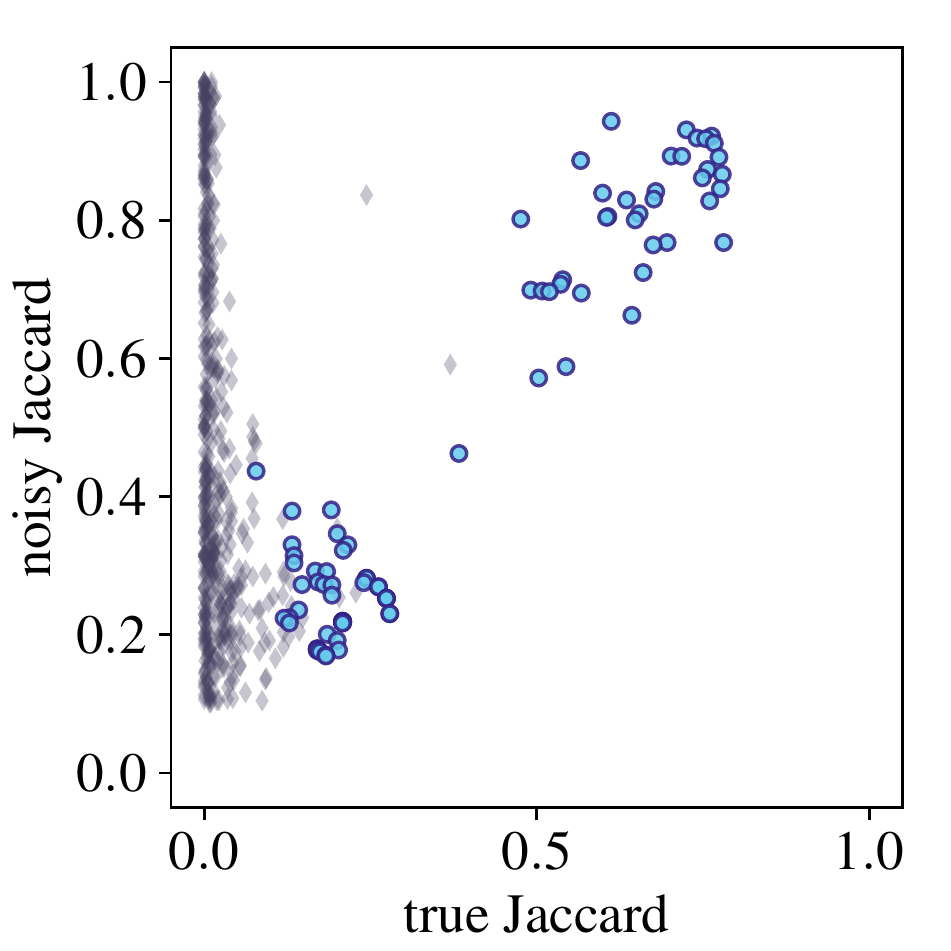} \\
    \includegraphics[width=\linewidth]{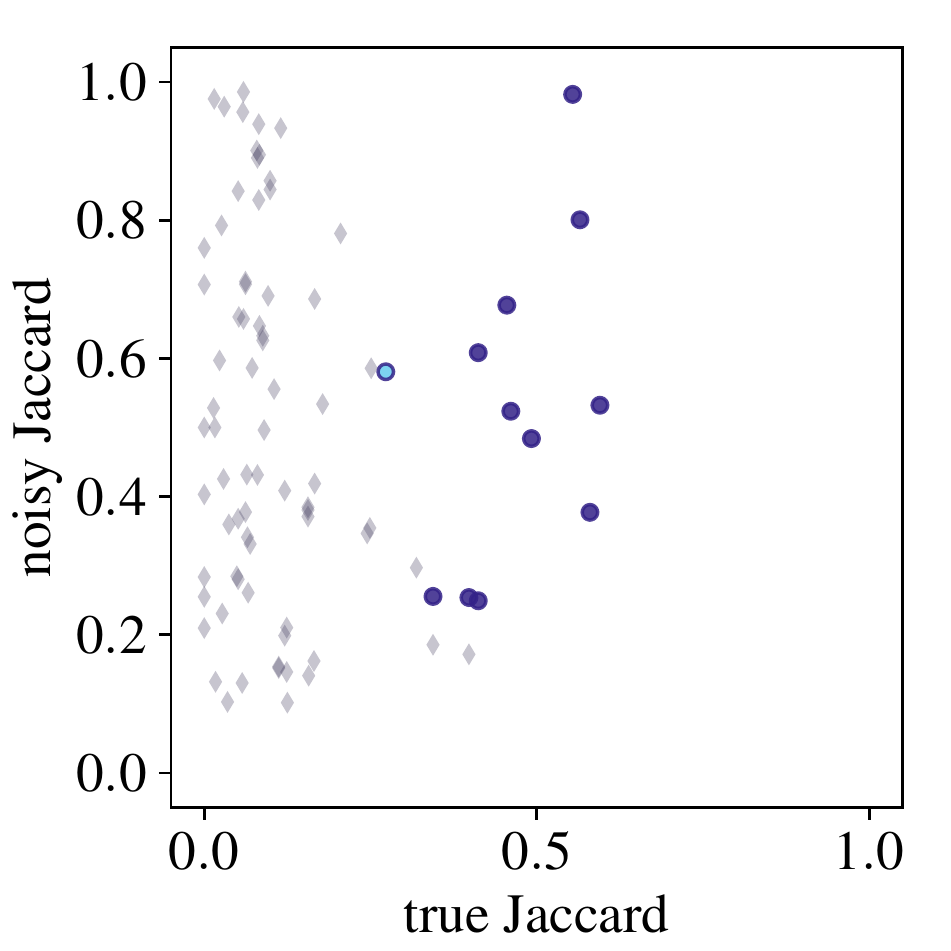} &
    \includegraphics[width=\linewidth]{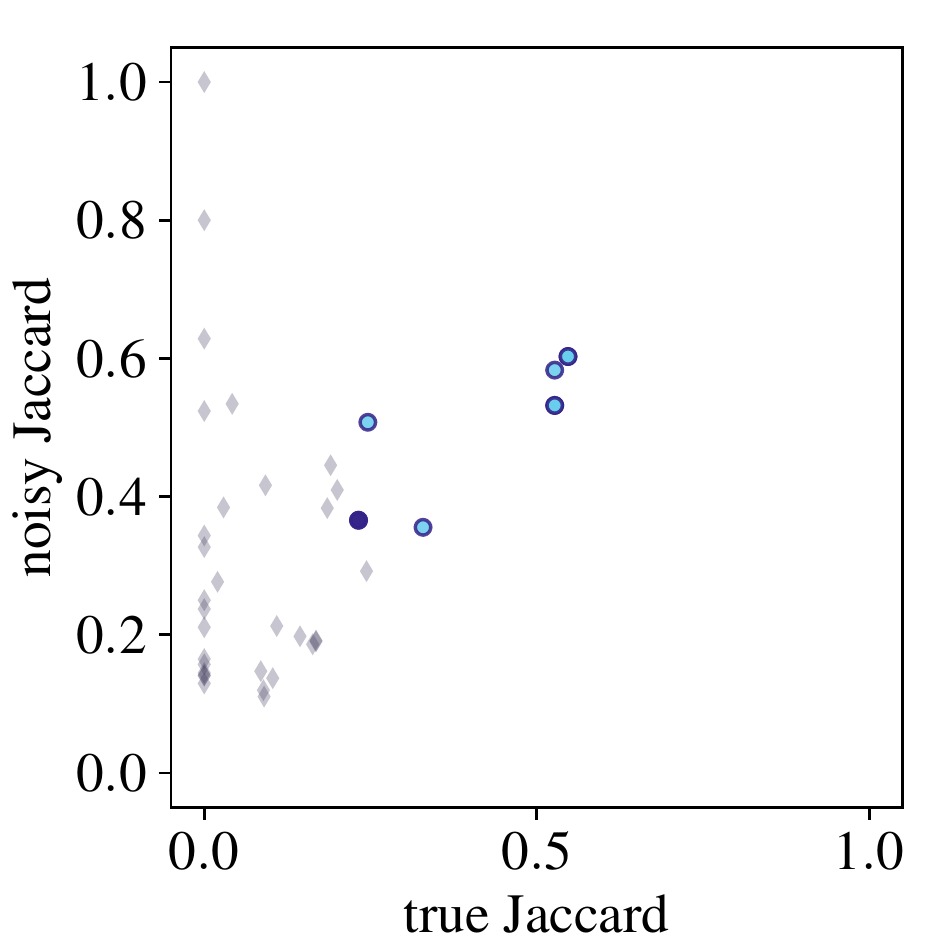} &
                                                               \includegraphics[width=\linewidth]{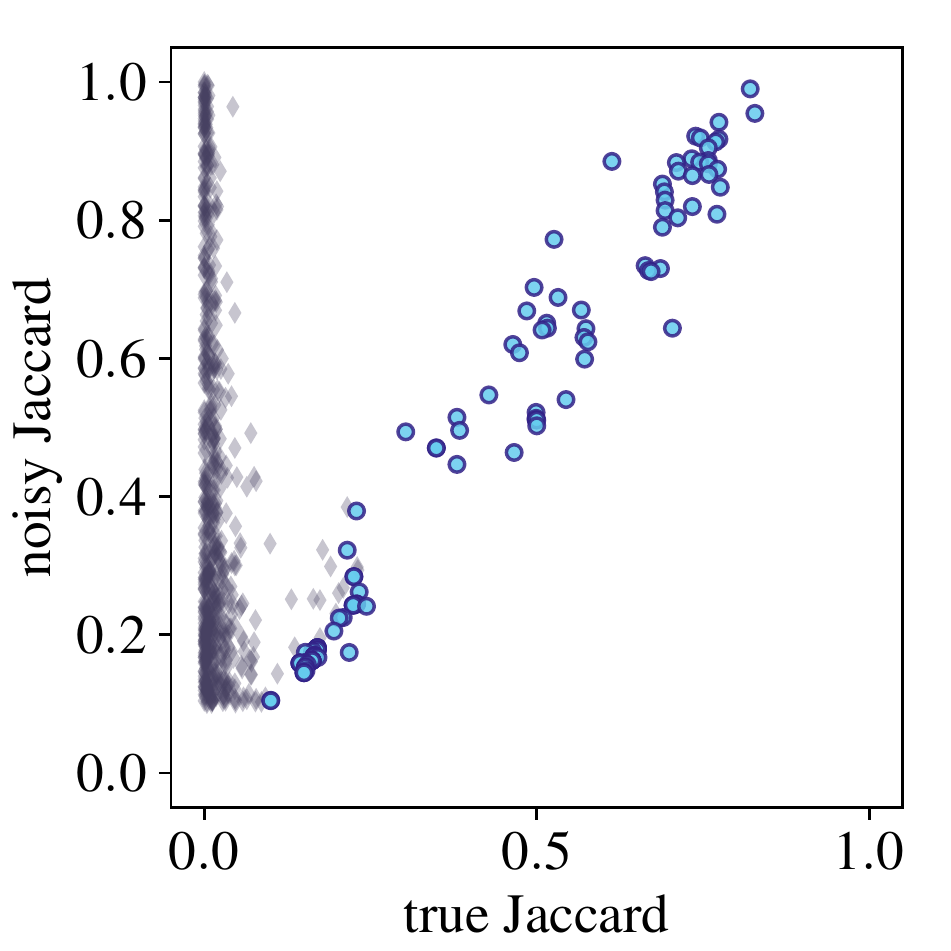} \\
    \includegraphics[width=\linewidth]{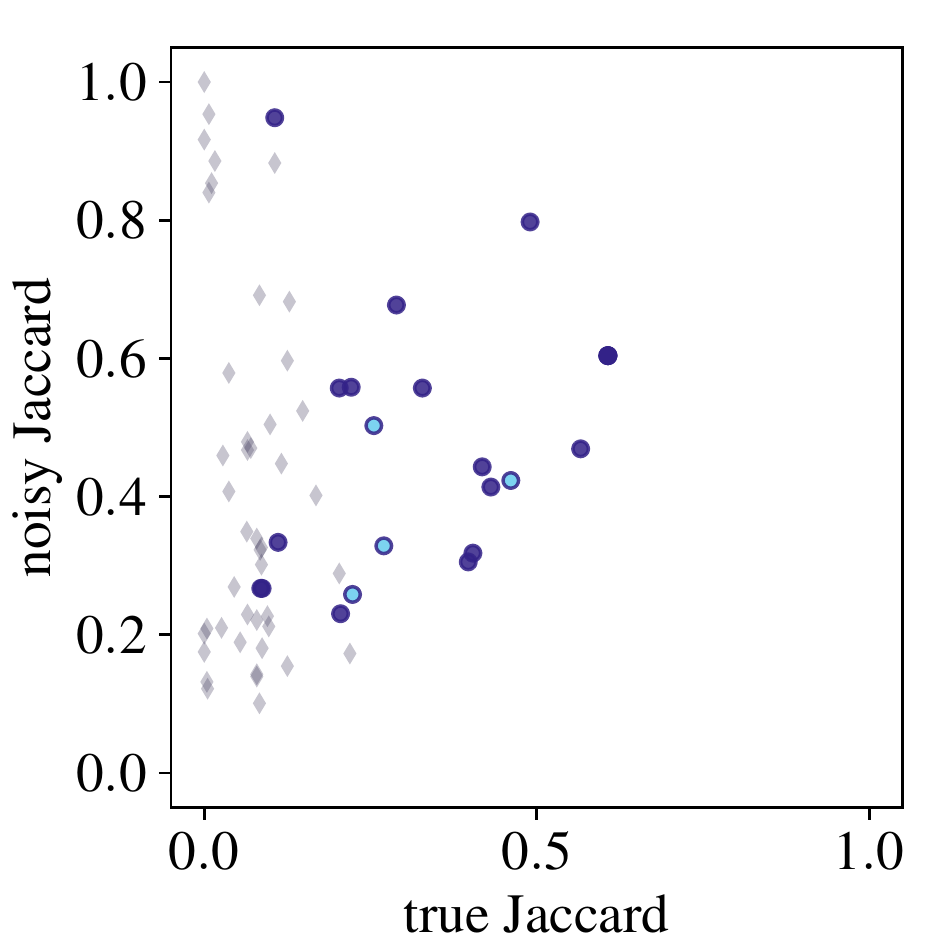} &
    \includegraphics[width=\linewidth]{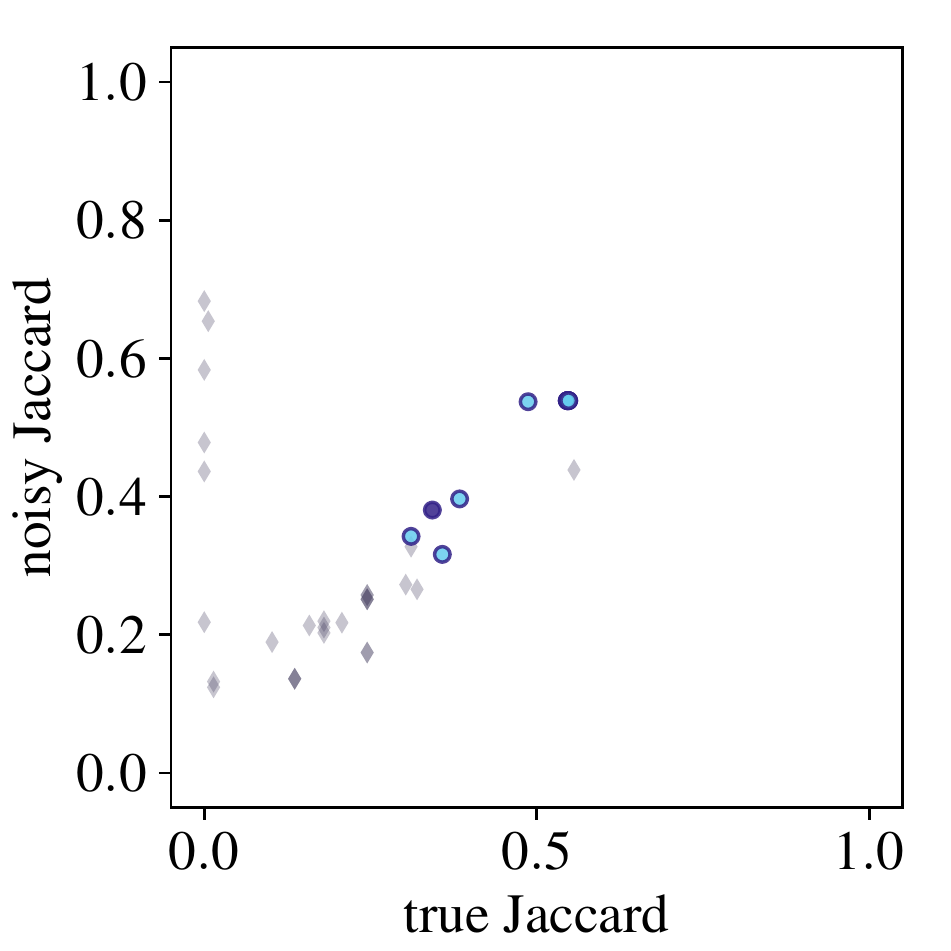} &
    \includegraphics[width=\linewidth]{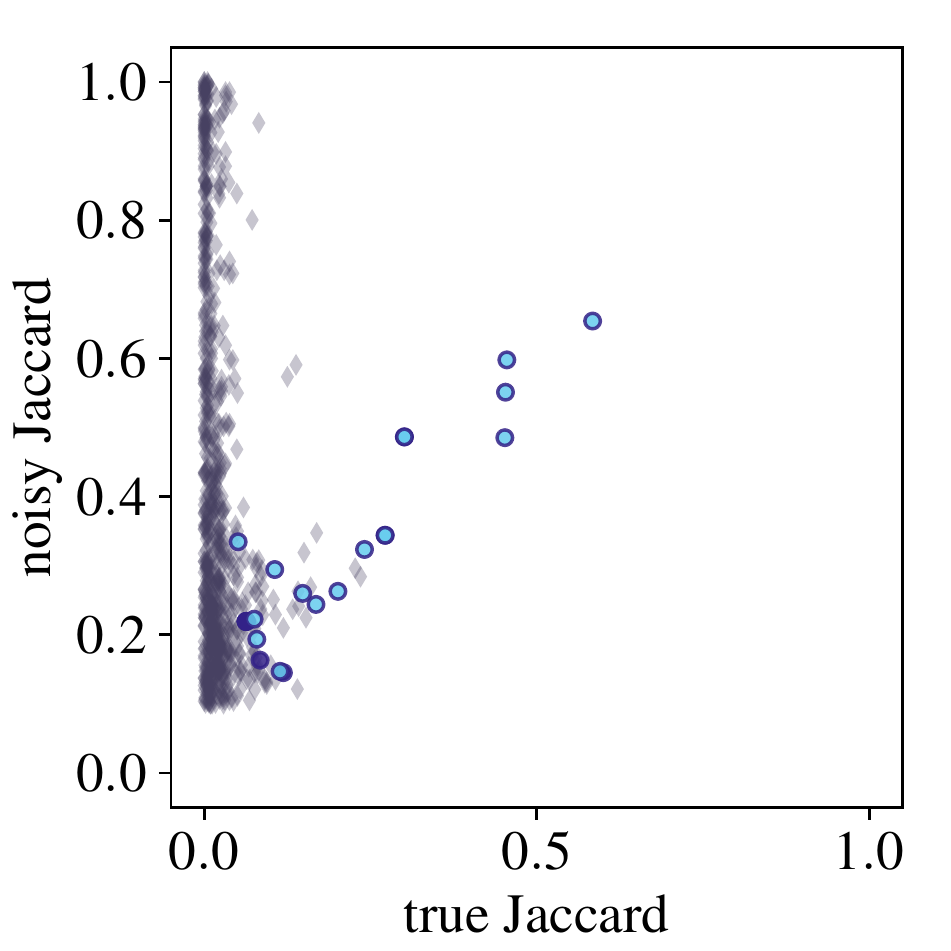} 
  \end{tabularx}
  \caption{\textbf{Impact of differential privacy on the quality of redescriptions for the three proposed differentially private algorithms.} Reported vs.\ true Jaccard indices for \sampling (top row), \expmech (middle row), and \ownalgo (bottom row). Light gray diamonds are redescriptions that are filtered out due to small noisy support. Dark blue dots are insignificant redescriptions and lighter blue dots are significant.}
  \label{fig:exp:scatter1}
\end{figure*}

\begin{figure*}[tbp]
  \centering
  \begin{tabularx}{\textwidth}{@{}X@{}X@{}X@{}X@{}}
    \multicolumn{1}{c}{\medicX{2}{1}} & \multicolumn{1}{c}{\medicX{2}{4}} & \multicolumn{1}{c}{\medicX{2}{8}} & \multicolumn{1}{c}{\medicX{2}{16}} \\
    \includegraphics[width=\linewidth]{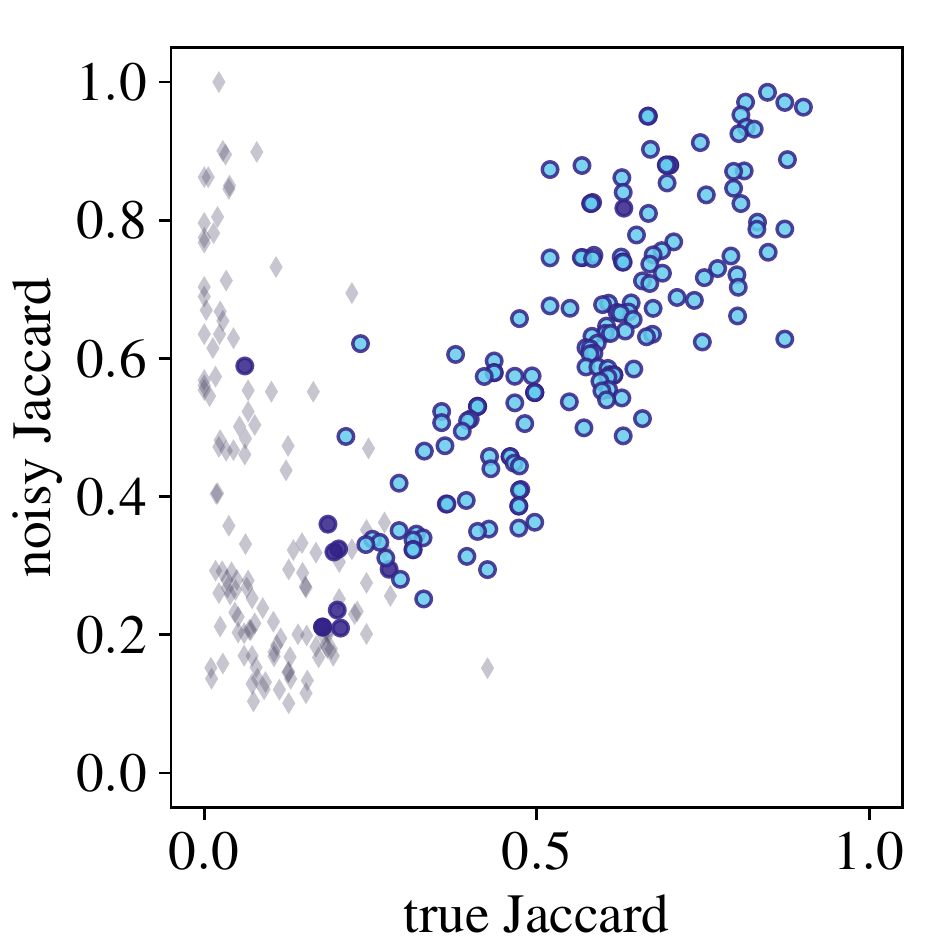} &
    \includegraphics[width=\linewidth]{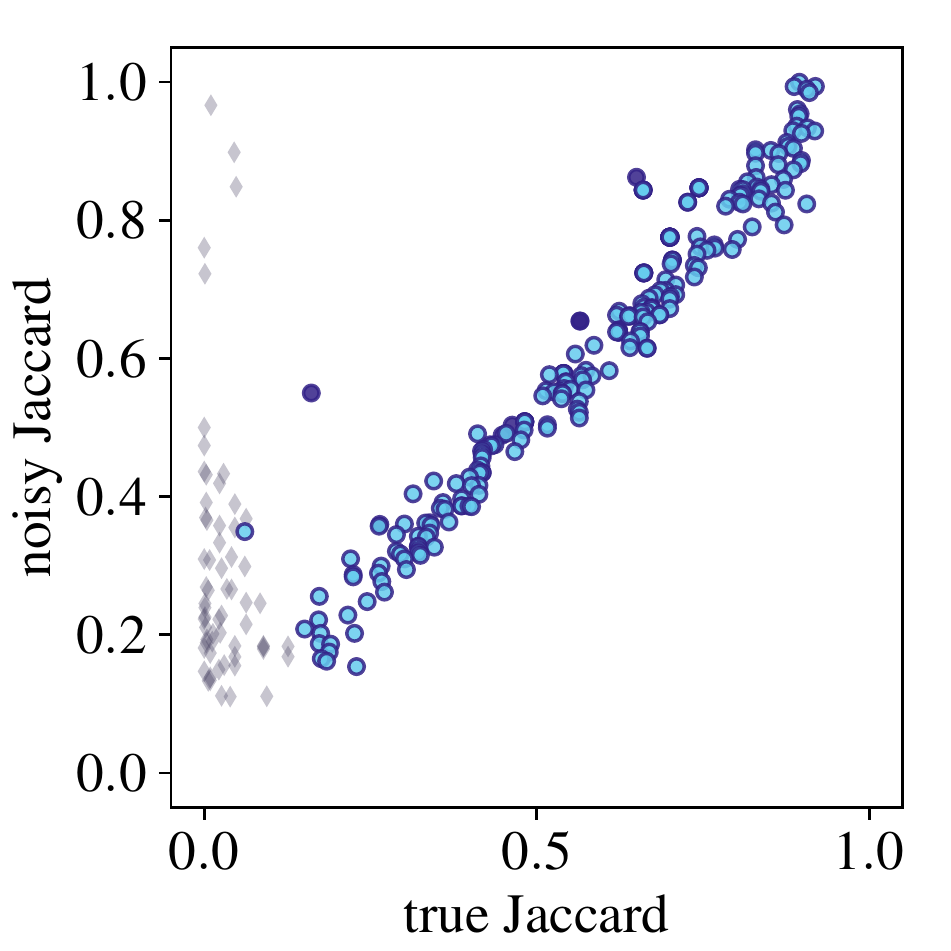} &
    \includegraphics[width=\linewidth]{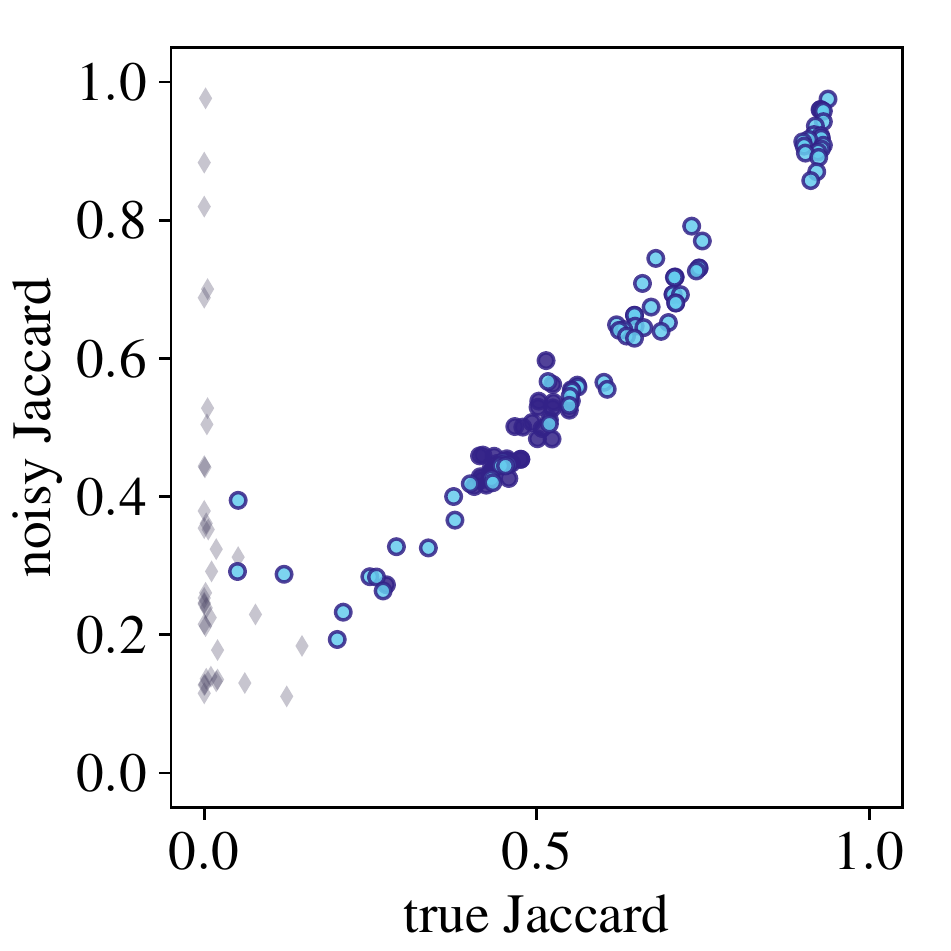} & 
    \includegraphics[width=\linewidth]{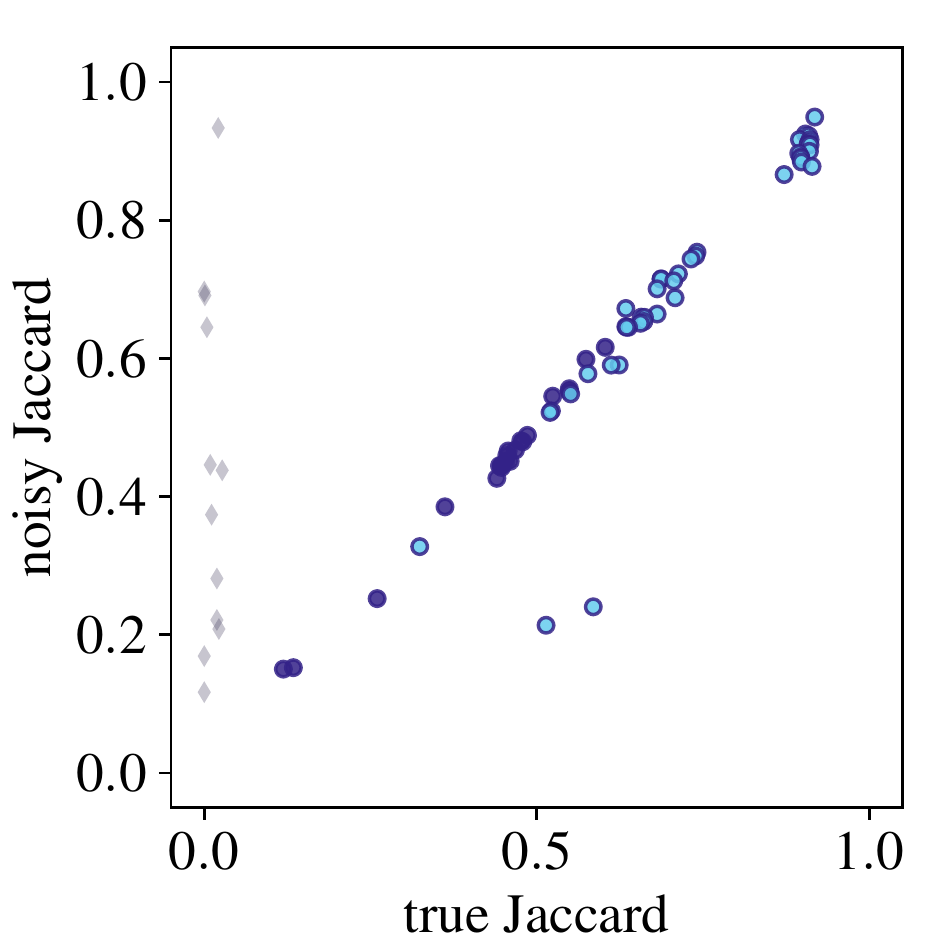}  \\
    \includegraphics[width=\linewidth]{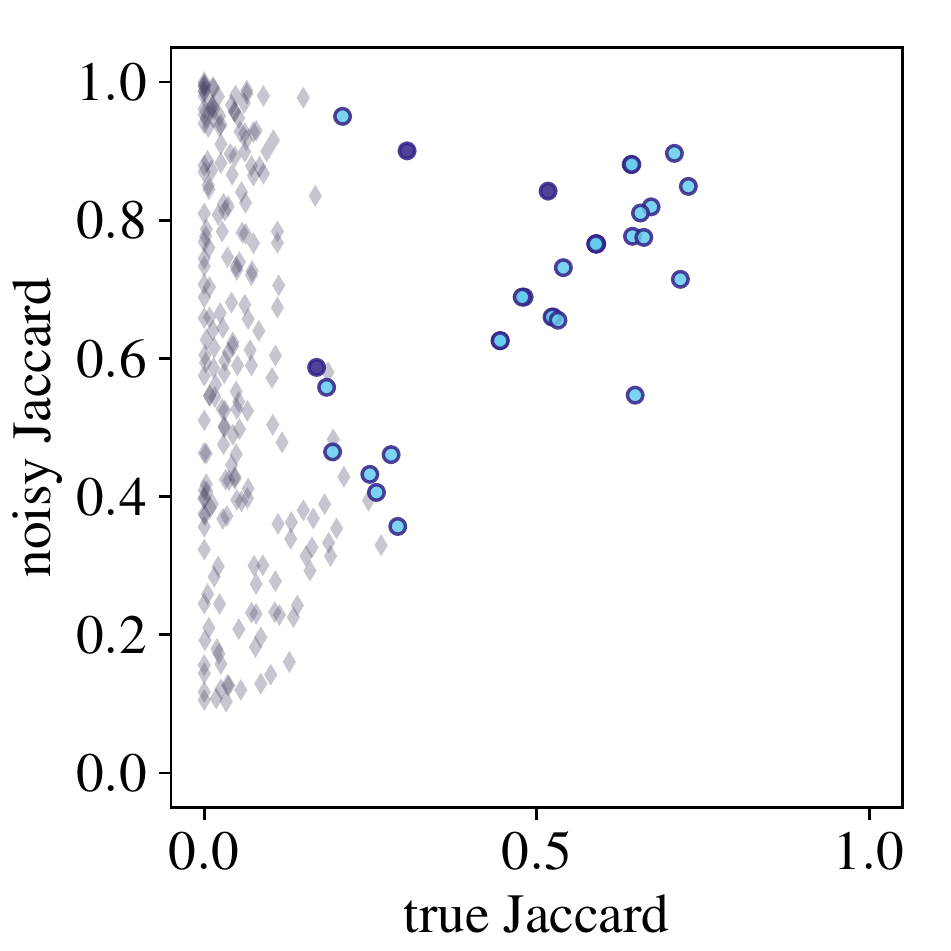} &
    \includegraphics[width=\linewidth]{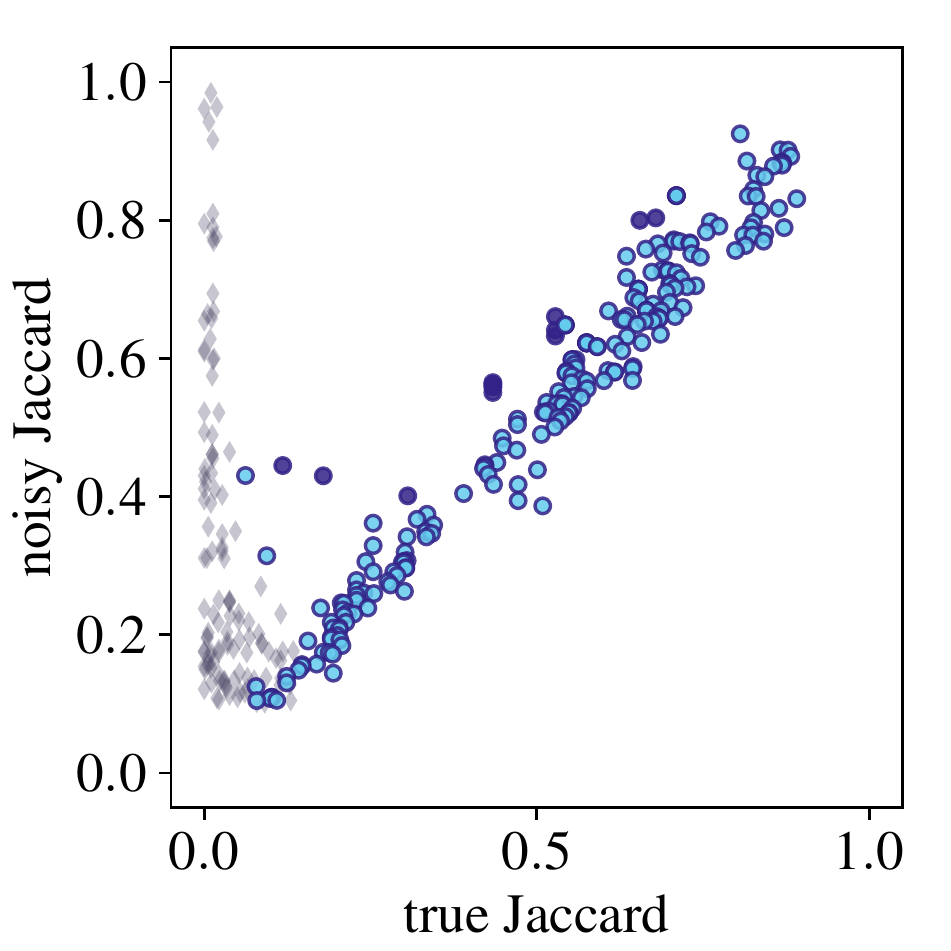} &
    \includegraphics[width=\linewidth]{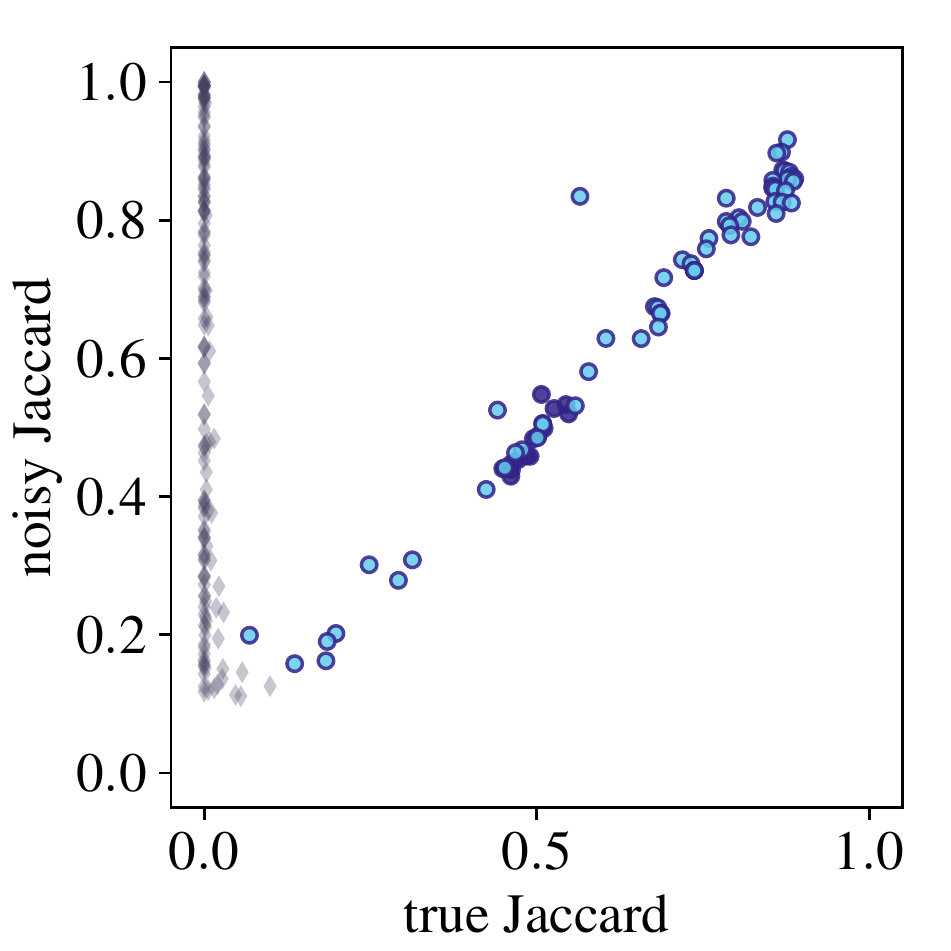} & 
    \includegraphics[width=\linewidth]{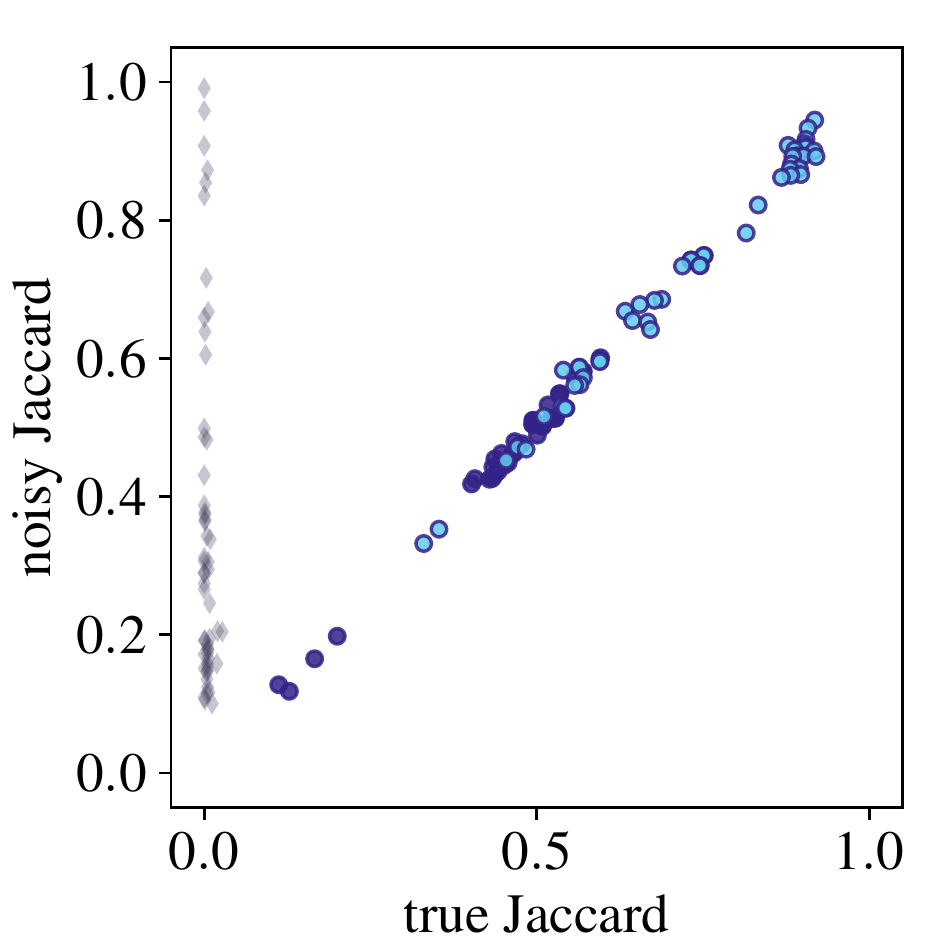}  \\
    \includegraphics[width=\linewidth]{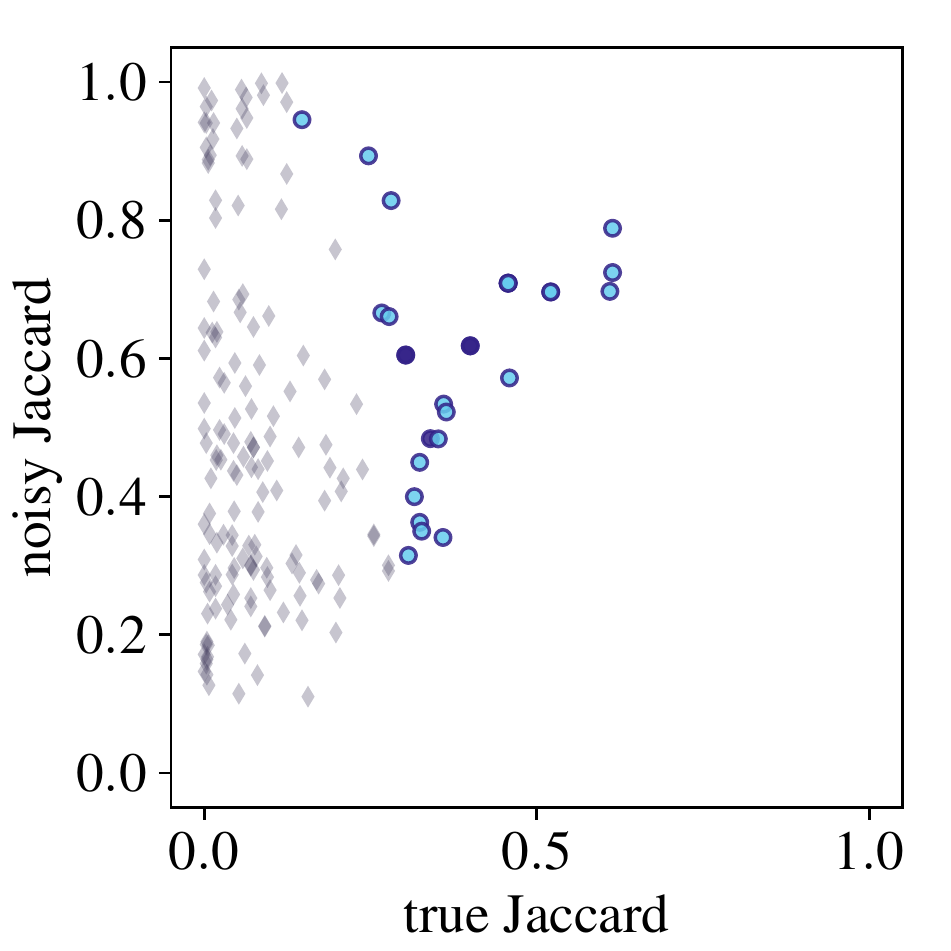} &
    \includegraphics[width=\linewidth]{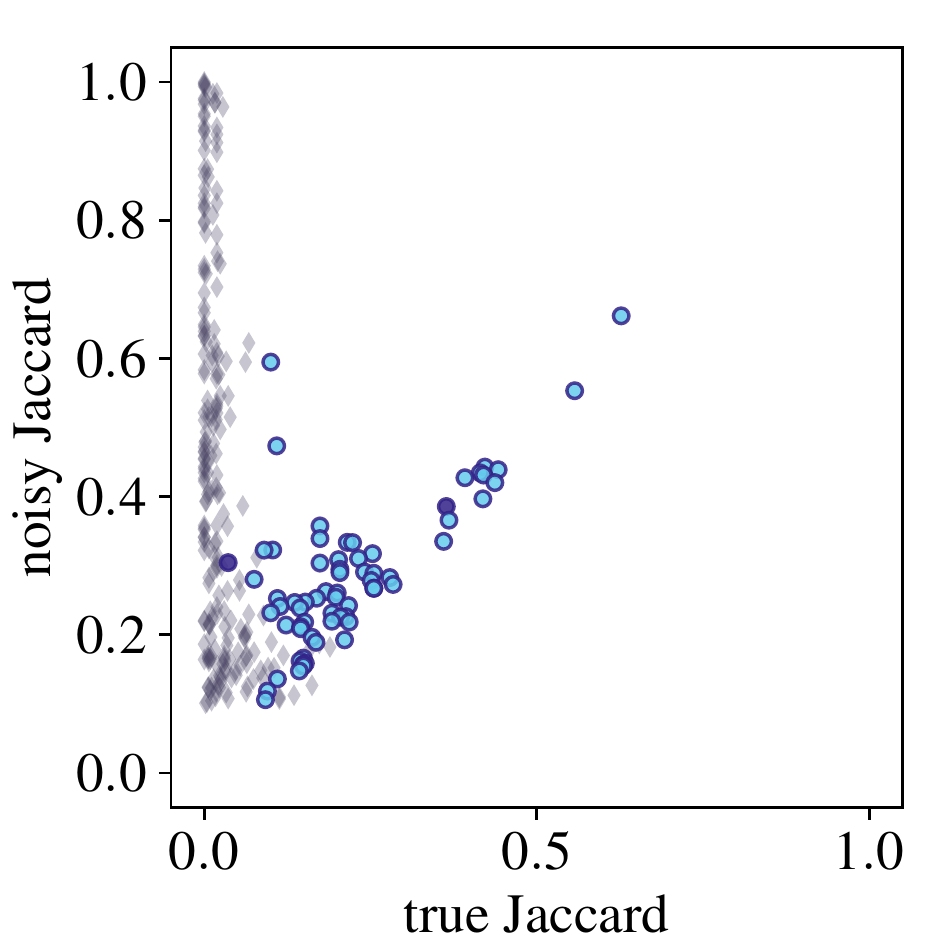} &
    \includegraphics[width=\linewidth]{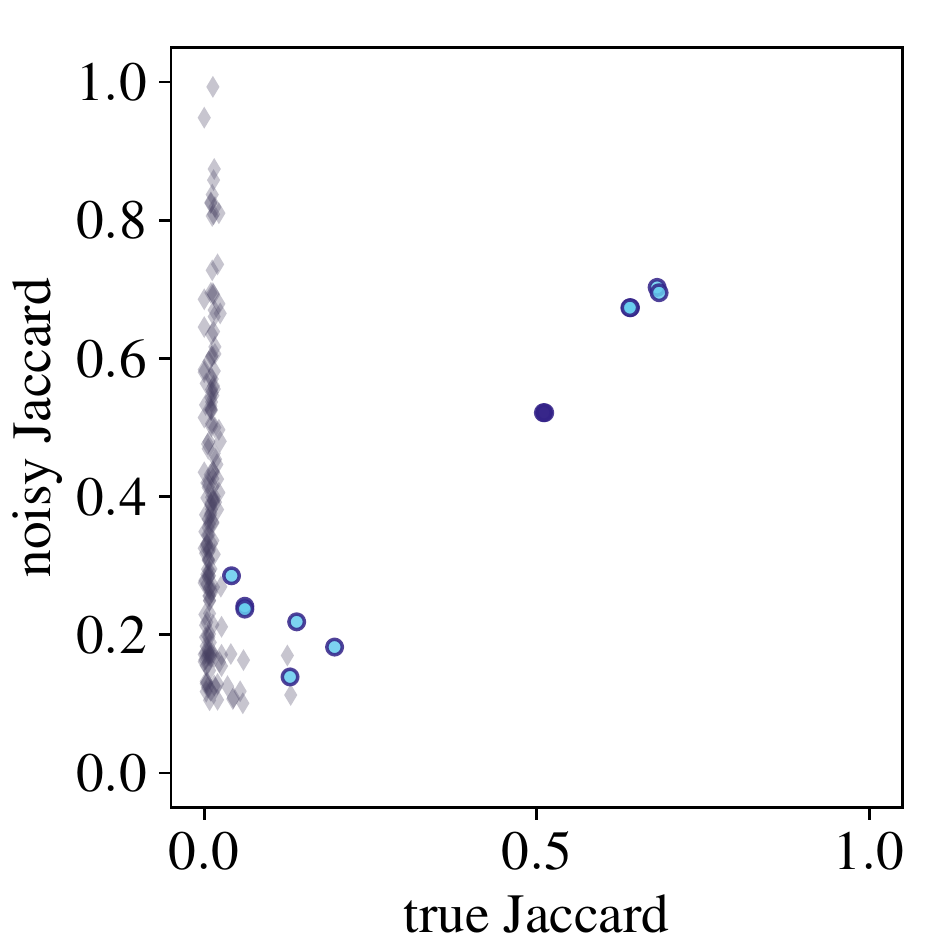} & 
    \includegraphics[width=\linewidth]{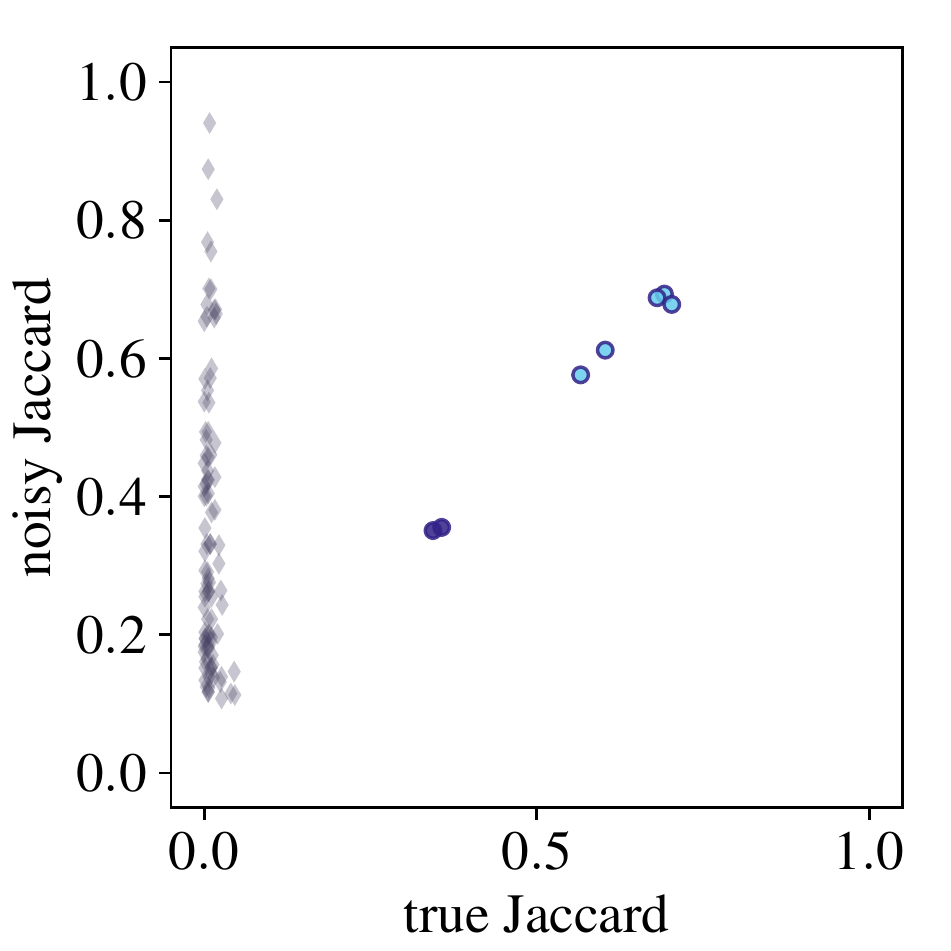} 
  \end{tabularx}
  \caption{\textbf{Impact of differential privacy on the quality of redescriptions for the three proposed differentially private algorithms.} See caption of Figure~\ref{fig:exp:scatter1}}
  \label{fig:exp:scatter}
\end{figure*}

The effects of the added noise on the reported Jaccard indices for the three developed algorithms can be observed in the scatter plots in  Figures~\ref{fig:exp:scatter1} and \ref{fig:exp:scatter}. Here, the $y$-axis shows the differentially private Jaccard indices reported by the differentially private algorithms, while the $x$-axis shows the true Jaccard indices computed with full access to the data.

Because of the noisy counting, the redescriptions with low (noisy) support can have very misleading quality measures. Hence, we recommend that the user prunes out all redescriptions with reported support below a pre-defined threshold (see Section~\ref{sec:redprun} for an explanation of the pruning strategy and Appendix~\ref{sec:param-used-exper} for the thresholds we used). The redescriptions pruned this way are depicted as light grey diamonds. The redescriptions not pruned are the larger blue dots. The colour of the dot indicates whether the true (i.e.\ not noisy) \pvalue is above \num{0.01} (dark blue) or below it (light blue). The user, with no access to the true data, can separate the diamonds from the dots, but cannot separate the dark from the light blue dots. 

We can see that for all approaches, the scatter plot slowly turns into a straight diagonal line as data size increases. This is due to the fact that noisy counts become more accurate as more data is available. Redescription accuracy on the \nerdy dataset is overestimated for all approaches, which is expected for such a small dataset. As it can be seen from Table~\ref{tab:corrcoef}, \ownalgo has the largest and most significant correlation between the noisy and true Jaccard indices on the \nerdy and \mammals datasets, \expmech has the highest correlation on the \mimic and \medicX{2}{16} datasets, and  \sampling on the \medicX{2}{1}, \medicX{2}{4}, and \medicX{2}{8} datasets. \ownalgo produced a smaller number of significant redescriptions on \mammals and on the larger datasets. 

\begin{table}
  \centering
  \caption{Spearman correlation coefficient ($\rho$) and statistical significance ($p$) of correlation between noisy and real Jaccard index for pruned redescriptions reported by different approaches.}
  \label{tab:corrcoef}
  \begin{tabular}{@{}
    l
    S[table-format=1.2]
    S[table-format=<1.2]
    S[table-format=1.2]
    S[table-format=<1.2]
    S[table-format=1.2]
    S[table-format=<1.2]
    @{}}
    \toprule
    Dataset & \multicolumn{2}{c}{\sampling} &  \multicolumn{2}{c}{\expmech}  & \multicolumn{2}{c}{\ownalgo} \\
    \cmidrule(r){2-3}\cmidrule(lr){4-5}\cmidrule(l){6-7}
    & {$\rho$} & {$p$} & {$\rho$} & {$p$} & {$\rho$} & {$p$} \\
    \midrule
    \nerdy & 0.03 & 0.9 & 0.32&0.31 &  0.44&0.04  \\
    \mammals & 0.82&<0.01 & 0.88&<0.01 &  0.90&<0.01\\
    \mimic & 0.78&<0.01 & 0.98&<0.01 &  0.75&<0.01  \\
    \medicX{2}{1} & 0.84&<0.01 & 0.56&<0.01 &  0.16&0.42   \\
    \medicX{2}{4} & 0.97&<0.01 & 0.97&<0.01 &  0.55&<0.01  \\
    \medicX{2}{8} & 0.98&<0.01 & 0.97&<0.01 &  0.77&<0.01  \\
    \medicX{2}{16} & 0.97&<0.01 & 0.99&<0.01 &  0.89&0.01  \\
    \bottomrule
  \end{tabular}
\end{table}

A very important observation is that the pruning strategy works. The grey diamonds to the left of the scatter plots correspond to false positives, that is, to redescriptions that differentially private algorithms return as valid redescriptions, but that have very low true Jaccard index value. On the other hand, there are very few diamonds towards the right, indicating that there are very few false negatives, that is, good redescriptions that are unnecessarily pruned. As the user does not have access to the true data and cannot compute the true Jaccard indices, it is of vital importance that this filtering works.

The first three rows of plots in Figures~\ref{fig:exp:bar} and \ref{fig:exp:bar1} show the distribution of the true Jaccard indices, with the number of redescriptions having true Jaccard index essentially zero indicated in the top-left corner. These plots make it clearer that the pruned redescriptions (light grey) are those with lower true Jaccard index values, while the non-pruned ones have higher Jaccard index values.

\subsection{Comparison to Other Algorithms}

\begin{figure*}[tbp]
  \centering
  \begin{tabularx}{0.9\textwidth}{@{}X@{}X@{}X@{}}
    \multicolumn{1}{c}{\nerdy}  & \multicolumn{1}{c}{\mammals} & \multicolumn{1}{c}{\mimicX{Ter}} \\
    \includegraphics[width=\linewidth]{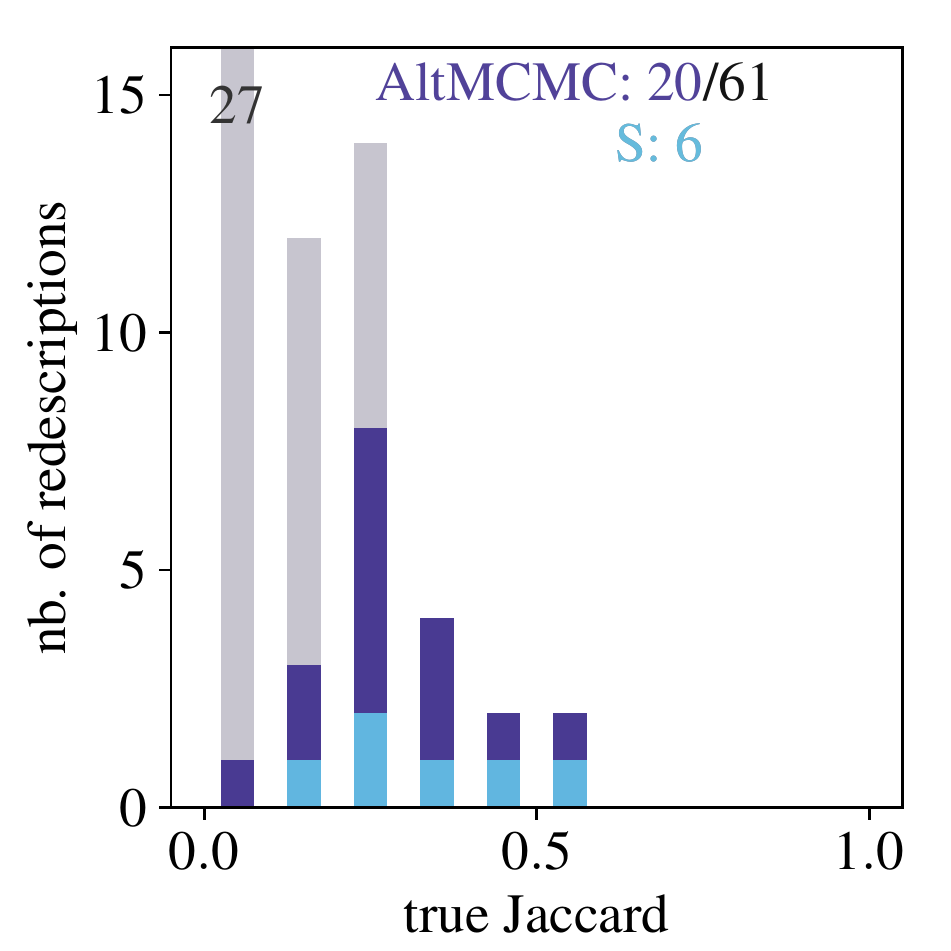} &
    \includegraphics[width=\linewidth]{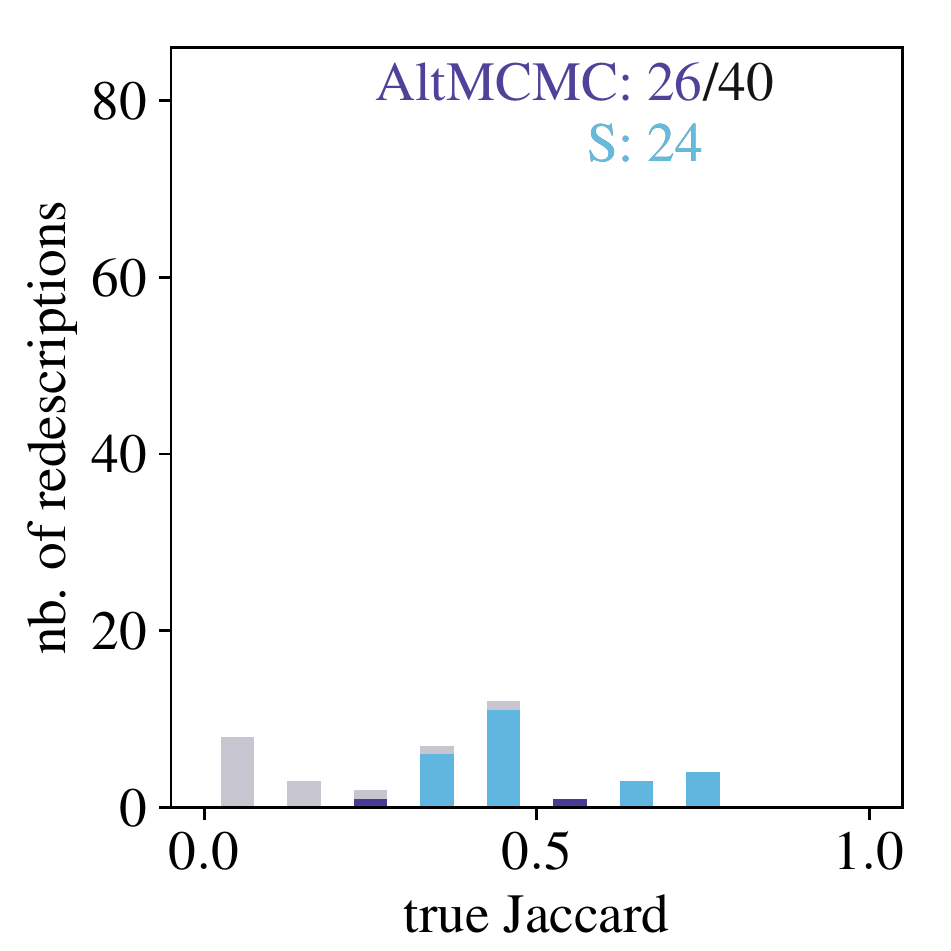} &
    \includegraphics[width=\linewidth]{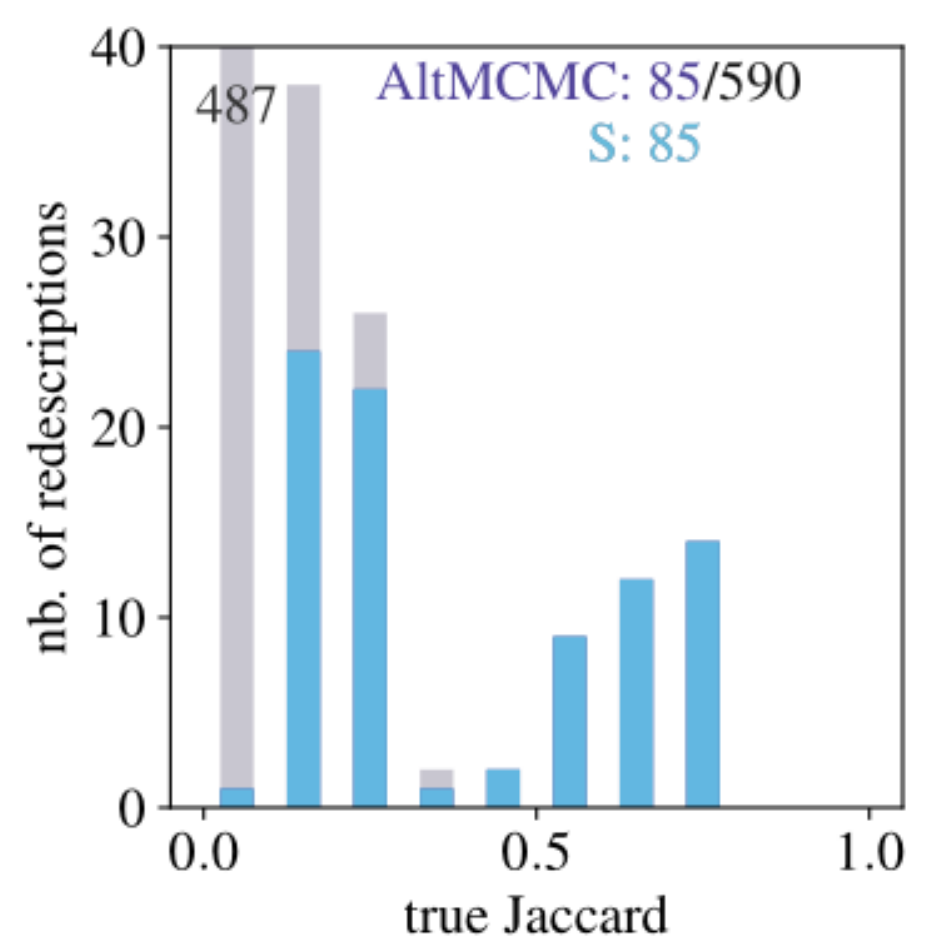} \\
   \includegraphics[width=\linewidth]{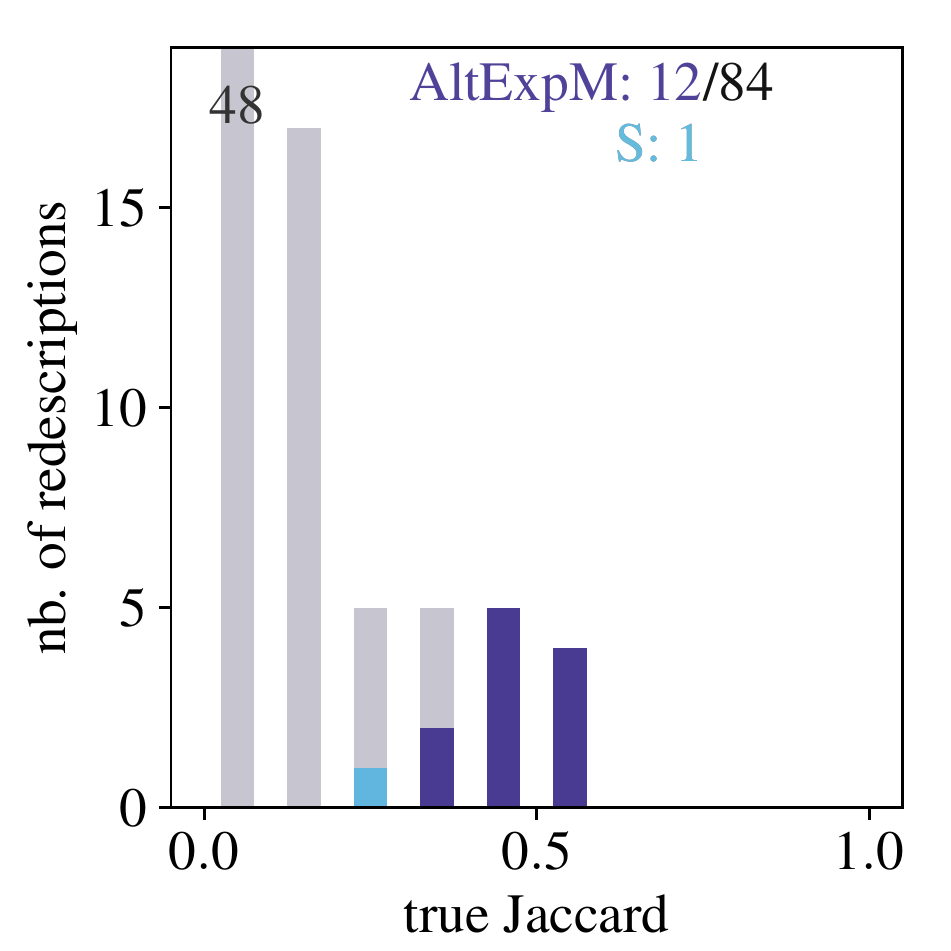} &
    \includegraphics[width=\linewidth]{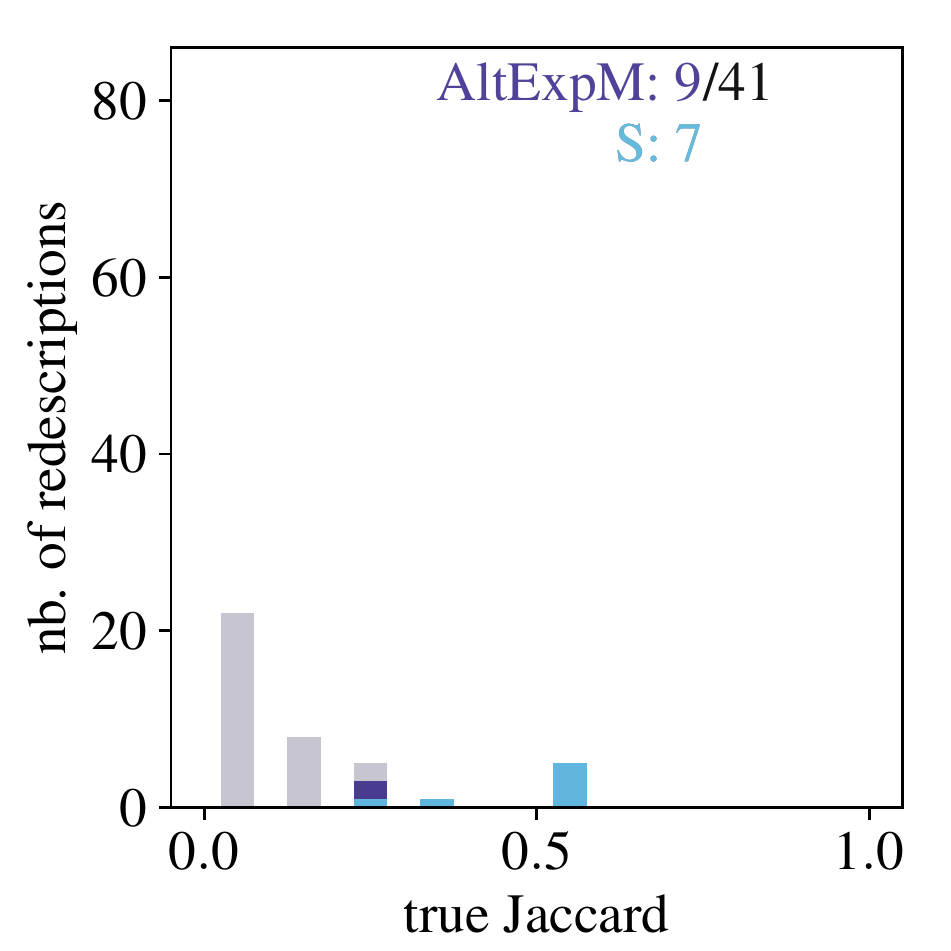} &
    \includegraphics[width=\linewidth]{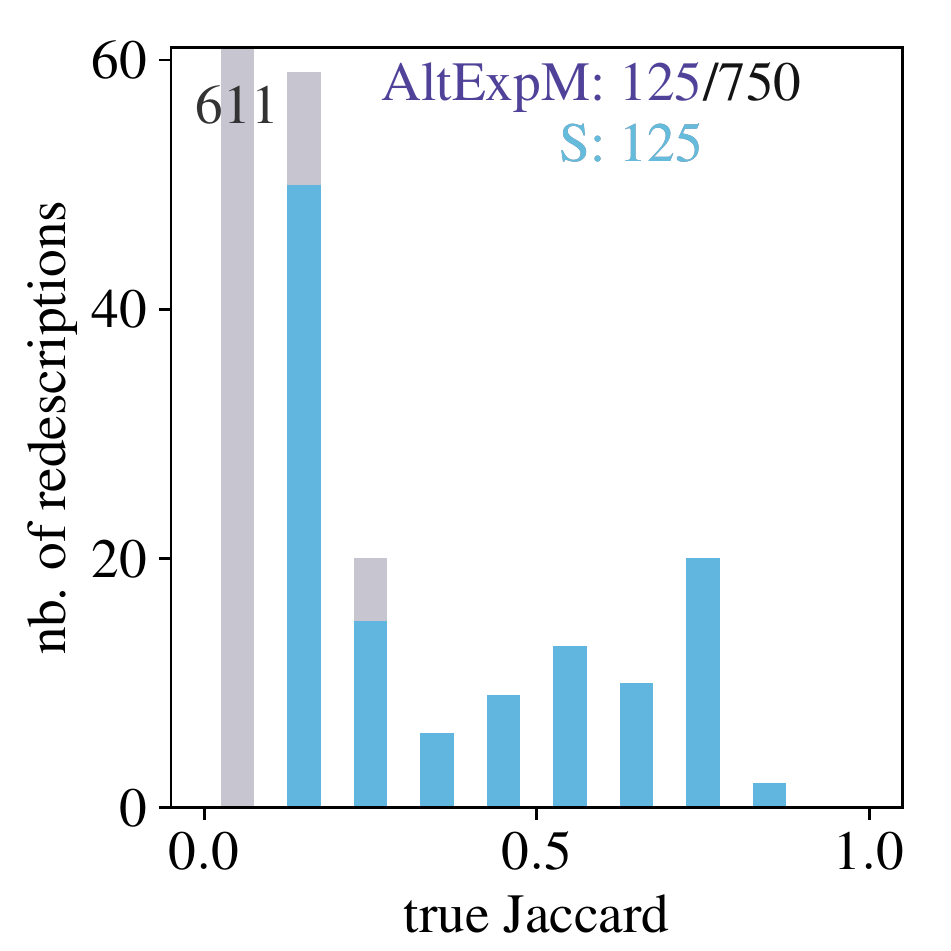} \\
    \includegraphics[width=\linewidth]{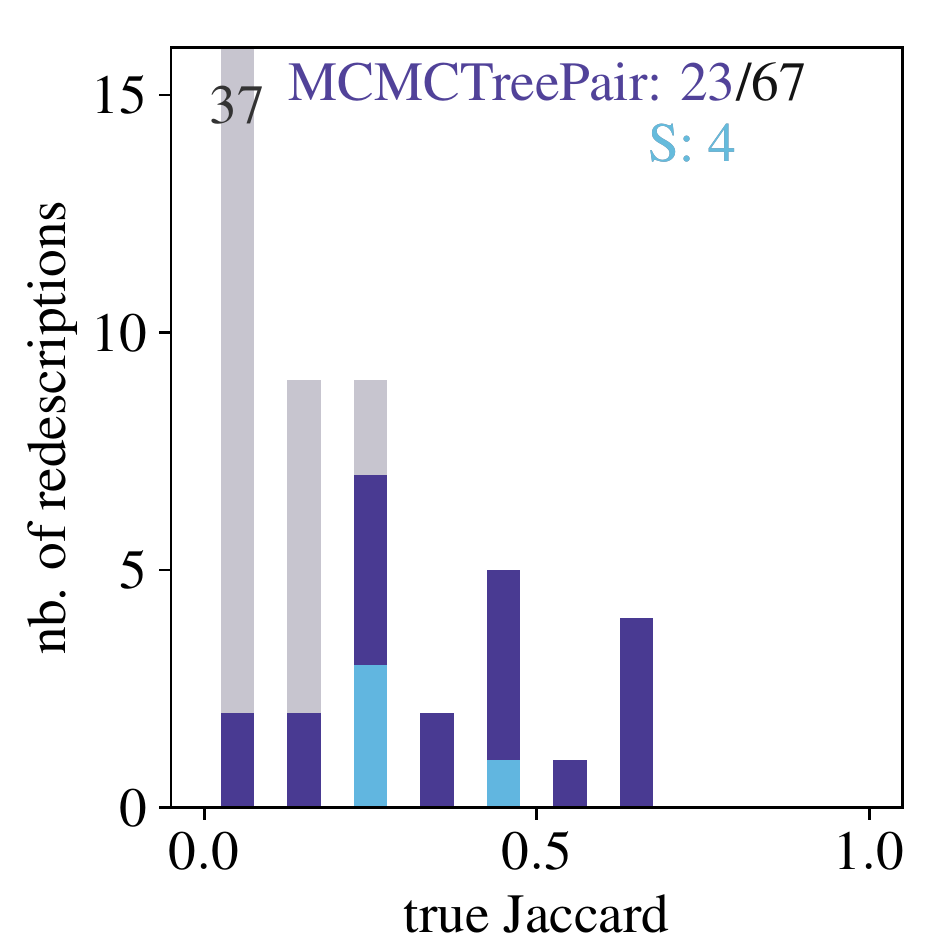} &
    \includegraphics[width=\linewidth]{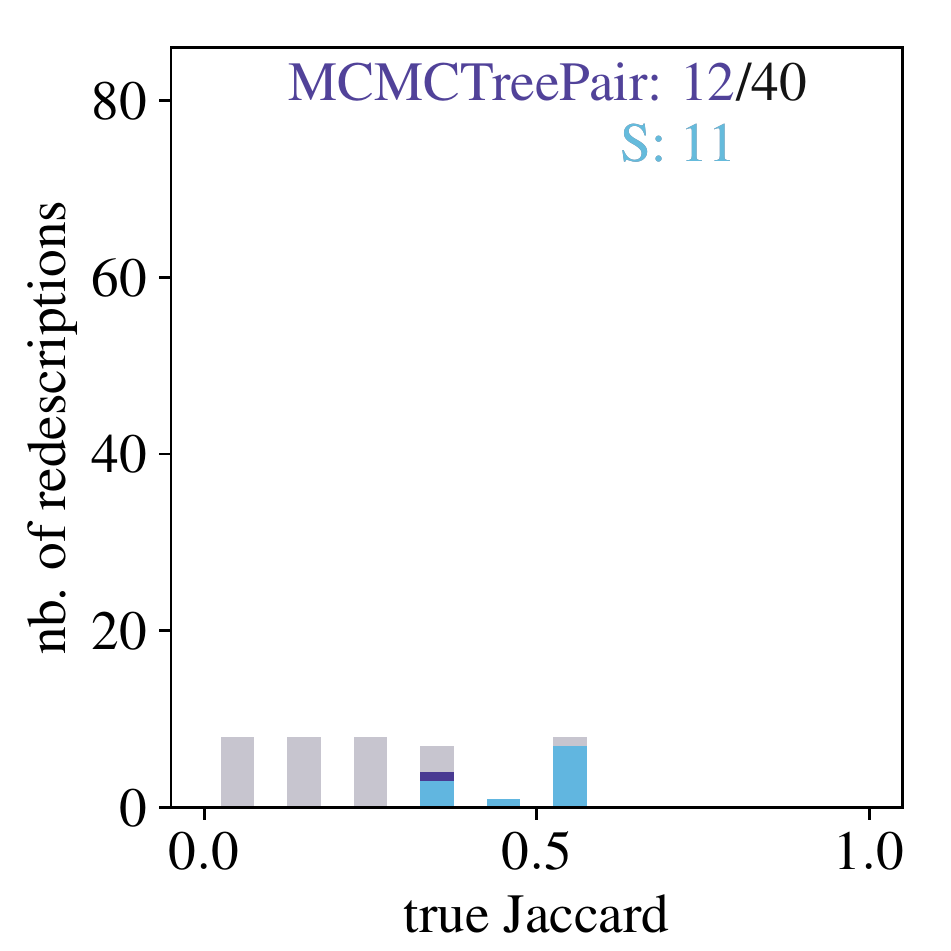} &
    \includegraphics[width=\linewidth]{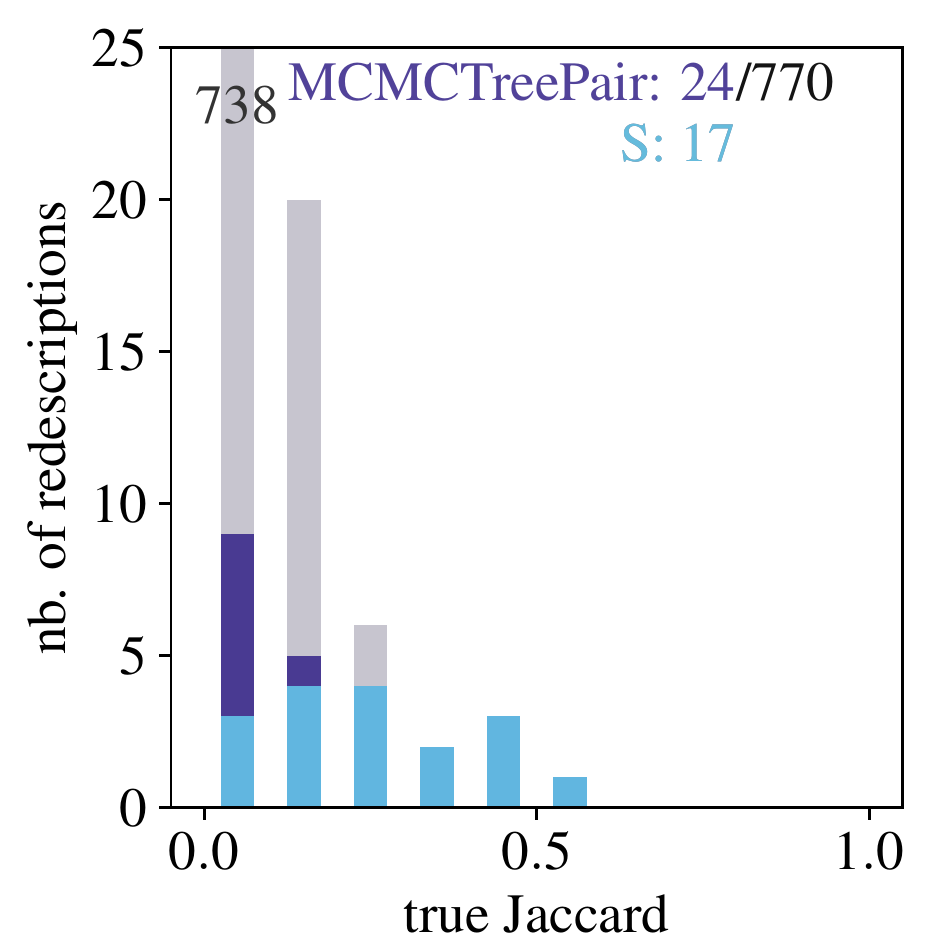} \\
    \includegraphics[width=\linewidth]{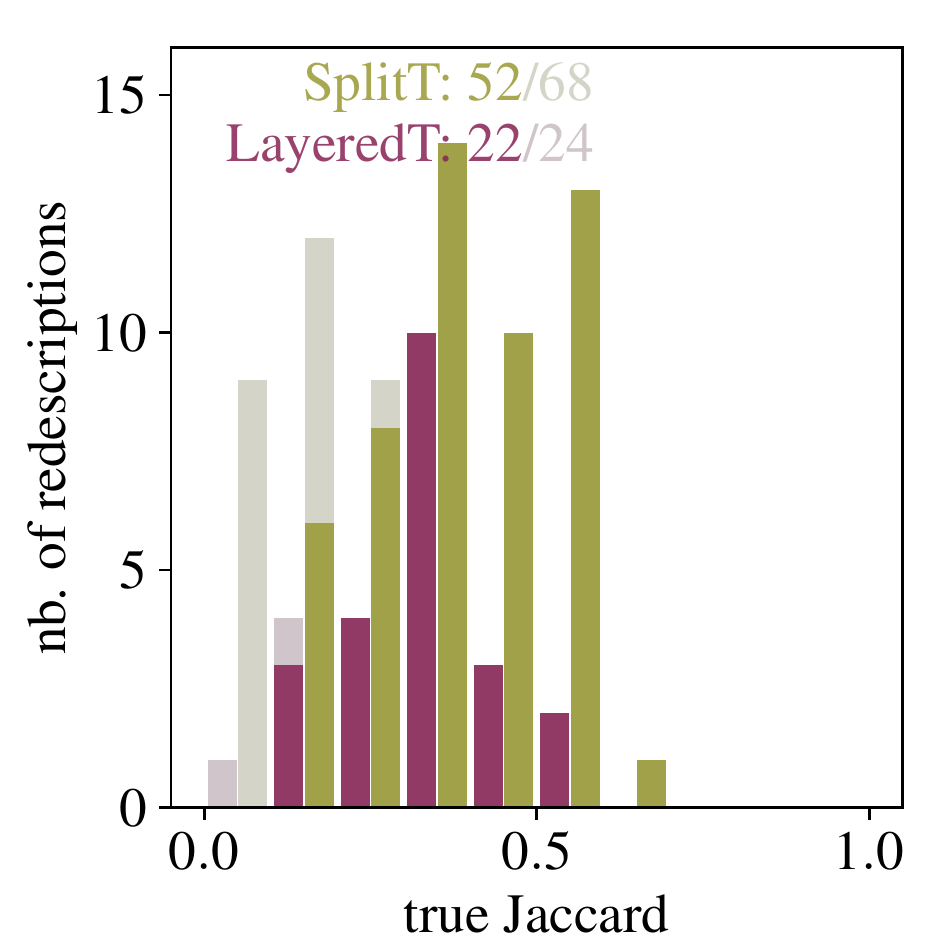} &
    \includegraphics[width=\linewidth]{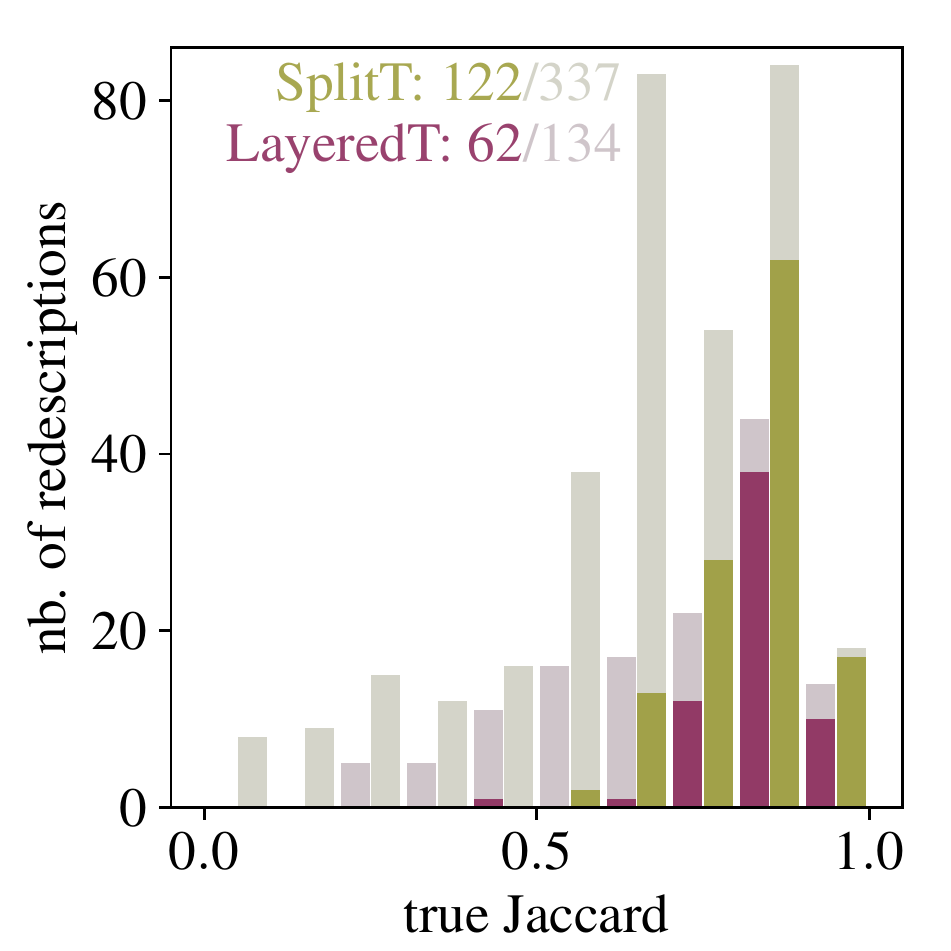} &
    \includegraphics[width=\linewidth]{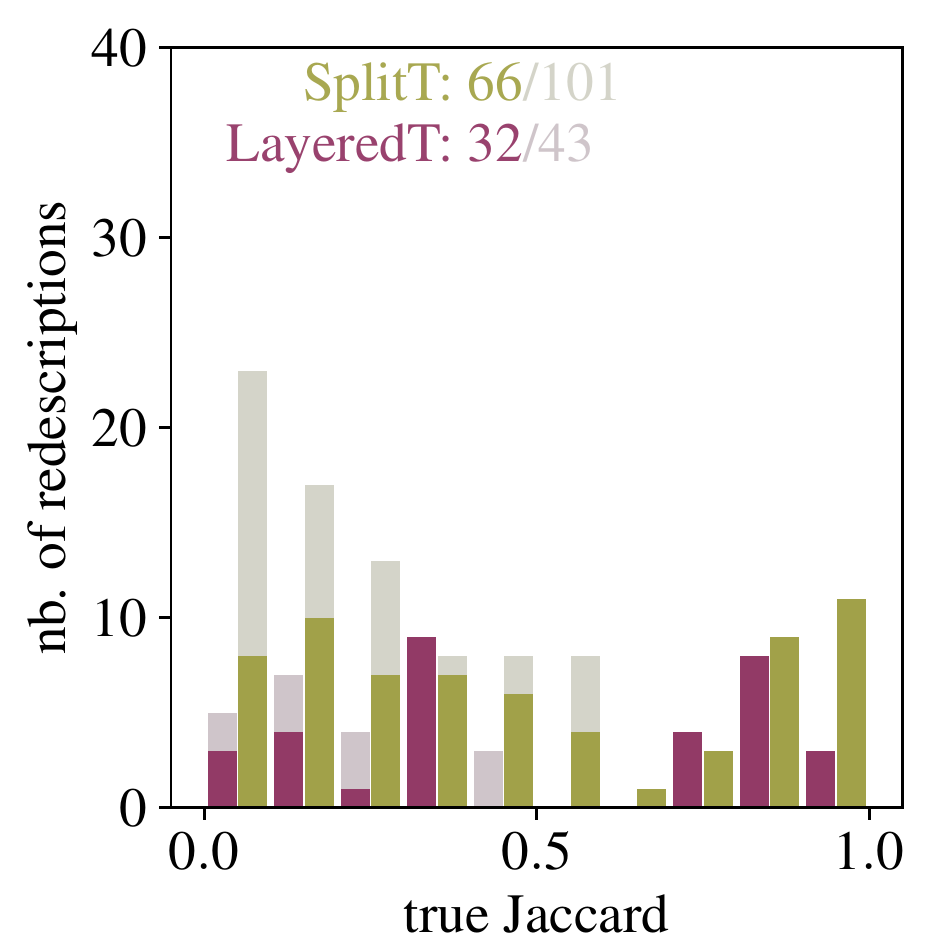} 
  \end{tabularx}
  \caption{\textbf{Impact of differential privacy on the quality of redescriptions and comparison to other algorithms} using the \nerdy, \mammals, and \mimic datasets. Distribution of true Jaccard indices for \sampling (top row), \expmech (second row), and \ownalgo (third row) and for the other non-differentially private algorithms (bottom row). 
   In the top three rows, light grey bars correspond to redescriptions that are pruned due to low noisy support, dark blue bars to statistically insignificant redescriptions, and light blue bars to significant redescriptions. For the bottom row, the greyed bars indicate insignificant redescriptions and coloured significant ones. The corresponding numbers are given at the top. Where relevant, the numbers in the top-left corner of bar plots indicate the total number of redescriptions that  have true Jaccard index value essentially $0$.}
  \label{fig:exp:bar}
\end{figure*}

\begin{figure*}[tbp]
  \centering
  \begin{tabularx}{\textwidth}{@{}X@{}X@{}X@{}X@{}}
    \multicolumn{1}{c}{\medicX{2}{1}} & \multicolumn{1}{c}{\medicX{2}{4}} & \multicolumn{1}{c}{\medicX{2}{8}} & \multicolumn{1}{c}{\medicX{2}{16}} \\
    \includegraphics[width=\linewidth]{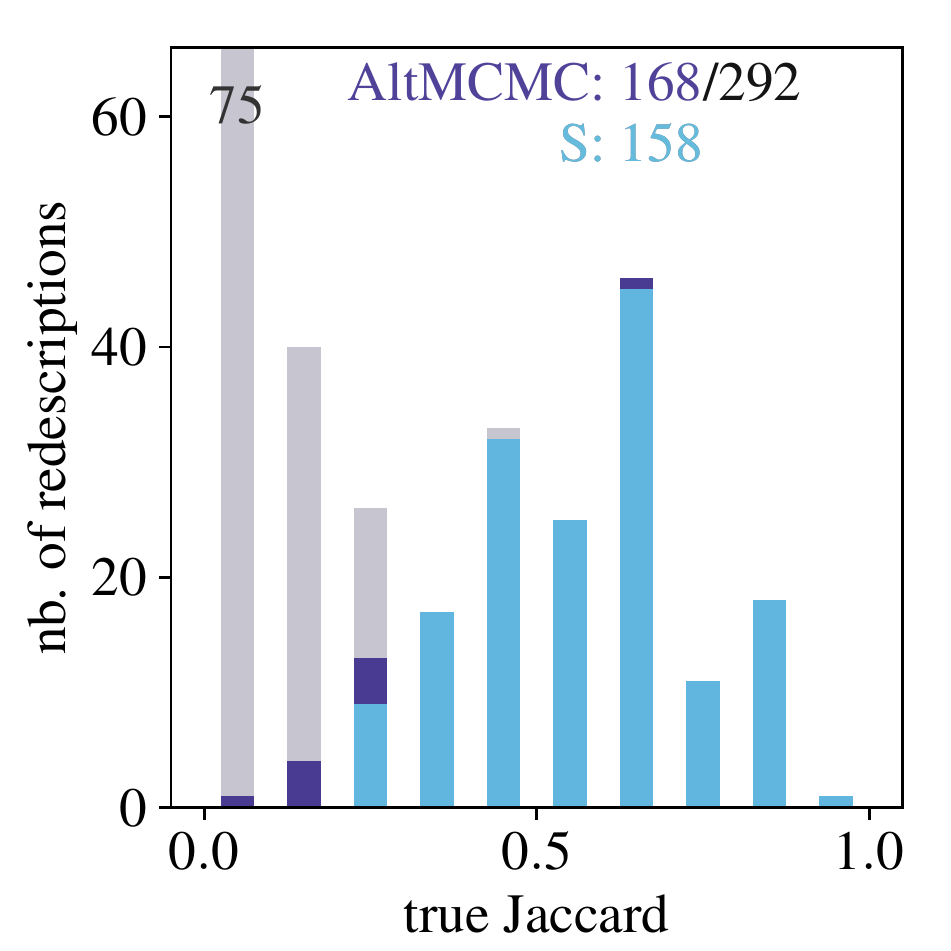} &
    \includegraphics[width=\linewidth]{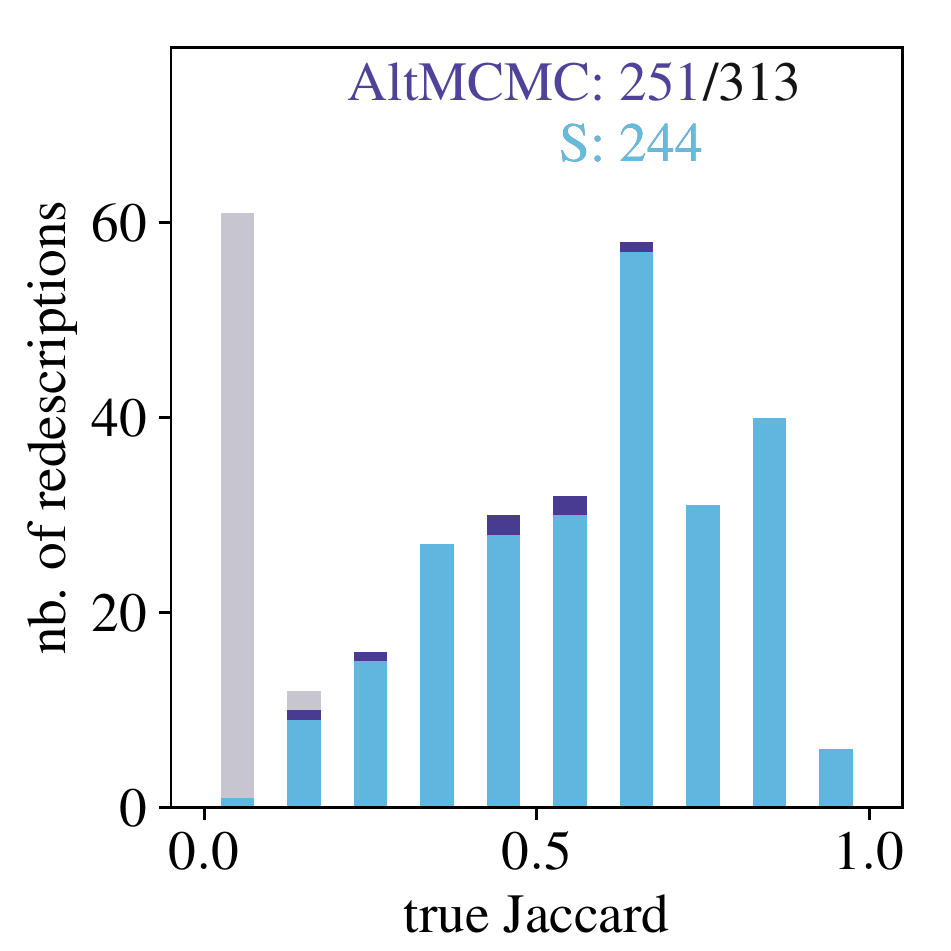} &
    \includegraphics[width=\linewidth]{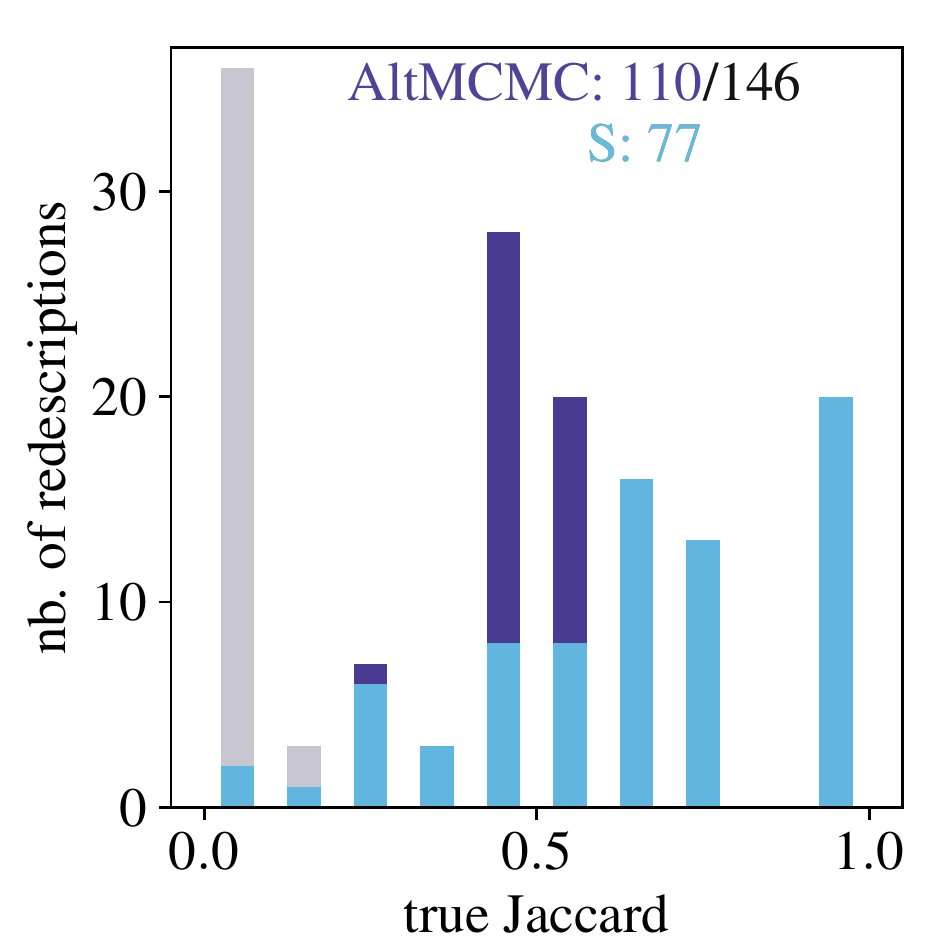} &
    \includegraphics[width=\linewidth]{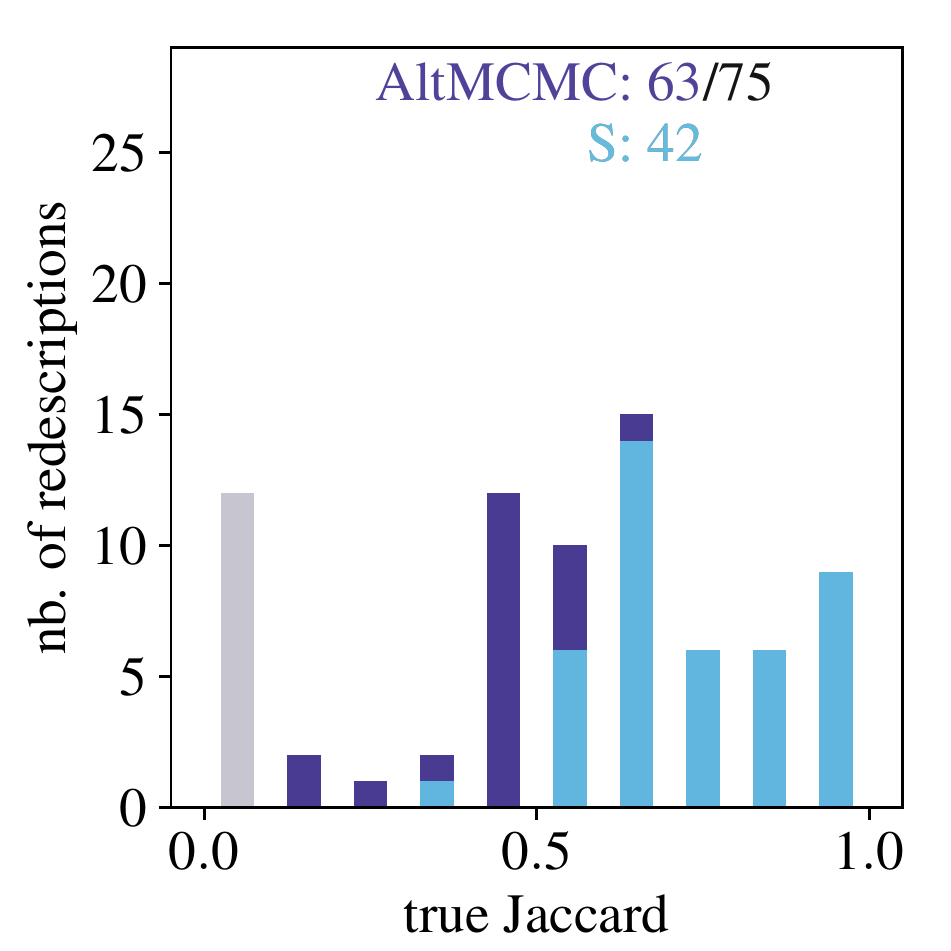}\\
    \includegraphics[width=\linewidth]{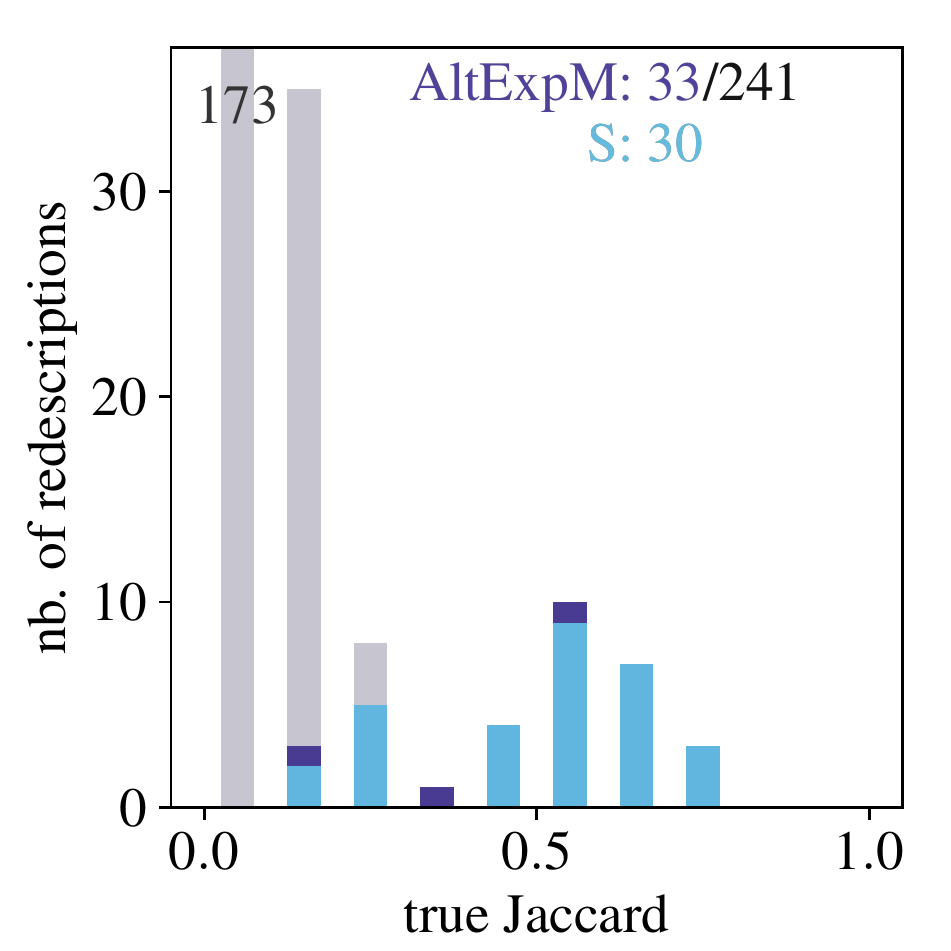} &
    \includegraphics[width=\linewidth]{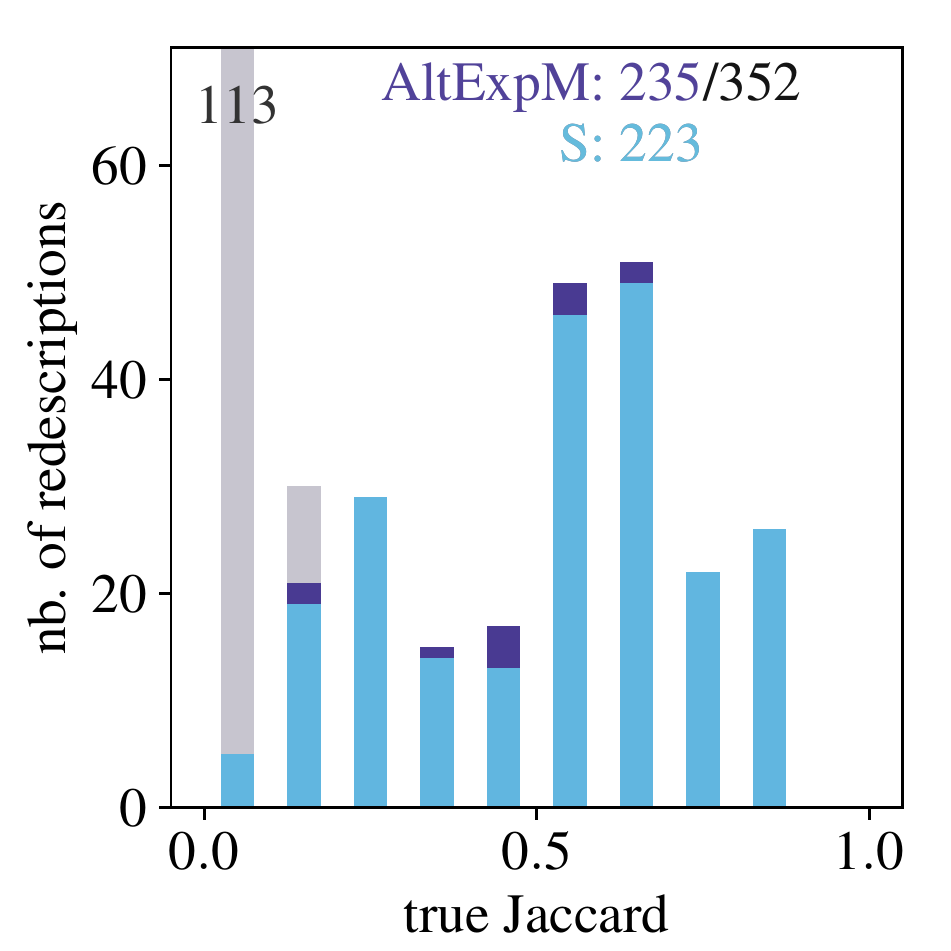} &
    \includegraphics[width=\linewidth]{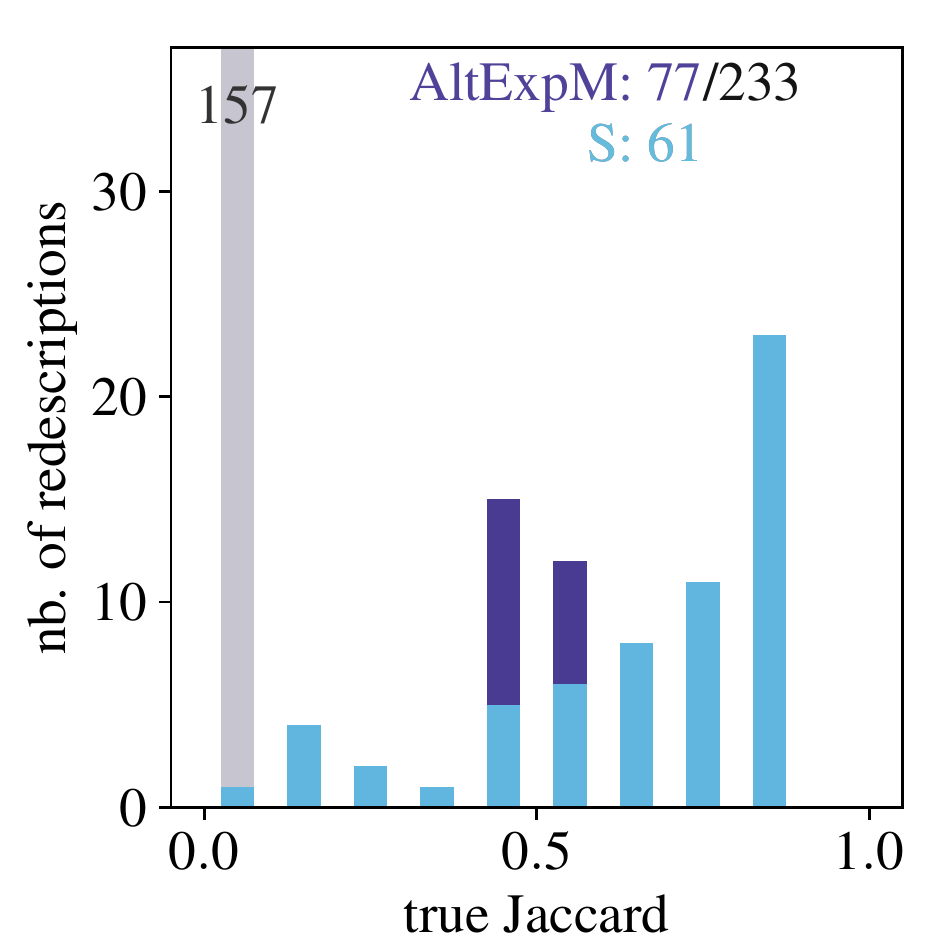} &
    \includegraphics[width=\linewidth]{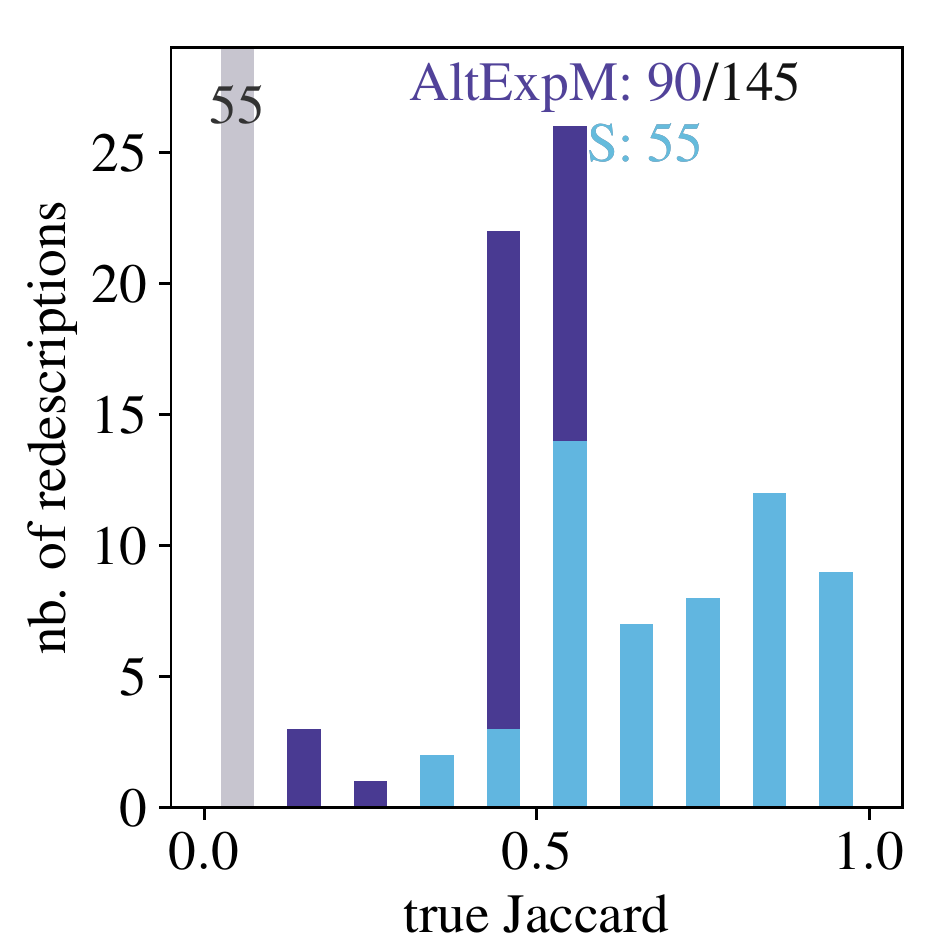} \\
    \includegraphics[width=\linewidth]{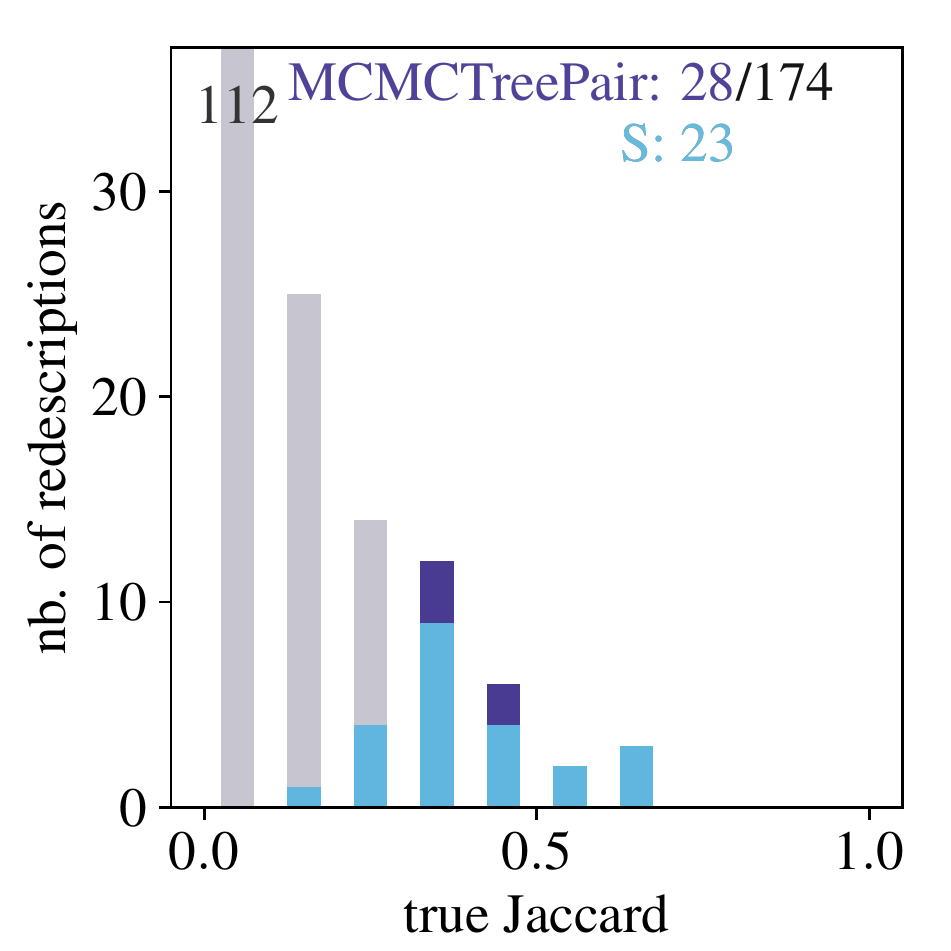} &
    \includegraphics[width=\linewidth]{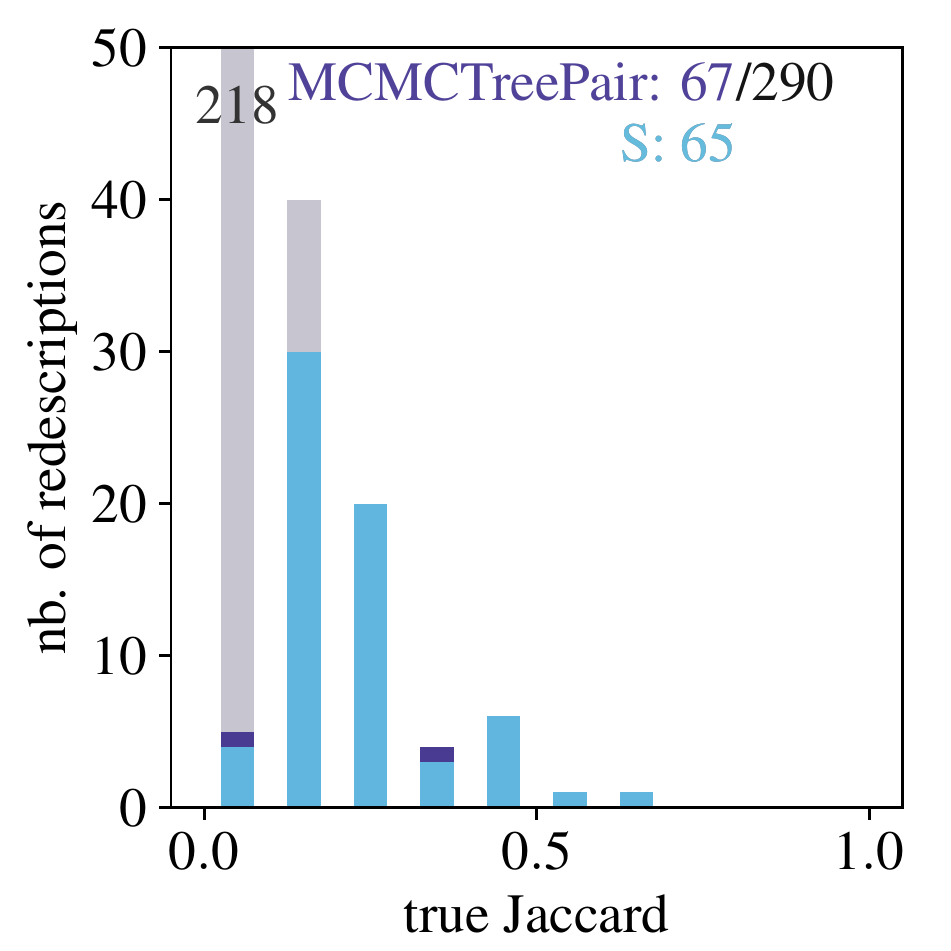} &
    \includegraphics[width=\linewidth]{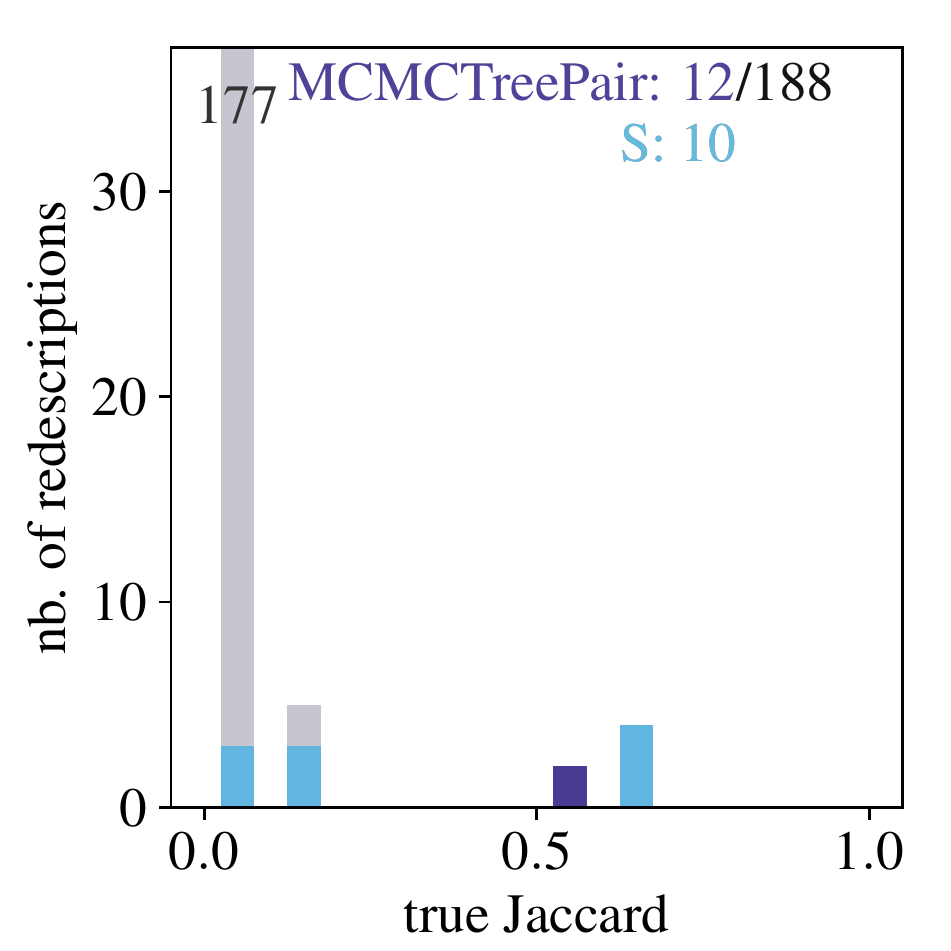} &
    \includegraphics[width=\linewidth]{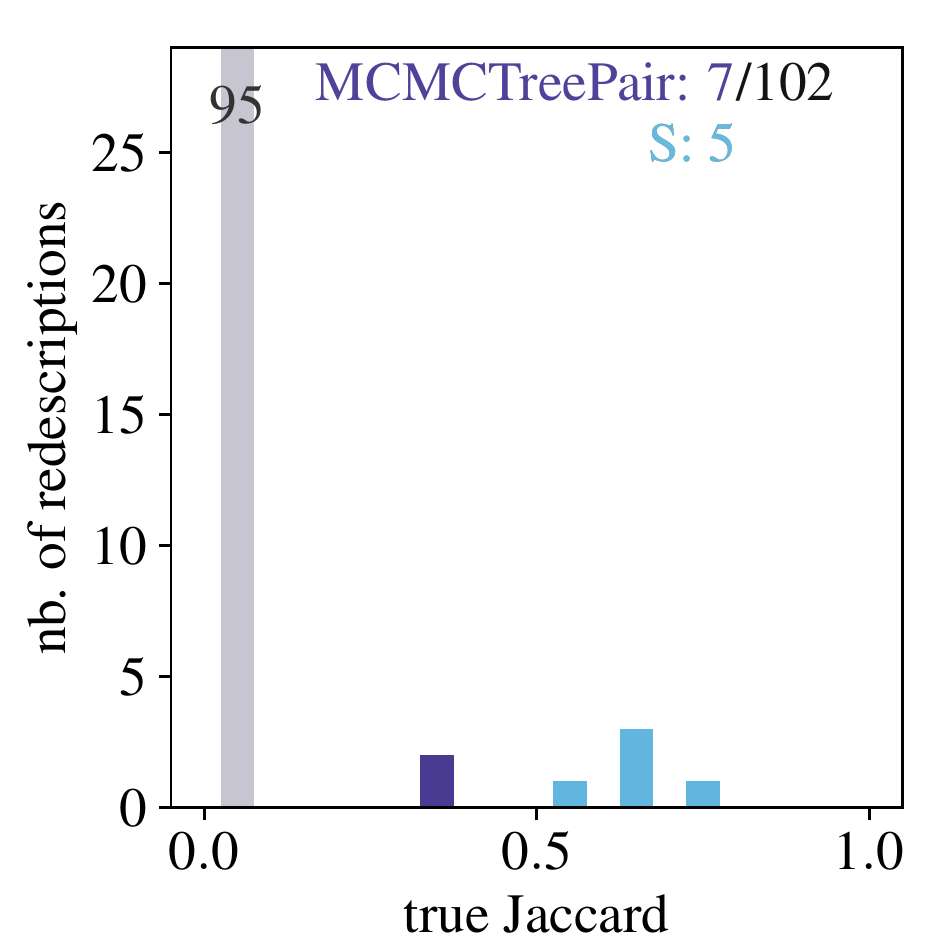} \\
    \includegraphics[width=\linewidth]{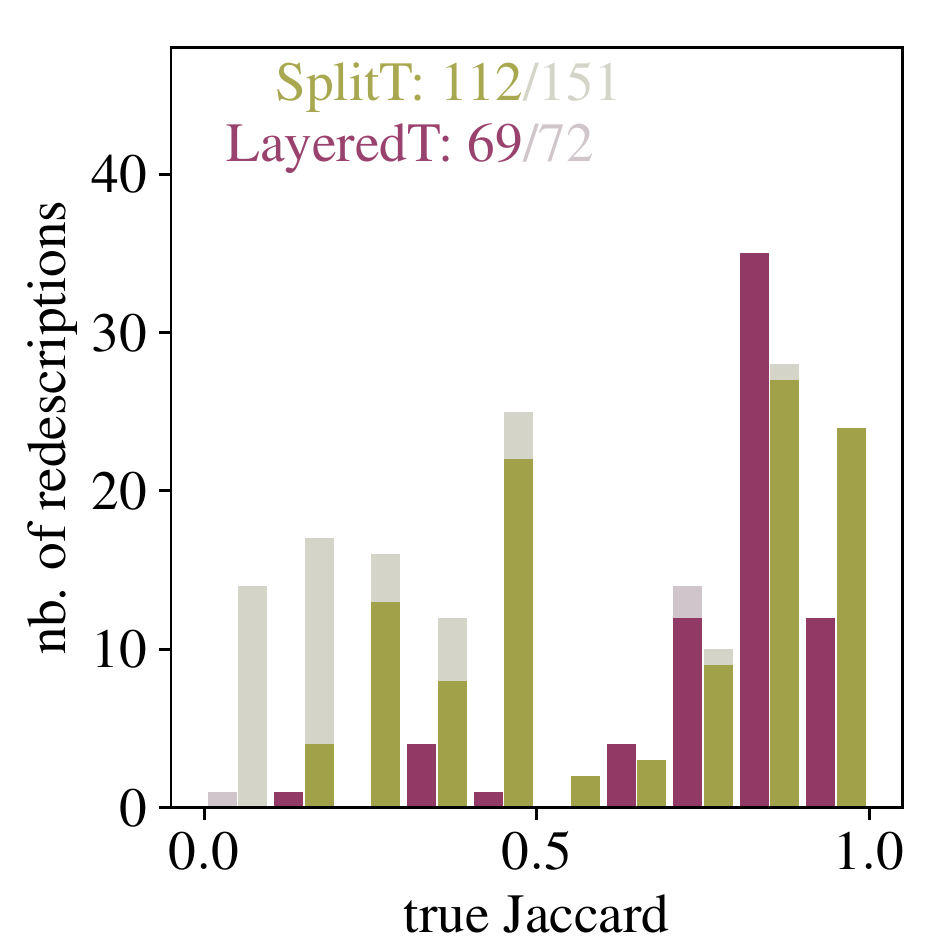} &
    \includegraphics[width=\linewidth]{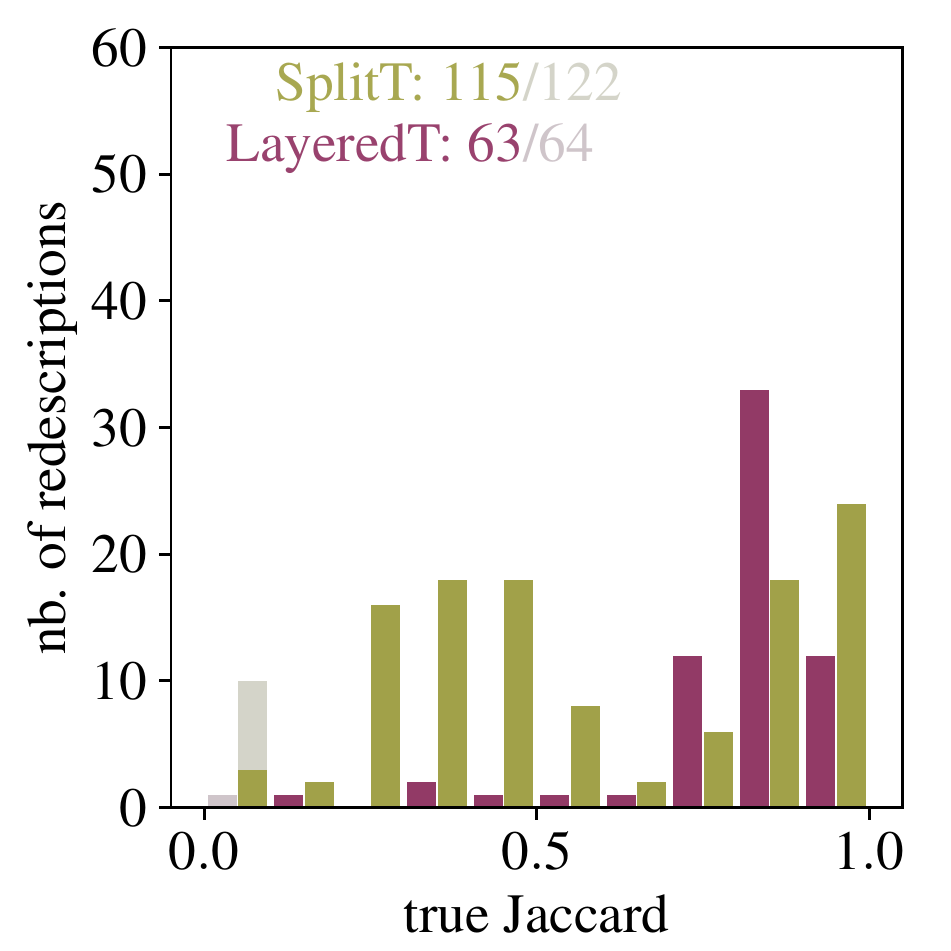} &
    \includegraphics[width=\linewidth]{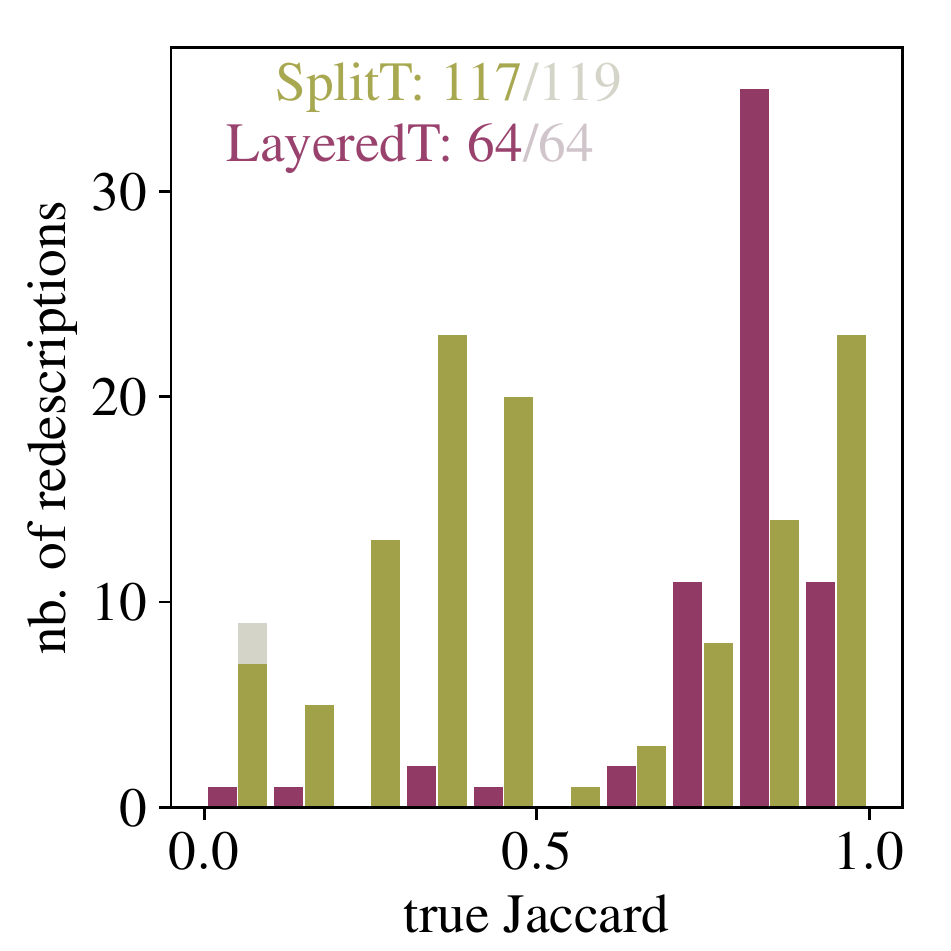} &
    \includegraphics[width=\linewidth]{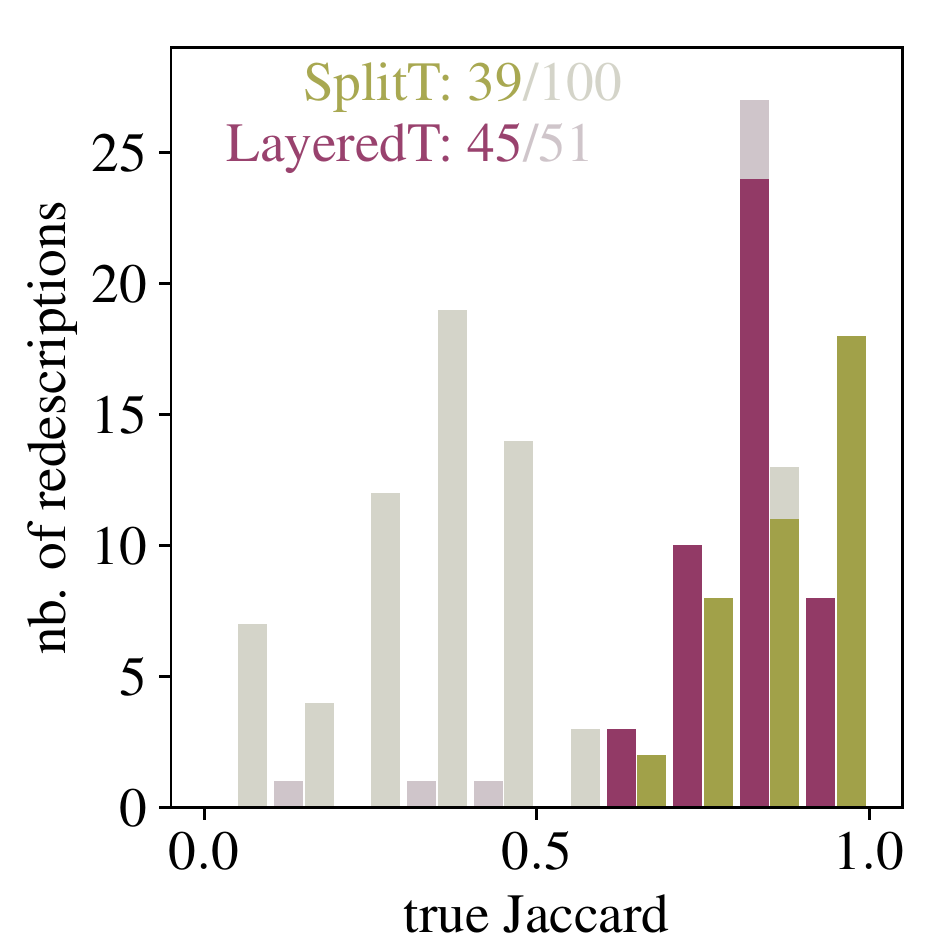}
  \end{tabularx}
  \caption{\textbf{Impact of differential privacy on the quality of redescriptions and comparison to other algorithms} using \medicare data. See caption of Figure \ref{fig:exp:bar} for details.}
  \label{fig:exp:bar1}
\end{figure*}

The comparison of \ownalgo, \expmech, and \sampling to \splittrees and \layeredtrees can be seen by comparing the first three rows of plots of Figures~\ref{fig:exp:bar} and \ref{fig:exp:bar1} to the last row. To make a fair comparison, we use only redescriptions with high enough support (non-filtered) for differentially private algorithms. Filtering for  \splittrees and \layeredtrees was performed by removing all redescriptions with true support smaller than the minimum support redescription retained for the \ownalgo algorithm.

As it can be seen from Figures~\ref{fig:exp:bar} and \ref{fig:exp:bar1}, differentially private algorithms mostly return a smaller number of redescriptions with the highest accuracy (Jaccard index in interval $[0.8,1.0]$) and \ownalgo returns fewer redescriptions overall (as noted earlier). 
One reason why differentially private algorithms might find fewer highly accurate redescriptions, except the obvious reason related to the introduced noise, is the maximum support constraint, that prevents the algorithms from creating redescriptions with (noisy) support larger than \SI{80}{\percent} of the dataset size. This bound is meant to prevent the creation of tautological statements and redescriptions describing almost all entities contained in the dataset. While the same bound is used in non-private algorithms, the added noise can lead the private algorithms to erroneously prune some redescriptions. 

These experiments demonstrate that when the proposed differentially private algorithms report good redescriptions with high quality and high-enough support, the user can be rather confident that she will find good results also from the true data, should she get a permission to study it. 

\subsection{Evaluation of the Properties of the Different Algorithms}

Here we provide a in-depth comparative analysis of the different algorithms. We will see that \sampling and \expmech perform very well and significantly outperform \ownalgo on medium-sized and large datasets when a very large budget is available or on large datasets with smaller budget. The  \ownalgo algorithm has the best performance on small datasets or when small budget is available on small and medium-sized datasets. The performance of  \sampling and \expmech increases with the increase of the data size.  

The \expmech and \sampling algorithms have higher dependence on the initialization procedure than \ownalgo.  When used in such a way that only one initial attribute is selected to start the alternations ($\InitTrials = 1$, the setting giving the largest budget for tree construction), a poor choice of initial attribute might produce an extremely bad or even empty redescription set. Since an analysis of the distribution of the attribute values is required to reduce the risk that this happens, these algorithms require either significantly pre-processed data (rarely available in real-world scenarios) or some more involved strategy to choose the initial attribute. Our experimental results show that such critical failure of \expmech and \sampling mostly happens with a probability less than $0.1$ (for more details, see Appendix \ref{sec:add-exp-res}).

In the absence of a good strategy to choose the initial attribute, one solution is to randomly select a larger number of initial attributes and reduce the number of alternations. Thus, we also test the performance of these two algorithms in a setting at the opposite end of the spectrum, where $\InitTrials = k$ and $\RMIter = 1$, to maximize the chances of choosing one good initial attribute, while keeping the tree-construction budget similar to the previous setting. In this setting, $k$ different initial attributes are chosen and for each one a pair of trees is created from which redescriptions are generated. The probability that algorithms with this parameter setting will choose a good initial candidate is now equal to that of the \ownalgo (thus we call this parameter setting \emph{stable} with respect to the choice of initial attributes). \sampling and \expmech allow different tradeoffs between $\InitTrials$ and $\RMIter$ parameters; further study of these tradeoffs is left for future work.

We test \ownalgo with a budget redistribution weight $\omega = 0.1$ (standard) and $\omega = 0.5$ (\emph{balanced} setting, denoted by ``b'').

\sisetup{detect-weight=true,detect-inline-weight=math}
\begin{table}
  \centering
  \caption{Percentage of significant redescriptions (out of all remaining redescriptions after filtering) produced by each approach on $7$ different datasets. $N$ -- \nerdy , $M$ -- \mammals, $MI$ -- \mimic , $C2\_1$ -- \medicX{2}{1}, $C2\_4$ -- \medicX{2}{4}, $C2\_8$ -- \medicX{2}{8}, and $C2\_16$ -- \medicX{2}{16}. Approaches  with initial budget of $1$ and $0.1$ are run $10$ times. Approaches with initial budget of $0.01$ are run $100$ times. Percentages are computed on the pooled redescription sets. The best results with initial budget of  $1$, $0.1$, and $0.01$ are highlighted. ``S'' denotes the stable parameter setting and ``b'' the balanced redistribution weight.  }
  \label{tab:percSig}
  \begin{tabular}{@{}l
    S[table-format=1.2]
    S[table-format=2.1]
    S[table-format=2.1]
    S[table-format=2.1]
    S[table-format=2.1]
    S[table-format=2.1]
    S[table-format=2.1]
    S[table-format=2.1]
    @{}
    }
    \toprule
    Approach & {Budget} & {$N$} & {$M$} &  {$MI$} & {$C2\_1$} & {$C2\_4$} & {$C2\_8$} & {$C2\_16$}  \\
    \midrule
    \sampling & 1 & \bfseries  13.5 & \bfseries 91.1 & \bfseries 99.9 & \bfseries 94.6 & 95.0 & 75.0 &  63.4 \\
    \expmech & &   10.9 & 87.8 & 98.2 & 92.2 & \bfseries 99.19 & \bfseries 85.7 & \bfseries 64.1 \\
     \sampling S & &   7.8 & 82.8 & 90.1 & 48.5 & 85.6 & 83.8 & 42.0 \\
    \expmech S & &  9.2 & 75.0 & 88.5 & 79.3 & 93.7 &  83.8 & 57.7 \\
    \ownalgo & &  7.4 & 74.3 & 75.6 & 61.8 & 95.1 & 85.1 & 34.4  \\
    \ownalgo b & &  6.5 &  84.1 & 85.9 & 80.6 & 82.5 & 73.3 & 28.0   \\[0.5ex]
    \sampling & 0.1 & 7.5 &  17.8  & 42.2 & 10.2 & 34.9 & 46.5 & \bfseries 64.4  \\
    \expmech & & 6.8 &   22.7 & 43.4 & 17.8 & \bfseries 54.9 & \bfseries 65.2 & 42.8   \\
     \sampling S & & 7.0 & 17.1  & 38.8 & 11.1 & 29.3 & 40.9 & 43.7  \\
    \expmech S & & 8.7 &  25.4 & 35.4 & 17.5 &  48.4 & 49.1 & 57.7   \\
    \ownalgo & & 4.4 & \bfseries 32.7  & \bfseries 49.5 & 20.0 & 54.1 & 42.3 & 42.5  \\
    \ownalgo b & & \bfseries  10.0 & 21.2  & 47.2 & \bfseries 23.3 & 48.8 & 39.3 & 41.1  \\[0.5ex]
     \sampling & 0.01 & 13.5 &  28.0  & 18.6 & 20.7 & 19.4 & 19.5 & 28.9  \\
    \expmech & & 18.3 &  28.3 & \bfseries 20.0 & 22.1 & 24.3 & \bfseries 27.0 & \bfseries 36.0   \\
      \sampling S & & 13.0 & 27.7  & 18.9 & 16.5 & 20.0 & 18.3 & 22.2  \\
    \expmech S & & \bfseries 19.0 &  \bfseries 29.3 & 19.4 & \bfseries 23.8 & \bfseries 24.6 & 24.3 & 28.9   \\
    \ownalgo & & 14.6 & 27.3  & 18.9 & 22.6 & 21.0 & 26.1 & 28.9  \\
    \ownalgo b & & 14.3 & 25.4  & 18.6 & 21.1 & 19.6 & 25.7 & 29.1  \\
    \bottomrule
  \end{tabular}
\end{table}

The statistical significance of differences in pooled distribution between approaches is measured using the Mann--Whitney U test \citep{MWUTest}.
As it can be seen from Table \ref{tab:percSig}, when an initial budget of $1.0$ is used, the \sampling and \expmech algorithms have the largest percentage of significant redescriptions on all datasets. This changes when $10$ runs with budget $0.1$ are used, where \ownalgo has the best results on \nerdy, \mammals, \mimic, and \medicX{2}{1} but \sampling and \expmech produce the largest percentage of significant redescriptions on \medicX{2}{4}, \medicX{2}{8}, and \medicX{2}{16}. When $100$ runs with initial budget of $0.01$ are used, the \expmech algorithm produces the highest percentage of significant redescriptions out of all filtered redescriptions on all datasets.

A better insight into the overall performance of the presented methodologies can be obtained from Table~\ref{tab:statSig}. The methodology with the highest median of true accuracy of filtered redescriptions in a pooled set is compared to all other approaches. The pooled sets contain redescriptions produced in all runs by some algorithm with a chosen initial budget. 

These results demonstrate that \ownalgo outperforms all methodologies with all values of initial budget on the \nerdy dataset. Due to the small size of this dataset, the difference is not always significant when a smaller initial budget is used. As data size increases, \sampling and \expmech have significantly better accuracy with a budget of $1.0$, but already with budget of $0.1$, \ownalgo outperforms other approaches on all datasets up to \medicX{2}{8}. On relatively large datasets, namely \medicX{2}{8} and \medicX{2}{16}, \sampling and \expmech significantly outperform \ownalgo if they are not stabilized, whereas \ownalgo outperforms other algorithms with lower initial budgets ($0.1$ or $0.01$) when other approaches are stabilized.

\begin{table}
  \centering
  \caption{Statistical significance of the hypothesis that distributions of \emph{Jaccard indices} are equal between the method with the largest distribution median (denoted Best) and other approaches (denoted Comp$_i$). One-sided Mann--Whitney U test. ``Stable'' indicates if \sampling ($S$) and \expmech ($EM$) with \emph{stable} parameter setting has been used. \ownalgo is denoted by $M$ and $Mb$ denotes the balanced variation thereof.}
  \label{tab:statSig}
  \begin{tabular}{
    @{}
    l
    S[table-format=1.2]
    c
    c
    R
    @{:}
    S[table-format=<1.4, round-mode=places, round-precision = 4, round-minimum=0.0001, exponent-mode=fixed, fixed-exponent=0]
    R
    @{:}
    S[table-format=<1.4, round-mode=places, round-precision = 4, round-minimum=0.0001, exponent-mode=fixed, fixed-exponent=0]
    R
    @{:}
    S[table-format=<1.4, round-mode=places, round-precision = 4, round-minimum=0.0001, exponent-mode=fixed, fixed-exponent=0]
    @{}
    }
    \toprule
    Dataset & {$\varepsilon$} & Stable &  Best & \multicolumn{2}{c}{Comp$_1$} & \multicolumn{2}{c}{Comp$_2$} & \multicolumn{2}{c}{Comp$_3$}\\
    \midrule
    \nerdy & 1.0 & No & $M$ & S & 0.09 & EM & 0.06 & Mb & 0.21 \\
    & 1.0 & Yes & $M$ & S & 0.64 & EM & 0.05  & Mb & 0.21\\
    & 0.1 & No & $Mb$ & S & 0.03 & EM & 0.23  & M & 0.17\\
    & 0.1 & Yes & $Mb$ & S & 0.31 & EM & 2.9e-6  & M & 0.17\\
    & 0.01 & No & $M$ & S & 2.8e-3 & EM & 0.09  & Mb & 0.02\\
    & 0.01 & Yes & $M$ & S & 0.07 & EM & 0.01  & Mb & 0.02\\[.5em]
    \mammals & 1.0 & No & $S$ & EM & 5.9e-8 & M & 3.9e-9 & Mb & 4.4e-8 \\
    & 1.0 & Yes & $S$ & EM & 3.3e-4 & M & 1.5e-4 & Mb & 9.4e-3\\
    & 0.1 & No & $M$ & S & 5.8e-3 & EM & 0.05 & Mb & 0.69\\
    & 0.1 & Yes & $M$ & S & 4.1e-4 & EM & 3.6e-5 & Mb & 0.69\\
    & 0.01 & No & $Mb$ & S & 0.78 & EM & 1.5e-3 & Mb & 0.31\\
    & 0.01 & Yes & $Mb$ & S & 0.38 & EM & 9.1e-7 & Mb & 0.31\\[.5em]
    \mimic & 1.0 &  No & $S$ & EM & 9.8e-9 & M & <2.2e-16 & Mb & <2.2e-16  \\
    & 1.0 &  Yes & $S$ & EM & 8.3e-6 & M & <2.2e-16 & Mb & 4.9e-8\\
    & 0.1 &  No & $Mb$& S & 0.07 & EM & 1.6e-6 & M & 0.02\\
    & 0.1 &  Yes & $Mb$& S & 8.2e-14 & EM & 6.0e-7 & M & 0.02\\
    & 0.01 &  No & $M$& S & <2.2e-16 & EM & <2.2e-16 & Mb & 0.7\\
    & 0.01 &  Yes & $M$& S & <2.2e-16 & EM & <2.2e-16 & Mb & 0.7\\[.5em]
    \medicX{2}{1} & 1.0 & No & $S$ & EM & <2.2e-16 & M & <2.2e-16  & Mb & <2.2e-16  \\
    & 1.0 & Yes & $S$ & EM & 1.4e-5 & M & 3.4e-5 & Mb & 0.02\\
    & 0.1 & No & $Mb$ & S & 6.7e-14 & EM & 2.3e-4 & M & 0.11\\
    & 0.1 & Yes & $Mb$ & S & <2.2e-16 & EM & 6.8e-9 & M & 0.11\\
    & 0.01 & No & $Mb$ & S & <2.2e-16 & EM & <2.2e-16 & M & 0.19\\
    & 0.01 & Yes & $Mb$ & S & <2.2e-16 & EM & <2.2e-16 & M & 0.19\\[.5em]
    \medicX{2}{4} & 1.0 & No  & $S$ & EM & 0.39 & M & <2.2e-16 & Mb & <2.2e-16   \\
    & 1.0 & Yes & $S$ & EM & 0.62  & M & <2.2e-16 & Mb & 2.6e-15\\
    & 0.1 & No & $M$ & S & 5.9e-6 & EM & 4.2e-6 & Mb & 0.18\\
    & 0.1 & Yes & $M$ & S & <2.2e-16 & EM & <2.2e-16 & Mb & 0.18\\
    & 0.01 & No & $M$ & S & < 2.2e-16 & EM & 8.8e-11 & Mb & 0.56\\
    & 0.01 & Yes & $M$ & S & <2.2e-16 & EM & <2.2e-16 & Mb & 0.56\\[.5em]
    \medicX{2}{8} & 1.0 &  No & $EM$ & S & 5.4e-11 & M & <2.2e-16 & Mb & <2.2e-16  \\
    & 1.0 & Yes & $S$ & EM & 0.05 & M & 1.7e-7 & Mb & 2.9e-6\\
    & 0.1 & No & $S$ & EM & <2.2e-16 & M & < 2.2e-16 & Mb & < 2.2e-16\\
    & 0.1 & Yes & $Mb$ & S & <2.2e-16 & EM & 5.2e-9 & M & 8.3e-3\\
    & 0.01 & No & $EM$ & S & <2.2e-16 & M & 1.1e-9 & Mb & 0.16\\
    & 0.01 & Yes & $Mb$ & S & <2.2e-16 & EM & 1.6e-3 & M & 5.3e-7\\[.5em]
    \medicX{2}{16} & 1.0 & No  & $S$ & EM & 8.8e-5 & M & <2.2e-16 & Mb & <2.2e-16 \\
    & 1.0 & Yes & $EM$ & S & 0.02 & M & 2.4e-8 & Mb & 1.7e-5\\
    & 0.1 & No & $S$ & EM & <2.2e-16 & M & <2.2e-16 & Mb & <2.2e-16\\
    & 0.1 & Yes & $EM$ & S & 1.2e-3 & M & 2.0e-3 & Mb & 2.0e-3\\
    & 0.01 & No & $EM$ & S & 6.7e-8 & M & 5.3e-13 & Mb & 2.1e-7\\
    & 0.01 & Yes & $Mb$ &S& < 2.2e-16 & EM & 0.35 & M & 0.02\\
    \bottomrule
  \end{tabular}
\end{table}

Results presented in Table \ref{tab:statSig}  show that every presented approach has its merits and can be effectively used under different circumstances. The \sampling and \expmech algorithms are best used with high initial budgets and on medium-sized and large datasets, whereas the \ownalgo algorithm should be used on smaller datasets and with smaller initial budget. The \emph{stable} parameter setting often significantly reduces the accuracy of redescriptions produced by the \sampling and \expmech algorithms, but it averts the undesirable outcome where no redescriptions are produced.

\begin{table}
  \centering
  \caption{Statistical significance of the hypothesis that distributions of \emph{absolute value of difference} between noisy and real Jaccard indices are equal between the method with the smallest distribution median (denoted Best) and other approaches (denoted Comp$_i$). One-sided Mann--Whitney U test. ``Stable'' indicates if \sampling ($S$) and \expmech ($EM$) with \emph{stable} parameter setting has been used. \ownalgo is denoted by $M$ and $Mb$ denotes the balanced variation thereof.}
  \label{tab:statSigSt}
  \begin{tabular}{
    @{}
   l
    S[table-format=1.2]
    c
    C
    R
    @{:}
    S[table-format=<1.4, round-mode=places, round-precision = 4, round-minimum=0.0001, exponent-mode=fixed, fixed-exponent=0]
    R
    @{:}
    S[table-format=<1.4, round-mode=places, round-precision = 4, round-minimum=0.0001, exponent-mode=fixed, fixed-exponent=0]
    R
    @{:}
    S[table-format=<1.4, round-mode=places, round-precision = 4, round-minimum=0.0001, exponent-mode=fixed, fixed-exponent=0]
    @{}
    }
    \toprule
    Dataset & {$\varepsilon$} & Stable &  \text{Best} & \multicolumn{2}{c}{Comp$_1$} & \multicolumn{2}{c}{Comp$_2$} & \multicolumn{2}{c}{Comp$_3$}\\
    \midrule
    \nerdy & 1.0 & No & M & S & 0.16 & EM & 0.08  & Mb & 0.013 \\
    & 1.0 & Yes & M & S & 0.25 & EM & 0.013  & Mb & 0.013\\
    & 0.1 & No & Mb & S &  <2.2e-16 & EM &  0.37  & M & 0.42\\
    & 0.1 & Yes & S & EM &  0.01 & M &  0.03  & Mb & 0.58\\
    & 0.01 & No & M & S &  0.07 & EM &  0.01  & Mb & 0.02\\
    & 0.01 & Yes & S & EM &  0.49 & M &  0.44  & Mb & 0.19\\[.5em]
    \mammals & 1.0 & No & M & S & 0.003 & EM & 0.04 & Mb & 0.16 \\
    & 1.0 & Yes & M & S & 2.2e-6 & EM & 4.0e-5 & Mb & 0.16\\
    & 0.1 & No & Mb & S & 0.15 & EM & 0.007 & M & 0.29\\
    & 0.1 & Yes & Mb & S & 6.6e-5 & EM & 0.002 & M & 0.29\\
    & 0.01 & No & M & S & 0.04 & EM & 8.6e-7 & Mb & 0.18\\
    & 0.01 & Yes & M & S & 0.1 & EM & 1.7e-6 & Mb & 0.18\\[.5em]
    \mimic & 1.0 &  No & Mb & S & < 2.2e-16 & EM & 2.1e-6 & Mb & <2.2e-16  \\
    & 1.0 &  Yes & Mb & S & <2.2e-16 & EM & <2.2e-16 & Mb & <2.2e-16\\
    & 0.1 &  No & M& S & 0.63 & EM & 8.4e-5 & Mb & 0.16\\
    & 0.1 &  Yes & M& S & 9.8e-11 & EM & 1.9e-7 & Mb & 0.16\\
    & 0.01 &  No & M& S & < 2.2e-16 & EM & < 2.2e-16 & Mb & 0.1\\
    & 0.01 &  Yes & M& S & 0.005 & EM & 1.7e-14 & Mb & 0.1\\[.5em]
    \medicX{2}{1} & 1.0 & No & S & EM & <2.2e-16 & M & 6.7e-6  & Mb & 0.0004  \\
    & 1.0 & Yes & Mb & S & 3.3e-10 & EM & <2.2e-16 & M & 0.24\\
    & 0.1 & No & Mb & S & 1.0e-15 & EM & 0.0002 & M & 0.01\\
    & 0.1 & Yes & Mb & S & <2.2e-16 & EM & 1.1e-10 & M & 0.01\\
    & 0.01 & No & M & S & 2.8e-8 & EM & 3.5e-15 & Mb & 0.24\\
    & 0.01 & Yes & M & S & 7.6e-7 & EM & <2.2e-16 & Mb & 0.24\\[.5em]
    \medicX{2}{4} & 1.0 & No  & S & EM & 0.71 & M & 3.3e-7 & Mb & 0.0009   \\
    & 1.0 & Yes & M & S &  2.4e-16  & EM & 6.1e-16 & Mb & 0.13\\
    & 0.1 & No & M & S & 1.7e-8 & EM & 0.0001 & Mb & 0.001\\
    & 0.1 & Yes & M & S & <2.2e-16 & EM & <2.2e-16 & Mb & 0.001\\
    & 0.01 & No & M & S & <2.2e-16 & EM & 0.36 & Mb & 8.9e-6\\
    & 0.01 & Yes & M & S & <2.2e-16 & EM & 3.0e-5 & Mb & 8.9e-6\\[.5em] 
    \medicX{2}{8} & 1.0 &  No & EM & S & 0.08 & M & 4.6e-9 & Mb & 1.4e-14  \\
    & 1.0 & Yes & M & S & 0.02 & EM & 1.0e-4 & Mb & 0.19\\
    & 0.1 & No & S & EM & <2.2e-16 & M & < 2.2e-16 & Mb & 2.4e-15\\
    & 0.1 & Yes & Mb & S & <2.2e-16 & EM & 1.1e-6 & M & 0.002 \\
    & 0.01 & No & EM & S & <2.2e-16 & M & <2.2e-16 & Mb & <2.2e-16\\
    & 0.01 & Yes & EM & S & <2.2e-16 & M & <2.2e-16 & Mb & <2.2e-16\\[.5em]
    \medicX{2}{16} & 1.0 & No  & EM & S & 0.0003  & M & 1.2e-8 & Mb & 3.1e-5 \\
    & 1.0 & Yes & EM & S & 0.31 & M & 0.007 & Mb & 0.17\\
    & 0.1 & No & S & EM & < 2.2e-16 & M & < 2.2e-16 & Mb & < 2.2e-16\\
    & 0.1 & Yes & EM & S & 3.3e-16 & M & < 2.2e-16 & Mb & < 2.2e-16\\
    & 0.01 & No & EM & S & < 2.2e-16 & M &< 2.2e-16 & Mb & < 2.2e-16\\
    & 0.01 & Yes & EM & S &< 2.2e-16 & M & < 2.2e-16 & Mb & 2.8e-5\\
    \bottomrule
  \end{tabular}
\end{table}

Table~\ref{tab:statSigSt} shows the statistical significance of the absolute difference between noisy and true redescription accuracy. It demonstrates that \ownalgo has the lowest difference in the majority of cases on the \nerdy, \mammals, \mimic, \medicX{2}{1}, and \medicX{2}{4} datasets, whereas \sampling as well as, in most cases, \expmech have the lowest difference on the larger \medicX{2}{8} and \medicX{2}{16} datasets.

The distributions of redescription accuracy and absolute difference between noisy and real redescription accuracy corresponding to the results presented in Tables \ref{tab:statSig} and \ref{tab:statSigSt} are provided in Appendix~\ref{sec:add-exp-res}.

\subsection{Scalability}
\label{sect:exp:scalability}

\begin{figure*}[tbp]
  \centering
  \includegraphics[width=\textwidth]{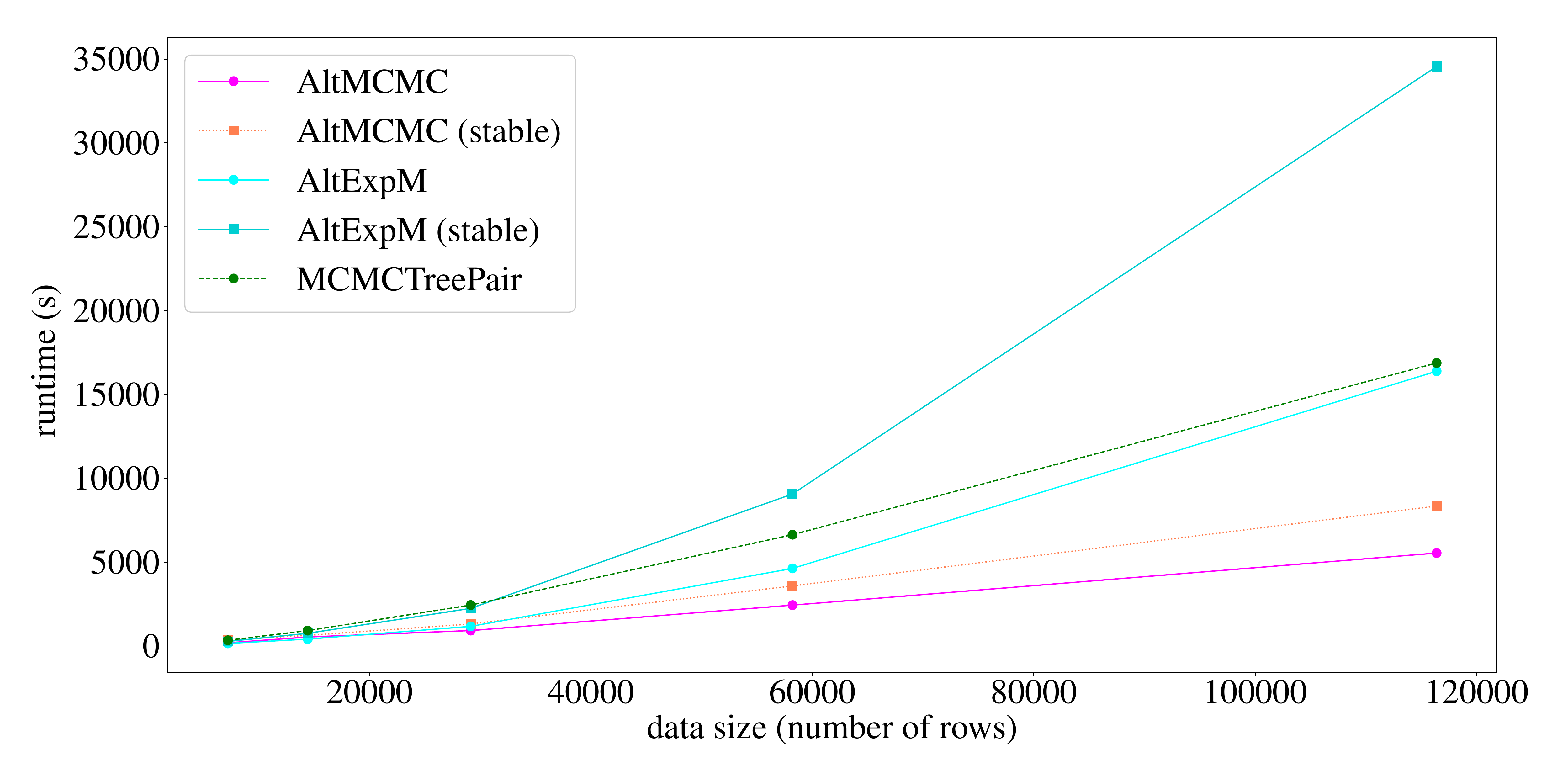}  
  \caption{\textbf{Runtime vs. data size}. Average runtime in total CPU seconds across $10$ runs.}
  \label{fig:exTimes}
\end{figure*}

To test scalability, we ran all proposed differentially private algorithms on the \medicX{2}{k} datasets for $k=\{1,2,4,8,16\}$. We run each algorithm $10$ times on each dataset. \sampling and \expmech were run with both tested parameter settings ($\InitTrials = 1$, $\RMIter = k$ and $\InitTrials = k$, $\RMIter = 1$). 
We report the total CPU time in seconds required to perform redescription mining in Figure~\ref{fig:exTimes}. The time required to load the input data is excluded to eliminate the effects of I/O and caching.

Algorithms scaled relatively well with data size (even single-threaded) and never used more than \SI{25}{\giga\byte} of main memory. Our experiments show that stabilization increases the execution times of the algorithms. This is the consequence of the fact that a larger number of trees need to be built and that the initial binning of an attribute needs to be repeated multiple times. \sampling has the lowest execution times, whereas the stabilized version of \expmech has the largest execution times. Variation in execution times increases for all approaches with the increase of data size. The increase is expectedly much larger for approaches based on sampling, since the involved random choices of attributes can drastically change the structure of the tree or  tree-pair being build, potentially leading to rebuilding larger parts of these trees (see Appendix \ref{sec:add-exp-res} for more details). 

As it can be seen from Figure~\ref{fig:exTimes}, all algorithms finish in less than $10$ hours on the largest dataset containing  \num{115200} entities.

\subsection{Discussion}
\label{sec:discussion}

The performed experimental evaluation shows that the presented differentially private redescription mining algorithms discover useful redescriptions on datasets of various sizes. Due to introduced noise, all presented differentially private approaches produce a number of redescriptions with very inaccurate estimates of the Jaccard index and \pvalue. Yet, as demonstrated, these can be detected and removed with a simple filter on the noisy support size. 

Differentially private redescription mining algorithms discover comparable number of redescriptions to the non-differentially private approaches on medium-sized and large datasets, whereas this number is smaller on datasets of smaller size. The overall accuracy of differentially private redescription mining approaches is also reduced compared to non-differentially private approaches due to the noise required to ensure privacy of individual records. 

The \expmech and \sampling differentially private redescription mining algorithms perform very well on datasets of medium and large size if: a) a maximal budget is used to perform the experiment (thus only one experiment overall can be performed) and  b) the data is carefully filtered and pre-processed to ensure successful initialization. With suitably adjusted parameters, requirement b) can be alleviated or completely removed at the cost of a significant drop in the overall accuracy. With the adjusted parameters, \expmech and \sampling perform very similarly to the \ownalgo algorithm. 

If the data size is relatively small, or it is required to allow several experiments to be performed on the data, the experimental results show increasing benefits of using the \ownalgo algorithm.

\section{Conclusions}
\label{sec:conclusions}

Differential privacy offers strong guarantees to protect the privacy of the data of every individual covered by a database. Unfortunately, these guarantees come at the cost of limited availability of information about the data set and reduced accuracy of the obtained results.

Despite this, such systems allow performing exploratory data analysis on privacy-sensitive data with the aim of generating new hypotheses. Preliminary results can be obtained quickly, safely, and cost-effectively, which can inform the decision to request a full access to the data (involving privacy agreements, potential data-use fees etc.).

Pattern-based data mining methods, such as redescription mining, are especially sensitive to the noise required by differential privacy. Designing effective, differentially private pattern mining algorithms is even harder than, say, classification methods. Nonetheless, our experiments show that the proposed differentially private redescription mining algorithms work: they can produce high quality redescriptions many of which are statistically significant. %

The proposed methods cannot be used on arbitrary data without care, however. The input data might require pre-processing (by the data owner), algorithmic parameters need to be set, and, most importantly, noisy support filtering must be performed to achieve reasonably good performance. 

Approaches based on the MCMC sampling of decision trees and the exponential mechanism (\sampling and \expmech) have the potential of returning very accurate sets of redescriptions with over \SI{90}{\percent} of the reported redescriptions (after noisy support filtering) being in fact significant. However, they are highly dependent on the initialization strategy or data preprocessing (which entirely depends on the data provider). Stabilizing these approaches in such a way that more initial attributes are chosen has a negative impact on the accuracy and significance of the produced redescriptions, as well as on the absolute difference between their noisy and true accuracies

The \ownalgo algorithm produces overall fewer redescriptions and these have lower accuracy compared to the results of \expmech and \sampling when a budget of $1.0$ is used. On the other hand, it is more stable, makes smaller errors in estimating true redescription accuracy, and is especially suited for use on smaller datasets or with smaller budgets, both common constraints in many real-world applications.

Tree-based algorithms are only one family of redescription mining algorithms. An interesting topic for future research is to study whether other types of algorithms could work equally well -- or even better -- than the proposed tree-based differentially private redescription mining algorithms.

\subsection*{Acknowledgements}

Part of this work was done while the first author was with University of Eastern Finland. The authors would like to thank Esther Galbrun for her help with experiments involving the \texttt{python-clired} package and for her helpful comments about the manuscript. We would also like to thank Pekka Hänninen for developing scripts to transform the data to the \texttt{.arff} format.

\appendix
\section{Correctness of Transition Probabilities}
\label{sect:appendix:correctness}

Given a set of objects $\mathcal{X}$ and a scoring function $q\colon(\mathcal{X}^n\times \mathbb{R})\to \mathbb{R}$, the main idea behind the simulation of the exponential mechanism based on MCMC 
sampling is to let the probability of a transition from object $O_1$ to object $O_2$ be $\exp\big(\frac{\pbdg\cdot q(O_2)}{2\cdot \Delta q}\big) \big/ \exp\big(\frac{\pbdg\cdot q(O_1)}{2\cdot \Delta q}\big)$, where $\Delta q$ represents the sensitivity of the quality function $q$. Then, upon convergence, the object $O$ is chosen with probability
\[\frac{\exp\big(\frac{\pbdg\cdot q(O)}{2\cdot \Delta q}\big)}{\sum_{O'\in \mathcal{O}} \exp\big(\frac{\pbdg\cdot q(O')}{2\cdot \Delta q}\big)}\;,\] which corresponds to the probabilities assigned by the exponential mechanism. 

The proposed algorithm utilizes Metropolis--Hastings algorithm \citep{Metropolis1953, Hastings} using the symmetric proposal distribution. This can be seen from the construction of the transitions. First, the algorithm chooses a node $n$ uniformly at random among the nodes in the pair of trees, i.e.\ from $N_i(T_L)\cup N_i(T_R)$. Then, depending on the tree from which the node was chosen, the new node $n'$ is chosen uniformly at random among all possible attribute--split choices from the corresponding view, i.e.\ from $\mathcal{S}(\Table_{\aview(n)})$. Thus, the overall probability is $(\abs{N_i(T_L)\cup N_i(T_R)} \cdot \abs*{\mathcal{S}(\Table_{\aview(n)})})^{-1}$. The original pair can be obtained with the same probability starting from the newly constructed tree pair. 

It follows from the construction of initial pairs of trees and positive transition probabilities that every possible pair of trees of a predefined depth $d$ can be reached with a positive probability. The algorithm accepts the MCMC transition from a pair of trees $(T_{L},\ T_{R})$ to a pair $(T'_{L},\ T'_{R})$ with probability $\min\{1,  z/z'\}$, where $z = \exp\big(\frac{\varepsilon\cdot \score(T_{L},\ T_{R})}{2\cdot \Delta \score}\big)$ and $z' = \exp\big(\frac{\varepsilon \cdot \score(T'_{L},\ T'_{R})}{2\cdot \Delta \score}\big)$. This follows the idea described above, thus upon convergence, the tree-pair $(T_L, T_R)$ is chosen with probability
\[
\frac{\exp(\pbdg\cdot \score(T_L,\ T_R)/(2\cdot \Delta \score))}{\sum_{(T'_L, T'_R)\in \mathit{AllTreePairs}_d} \exp(\pbdg \cdot \score(T'_L, T'_R)/(2\cdot \Delta \score))}\;.
\]

\section{Proof of Lemma~\ref{lemma:score_sensitivity}}
\label{sect:appendix:sensitivity}

First, note that the sensitivity of counting queries is $1$ \citep[see][]{DworkFoundations}.
Recall from Equation \ref{eq:quality} that, given a decision tree $T$ and a target variable $\classvar$, the quality of the subtree rooted in node $n$ is defined as
\begin{equation}
  \label{eq:quality-app}
  \quality(n) = 
     \begin{cases}
       \sum_{m \in C(n)} \quality(m)  & \text{if }n \in N_i(T)\;,\\
       \frac{\tau_n}{\tau} \sum_{c \in \classvar} (\frac{\tau_{n,c}}{\tau_n})^2 & \text{if }n\in N_l(T)\;,
     \end{cases}
   \end{equation}
   where $C(n)$ denotes the set of children of $n$, while $\tau$, $\tau_n$, and $\tau_{n,c}$ denote respectively the total number of entities in the dataset, the number of entities in $n$, and the number of entities of class $c$ in $n$.

Since $\tau_{n,c} \geq 0$ for all $n$ and $c$, we have that
$\sum_{c \in \classvar} \tau_{n,c}^2 \leq \big(\sum_{c \in \classvar} \tau_{n,c}\big)^2$.
Furthermore, since $\sum_{c \in \classvar} \tau_{n,c} = \tau_n$, we have that
\[
\sum_{c \in \classvar} \Bigl(\frac{\tau_{n,c}}{\tau_n}\Bigr)^2 \leq \frac{\bigl(\sum_{c \in \classvar} \tau_{n,c}\bigr)^2}{\tau_{n}^2} = 1,\quad \text{that is}\quad \sum_{c \in \classvar} \Bigl(\frac{\tau_{n,c}}{\tau_n}\Bigr)^2 \in [0,1].
\]

Since $0 \leq \tau_n/\tau \leq 1$ and $\sum_{n \in N_l(T)} \tau_{n}/\tau = 1$, the sum of scores of leaves is a convex sum, and hence takes value between $0$ and $1$, so that $\quality(n) \in [0,1]$ and
$\quality(\treeroot(T)) = \sum_{n \in N_l(T)} \quality(n) \in [0,1]$, in particular.

Therefore, the measure used to evaluate a pair of trees, defined in \eqref{eq:score} as 
\[
\score(T_L,T_R) = \frac{\quality(\treeroot(T_L))\cdot \big(1+ \quality(\treeroot(T_R))\big)}{2}\;,
\]
clearly also takes value in the unit interval. Thus, the sensitivity of $\score(T_L,\ T_R)$ equals $1$.

\section{Negated Queries in the Presence of Missing Values}
\label{sect:appendix:missing}

\begin{figure*}[tbp]
  \centering
    \includegraphics[width=.9\linewidth]{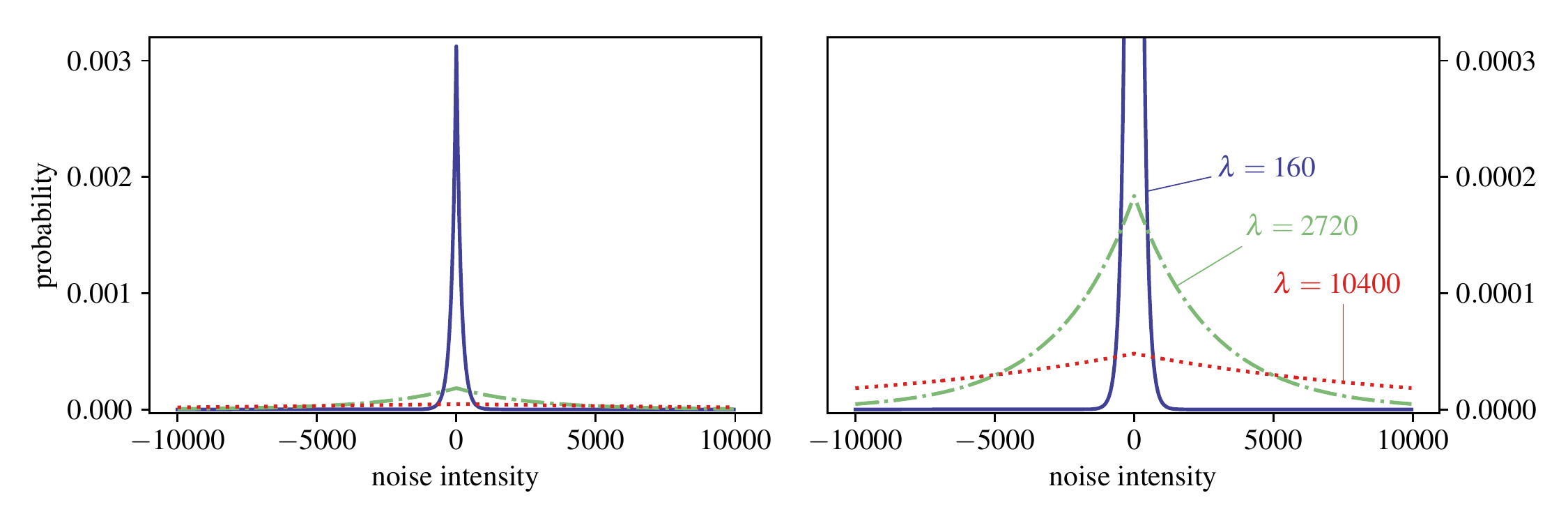}
  \caption{\textbf{Laplace noise distributions.} The expected Laplace noise distribution used to compute base query support sizes when $40$ iterations are used and trees of maximal depth $4$ or $6$ are built. Left: Illustration with full $y$-range. Right: Zoomed-in view. The $\lambda$ parameter, that determines the standard deviation $\sqrt{2}\cdot \lambda$ of this distribution, equals $4\cdot 40 = 160$ to compute queries when negations are computed heuristically (in the presence of missing values), $4\cdot 40 \cdot 17 = 2720$ for queries (if equal budget distribution is used) when they are computed exactly using trees of maximal depth $4$ and $4\cdot 40 \cdot 65 = 10400$ when they are computed exactly using trees of maximal depth $6$.}
  \label{fig:lapDist}
\end{figure*} 

When using heuristic noisy counts, $\abs{\supp(\neg\,\query_m)} \approx \sum_{l \in I_L\setminus \{m\}}\abs{\supp(n_l)}$, every counted entity truly belongs to the support of a query, however the counting process potentially leaves out some entities. Since the negations of leaf nodes can be highly overlapping, computing these counts exactly requires increasing the budget by a factor of $2^d$, which results in prohibitive additional noise. This is illustrated in Figure~\ref{fig:lapDist}. Using the proposed heuristic, setting the Laplace parameter $\lambda = \num{160}$ (regardless of the tree depth) results in a distribution where the noise is tightly concentrated around the origin. Using exact counting with tree depth $4$, we have to set the Laplace parameter to $\lambda = \num{2720}$, resulting in much less concentrated distribution causing ultimately significantly larger noise. Using exact counting with tree depth $d=6$ and letting the Laplace parameter $\lambda = \num{10400}$ results in extremely wide distribution, causing prohibitive amount of noise.

It is important to notice that the support size of the negated query depends on the sum of support sizes of all other leaves of the three (which have added noise with a deviation of $\sqrt{2}\cdot \lambda$). Thus, the noise accumulates as 
$\sigma_{total} = \sum_{l \in I_L\setminus \{m\}} \sigma_l$ and we can assume the very conservative estimate $\sigma_{total}\leq (2^{d}-1)\cdot \sqrt{2}\cdot \lambda$. The Laplace distribution used to produce noise is centred around $0$ (which is the expected value) thus we can assume more realistically that the expected introduced noise will be closer to $\sigma_{total}\leq ((2^{d}-1)\cdot \sqrt{2}\cdot \lambda)/2$.

Given $\lambda = 160$ (as above when negations are computed heuristically), for $d = 4$ we have $\sigma_{total} \leq 1810$ and for $d = 6$ it follows $\sigma_{total} \leq 7241$. Our simulation experiments show that out of $1 000 000$ samples of size $2^4-1 = 15$ generated from the \texttt{Laplace}($0$,$160$), only $40741$ (\SI{4.1}{\percent}) had the absolute value of a sum larger than $1810$ and only $65$ had a sum larger than $\sqrt{2}\cdot 2720$. The same experiment for samples of size $2^6-1 = 63$ showed that only $89$ samples ($0.009$\%) had the absolute values of a sum larger than $7241$ from which $0$ had the absolute value of a sum larger than $\sqrt{2}\cdot 10400$.

Thus, we conclude that the use of the proposed heuristic solution is justified in differentially private redescription mining for small and medium-sized data and for data with small or medium amount of missing values.

\section{Asymptotic Time Complexity}
\label{sec:time_complexity}

We use $\Table$ to denote the input data, $\attrs(\Table)$ to denote the set of attributes of the input data, $S_p$ the number of splitting points, attribute-value pairs in the data, $\MCIter$ to denote the number of MCMC iterations, and $\MCIter'_{\texttt{algo}}$ to denote the number of MCMC iterations in which the root node of a tree has been changed during execution of the \texttt{algo} algorithm ($\MCIter'_{\texttt{algo}}\geq 0$ and $\MCIter'_{\texttt{algo}}\ll\MCIter$).

The time complexity of the tree construction with exponential mechanism is $O\bigl( \abs{\Table} (\abs{\attrs(\Table)} + \abs{S_p})\bigr)$. The first step is to compute all possible splitting points, which requires iterating over the whole dataset for all attributes. Tree construction, at each level, includes iterating over all splitting points which contain only non-redundant attribute values. For all levels except the root node, the splitting point is evaluated for a subset of data $\Table'$. 
When MCMC sampling is used, the time complexity is $O\bigl(\abs{\Table} (\MCIter'_{\sampling}+1 + \abs{\attrs(\Table)})\bigr)$. The first step in tree construction is to create splitting points. The next step is to create a random tree of depth $d$, which can be done in time $O(\abs{\Table})$. Then $\MCIter$ iterations iteratively change the input tree. Iterating over all entities of the dataset is required in only $\MCIter'$ iterations.

The time complexity of the tree-pair construction in \ownalgo is $O\bigl(\abs{\Table} (\MCIter'_{\ownalgo}+1 + \abs{\attrs(\Table_L\cup \Table_R)}) \bigr)$. Constructing tree-pairs necessitates creating all splitting points for both data tables. First, a random pair of trees of depth $d$ is created (doable in $O(\abs{\Table})$ time); next, MCMC iterations iteratively change the input pair of trees. Iterating over all entities of the dataset is required in only $\MCIter'_{\ownalgo}$ iterations.

Redescription generation has an average time complexity of $O(\abs{\Table})$ and worst-case time complexity of $O(\abs{\Table}^2)$.

Since $\InitTrials$ and  $\RMIter$ represent constants for average time complexity, the overall average time complexity of the presented redescription mining algorithms are equal to the complexity of underlying tree or tree-pair construction: 
\begin{flalign*}
&O\bigl(\abs{\Table}\cdot (\abs{\attrs(\Table)} + \abs{S_p})\bigr) & &\text{for \expmech algorithm,} \\
&O\bigl(\abs{\Table}\cdot (\MCIter'_{\sampling}+1 + \abs{\attrs(\Table)})\bigr) & &\text{for \sampling and} \\
&O\bigl(\abs{\Table}\cdot (\MCIter'_{\ownalgo}+1 + \abs{(\attrs(\Table_L\cup \Table_R)})\bigr) & &\text{for \ownalgo.}
\end{flalign*}

The maximum number of initial trials $\InitTrials$ is at most the number of attributes in the data whereas the $\RMIter$ can be an arbitrary number (although it is typically a constant), thus the worst-case time complexity of the approaches are:
\begin{gather*}
  \begin{multlined}[b]
    O\big(\abs{\attrs(\Table_L)\cup \attrs(\Table_R)}\cdot 
    (\abs{\Table}^2 \cdot \max\{\abs{\attrs(\Table_L)}, \abs{\attrs(\Table_R)}\})\big) \end{multlined}\; ,  \\
  \begin{multlined}[b][\textwidth]
    O\big(\abs{\attrs(\Table_L)\cup \attrs(\Table_R)} 
    \cdot (\abs{\Table}\cdot (\MCIter + \max\{\abs{\attrs(\Table_L)},\abs{\attrs{\Table_R}}\}) + \abs{\Table}^2)\big)\end{multlined}\; ,\\
  \shortintertext{and}
  \begin{multlined}[b][\textwidth]
    O\big(\abs{\attrs(\Table_L)\cup \attrs(\Table_R)} 
    \cdot (\abs{\Table}\cdot (\MCIter + \abs{\attrs(\Table_L) \cup \attrs(\Table_R)}) + \abs{\Table}^2)\big)\end{multlined} \; ,
\end{gather*}
respectively for \expmech, \sampling, and \ownalgo. In practice, the initial number of trials is typically set much smaller than the number of attributes, in order to spend the available budget in a rational manner. 

As it can be seen from these analyses, the \expmech algorithm has the highest worst-case computational complexity. Relating the average time complexity is more difficult, since it depends on the actual data.  However, for large data, it is safe to assume that $\MCIter'_{algo}\ll\abs{S_p}$ and $(\MCIter'_{algo}+ \attrs(\Table) )\ll\abs{S_p}$. The \ownalgo algorithm has worse computational complexity than \sampling since attributes from both data tables need to be used at each trial. The worst case for the \expmech algorithm is when all attributes have distinct values for all entities in the dataset, then creating a root node requires quadratic computation on the number of entities in the dataset for each attribute of the current view. This is not required in MCMC-based approaches, since split points are chosen at random at each MCMC iteration.

\section{Details on Experimental Evaluation and Datasets}
\label{sec:experiment_setup}

\subsection{Dataset Properties and Pre-Processing}
\label{sec:datas-prop-pre}

We prepared the \medicare dataset(s) based on another publicly available dataset, namely the Data Entrepreneurs Synthetic Public Use Files (DE-SynPUF)\footnote{\url{https://www.cms.gov/Research-Statistics-Data-and-Systems/Downloadable-Public-Use-Files/SynPUFs/DE_Syn_PUF}, accessed 9 June 2020} released by the Centers for Medicare and Medicaid Services (CMS). It consists of realistic synthetic records of beneficiaries and claims information for three consecutive years (2008--2010).
An entity in the dataset is a (synthetic) beneficiary.
As the left-hand side table, we collected the personal information (birth year, sex, race, etc.) and diagnoses received by the beneficiaries over the three years. ICD-9 diagnosis codes were mapped to the second level of the categorization scheme from the Clinical Classifications Software.\!\footnote{\url{https://www.hcup-us.ahrq.gov/toolssoftware/ccs/ccs.jsp}, accessed 9 June 2020} The right-hand side table contains the yearly total amounts in different reimbursement, deductible, co-insurance components, relevant to inpatient, outpatient, carrier, and prescription drug event claims.
As a result, one side of the data is mostly Boolean (indicators of different diagnoses) while the other is fully numeric.
The original data is divided into $20$ samples. We used Sample $2$, which we split further into $16$ segments based on the first digit of the hexadecimal identifier assigned to each beneficiary. Each segment contains roughly $7200$ beneficiaries. We can construct datasets of different sizes by merging segments together. We denote as \medicX{2}{k} the datasets containing the first $k$ segments from Sample $2$. For instance,  \medicX{2}{4} contains beneficiaries from Sample $2$ whose identifier starts with `0', `1', `2', or `3'.

\nerdy is a publicly available\footnote{\url{https://openpsychometrics.org/_rawdata/}, accessed 9 June 2020, data from 19 February 2016} collection of answers to an online psychological assessment questionnaire. One side of the data contains background information about the respondents (age, education level, etc.) and the other their answers to the questions on the $5$ point Likert scale. We removed all answers that contained missing values, since the \splittrees and \layeredtrees baseline algorithms cannot handle them. The \nerdy is significantly smaller than \medicare datasets, and it also differs from the others in that it contains categorical variables. We use it in part to evaluate the impact of noise on smaller datasets.

The \mimic dataset\footnote{\url{https://mimic.physionet.org/gettingstarted/overview/}, accessed 18 September 2020} is a prime example of health data where patient privacy is extremely important. It contains de-identified health information about patients who were treated at the Beth Israel Deaconess Medical Center (Boston, MA, US). Using this data, we constructed a dataset containing two views: the diagnoses view and the lab events view. The aim when mining this dataset is to find out what kind of lab events preceded and followed various diagnoses in the studied patients.

The dataset contains $808$ Boolean attributes in the diagnoses view ($1$ indicating positive diagnosis) and $260$ ternary attributes in the lab events view. The ternary attributes distinguish between normal lab test results (value $0$), abnormal lab test results (value $1$) and lab test that were not performed (value $-1$) for a given patient. With $\num{46065}$ entities, it is a medium-sized data.

The \mimic,
\medicare, and \nerdy datasets are examples of datasets that contain potentially sensitive information. Our last dataset, \mammals, is a collection of records about the distribution of land mammal species~\citep{mitchell-jones99atlas} and climate~\citep{hijmans05very} in Europe. This information is hardly sensitive, but we include this dataset as a point of reference, because it has been used in experiments in several redescription mining articles \citep[see, e.g.][]{galbrun12black,galbrun12siren,zinchenko15mining,kalofolias16from,mihelcic17framework,galbrun18mining}.

\subsection{Filtering NPAS Data}
\label{sec:filteringNPAS}
NPAS is a relatively small dataset (just $17$ and $47$ attributes for $1221$ entities) containing attributes of mixed type in the first view and numeric attributes in the second view. 
In order to obtain any useful redescriptions despite the introduced noise of the differential privacy mechanisms, we need to make sure the input data is properly prepared. Since this step can be performed on the side of a data provider, pre-processing and filtering data does not break any privacy constraints and the resulting redescriptions are differentially private. 

As a first filtering step, we removed all attributes containing only one value. Such attributes are useless in every data mining task, but are especially harmful in a scenario where differential privacy needs to be preserved. Using such attributes not only wastes budget but also causes the creation of many redescriptions with empty support, which are neither detected nor removed due to the introduced noise. 

As a second step, we merged some values of categorical attributes. Similarly to useless attributes, categorical values that occur very rarely can also cause the creation of many redescriptions of very poor quality. For a given attribute, we merged categorical values occurring in less than $50$ entities. As a consequence, the new category, $category_{new}$, must be interpreted as $category_{old1}\ \vee\ category_{old2}$.  

Redescriptions with smaller support are more affected by the introduced noise. This is expected, since the level of noise introduced depends on the algorithm parameters and not on the data size. For this reason, it is substantially harder to obtain a good differentially private redescription set on datasets containing fewer than $2000$ entities.

\subsection{Parameters Used in Experiments}
\label{sec:param-used-exper}

The maximum tree depth $d$ affects mostly the complexity of the resulting redescriptions and depends on the application. We used maximum depth of \num{4} for all experiments. This depth allows creating interpretable redescriptions (with queries containing $\leq 4$ attributes) and keeps the level of noise required to obtain negated queries at reasonable levels.
All algorithms use the Gini coefficient as the splitting criterion.

The minimum support of the redescriptions should be set small enough to find all interesting results, and depends somewhat on the data size. In the differentially private setting where information about the data is unavailable, the preferable setting is $\geq 10$ (since this will remove redescriptions with very small or very inaccurate support). If one knows that the used data is large (regardless of the real number of entities) this threshold can be increased to $100$ which will reduce even larger number of redescriptions that are very sensitive to noise.  The noisy counting also means that the Jaccard index computation can be very inaccurate for redescriptions with very small support. Hence, we mined redescriptions using the same relatively low minimum support for all algorithms (the only knowledge we used is if the data is \emph{small} or \emph{medium/large} which does not give out any sensitive information) and then pruned away those redescriptions returned by the \ownalgo that had too small (noisy) support. The minimum support for the redescriptions was \num{10} for \mammals and \nerdy and \num{100} for  \mimic and \medicare. In post-processing, we removed all redescriptions from the \ownalgo that had noisy support smaller than \num{2000} for  \mimic, \num{1000} for \medicare, \num{500} for \mammals and \num{200} for \nerdy. 
Obtaining proper post-processing filtering threshold is explained in Section \ref{sec:redprun}.

All algorithmic parameters used to perform the experiments are listed in Table \ref{tab:parameters}. \sampling and \expmech have additional parameter $\RMIter$. For these two algorithms we perform two experiments, the first uses $\InitTrials = 1$ and $\RMIter$ equal the number of initial trials from Table \ref{tab:parameters} and the second where $\InitTrials$ is as reported in Table \ref{tab:parameters} and $\RMIter = 1$. The second setting ensures equal probability of selecting a satisfactory initial attribute for all approaches.

\begin{table}
  \centering
  \caption{The parameters used to perform the experiments presented in figures of Section~\ref{sec:exper-eval} and Appendix~\ref{sec:add-exp-res} with the exception of experiments in Figures~\ref{fig:RAccB1}  and \ref{fig:RStabB1} which use $\varepsilon = 0.1$.}
  \label{tab:parameters}
  \begin{minipage}{0.9\textwidth}
    \begin{tabular}{
      @{}
    l
    S[table-format=5.0]
    S[table-format=1.3]
    S[table-format=1.0]
    S[table-format=2.0]
    S[table-format=1.1]
      S[table-format=1.1]
      @{}
    }
    \toprule
    Dataset          & {$\MCIter$} & {$\VarTh$} & {$d$} & {$\InitTrials$} & {$\omega$} & {$\varepsilon$} \\
    \midrule
    \medicX{2}{k} & 10000        & 0.005         & 4       & 20                   & 0.1               & 1.0 \\
    \nerdy            & 10000        & 0.005         & 4       & 4                     & 0.1               & 1.0 \\
    \mammals     & 10000         & 0.005        & 4        & 4                     & 0.1               & 1.0 \\
    \mimic           & 10000        & 0.005         & 4       & 20                    & 0.1              & 1.0 \\
    \midrule
  \end{tabular}\newline
  \textit{Redescription constraints}\newline
  \begin{tabular}{
    @{}
    l
    S[table-format=1.2]
    S[table-format=1.1]
    S[table-format=3.0]
    S[table-format=1.1]
    @{}
    }
    \midrule
    Dataset         & {$\Gamma.p_{\mathit{val}}$} & {$\Gamma.\mathit{minJ}$} & {$\Gamma.\mathit{minSupp}$} & {$\Gamma.\mathit{maxSupp}$ (\% of $\abs{\Table}$)} \\
    \midrule
    \medicX{2}{k} & 0.01         & 0.1             & 100           & 0.8 \\
    \nerdy            & 0.01         & 0.1             & 10             & 0.8 \\
    \mammals      & 0.01        & 0.1              & 10             & 0.8 \\
    \mimic           & 0.01         & 0.1              & 100           & 0.8 \\
    \bottomrule
  \end{tabular}
  \end{minipage}
\end{table}

One should use a high number of MCMC iterations ($\MCIter$ parameter) to ensure the obtained tree-pairs appropriately mimic the probability distribution of the exponential mechanism. We used \num{10000} MCMC iterations as a relatively high number for shallow trees (depth $\leq 4$). Determining the exact number of iterations for general trees requires knowledge about the depth of the tree, which depends on data size and experimental evaluation \citep{Xuanyu17}, neither of which is possible in the differentially private setting. The MCMC iteration termination variance threshold $\VarTh$ terminates MCMC iterations if the variance of the MCMC score in the previous $k$ iterations is smaller than this threshold (there is little change in the score, thus we may consider the process to have converged); $k = 500$ is used in all experiments. The $\VarTh$ should be set to a small fraction of a quality score value range. Here, we set it to \SI{0.5}{\percent} of the value range of the tree or tree-pair quality evaluation function.

The $\InitTrials$ and $\RMIter$ parameters are very important since more initial trials and iterations allow creating more redescriptions. Larger $\InitTrials$ allows using a larger number of different initial targets to start growing trees or tree-pairs, and larger $\RMIter$ allows using more alternations inside the \sampling and \expmech algorithms. However, increasing either of these parameters inevitably reduces the budget for individual tree or tree-pair construction. As it turns out, using $4$ iterations with high weight ($1-\omega=0.9$) for noisy counts in \ownalgo leads to fairly accurate redescriptions even on small datasets (such as \mammals and \nerdy). On datasets with a larger number of entities, the $\InitTrials$ parameter can be increased. The weight parameter $\omega$ allows balancing the budget between tree-pair creation and noisy counts. Using very low $\omega$ causes tree-pairs to be of lower quality, which results in a smaller number of produced redescriptions. At the same time, since this means allocating a larger portion of the budget for noisy counts, the remaining redescription statistics are computed more accurately.

The privacy budget $\varepsilon$ is set to $1$ if the algorithm is to be run only once on the data. No more runs would be permitted after that. Using a smaller privacy budget allows multiple exploratory runs on the data.

The impact of further algorithm parameters in $\Gamma$ is very limited on datasets with a small number of entities and for redescriptions with small support set size. Among these parameters, the most useful one is $\Gamma.\mathit{minSupp}$, which is used for redescription filtering (as explained earlier). On the other hand, the $\Gamma.\mathit{maxSupp}$ is typically set to $0.8$, to prevent various potentially uninteresting results such as tautologies and insignificant redescriptions (which are hard to detect internally due to noise). Calculating this requires first calculating the estimated data size. Since noisy counts are relatively accurate on the higher support size spectra, setting the maximum support to \SI{80}{\percent} dataset size should remove the majority of redescriptions with support equal to or very close to all entities in the dataset. The $p$-value threshold ($\Gamma.p_{\mathit{val}}$) is set to $0.01$ by default, but even stricter values might make sense in this scenario. The parameter $\Gamma.\mathit{minJ}$ is most affected by the introduced noise, and setting it to a very high value comes with the risk of eliminating many true, accurate redescriptions.

\section{Additional Experimental Results}
\label{sec:add-exp-res}

\subsection{Convergence Properties of MCMC Iterations Inside \ownalgo}
The MCMC iterations inside the \ownalgo algorithm are repeated until they reach a predefined threshold ($\num{100000}$ in this experiment, $\num{10000}$ in all other experiments) or if the variance of the last $\num{500}$ scores is less than the predefined variance threshold ($\num{0.0005}$ in this experiment, $\num{0.005}$ in all other experiments, i.e.\ \SI{0.05}{\percent} and \SI{0.5}{\percent} of the tree-pair quality score range, respectively). If either iteration or variance thresholds are reached, the MCMC iterations are terminated and the last obtained pair of trees is returned. 

\begin{figure*}[tbp]
  \centering
  \begin{tabular}{@{}c@{}c@{}c@{}}
    \includegraphics[height=3.1cm]{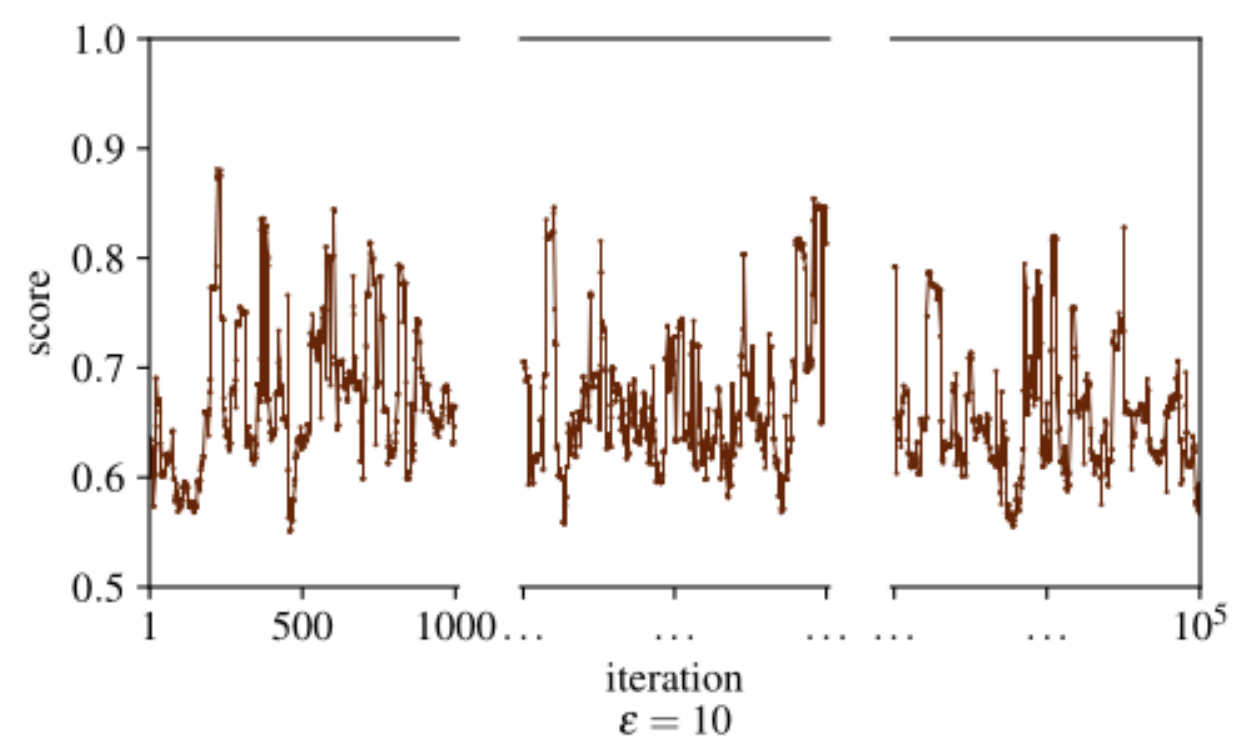} &
    \includegraphics[height=3.1cm]{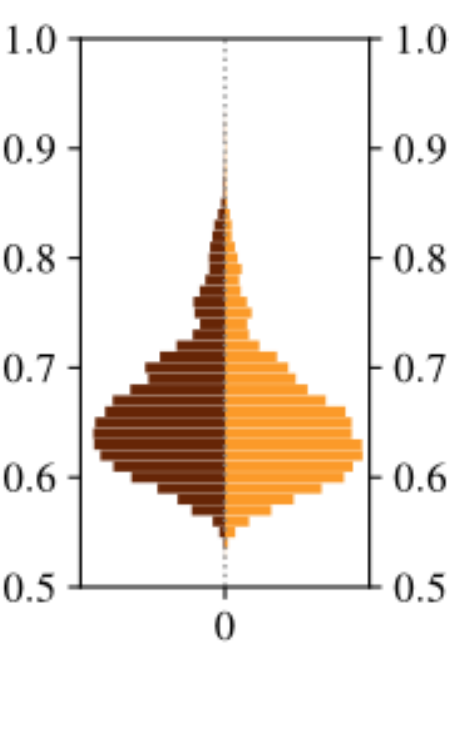} &
    \includegraphics[height=3.1cm]{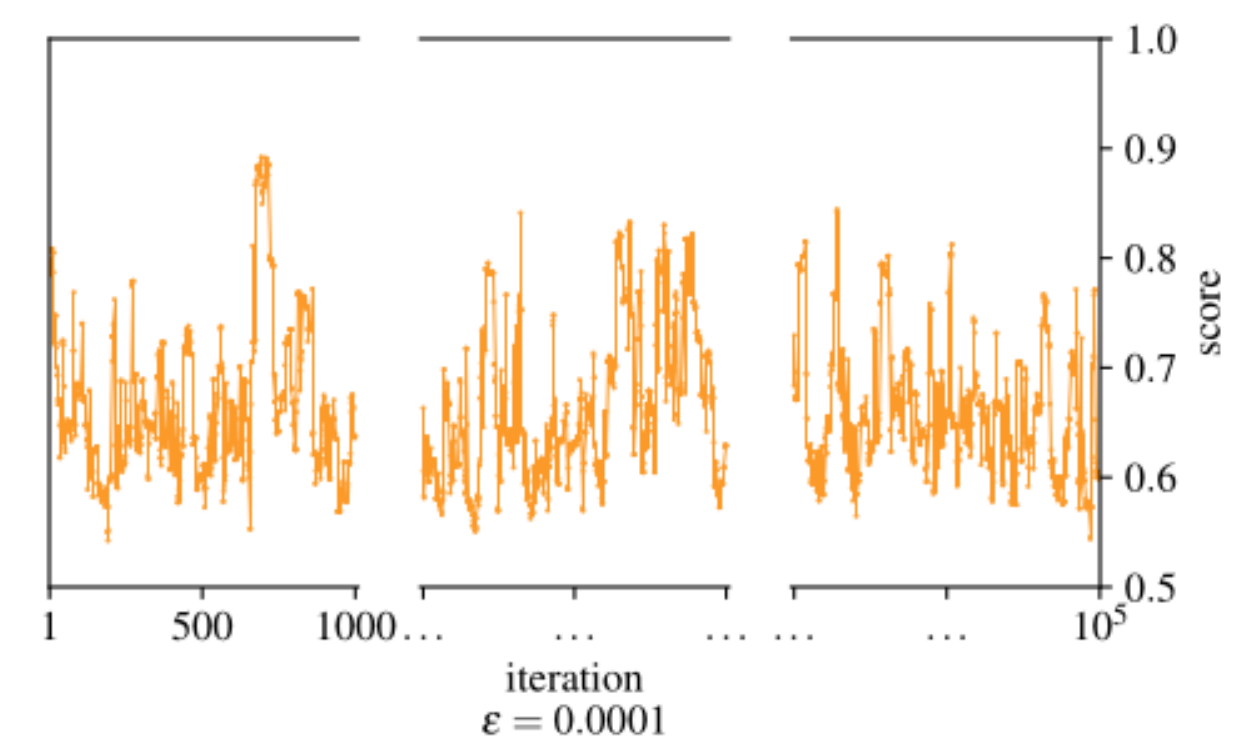}
  \end{tabular}
  \caption{\textbf{Value of MCMC scores across MCMC iterations.} The scores achieved during MCMC iterations of tree-pairs construction of the \ownalgo algorithm using the same initial targets with budgets $\varepsilon = 10$ (left) and $\varepsilon = 0.0001$ (right), for the first, mid and last \num{1000} iterations of \num{100000} total iterations. Corresponding histograms of the scores (center).}
  \label{fig:MCMCIterations}
\end{figure*}

Here, we inspect the distribution of tree-pair scores achieved during the MCMC iterations with $\varepsilon = 10$ and $\varepsilon = 0.0001$. Intuitively, using a larger budget should result in more accurate tree-pairs. Indeed, this happens, but manifests as a higher probability to obtain a tree-pair with higher score rather than by obtaining much larger scores on average. This occurs because values of the used score are in the unit interval, and the exponential mechanism gives very high probabilities even to tree-pairs of very low quality. As a result, during tree-pair construction good split points will generally be chosen, but bad choices will occasionally be incorporated into tree-pair construction as well. This results in an almost periodic behaviour of the quality score (see Figure \ref{fig:MCMCIterations}).

Using a larger budget indeed increases the overall score of tree-pairs. Histogram in \ref{fig:MCMCIterations} (center) indicates slightly better behaviour for $\varepsilon=10$, and indeed the average scores ($\pm$ standard deviation) a \num{0.668+-0.060} for $\varepsilon=10$ and \num{0.657+-0.058} for $\varepsilon=0.0001$. Using a one-sided Mann--Whitney test, we determined that scores achieved using the higher budget are significantly larger ($p<2\cdot 10^{-16}$) than scores achieved using the lower budget. Thus, indeed, it is more probable to obtain a good solution using a higher budget. This is reflected in the accuracy and the number of produced redescriptions (compare Figures \ref{fig:RAccB1} and \ref{fig:RAccB01}).

\subsection{Comparison of Differentially Private Redescription Mining Algorithms}
Percentages of successful executions (returns any redescription) for all differentially private approaches are presented in Tables \ref{tab:succexec} and \ref{tab:succexec1}. As it can be seen from these results, the \sampling and \expmech fail with probability up to $0.13$ when $100$ runs are used (which should provide relatively stable statistics).

\begin{table}
  \centering
  \caption{Number of executions (out of $10$) returning at least one redescription when $\InitTrials = 1$ and initial budget $\varepsilon = 0.1$.}
  \label{tab:succexec}
  \begin{tabular}{lS[table-format=3.0]S[table-format=3.0]S[table-format=3.0]}
    \toprule
    Dataset &{\sampling} &  {\expmech}  & {\ownalgo} \\
    \midrule
    \nerdy & 10/10 & 10/10 & 10/10  \\
    \mammals & 10/10 & 9/10 & 10/10 \\
    \mimic & 10/10 &  10/10 & 10/10  \\
    \medicX{2}{1} & 7/10 & 9/10   & 10/10   \\
    \medicX{2}{4} & 10/10 & 10/10  & 10/10  \\
    \medicX{2}{8} & 10/10 &  10/10 & 10/10  \\
    \medicX{2}{16} & 10/10 & 10/10  & 10/10  \\
    \bottomrule
  \end{tabular}
\end{table}

\begin{table}
  \centering
  \caption{Number of executions (out of $100$) returning at least one redescription when $\InitTrials = 1$ and initial budget $\varepsilon = 0.01$.}
  \label{tab:succexec1}
  \begin{tabular}{lS[table-format=3.0]S[table-format=3.0]S[table-format=3.0]}
    \toprule
    Dataset &{\sampling} &  {\expmech}  & {\ownalgo} \\
    \midrule
    \nerdy & 99/100 & 100/100 & 100/100  \\
    \mammals & 98/100 & 100/100 & 100/100 \\
    \mimic & 100/100 &  100/100 & 100/100  \\
    \medicX{2}{1} & 87/100 & 91/100   & 100/100   \\
    \medicX{2}{4} & 98/100 & 96/100  & 100/100  \\
    \medicX{2}{8} & 90/100 &  97/100 & 100/100  \\
    \medicX{2}{16} & 95/100 & 97/100  & 100/100  \\
    \bottomrule
  \end{tabular}
\end{table}

\begin{figure*}[tbp]
  \centering
  \begin{tabularx}{\textwidth}{@{}X@{}X@{}}
    \multicolumn{1}{c}{\nerdy}  & \multicolumn{1}{c}{\mammals}   \\
    \includegraphics[height=2.8cm]{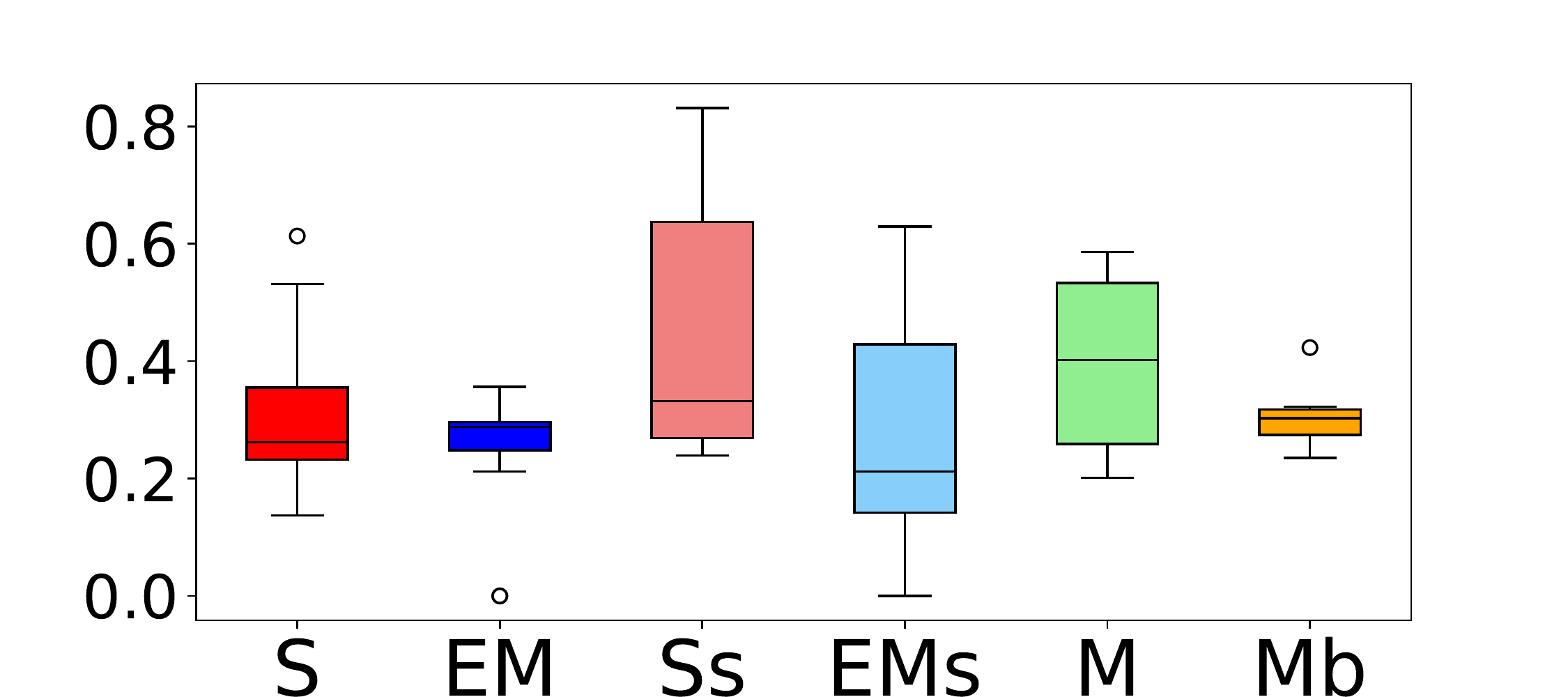} &
    \includegraphics[height=2.8cm]{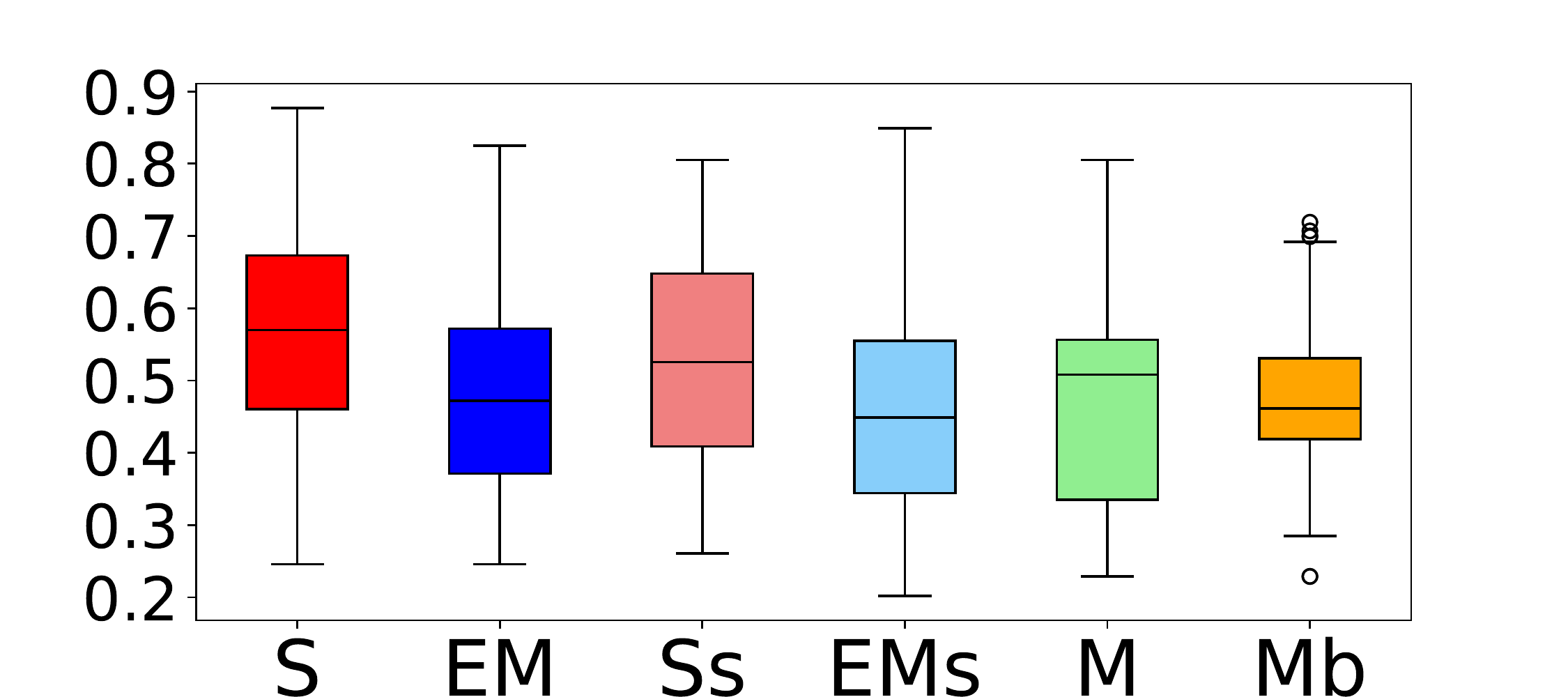} \\
     \multicolumn{1}{c}{\mimicX{Ter}} & \multicolumn{1}{c}{\medicX{2}{1}}  \\
     \includegraphics[height=2.8cm]{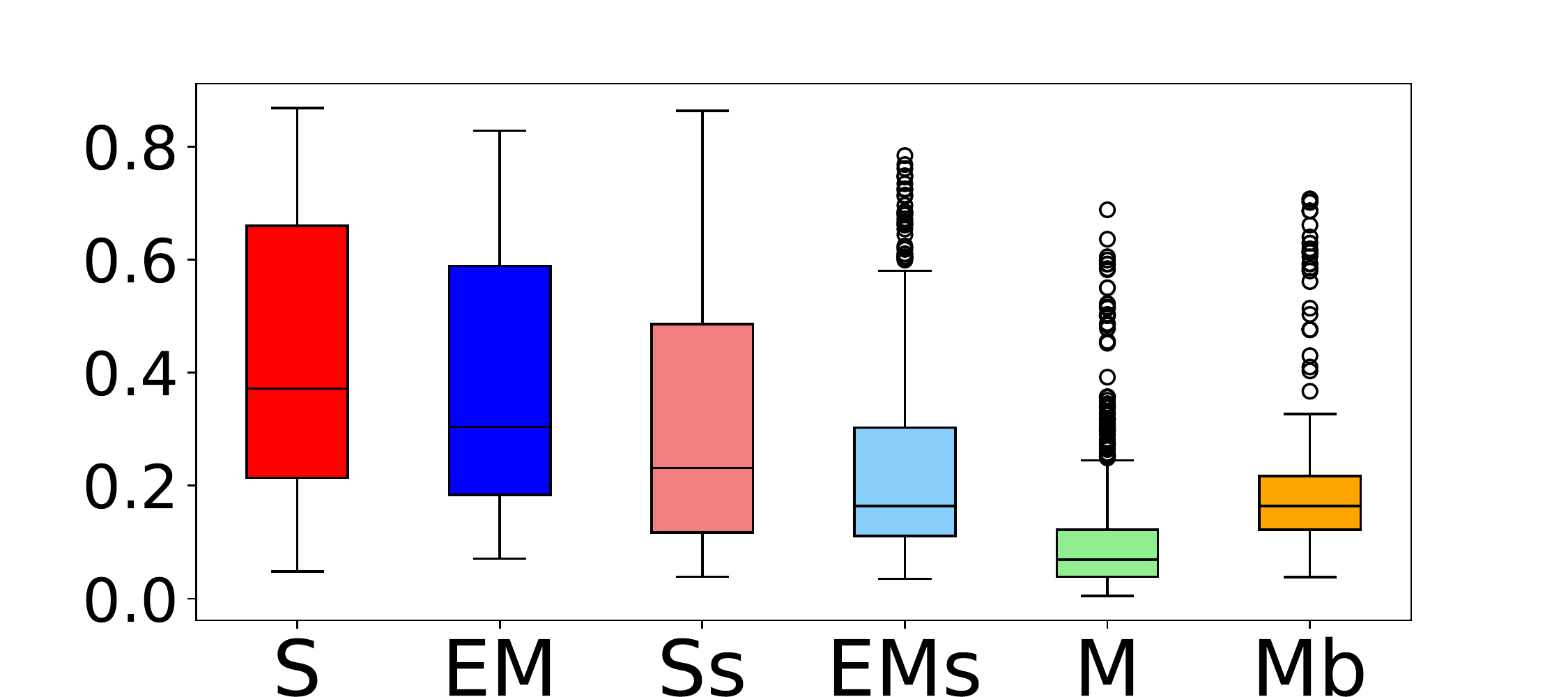} &
    \includegraphics[height=2.8cm]{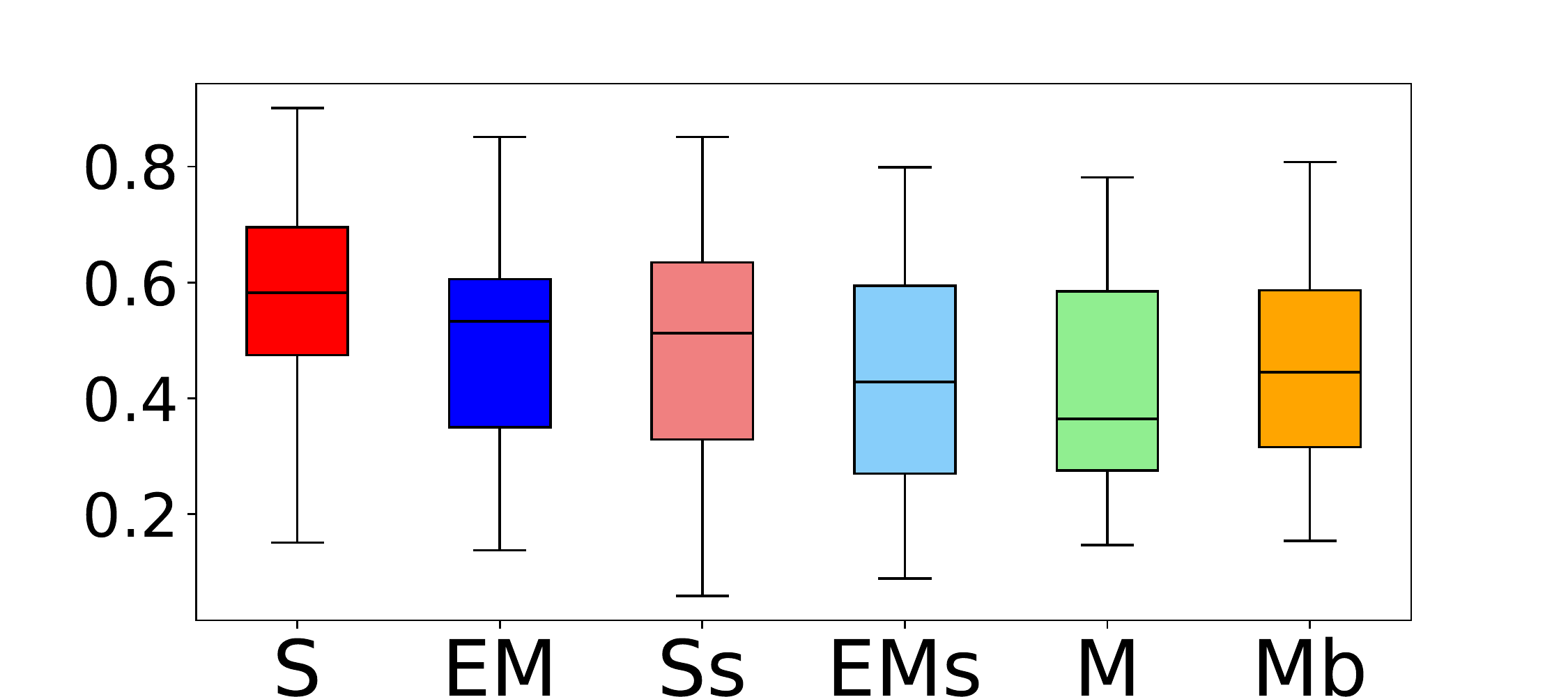} \\
     \multicolumn{1}{c}{\medicX{2}{4}} & \multicolumn{1}{c}{\medicX{2}{8}}  \\
     \includegraphics[height=2.8cm]{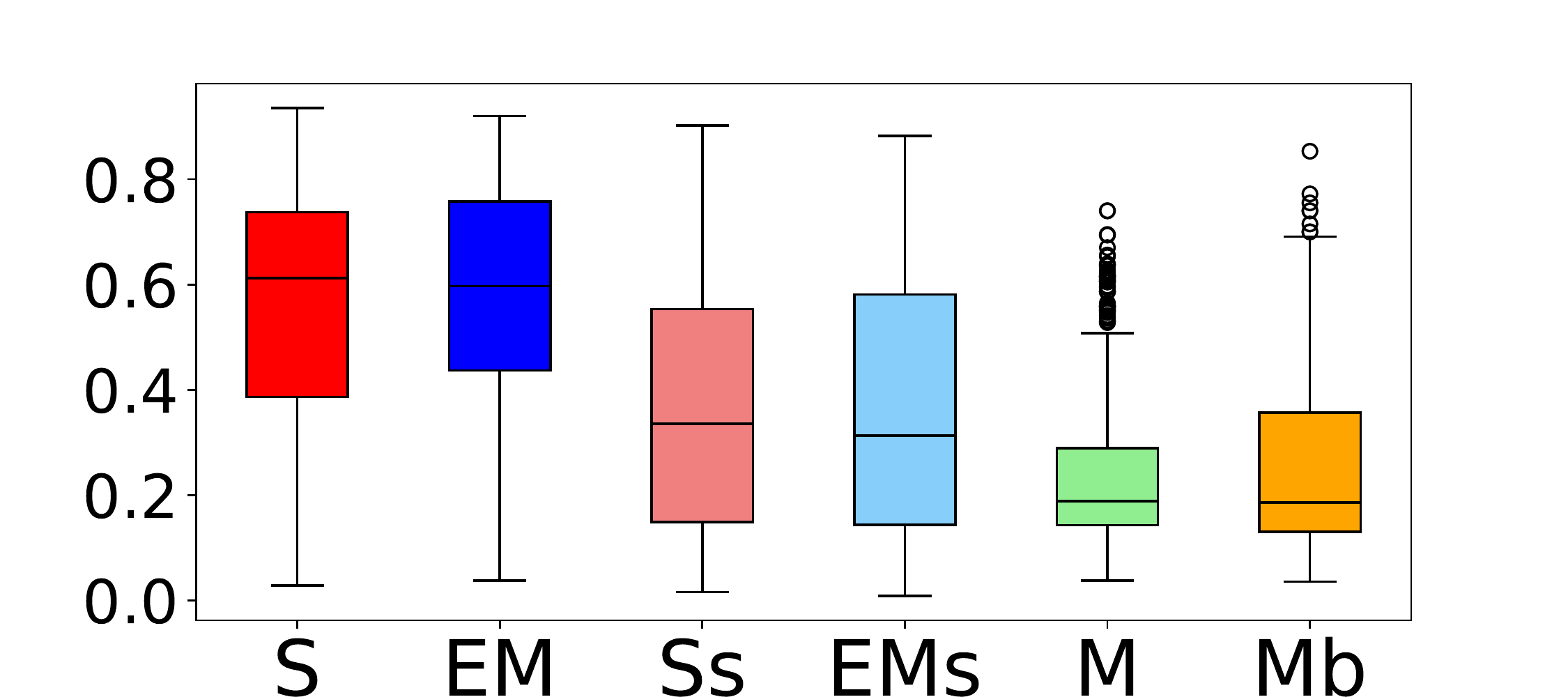} &
    \includegraphics[height=2.8cm]{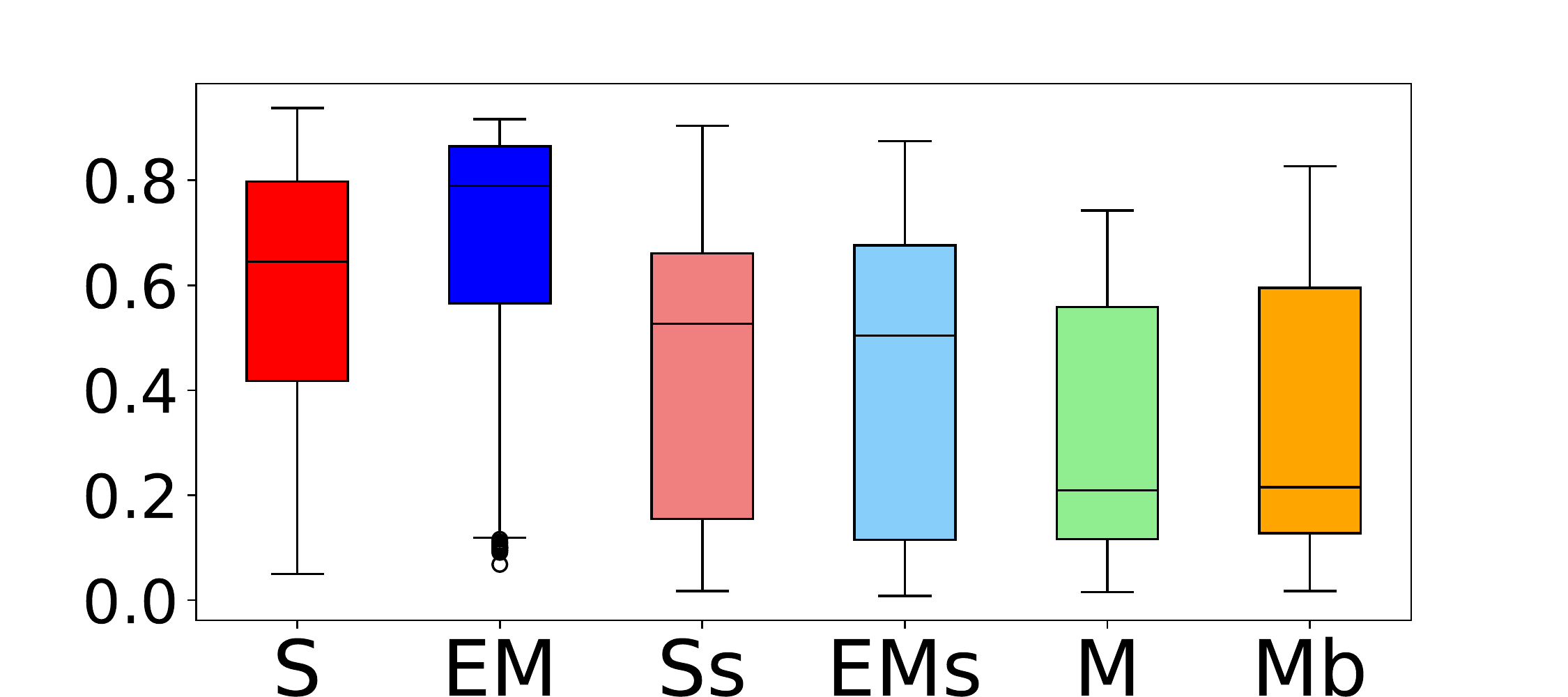}  \\
     \multicolumn{1}{c}{\medicX{2}{16}} &  \\
     \includegraphics[height=2.8cm]{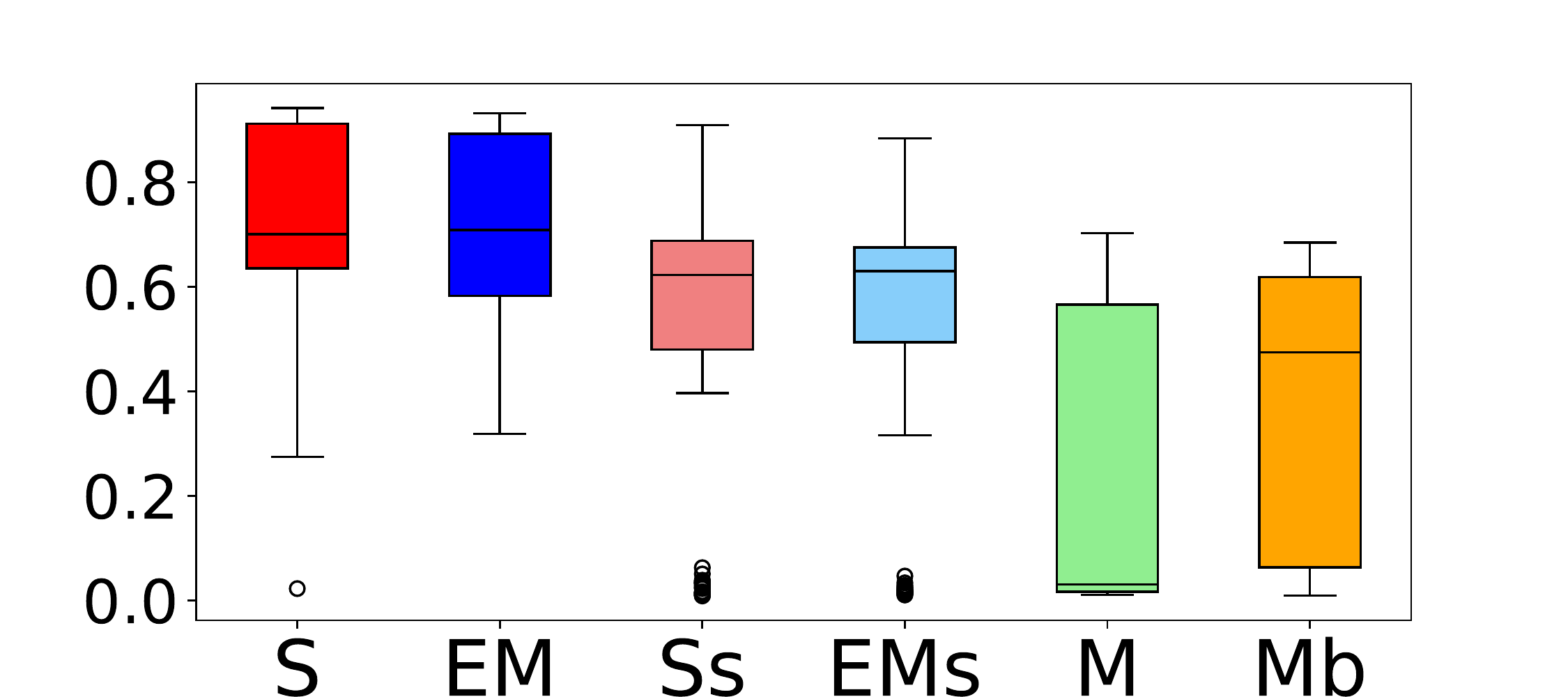} &
  \end{tabularx}
  \caption{Distribution of accuracy of filtered significant redescriptions produced by each differentially private approach in $10$ executions with budget $1$. $S$ -- \sampling, $EM$ -- \expmech, $M$ -- \ownalgo. $s$ denotes accuracy after stabilization and $b$ denotes $\omega = 0.5$ in the \ownalgo.}
  \label{fig:RAccB1}
\end{figure*}

\begin{figure*}[tbp]
  \centering
  \begin{tabularx}{\textwidth}{@{}X@{}X@{}}
    \multicolumn{1}{c}{\nerdy}  & \multicolumn{1}{c}{\mammals}   \\
    \includegraphics[height=2.8cm]{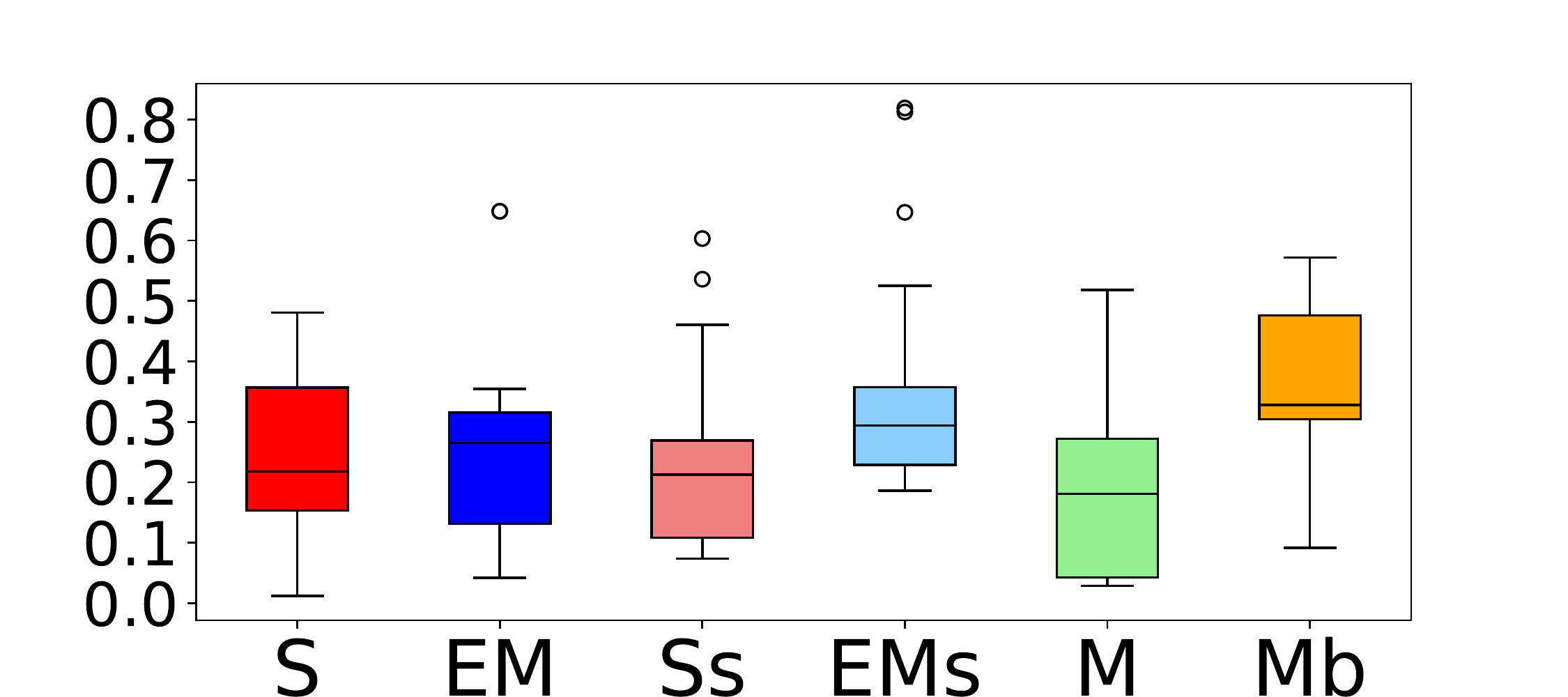} &
    \includegraphics[height=2.8cm]{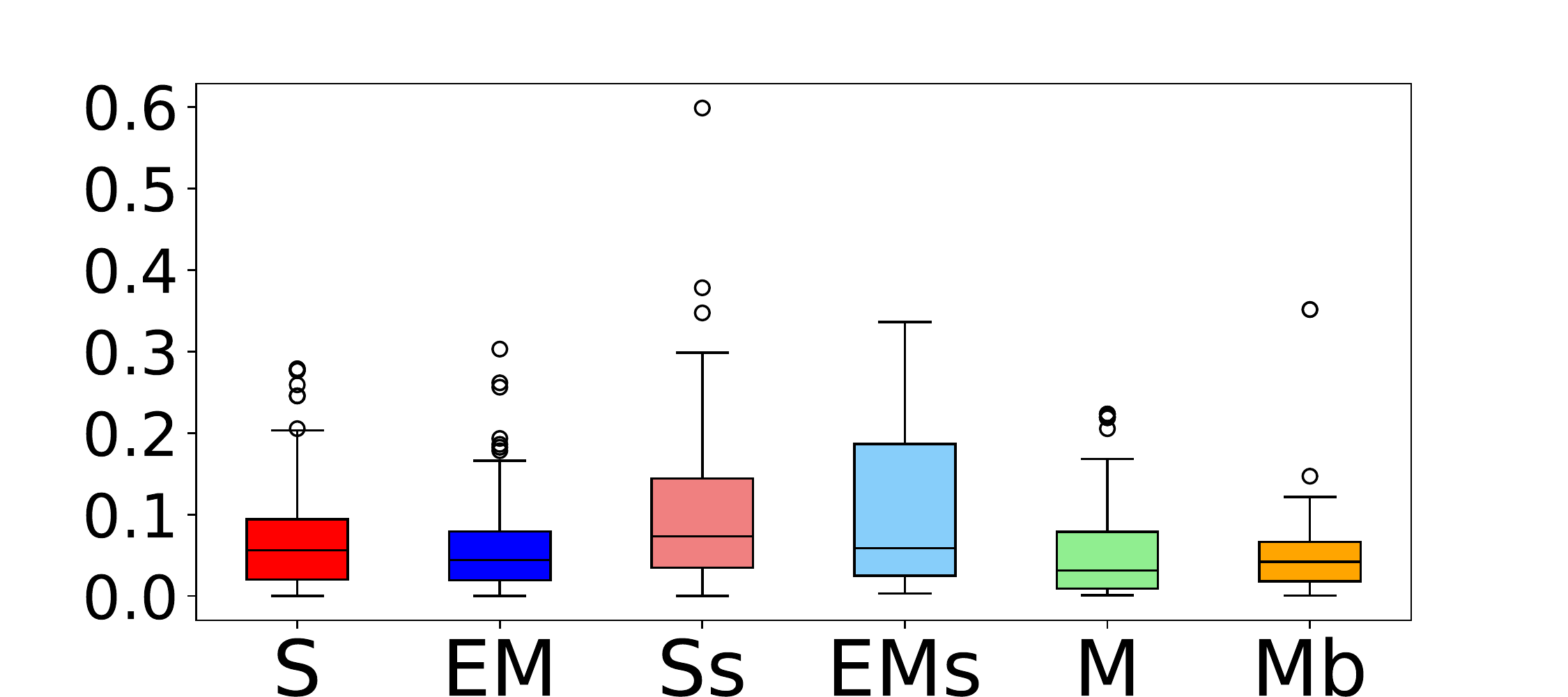} \\
     \multicolumn{1}{c}{\mimicX{Ter}} & \multicolumn{1}{c}{\medicX{2}{1}}  \\
     \includegraphics[height=2.8cm]{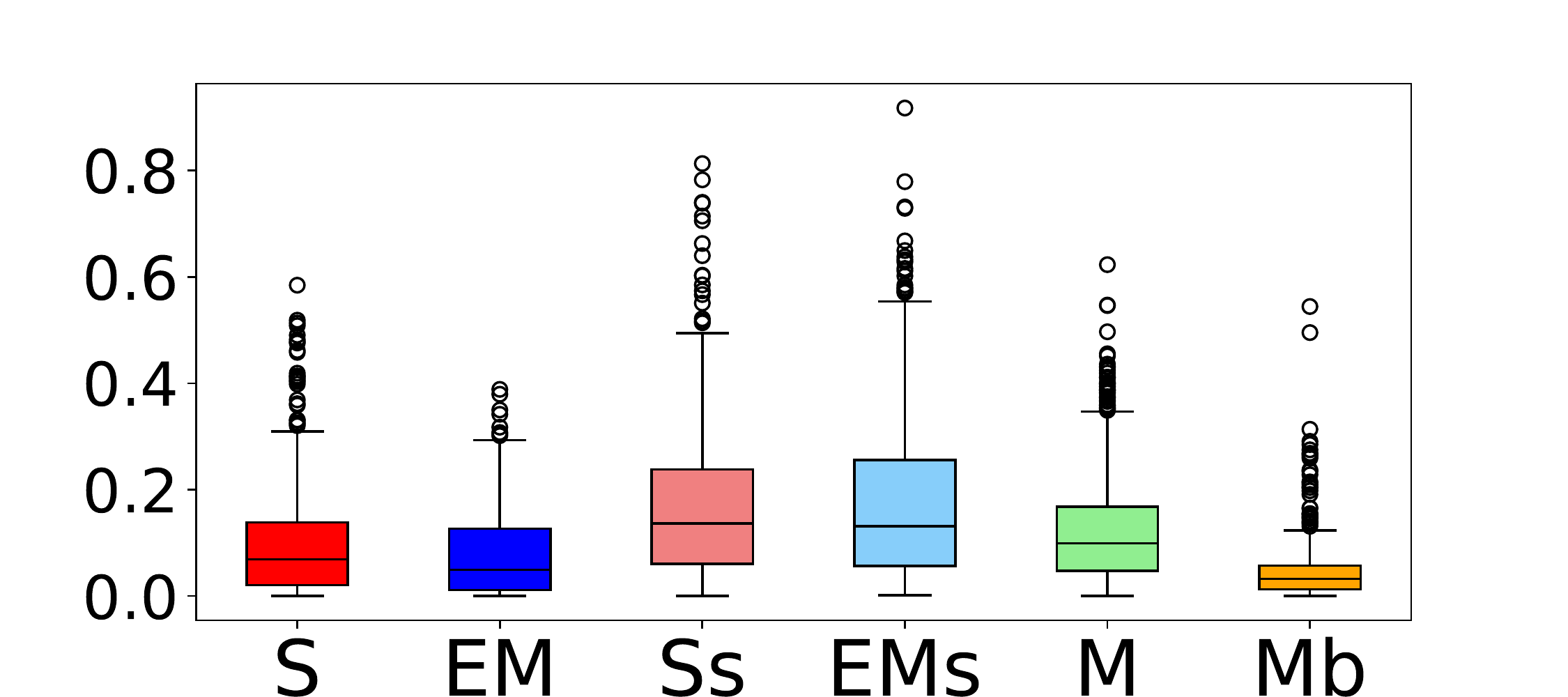} &
    \includegraphics[height=2.8cm]{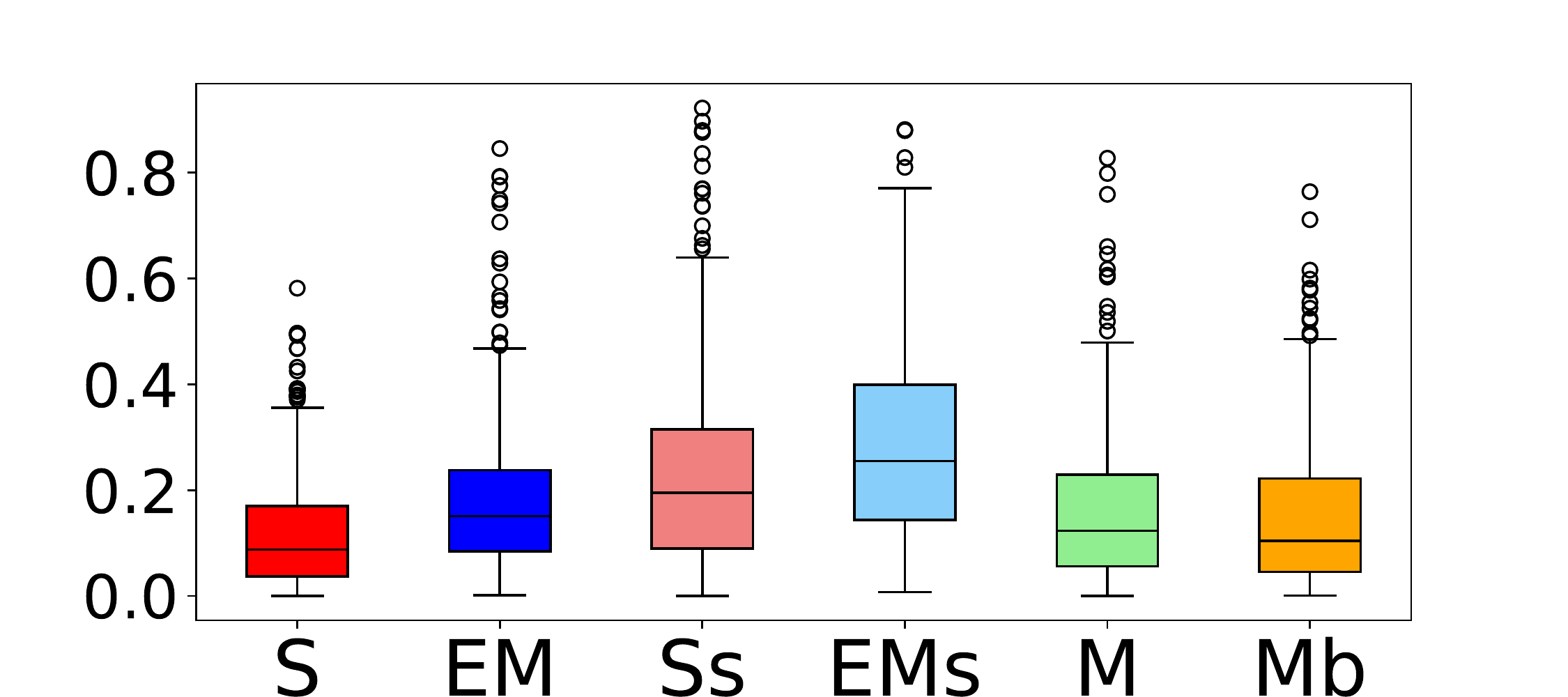} \\
     \multicolumn{1}{c}{\medicX{2}{4}} & \multicolumn{1}{c}{\medicX{2}{8}}  \\
     \includegraphics[height=2.8cm]{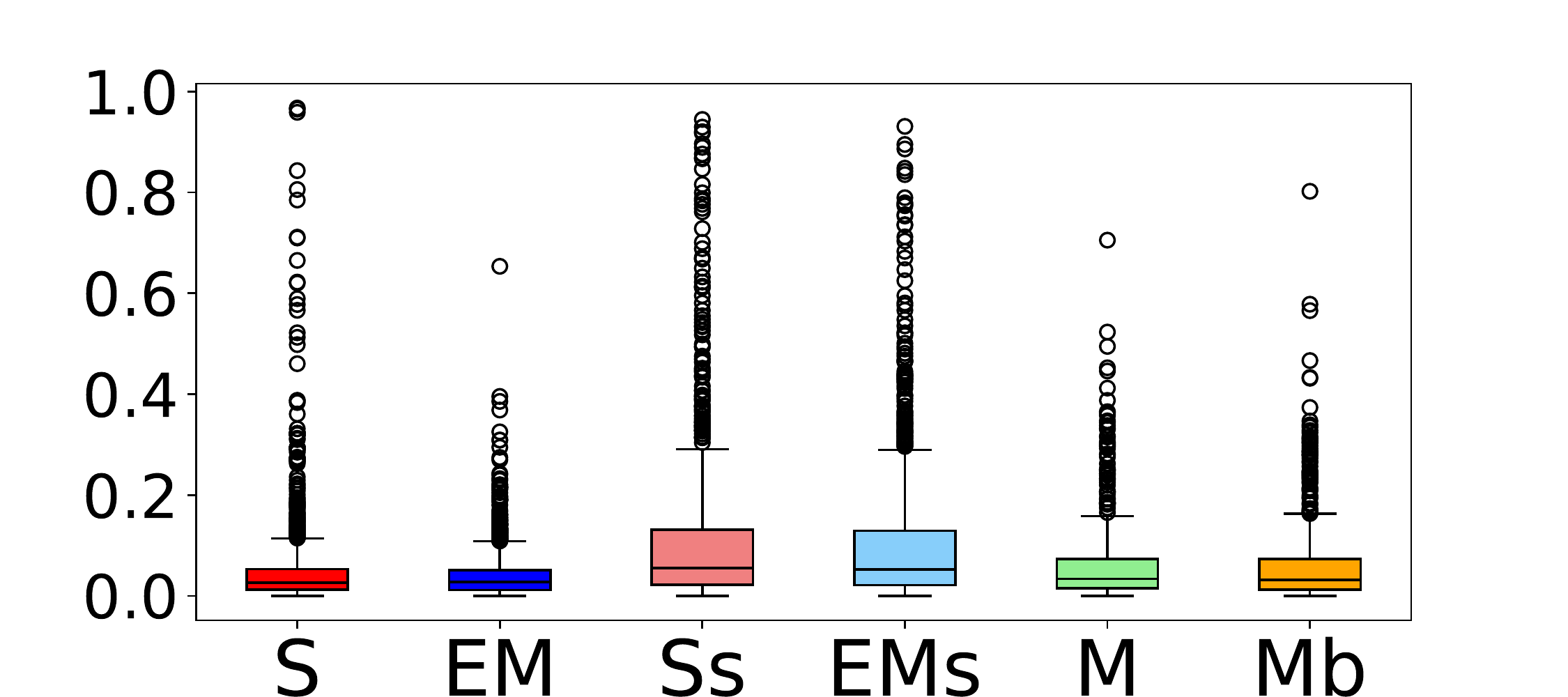} &
    \includegraphics[height=2.8cm]{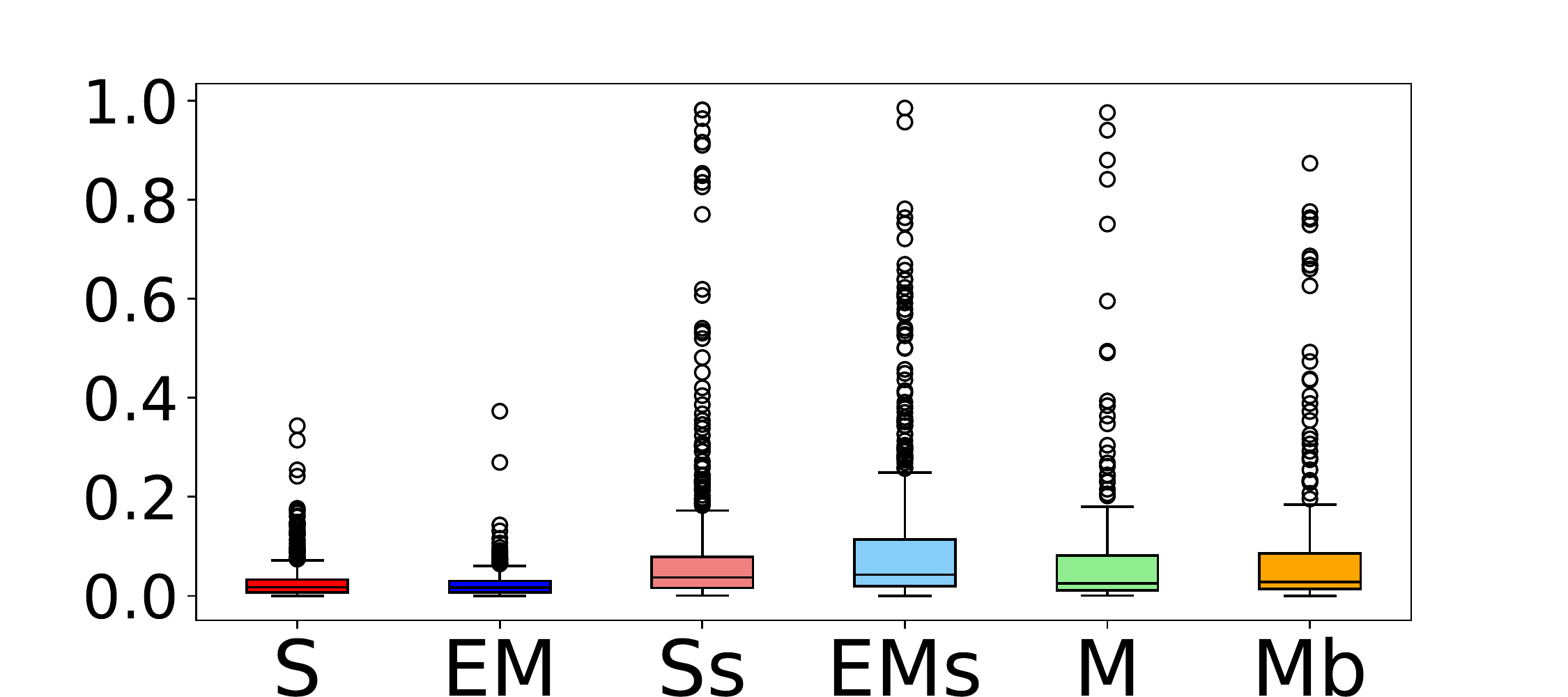}  \\
     \multicolumn{1}{c}{\medicX{2}{16}} &  \\
     \includegraphics[height=2.8cm]{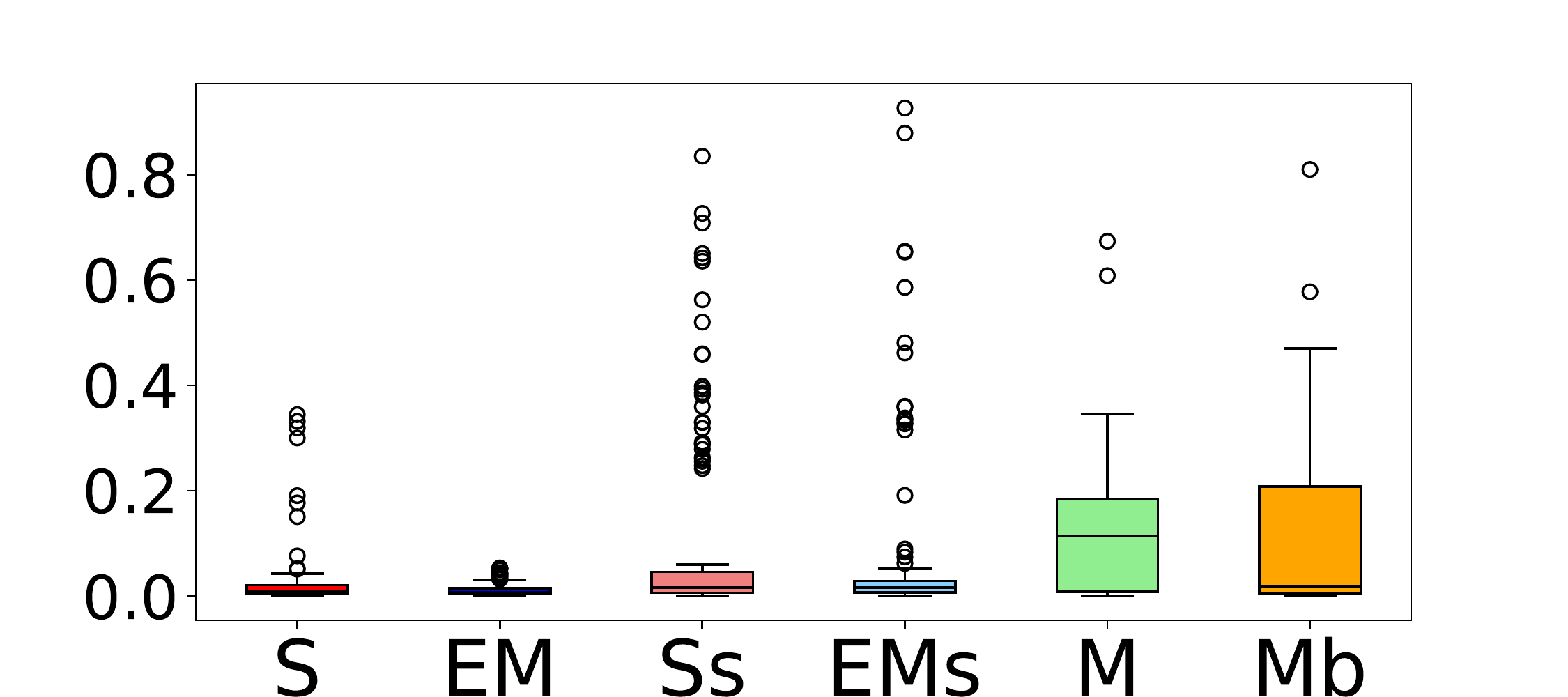} &
  \end{tabularx}
  \caption{Distribution of absolute difference between noisy and real redescription accuracy of filtered significant redescriptions produced by each differentially private approach in $10$ executions with budget $1$. Algorithm names are as in Figure \ref{fig:RAccB1}.}
  \label{fig:RStabB1}
\end{figure*}

Parallel boxplots in Figure \ref{fig:RAccB1} show pooled distributions of redescription accuracy for filtered, significant redescriptions obtained across $10$ executions with budget $1.0$. Regular setting denotes $\InitTrials = 1$ and, $\RMIter = k$ while stable ($s$) denotes setting $\InitTrials = k$ and $\RMIter = 1$. The second setting is denoted \emph{stable}, since there is substantially higher probability for the \sampling and \expmech with these parameters to choose at least one initial attribute that will yield creation of redescriptions.  We also present two different executions of the \ownalgo, with the budget redistribution weight $\omega = 0.1$ and $\omega = 0.5$ (balanced,  $b$). Distributions of absolute difference between noisy and real redescription accuracy ($\abs{J_{noisy}(R) - J_{real}(R)}$) in these settings can be seen in Figure \ref{fig:RStabB1} (lower numbers are better since they denote smaller variation between the noisy estimate and the real accuracy of produced redescriptions).

\begin{figure*}[tbp]
  \centering
  \begin{tabularx}{\textwidth}{@{}X@{}X@{}}
    \multicolumn{1}{c}{\nerdy}  & \multicolumn{1}{c}{\mammals}   \\
    \includegraphics[height=2.8cm]{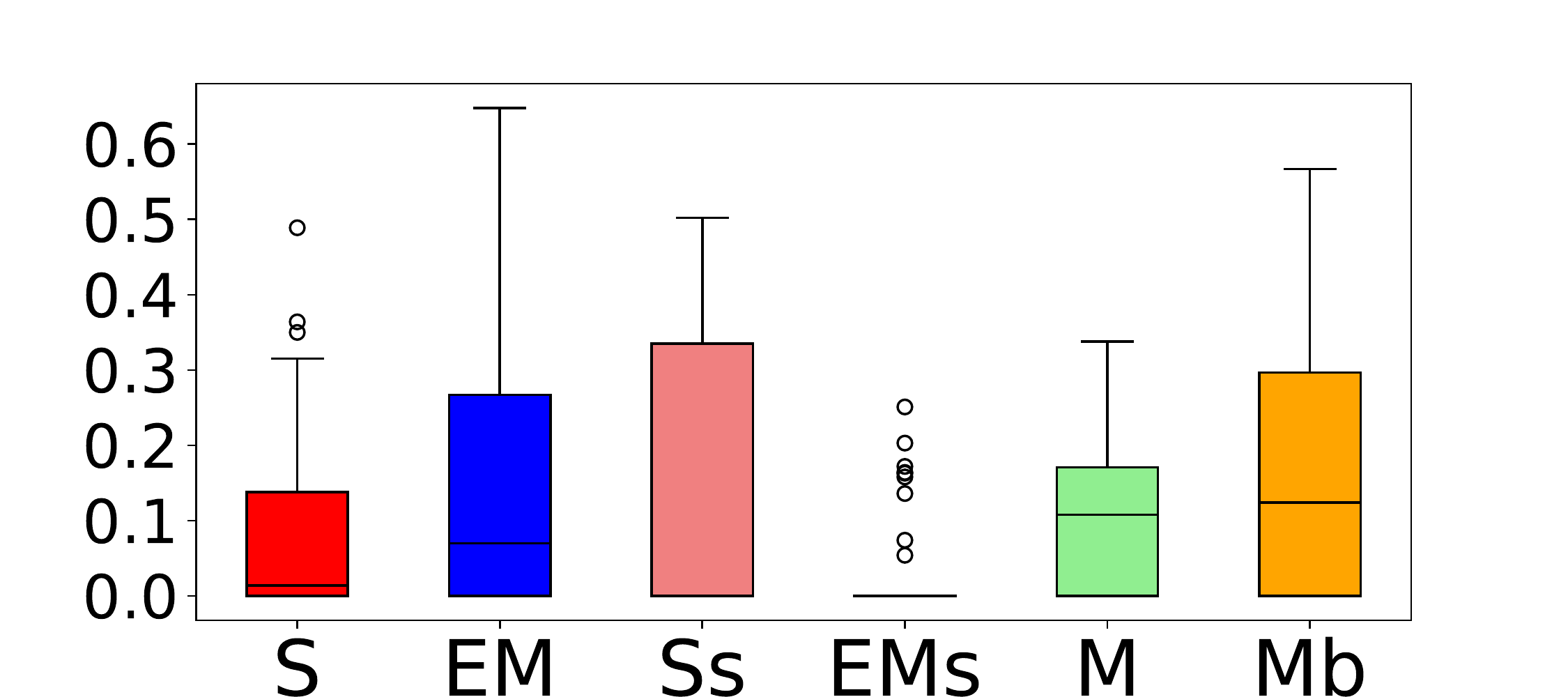} &
    \includegraphics[height=2.8cm]{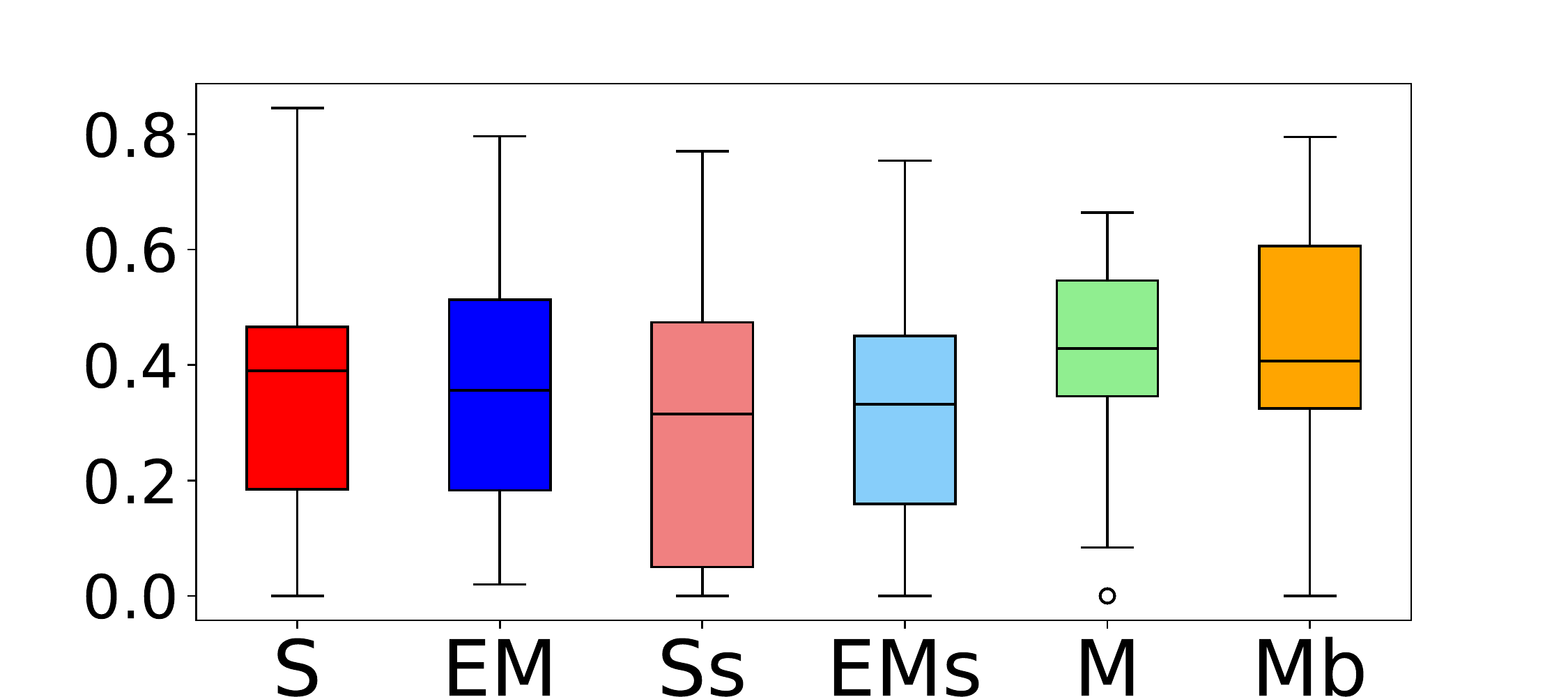} \\
     \multicolumn{1}{c}{\mimicX{Ter}} & \multicolumn{1}{c}{\medicX{2}{1}}  \\
     \includegraphics[height=2.8cm]{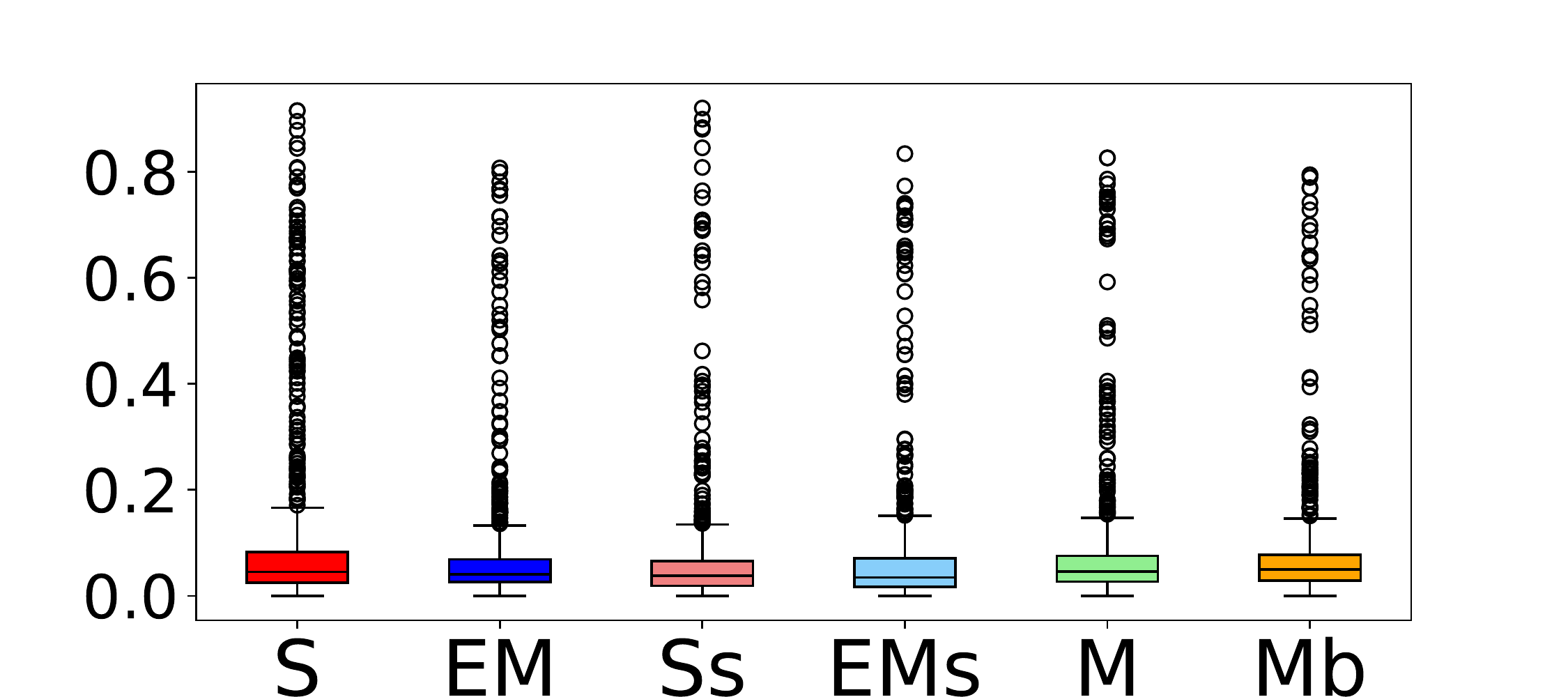} &
    \includegraphics[height=2.8cm]{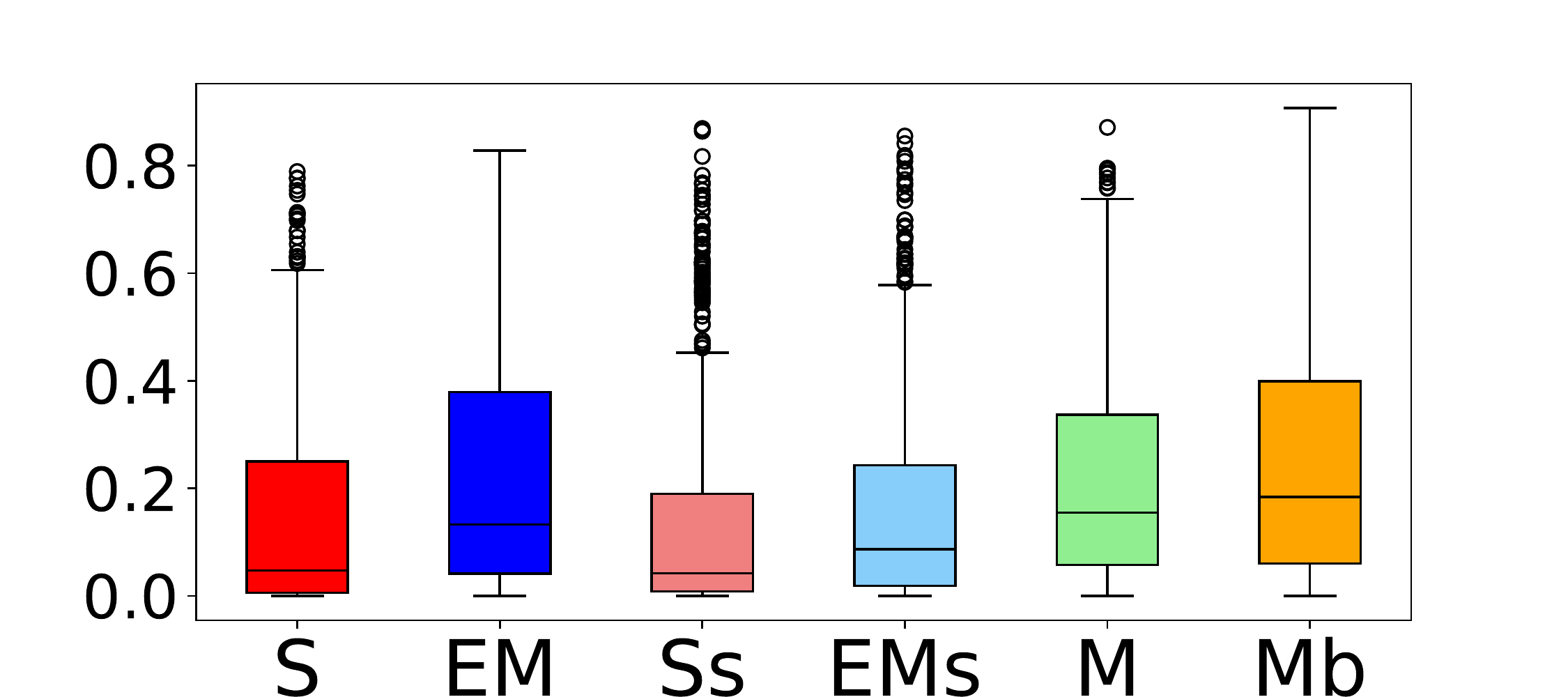} \\
     \multicolumn{1}{c}{\medicX{2}{4}} & \multicolumn{1}{c}{\medicX{2}{8}}  \\
     \includegraphics[height=2.8cm]{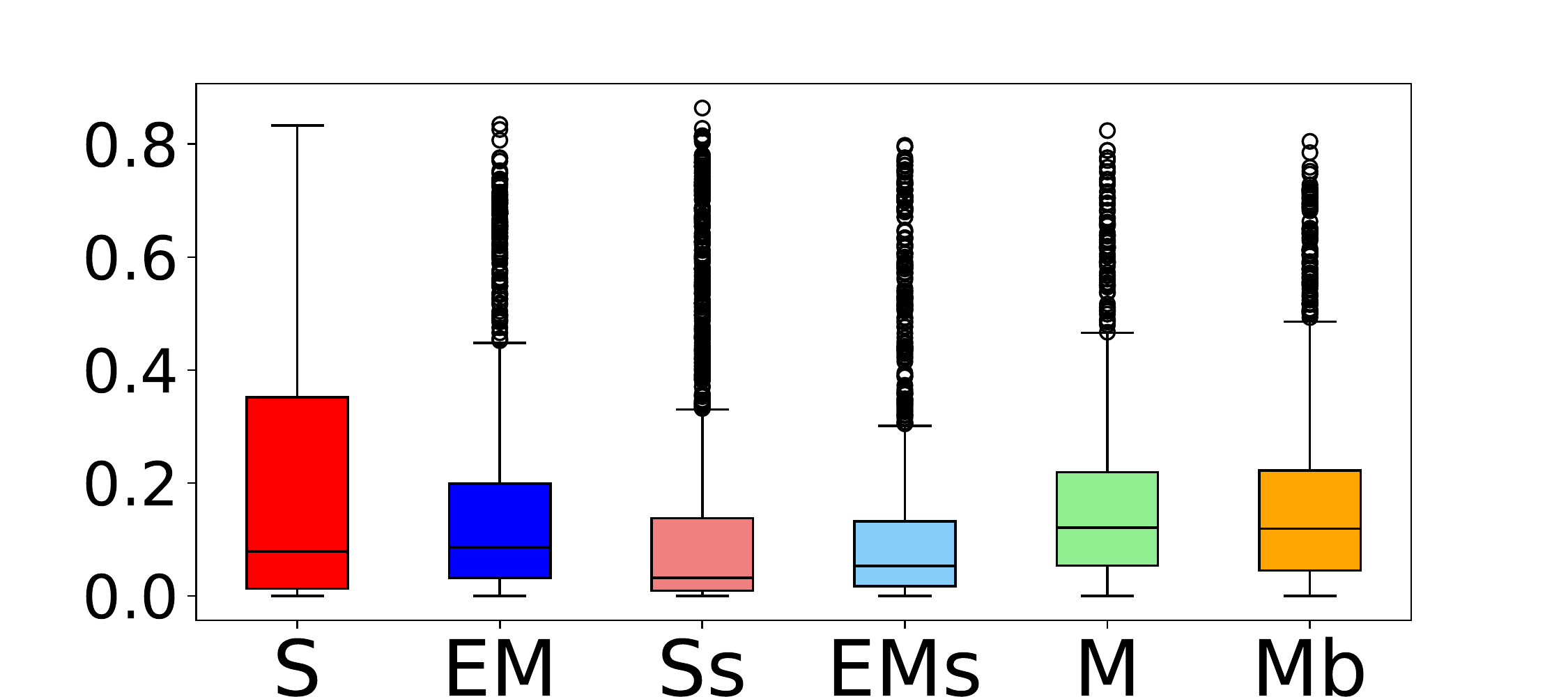} &
    \includegraphics[height=2.8cm]{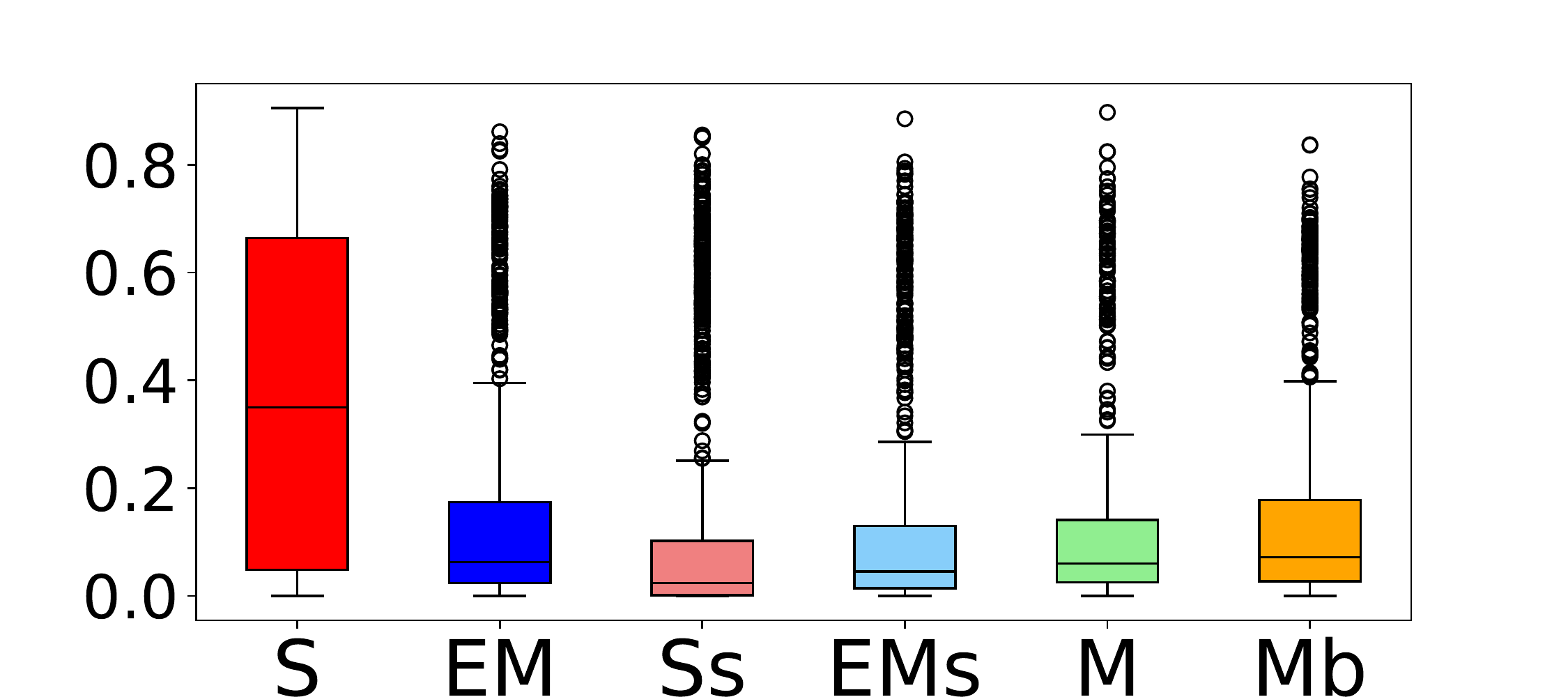}  \\
     \multicolumn{1}{c}{\medicX{2}{16}} &  \\
     \includegraphics[height=2.8cm]{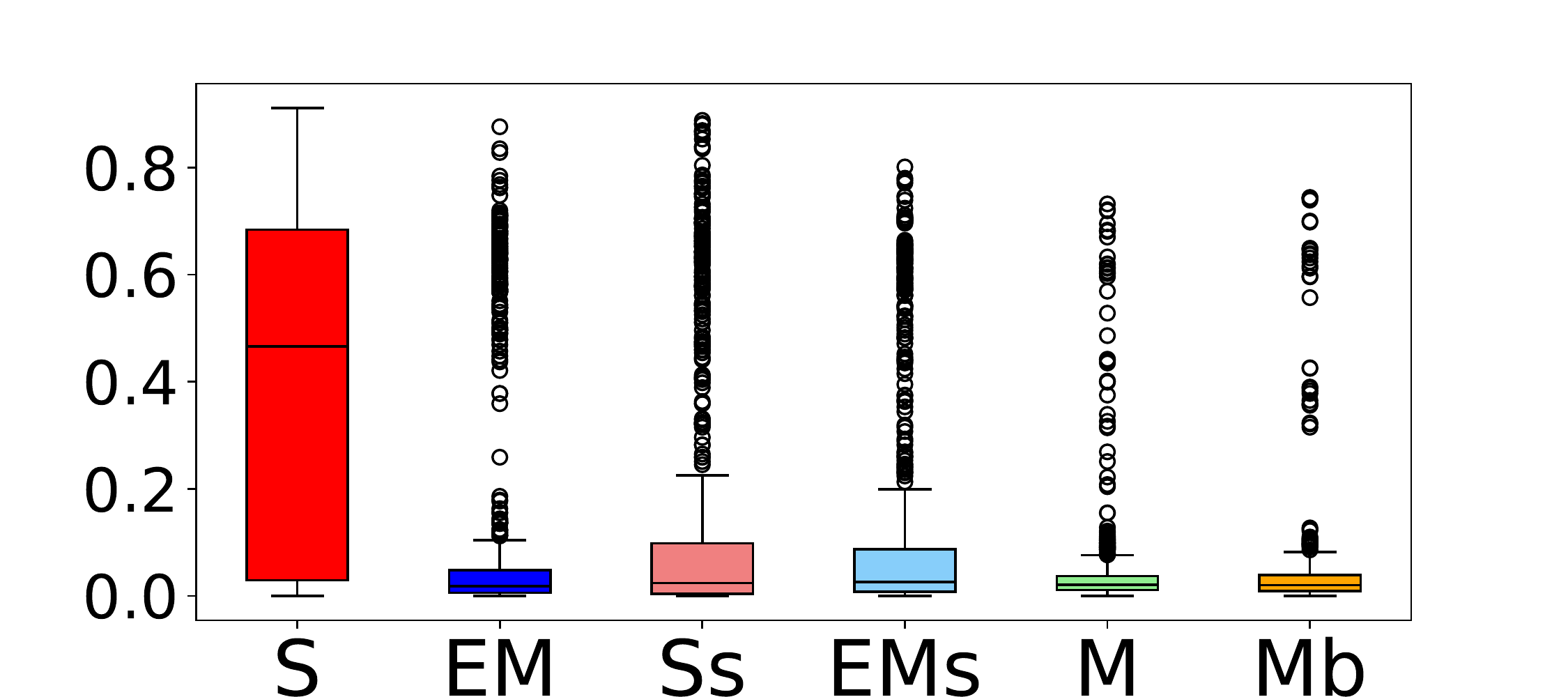} &
  \end{tabularx}
  \caption{Distribution of accuracy of filtered significant redescriptions produced by each differentially private approach in $10$ differentially private executions (each with budget $0.1$). Algorithm names are as in \ref{fig:RAccB1}.}
  \label{fig:RAccB01}
\end{figure*}

\begin{figure*}[tbp]
  \centering
  \begin{tabularx}{\textwidth}{@{}X@{}X@{}}
    \multicolumn{1}{c}{\nerdy}  & \multicolumn{1}{c}{\mammals}   \\
    \includegraphics[height=2.8cm]{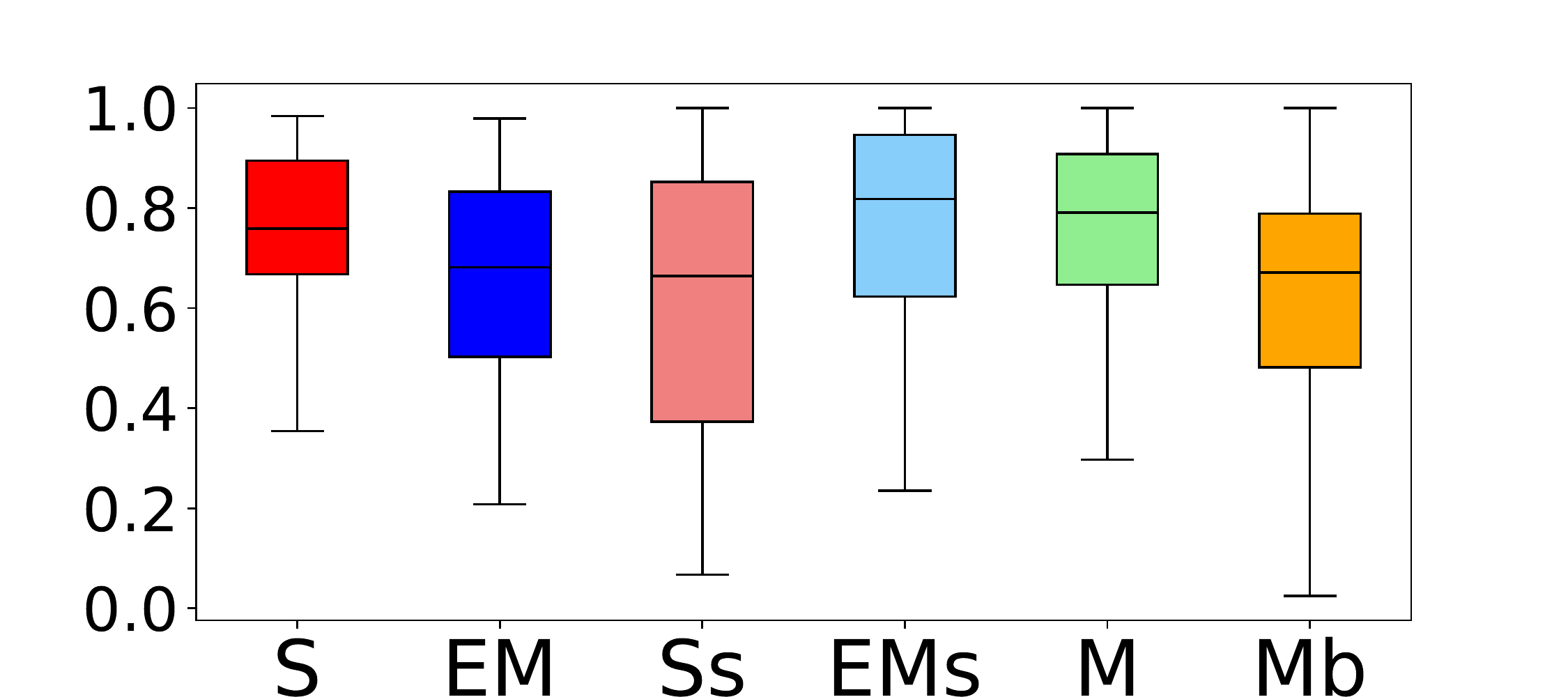} &
    \includegraphics[height=2.8cm]{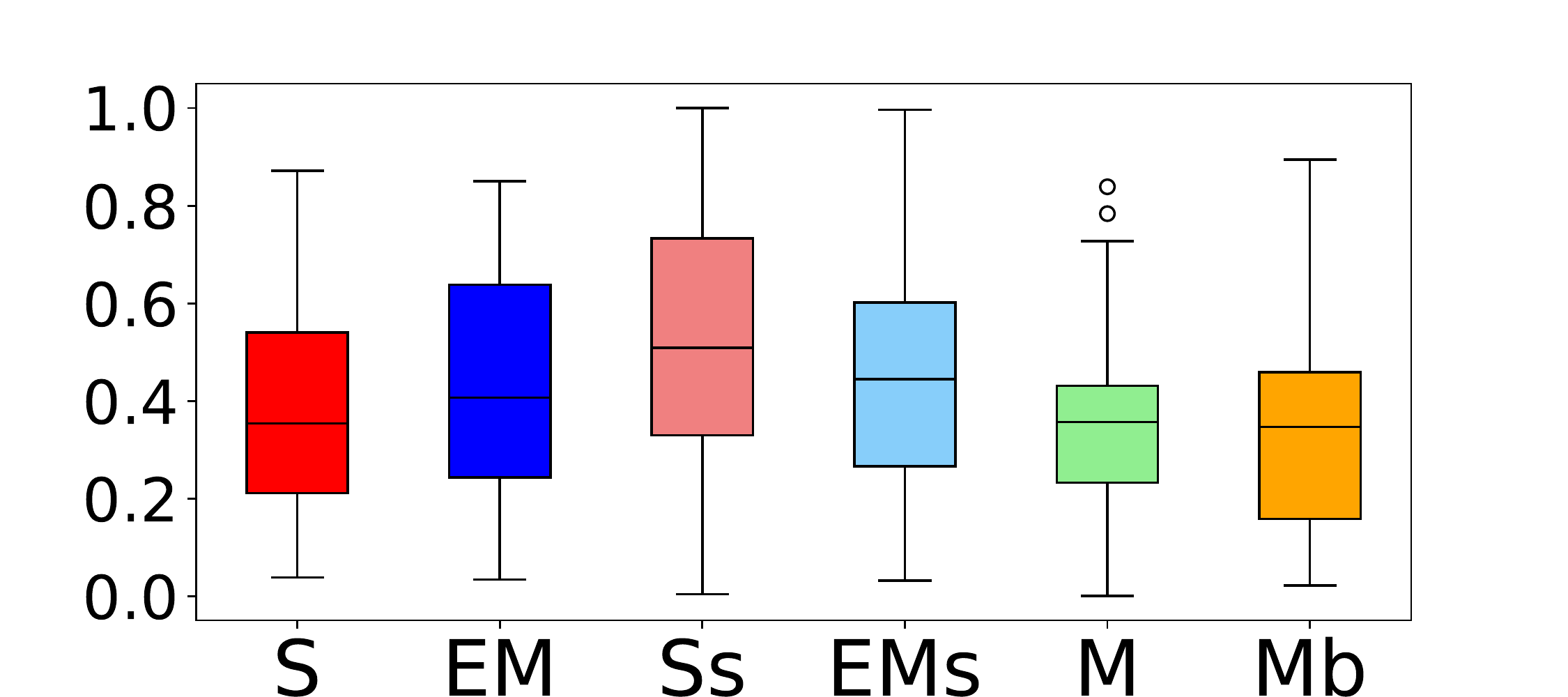} \\
     \multicolumn{1}{c}{\mimicX{Ter}} & \multicolumn{1}{c}{\medicX{2}{1}}  \\
     \includegraphics[height=2.8cm]{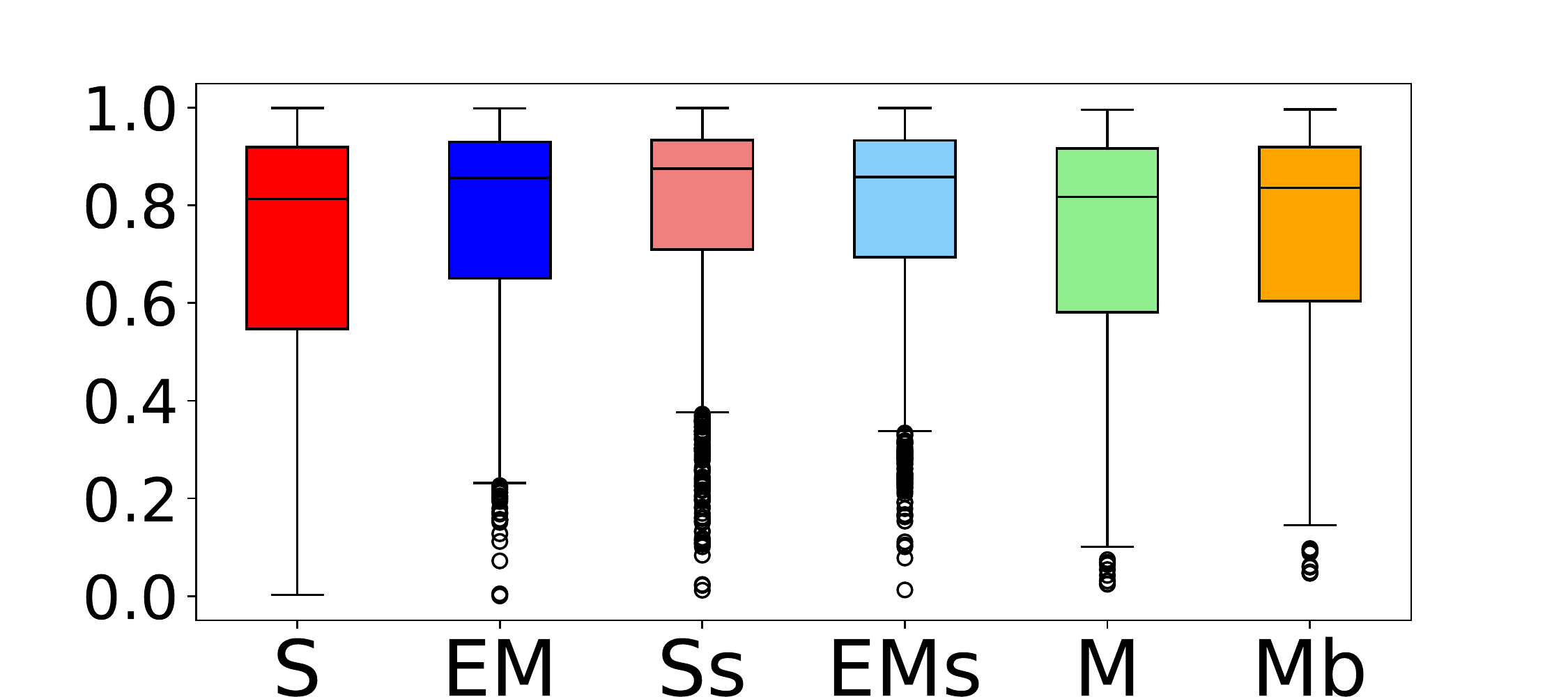} &
    \includegraphics[height=2.8cm]{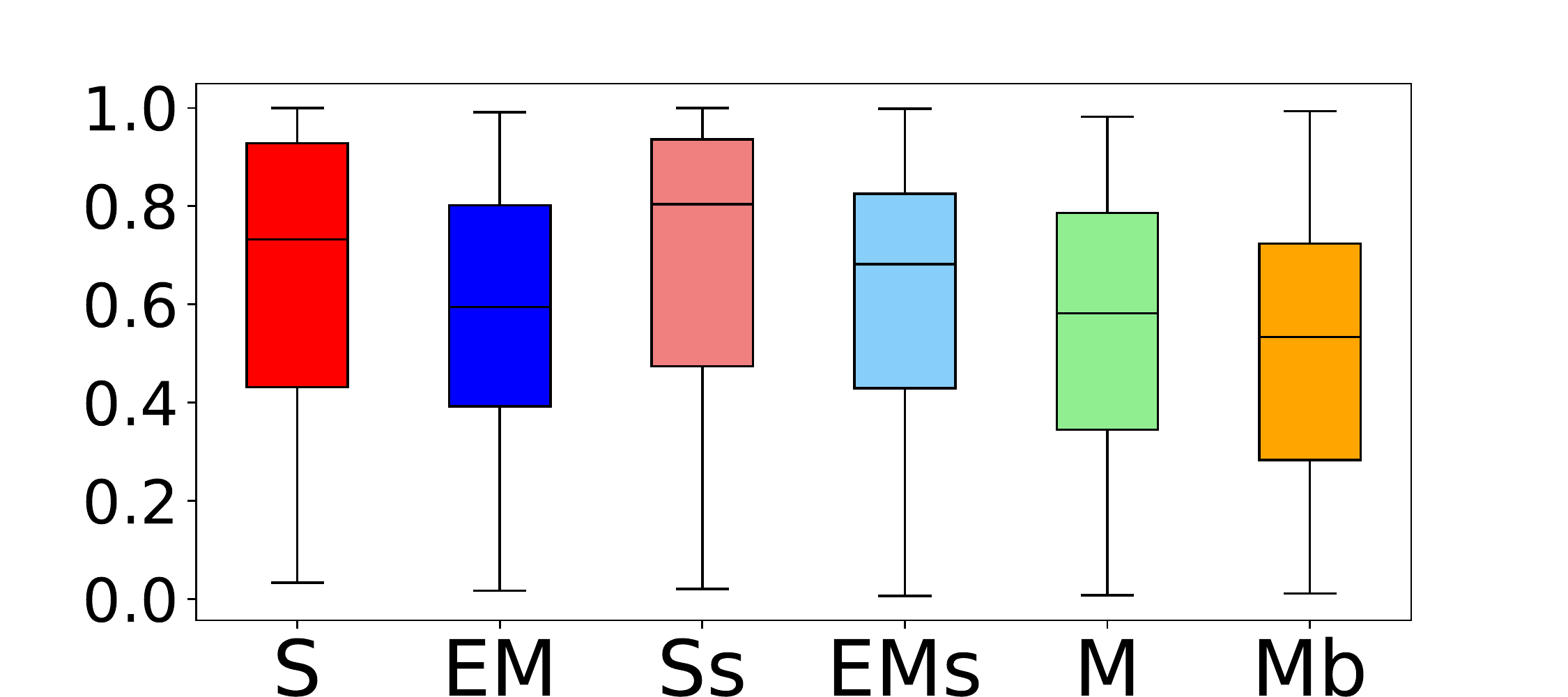} \\
     \multicolumn{1}{c}{\medicX{2}{4}} & \multicolumn{1}{c}{\medicX{2}{8}}  \\
     \includegraphics[height=2.8cm]{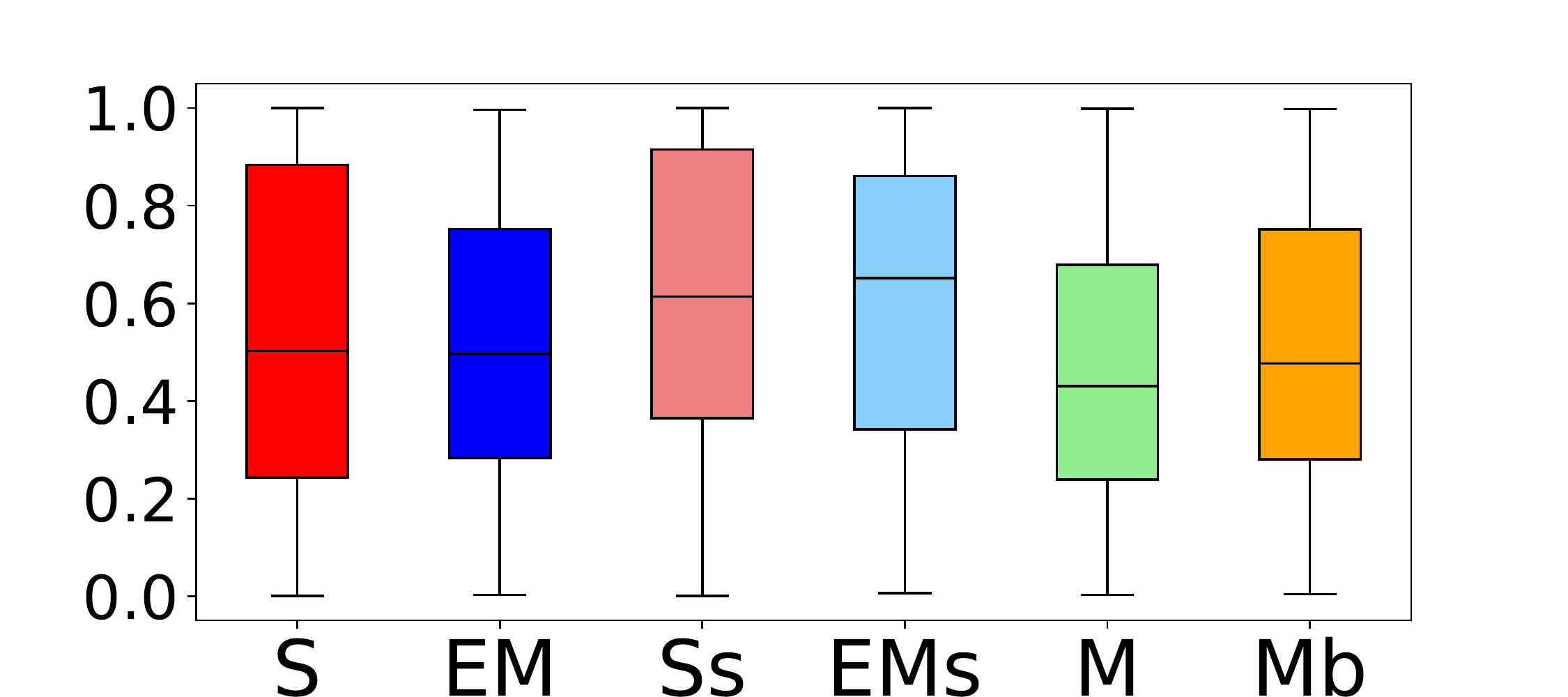} &
    \includegraphics[height=2.8cm]{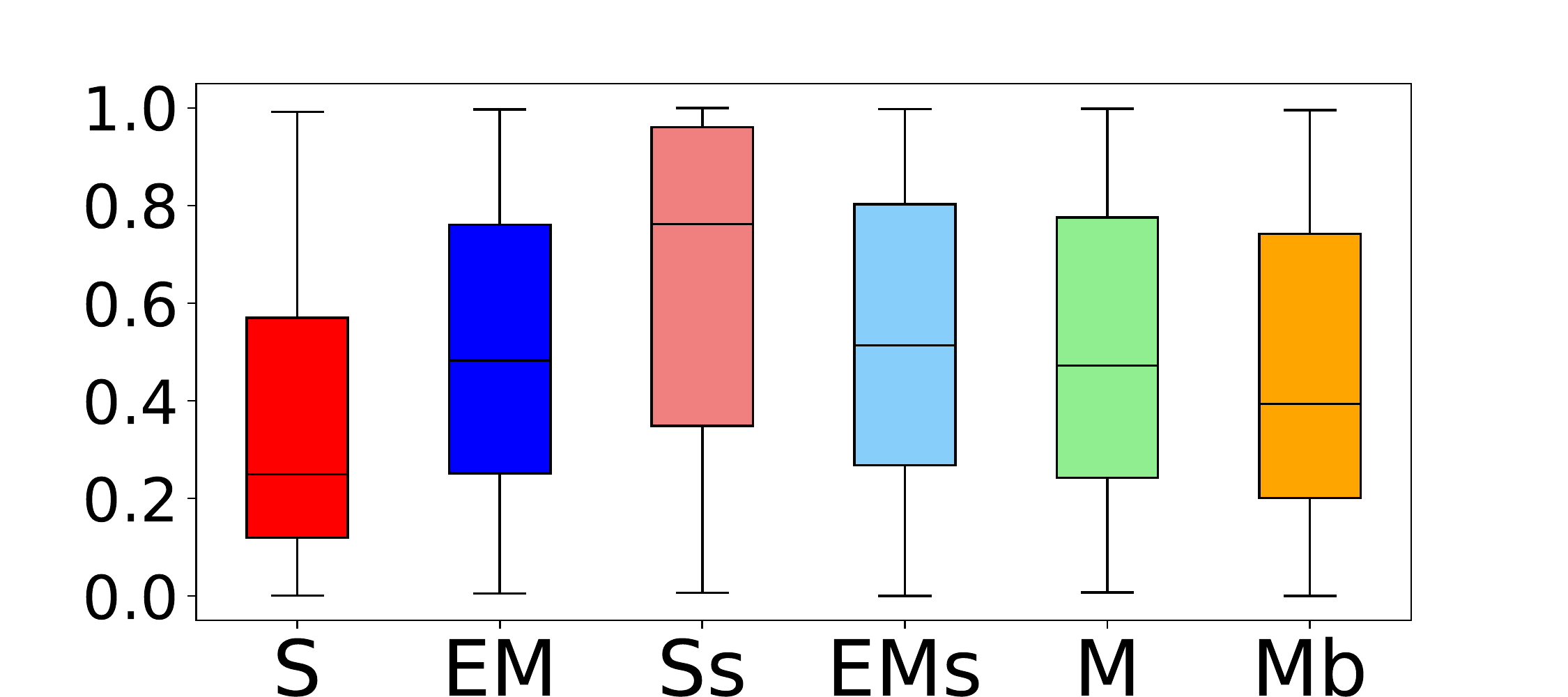}  \\
     \multicolumn{1}{c}{\medicX{2}{16}} &  \\
     \includegraphics[height=2.8cm]{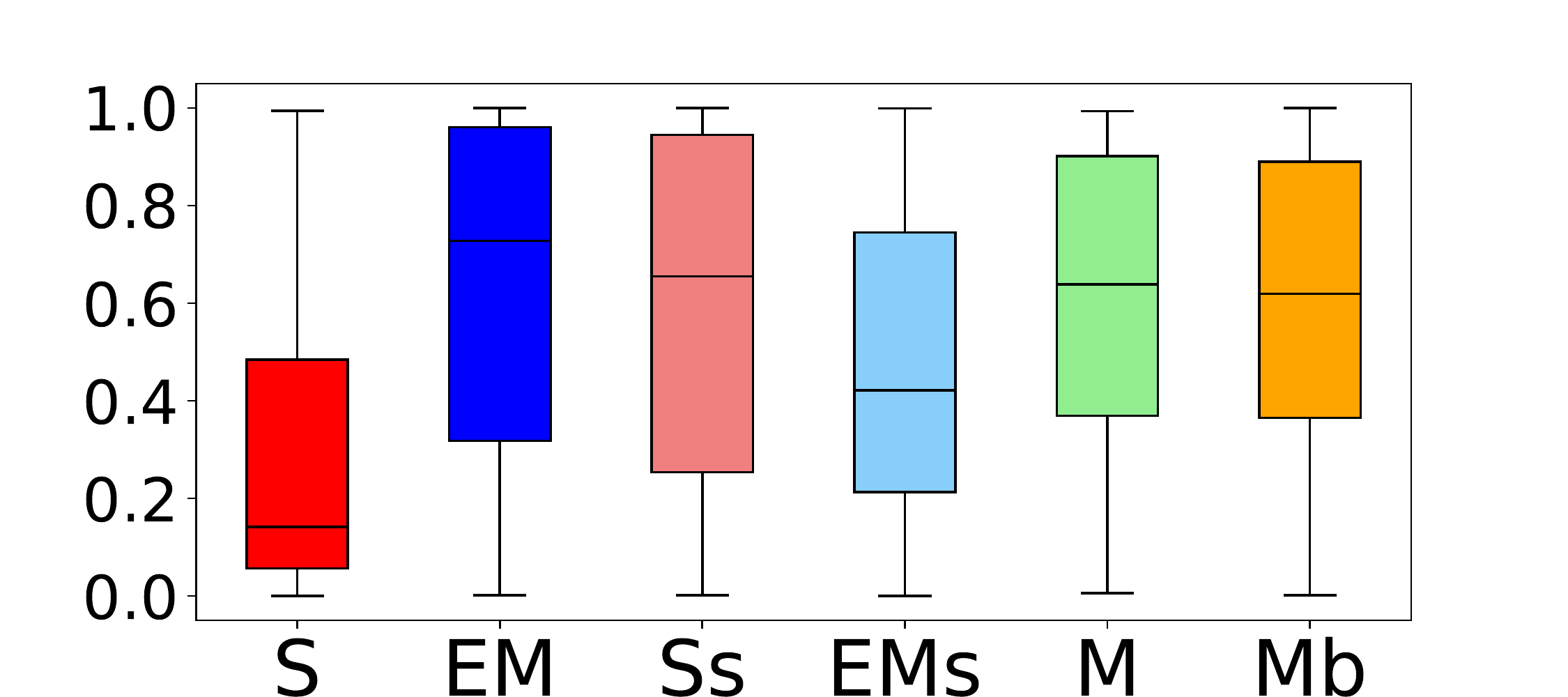} &
  \end{tabularx}
  \caption{Distribution of absolute difference between noisy and real redescription accuracy of filtered significant redescriptions produced by each differentially private approach in $10$ differentially private executions (each with budget $0.1$). Algorithm names are as in Figure \ref{fig:RAccB1}.}
  \label{fig:RStab0B1}
\end{figure*}

\begin{figure*}[tbp]
  \centering
  \begin{tabularx}{\textwidth}{@{}X@{}X@{}}
    \multicolumn{1}{c}{\nerdy}  & \multicolumn{1}{c}{\mammals}   \\
    \includegraphics[height=2.8cm]{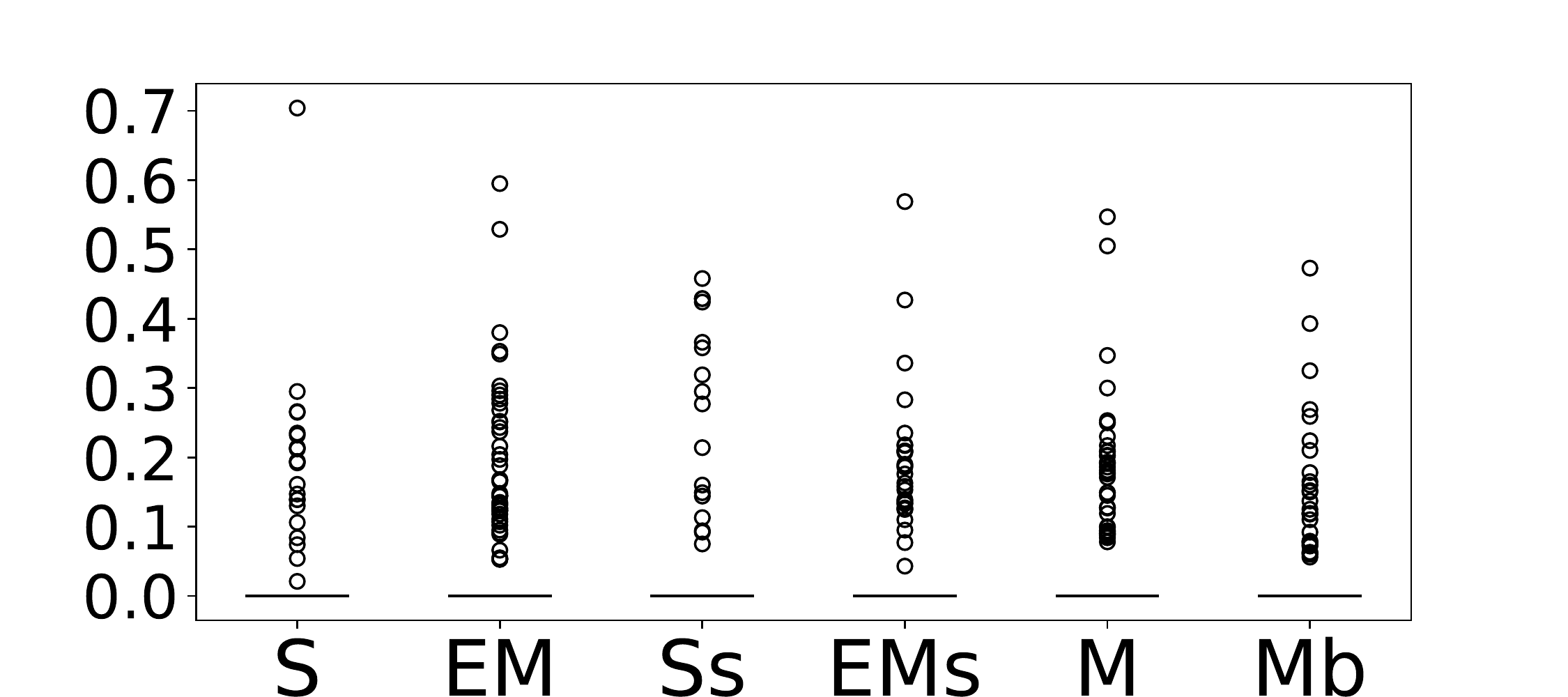} &
    \includegraphics[height=2.8cm]{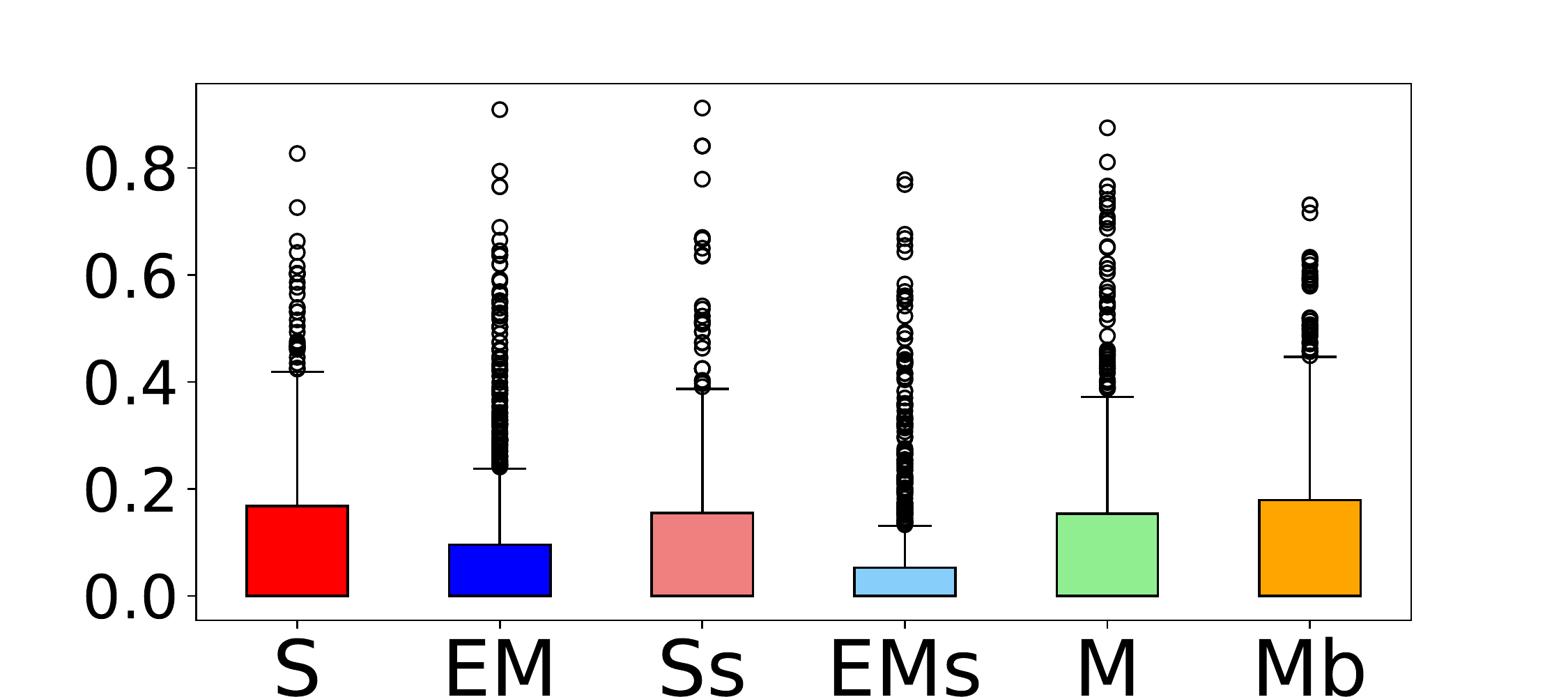} \\
     \multicolumn{1}{c}{\mimicX{Ter}} & \multicolumn{1}{c}{\medicX{2}{1}}  \\
     \includegraphics[height=2.8cm]{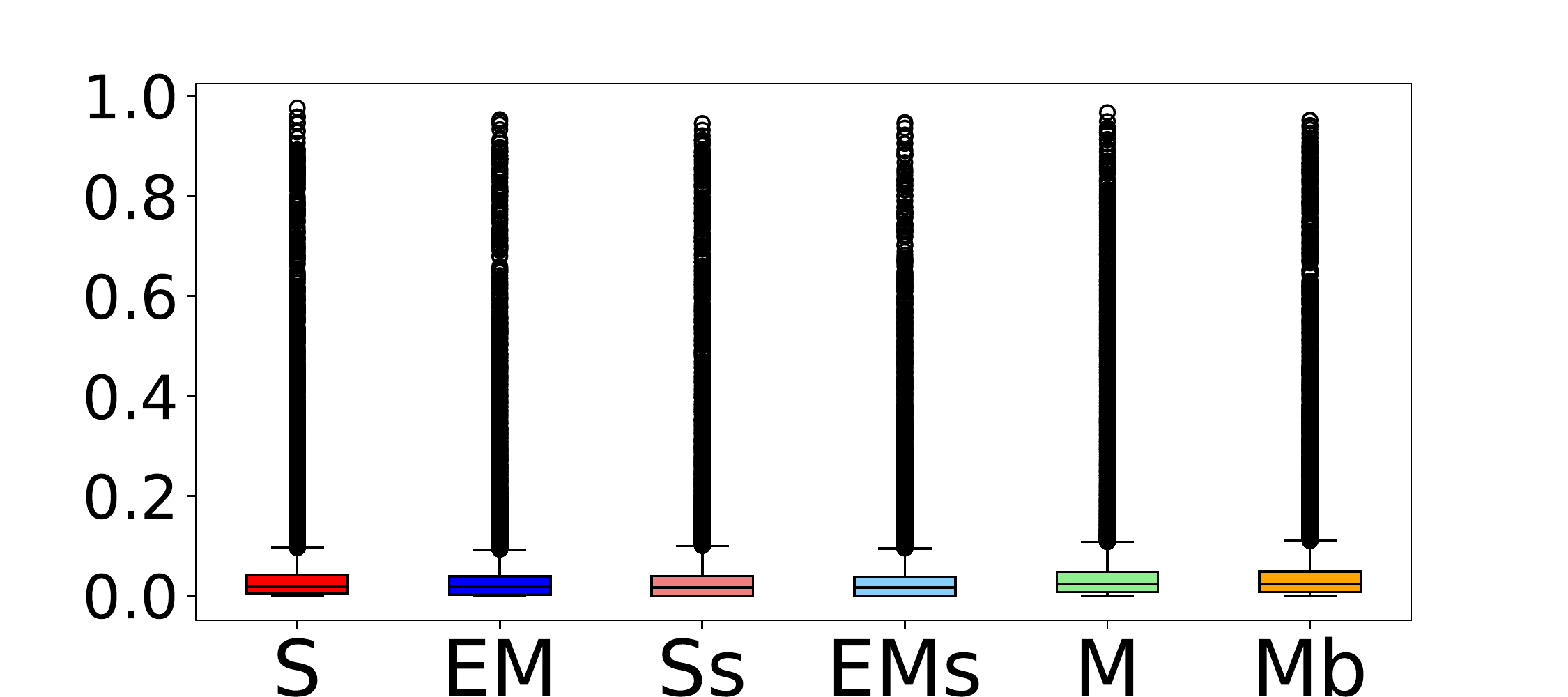} &
    \includegraphics[height=2.8cm]{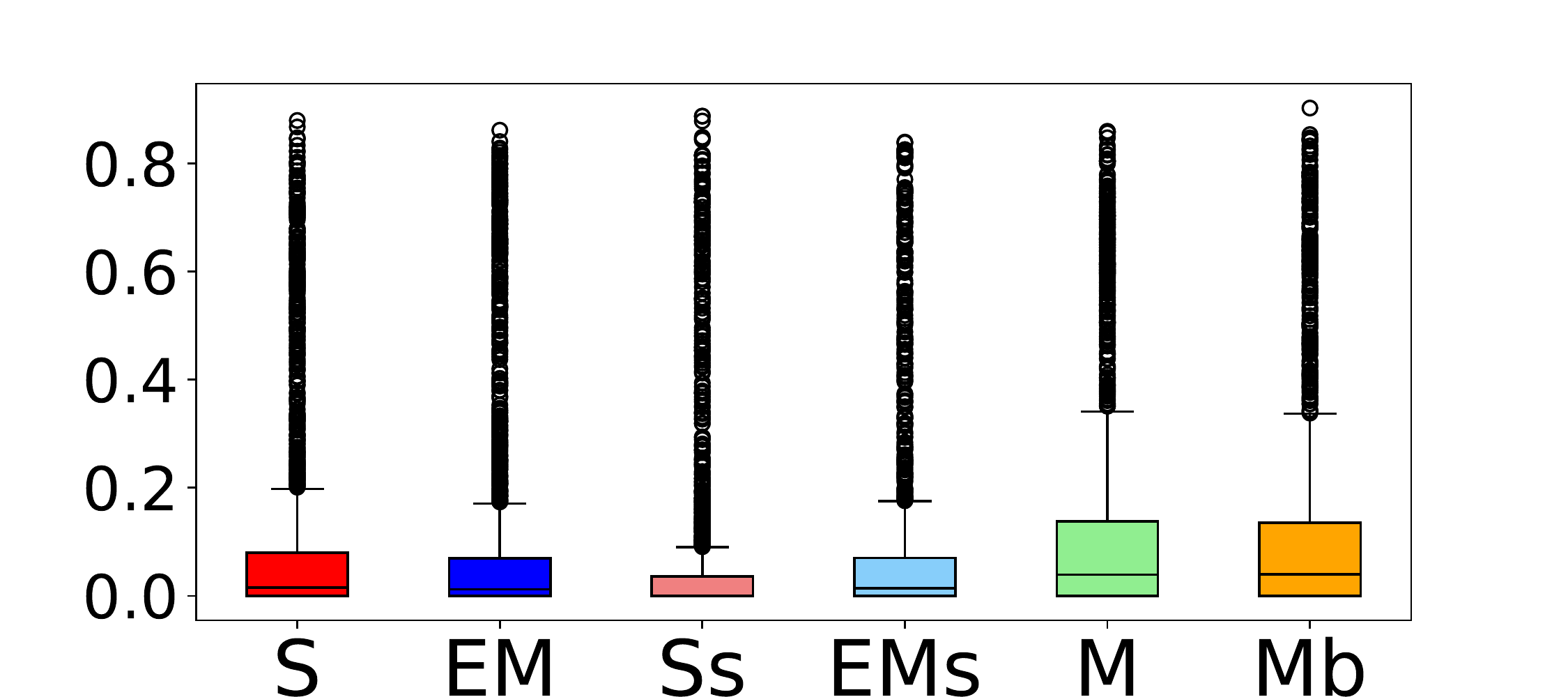} \\
     \multicolumn{1}{c}{\medicX{2}{4}} & \multicolumn{1}{c}{\medicX{2}{8}}  \\
     \includegraphics[height=2.8cm]{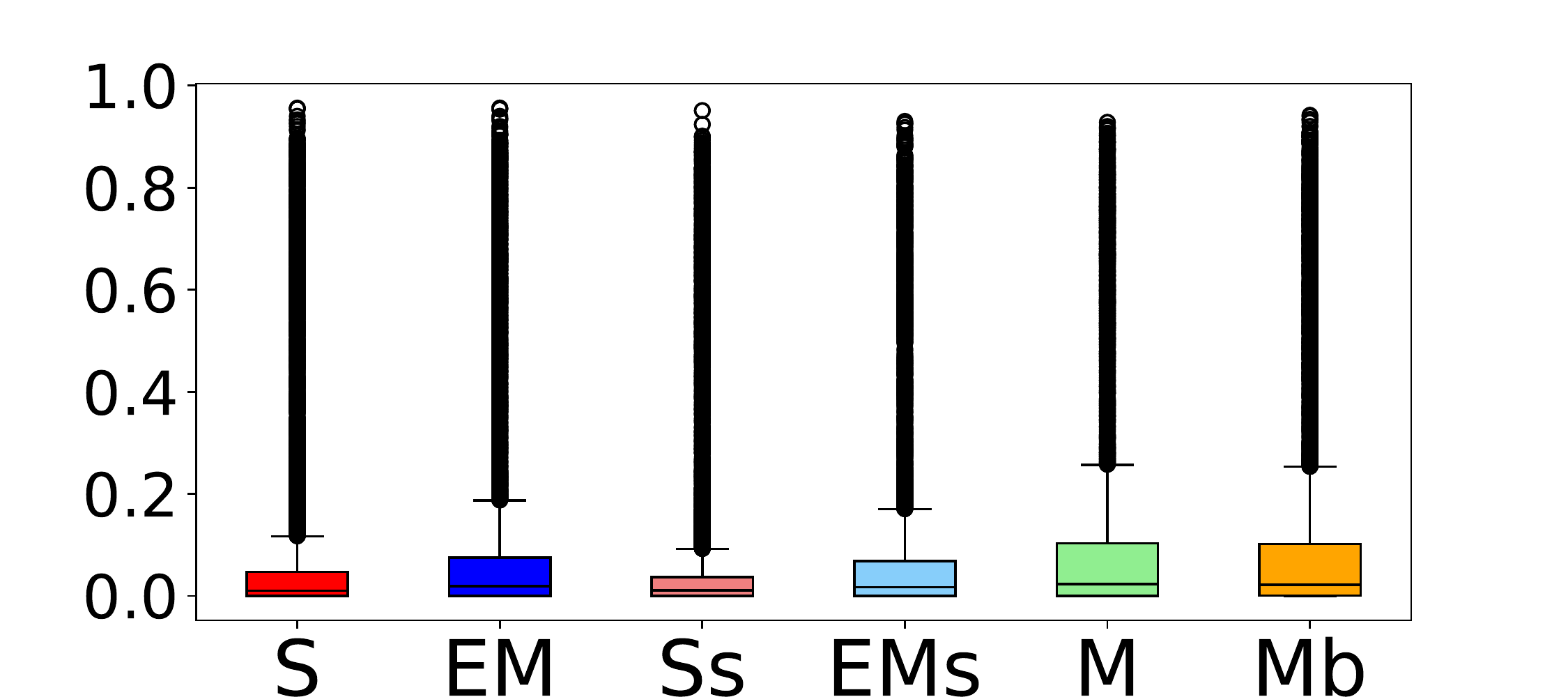} &
    \includegraphics[height=2.8cm]{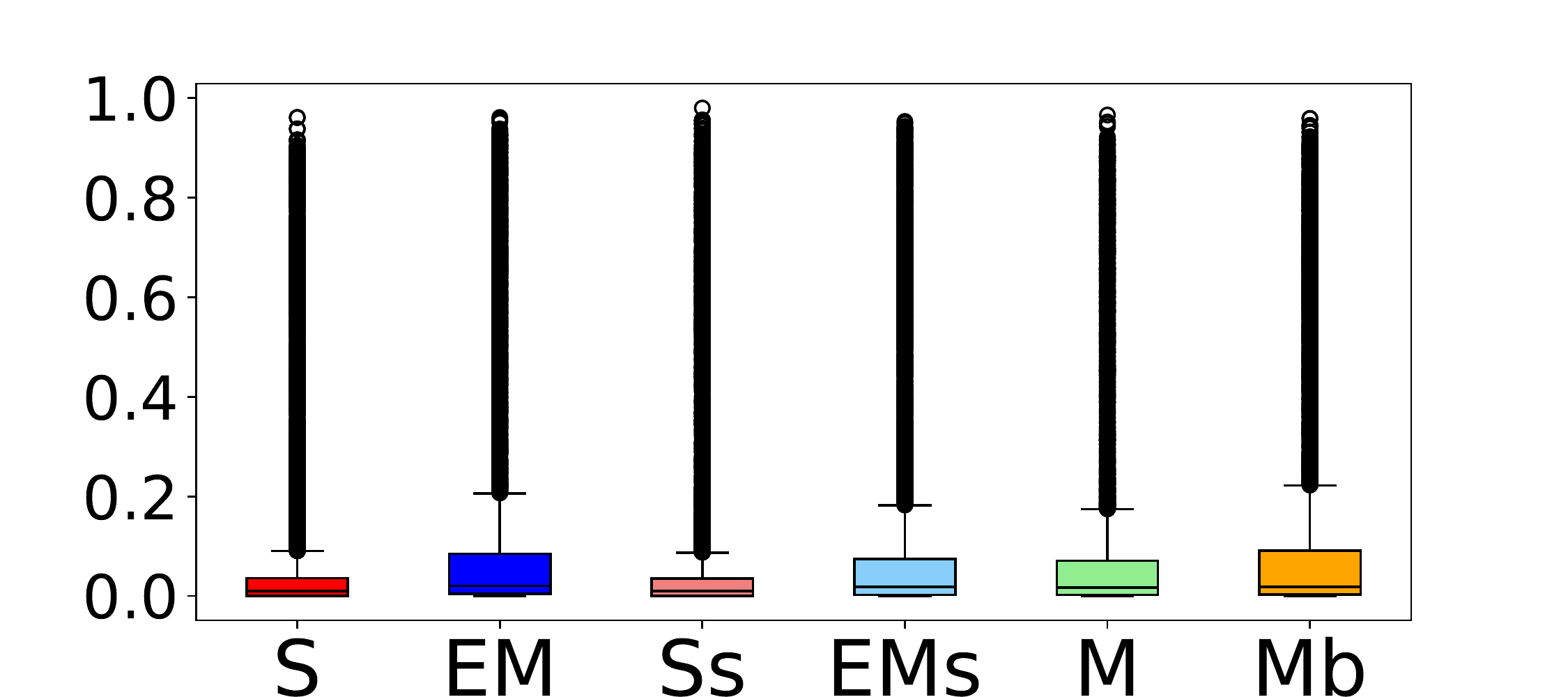}  \\
     \multicolumn{1}{c}{\medicX{2}{16}} &  \\
     \includegraphics[height=2.8cm]{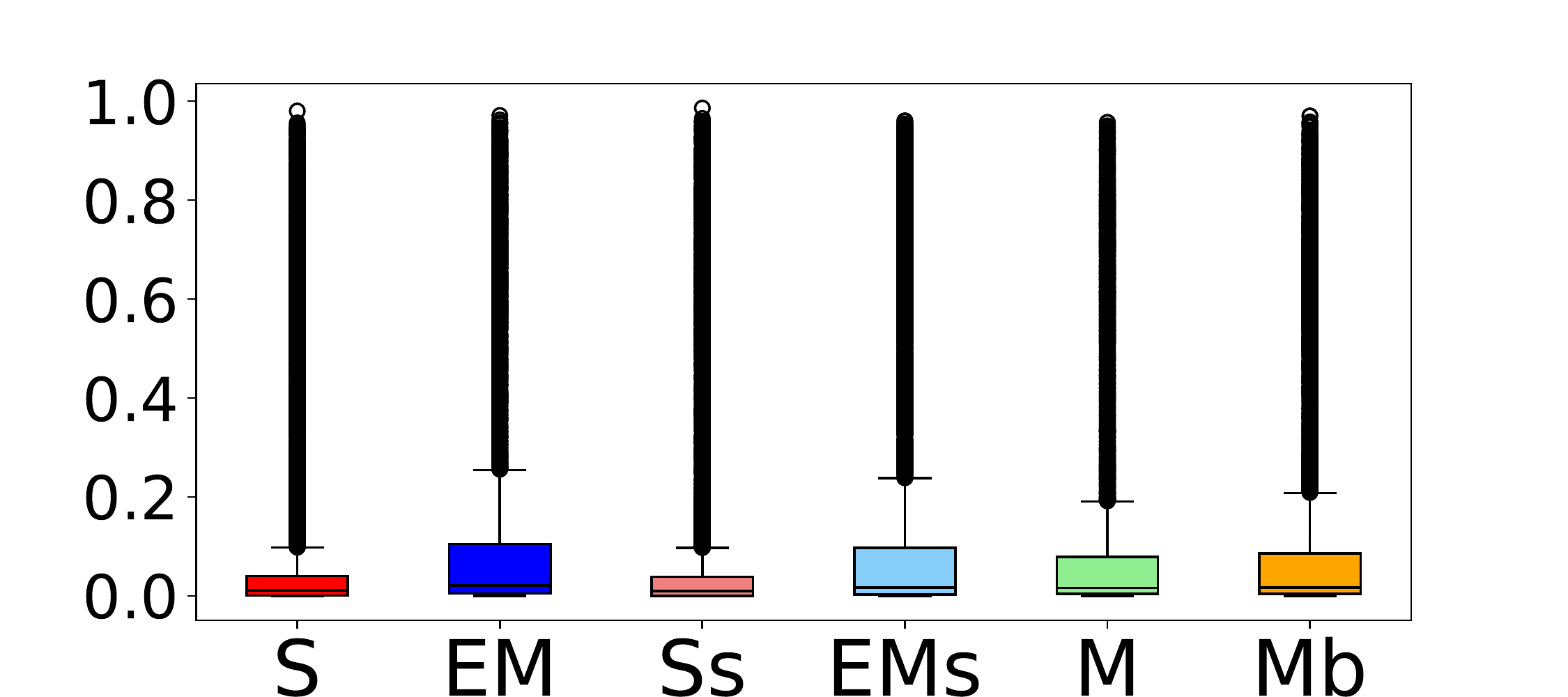} &
  \end{tabularx}
  \caption{Distribution of accuracy of filtered significant redescriptions produced by each differentially private approach in $100$ differentially private executions (each with budget $0.01$). Algorithm names are as in \ref{fig:RAccB1}.}
  \label{fig:RAccB001}
\end{figure*}

\begin{figure*}[tbp]
  \centering
  \begin{tabularx}{\textwidth}{@{}X@{}X@{}}
    \multicolumn{1}{c}{\nerdy}  & \multicolumn{1}{c}{\mammals}   \\
    \includegraphics[height=2.8cm]{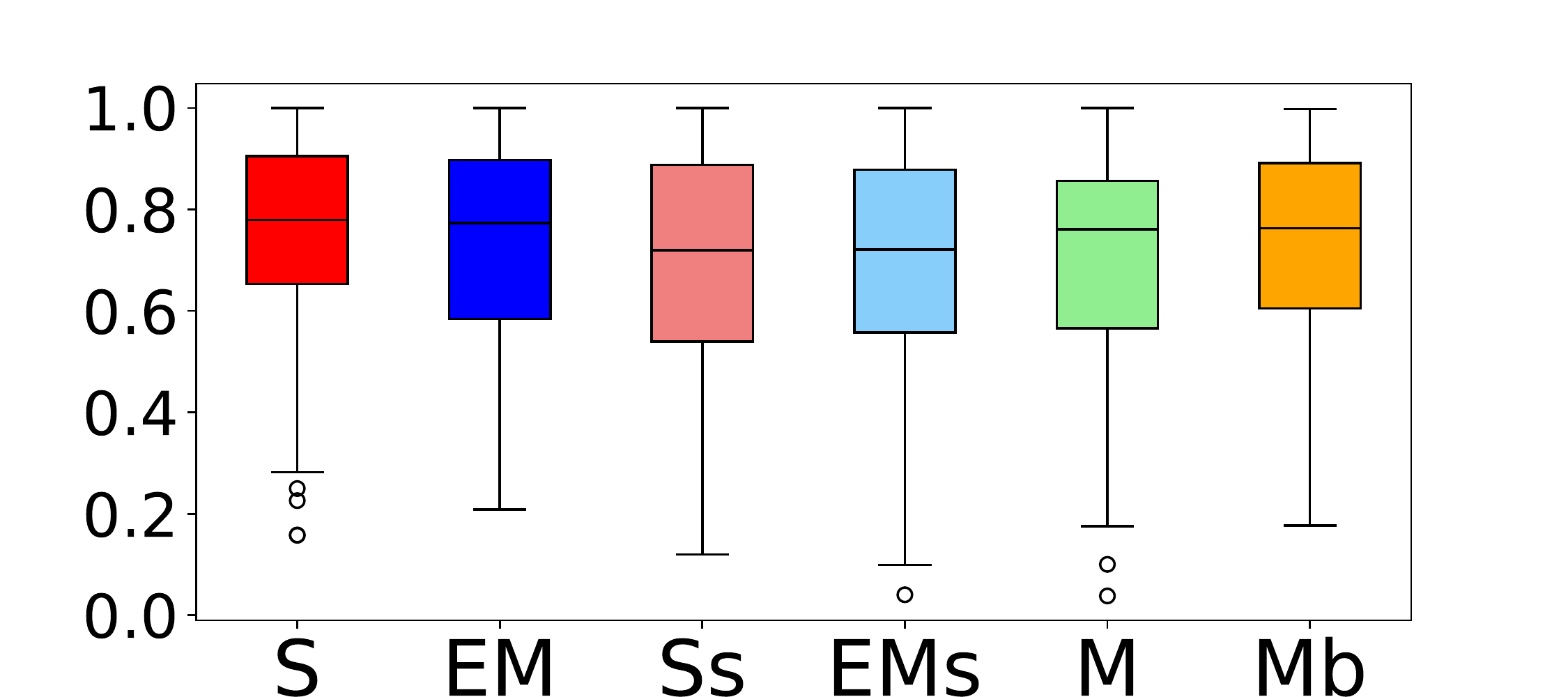} &
    \includegraphics[height=2.8cm]{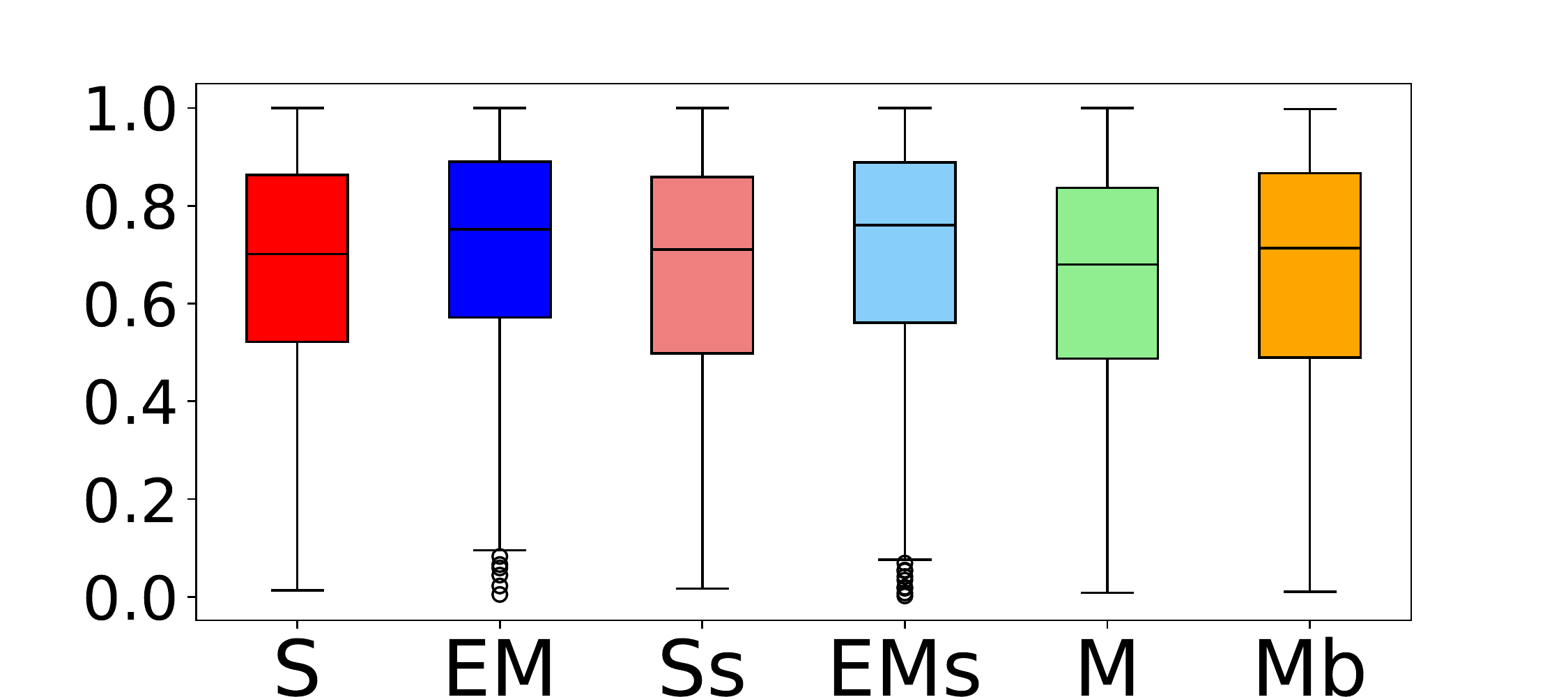} \\
     \multicolumn{1}{c}{\mimicX{Ter}} & \multicolumn{1}{c}{\medicX{2}{1}}  \\
     \includegraphics[height=2.8cm]{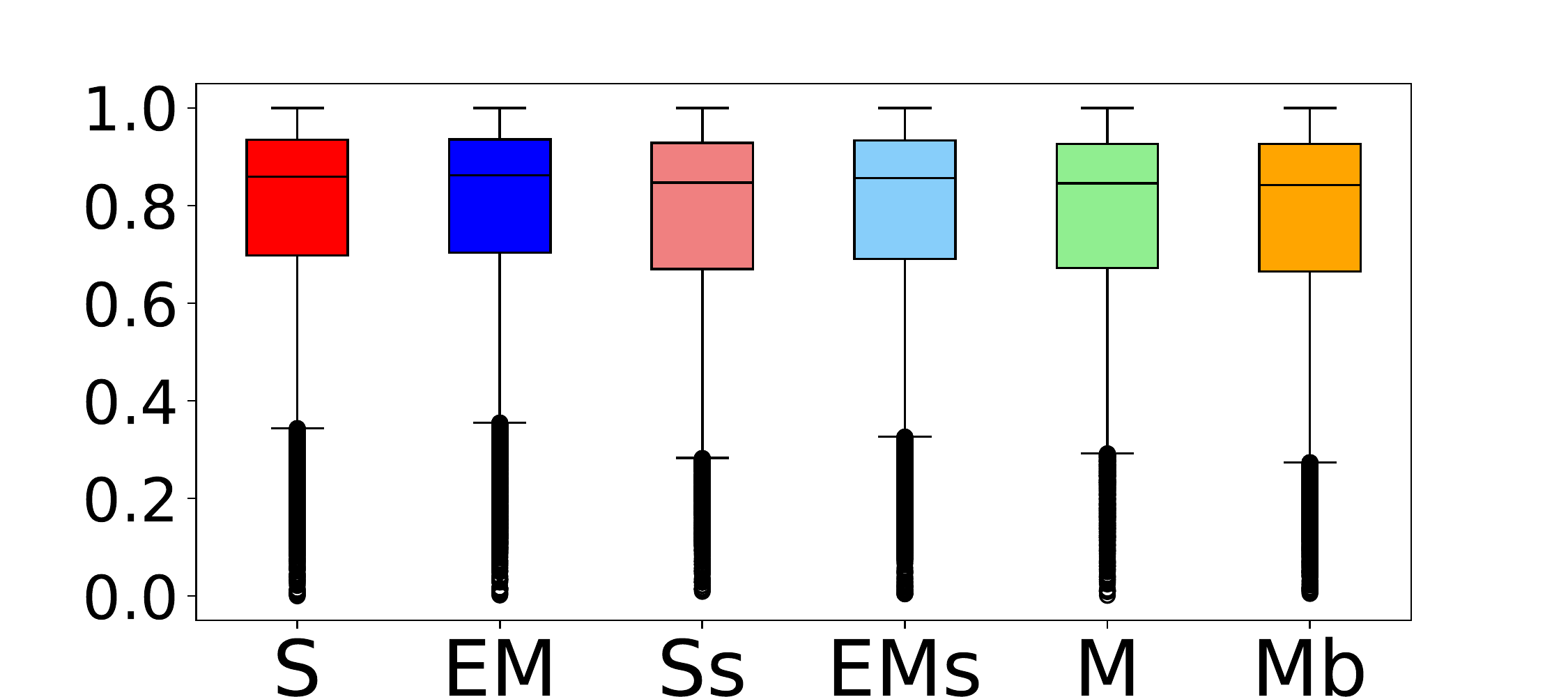} &
    \includegraphics[height=2.8cm]{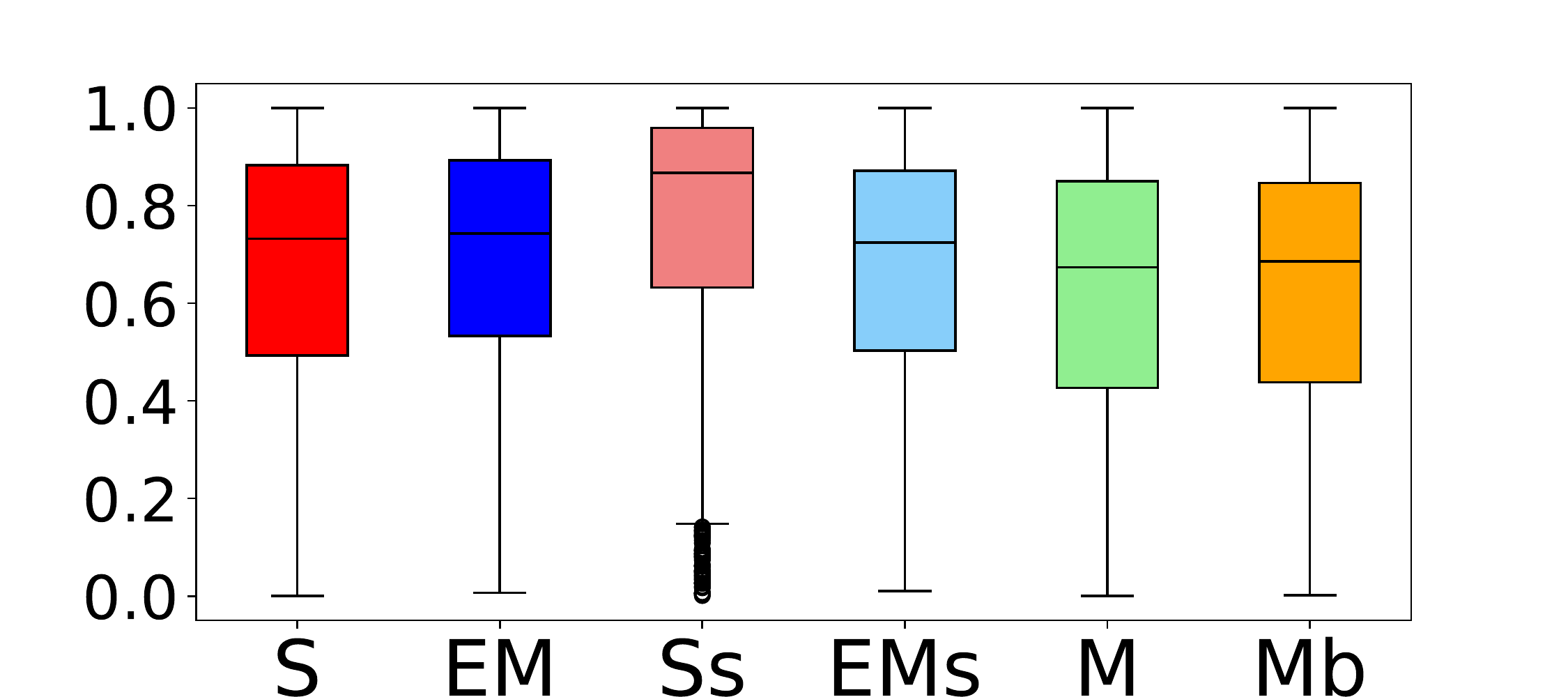} \\
     \multicolumn{1}{c}{\medicX{2}{4}} & \multicolumn{1}{c}{\medicX{2}{8}}  \\
     \includegraphics[height=2.8cm]{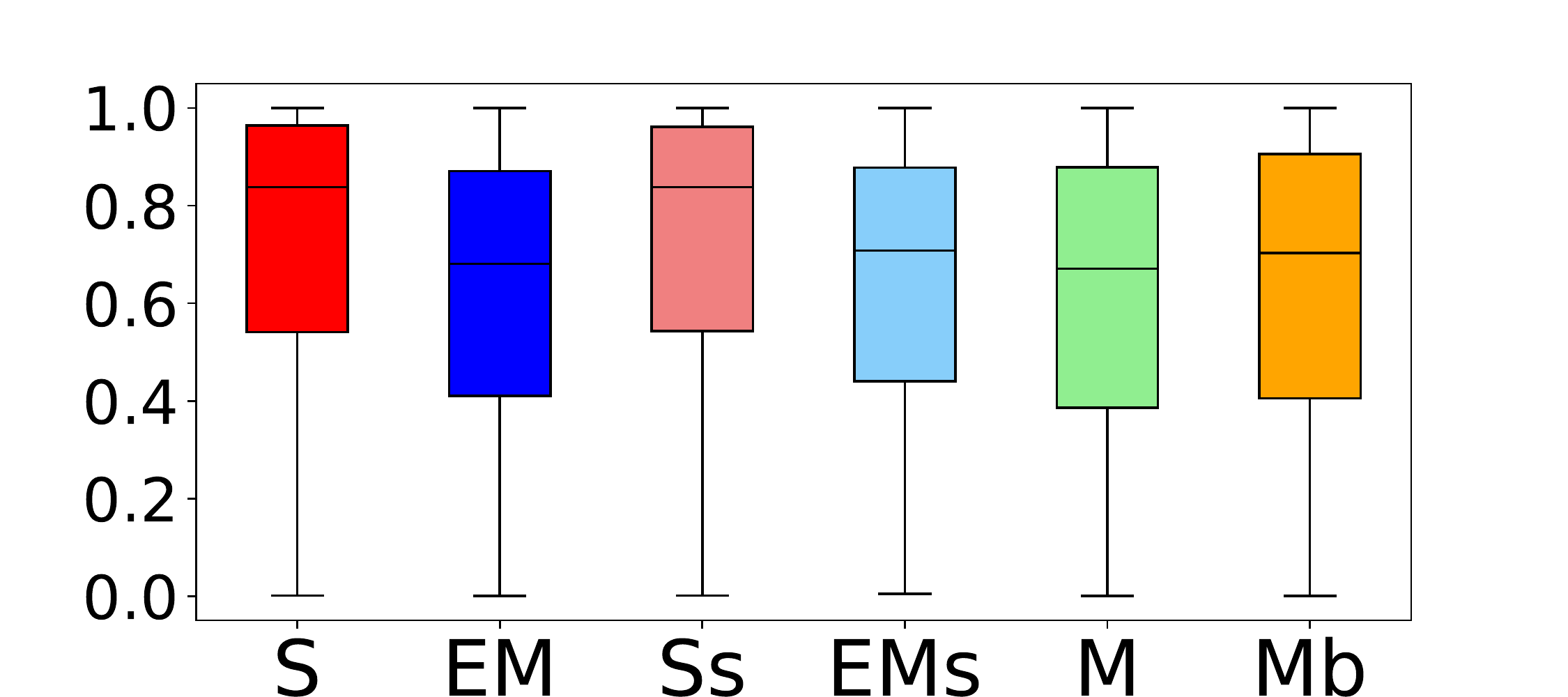} &
    \includegraphics[height=2.8cm]{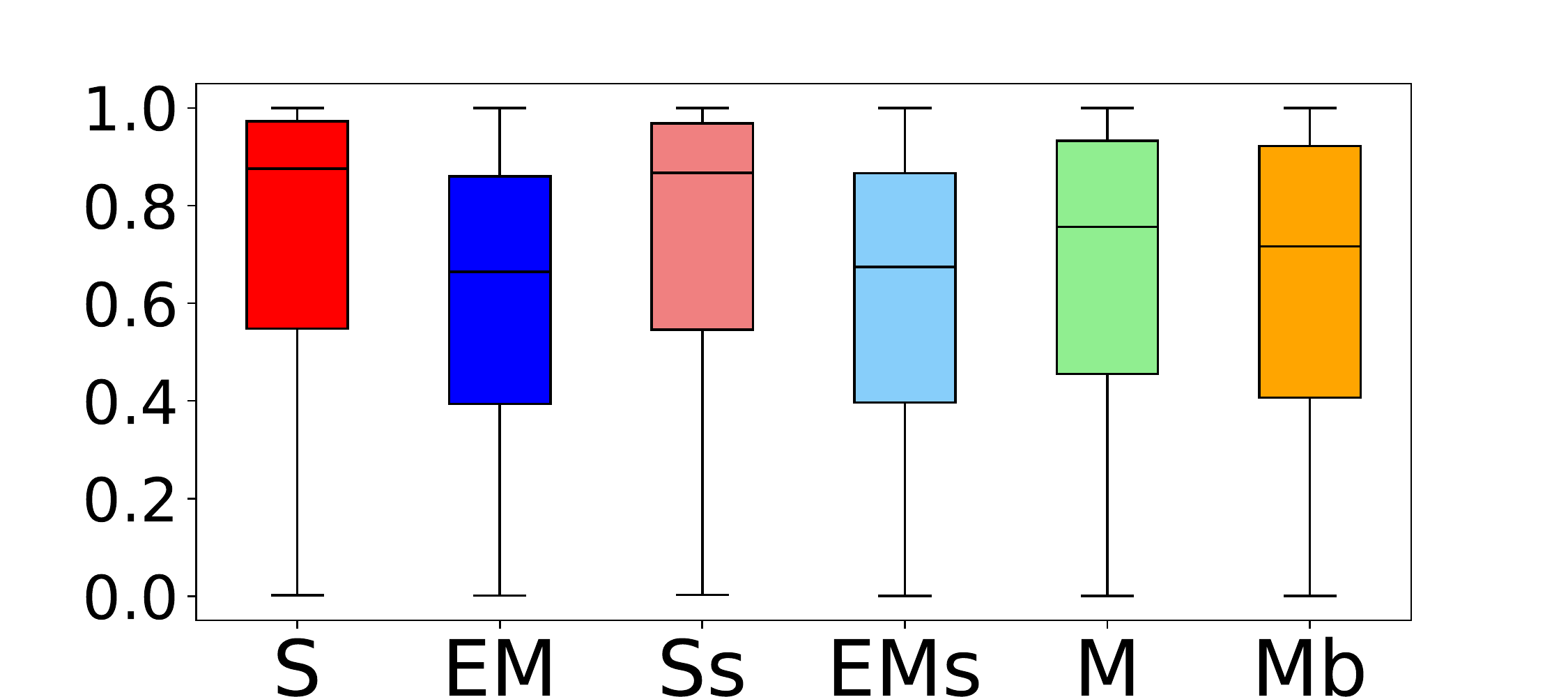}  \\
     \multicolumn{1}{c}{\medicX{2}{16}} &  \\
     \includegraphics[height=2.8cm]{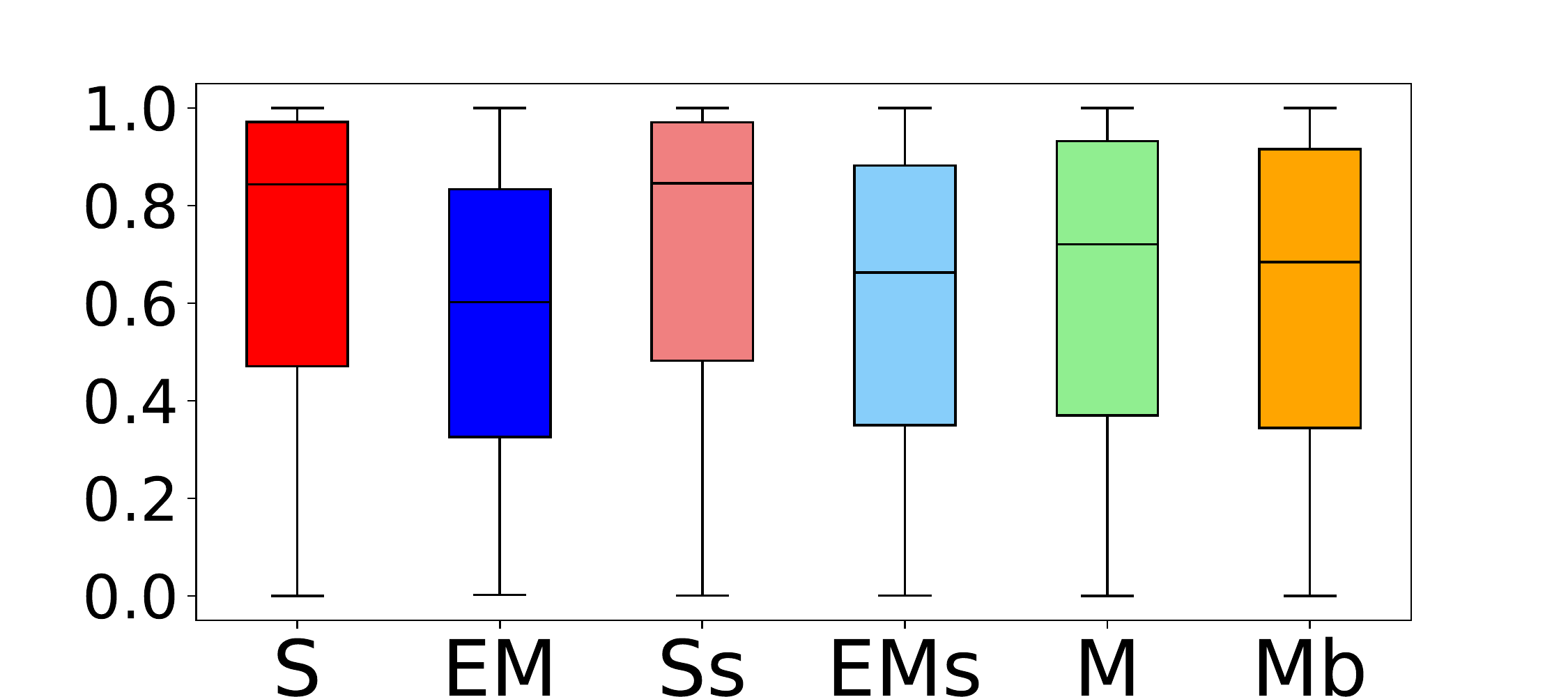} &
  \end{tabularx}
  \caption{Distribution of absolute difference between noisy and real redescription accuracy of filtered significant redescriptions produced by each differentially private approach in $100$ differentially private executions (each with budget $0.01$). Algorithm names are as in Figure \ref{fig:RAccB1}.}
  \label{fig:RStab00B1}
\end{figure*}

Perhaps the most noticeable thing from  Figures~\ref{fig:RAccB1} and \ref{fig:RStabB1} is a huge drop in redescription accuracy that occurs after stabilization of alternation-based approaches.    
Expectedly, as initial budget is lowered, the overall redescription accuracy decreases whereas the absolute difference between noisy and real redescription accuracy increases (see Figures \ref{fig:RAccB01}, \ref{fig:RAccB001}, \ref{fig:RStab0B1} and \ref{fig:RStab00B1}). This is the result of the increased noise in all algorithms, however as it can be seen from the results, the \ownalgo is the most resilient to the added noise from the $3$ presented approaches.   

\subsection{Execution Time Analyses}
\begin{figure*}[tbp]
  \centering
  \begin{tabularx}{\textwidth}{@{}X@{}X@{}}
    \multicolumn{1}{c}{\sampling}  & \multicolumn{1}{c}{\sampling (stable)}   \\
    \includegraphics[height=2.8cm]{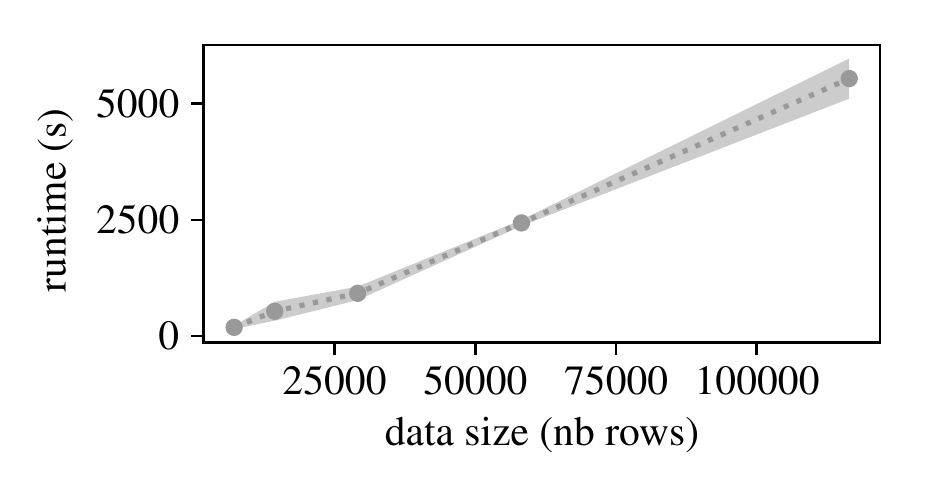} &
    \includegraphics[height=2.8cm]{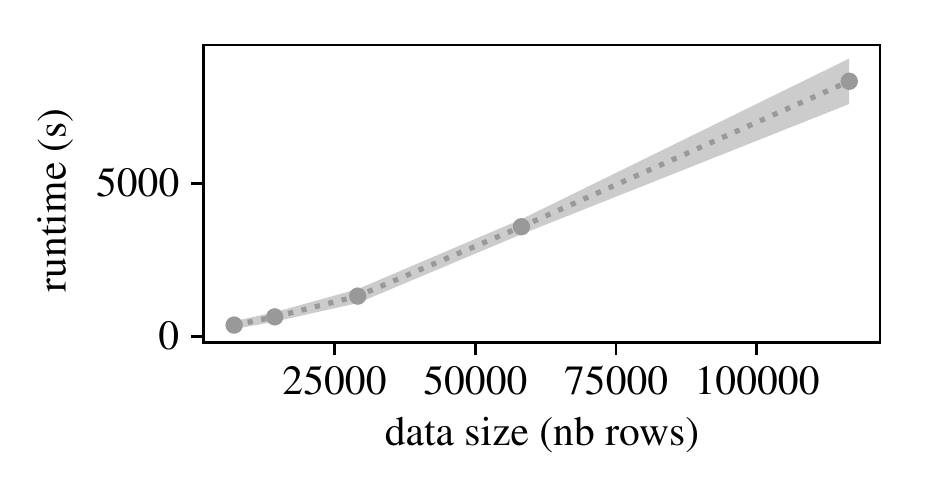} \\
     \multicolumn{1}{c}{\expmech} & \multicolumn{1}{c}{\expmech (stable)}  \\
     \includegraphics[height=2.8cm]{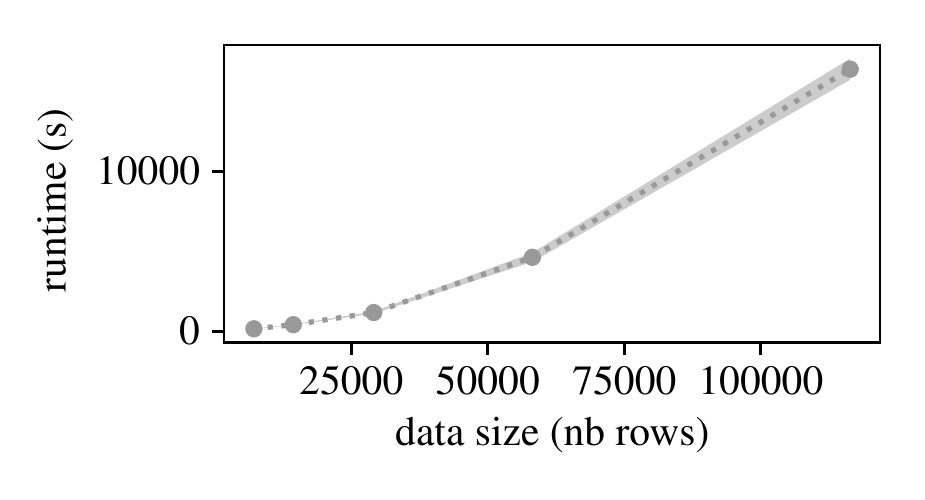} &
    \includegraphics[height=2.8cm]{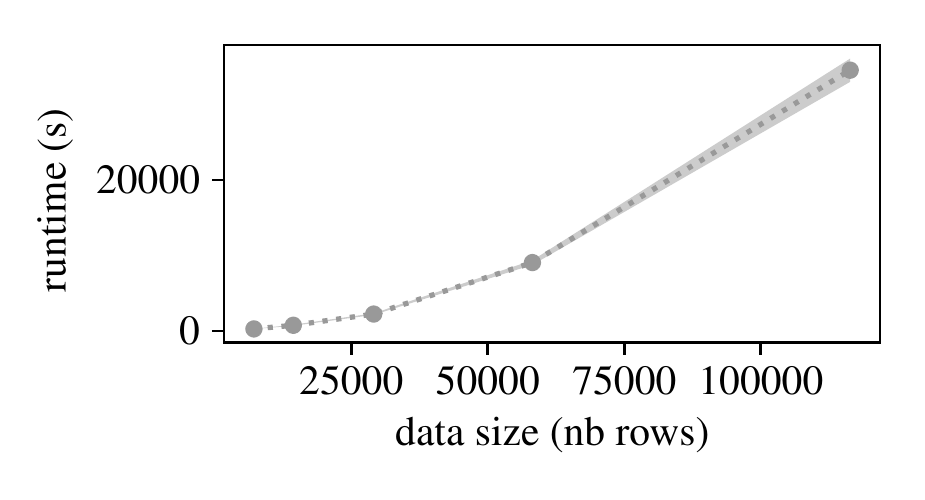} \\
     \multicolumn{1}{c}{\ownalgo} &  \multicolumn{1}{c}{\texttt{Comparative plot}}  \\
     \includegraphics[height=2.8cm]{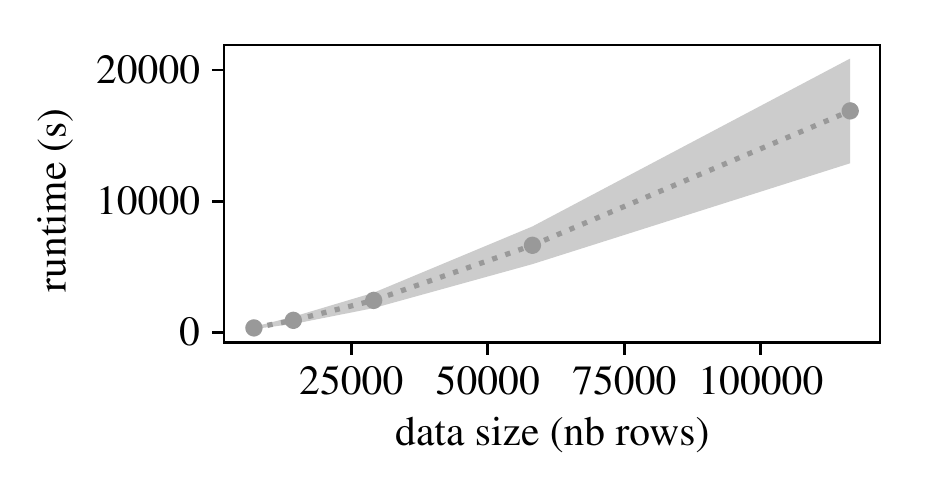} &
     \includegraphics[height=2.8cm]{runtime-lin_CMS-2-kComp}  \\
  \end{tabularx}
  \caption{\textbf{Runtime vs. data size}. Runtime in total CPU seconds. The shaded area indicates one standard deviation above and below the average.}
  \label{fig:AppexTimes}
\end{figure*}

Algorithm execution times are presented in Figure \ref{fig:AppexTimes}.

As it can be seen from Figure \ref{fig:AppexTimes}, the \sampling algorithm has the smallest executions time, followed by the \expmech (with similar execution time as the \ownalgo). As data size increases, so that variability in the execution times of MCMC-based approaches. The \expmech execution time across different runs changes only slightly on larger data. 

As already explained, the main cause of variability is the that depending on the initial randomly generated tree or a tree-pair, the MCMC iterations can converge in smaller or larger number of steps. The \expmech algorithm does not have such behaviour, which is reflected in almost identical execution time across runs.

\newcommand\textdot{\.}

\end{document}